\documentclass[prd,aps,nofootinbib,showpacs,10pt]{revtex4}
\usepackage{amsmath,graphicx,color,epsfig,float,textcomp}
\usepackage{epstopdf}

\newcommand{\myslash}[1]{/\kern-0.57 em {#1}}

\begin{document}
\title{Weak Decays of Doubly Heavy Baryons: ${\cal B}_{cc}\to {\cal B}_c V$}
\author{Li-Juan Jiang, Bei He, Run-Hui Li\footnote{E-mail: lirh@imu.edu.cn}}
\affiliation{ School of Physical Science and Technology, Inner Mongolia University, Hohhot 010021, China }
\begin{abstract}
The weak decays of a spin-$1/2$ doubly charm baryon (${\cal B}_{cc}$) to a spin-$1/2$ singly charm baryon (${\cal B}_c$) and a light vector meson ($V$) are studied under a phenomenological scheme. The contributions are classified into different topological diagrams, among which the short distance ones are calculated under the factorization hypothesis, and the long distance contributions are modelled as final-state interactions (FSIs) which are estimated with the one-particle-exchange model. In calculation the topological contributions tend to fall in a hierarchy. The branching fractions or decay widths are estimated, and it indicates that $\Xi_{cc}^+\to\Xi_c^+\pi^+\pi^-$ and $\Omega_{cc}^{+}\rightarrow\Xi_{c}^{+}K^-\pi^+$ can be used as candidate decays for searching $\Xi_{cc}^+$ and $\Omega_{cc}^+$. Some decays that are mainly activated by the long distance effects are found, observation on which in future experiments can help to understand the role of FSIs in charm baryon decays.
\end{abstract}
\pacs{13.30.-a,12.40.-y,12.15.-y,12.39.Fe}
 \maketitle
%=========================================================
\section{Introduction}\label{sec:introduction}
%=========================================================
The doubly heavy baryons with two heavy quarks ($b$ or $c$ quark), predicted by the quark model, had not been established in experiments for a long time, although they are allowed by the quantum chrodynamics theory. The SELEX collaboration used to report the discovery of doubly charmed baryon $\Xi_{cc}^+$ with an unexpected short lifetime and a relatively large production rate\cite{Mattson:2002vu,Ocherashvili:2004hi}. However, their results have not been confirmed by the following searches from FOCUS \cite{Ratti:2003ez}, BaBar \cite{Aubert:2006qw}, Belle \cite{Chistov:2006zj}, and LHCb \cite{Aaij:2013voa}, which becomes a longstanding puzzle in experiments. Out of the requirement of experimental searching, we investigated some weak decays of doubly charmed baryons and presented $\Xi_{cc}^{++} \to \Lambda_c^{+} K^- \pi^+ \pi^+$ and $\Xi_{cc}^{++} \to \Xi_c^+ \pi^+$ as candidates of discovery decay \cite{Yu:2017zst,Li:2018epz}. In 2017 the LHCb collaboration announced the discovery of $\Xi_{cc}^{++}$ with properties predicted by theorists via our first suggesting decay \cite{Aaij:2017ueg} , and recently they reported observation of $\Xi_{cc}^{++}$ via the second decay \cite{Aaij:2018gfl}.

Before the discovery, there are only a few literatures about the weak decays of doubly charmed baryons, although they are really essential to searching for the lowest lying doubly charmed baryon states. So far there are still very few methods to deal with the weak decays of doubly heavy baryons. One of these methods is based on $SU(3)$ symmetry \cite{Egolf:2002nk,Wang:2017azm,Shi:2017dto}, which can figure out some relationships among the amplitudes or branching fractions but is not capable of predicting their values. What theorists do intensively is the calculation of weak transition form factors under the framework of various quark models or sum rules \cite{Onishchenko:2000yp,Ebert:2004ck, Ebert:2005ip, Albertus:2006wb, Albertus:2012jt, Wang:2017mqp, Hu:2017dzi, Zhao:2018mrg, Xing:2018lre,Wang:2017jow,Wang:2018wfj,Shen:2016hyv,Li:2012cfa,Li:2018lxi}. The branching fractions of related semileptonic decays can be obtained easily with these form factors. However, semileptonic decays are not so ideal candidates for particle searching because of the missing energy problem in experiments caused by neutrinos. Therefore the study on nonleptonic weak decays, which is meaningful to both particle searches and research on dynamics of baryon decays, becomes necessary.

Unlike the situation of heavy meson decays, there is still no systematic factorization theories for heavy baryon decays because of the complicated dynamics. Therefore explorations at the beginning tends to be undertaken in phenomenological ways \cite{Li:2017ndo}. Based on some assumptions, a scheme is figured out for the calculation of baryon decays induced by a $c$ quark weak decay \cite{Yu:2017zst}. These decays are quite similar to charm meson decays, i.e. they take place at a relatively low energy scale. The releasing energy is not large enough to activate the perturbative theory, and it is commonly thought that nonperturbative dynamics plays an important role \cite{FSIs}. In our scheme, the short distance dynamics is calculated under the hypothesis of factorization just as authors did in Ref. \cite{Onishchenko:2000yp}. The long distance contributions are modeled as final-state interactions (FSIs) and calculated under the one-particle-exchange model, which will be introduced in details in section \ref{sec:analytic}.  Instead of accurate predictions, the goal of this work is to present the first step estimations. Therefore, in our calculation only the lowest lying baryon states, light pseudoscalar and vector mesons are considered in the FSIs. The excited states, which are thought to be unimportant or lack of data, are ignored.

Under the above scheme, we calculate the nonleptonic weak decays of a spin-$1/2$ doubly charmed baryon $\Xi_{cc}^{++}$, $\Xi_{cc}^+$ or $\Omega_{cc}$ to a spin-$1/2$ singly charmed baryon and a light vector meson. This decay mode is special in the meaning of doubly charm baryon searches. For one reason, $\Xi_{cc}^{++}$ is discovered through $\Xi_{cc}^{++} \to \Lambda_c^{+} K^- \pi^+ \pi^+$, which is firstly studied via $\Xi_{cc}^{++} \to \Sigma_c^{++} \bar K^{*0}$ \cite{Yu:2017zst}. For another reason, some neutral vector mesons can be reconstructed by charged pseudoscalar mesons, such as $\rho^0$ by $\pi^+\pi^-$, $\bar K^{*0}$ by $K^-\pi^+$, $K^{*0}$ by $K^+\pi^-$, $\phi$ by $K^+K^-$ and so on,which are preferred in the experiments because of the high detection efficiencies. This paper is organized as follows. In section \ref{sec:analytic} we discuss the dynamics, introduce the theoretical framework, specify calculation details with an example decay, and list its analytic expressions. In section \ref{sec:results} inputs are introduced, numerical results are gathered, discussions and analysis based on the results are presented. A brief summary is given in section \ref{sec:summary}. The expressions for each decay are all listed in appendix \ref{app:amps}, and the strong coupling constants used in our calculation are collected in appendix \ref{app:stcouplings}.

%=========================================================
\section{Theoretical Framework and Analytical Calculations}\label{sec:analytic}
%=========================================================
%-------------------------------
\subsection{Effective Hamiltonian and Topological Diagrams}\label{ssec:hamiltonian}
%-------------------------------
In this paper we focus on weak decays induced by the charge current $c \to s/d$. The contributing low energy effective Hamiltonian is given by
 \begin{eqnarray}
 {\cal H}_{eff} = \frac{G_{F}}{\sqrt{2}}
     \sum\limits_{q=d,s} V^{*}_{cq} V_{uD} \big[
     C_{1}({\mu}) O^{q}_{1}({\mu})
  +  C_{2}({\mu}) O^{q}_{2}({\mu})\Big] + \mbox{h.c.} ,
 \label{eq:hamiltonian}
\end{eqnarray}
where $D=s,d$ stands for a down type light quark, $V_{cq}$ and
$V_{uD}$ are Cabibbo-Kobayashi-Maskawa (CKM) matrix elements whose values are taken from the CKMfitter Group \cite{CKM}, $C_{1/2}(\mu)$ are the Wilson coefficients, and $G_F=1.166\times 10^{-5}\mbox{ GeV}^{-2}$ is the Fermi constant.
The local four-quark operators $O^q_{1/2}$ are given as
\begin{eqnarray}
  O^{q}_{1}=({\bar{u}}_{\alpha}D_{\beta} )_{V-A}
               ({\bar{q}}_{\beta} c_{\alpha})_{V-A},
    \ \ \ \
   O^{q}_{2}=({\bar{u}}_{\alpha}D_{\alpha})_{V-A}
               ({\bar{q}}_{\beta} c_{\beta} )_{V-A}
    \label{eq:operators}
\end{eqnarray}
with $\alpha$ and $\beta$ as color indices.
%++++++++++++++++++++++++++++++++++++++++++++++++++++++++++++++++++++++++++++
\begin{figure}[htp]
\begin{center}
\begin{minipage}{0.25\linewidth}
\centerline{\includegraphics[scale=0.5]{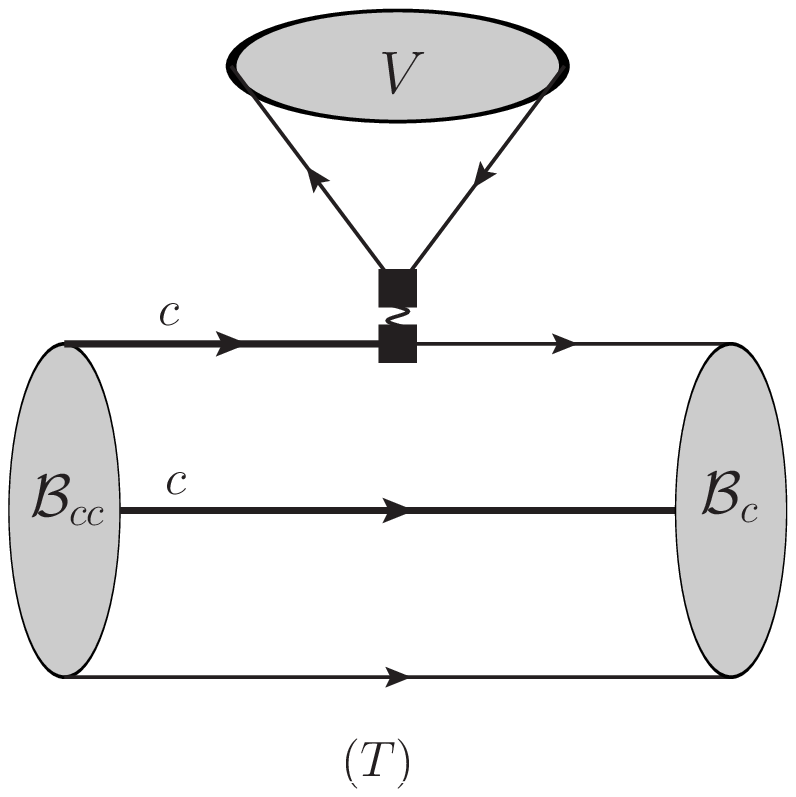}}
\centerline{\includegraphics[scale=0.5]{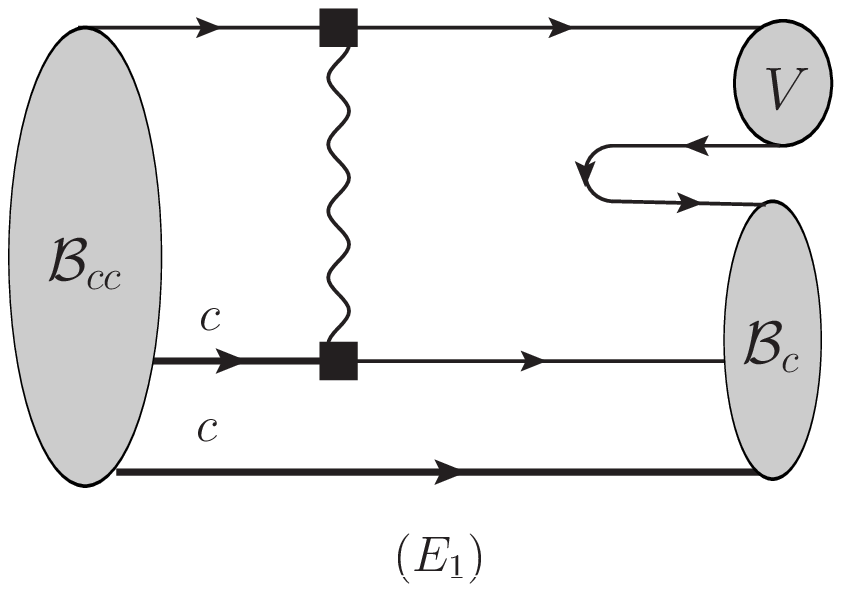}}
\end{minipage}
\begin{minipage}{0.25\linewidth}
\centerline{\includegraphics[scale=0.5]{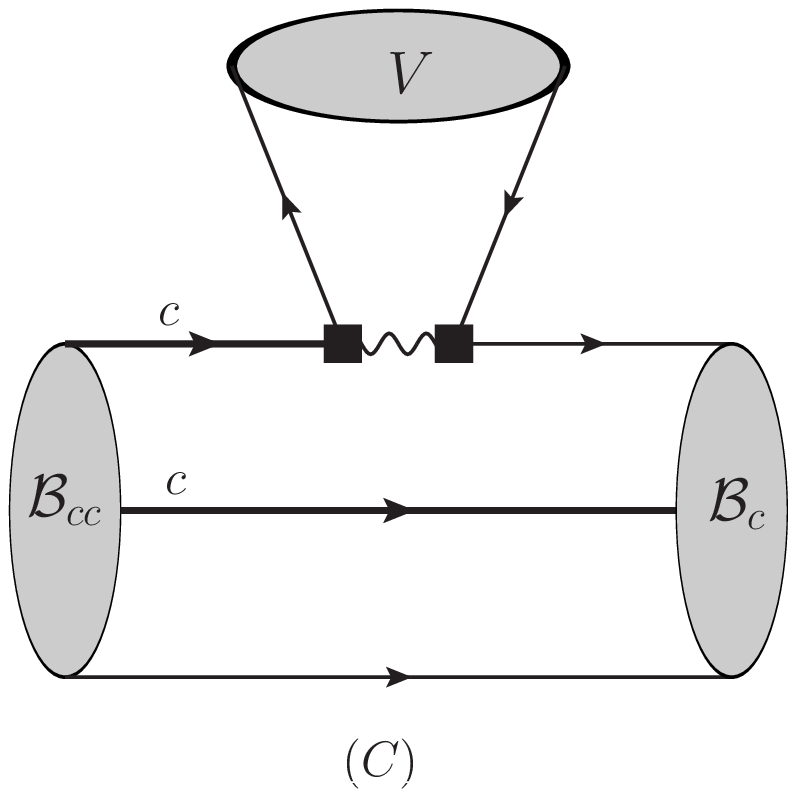}}
\centerline{\includegraphics[scale=0.5]{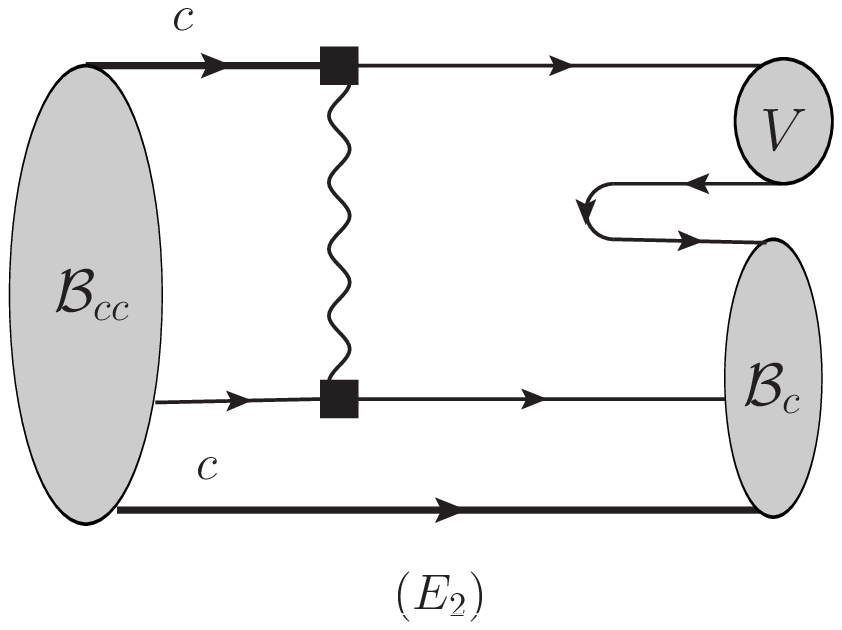}}
\end{minipage}
\begin{minipage}{0.25\linewidth}
\centerline{\includegraphics[scale=0.5]{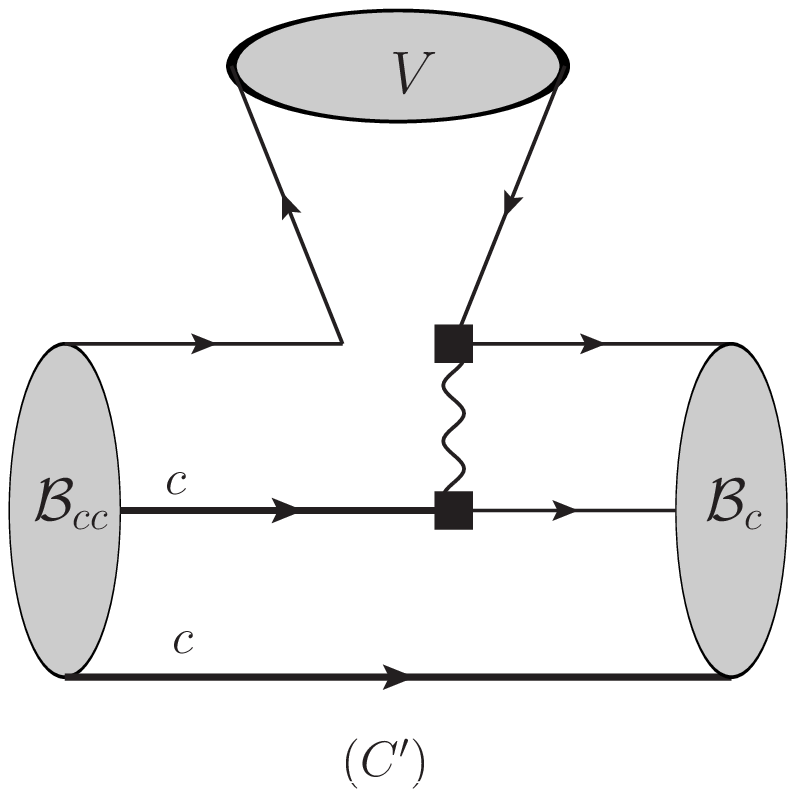}}
\centerline{\includegraphics[scale=0.5]{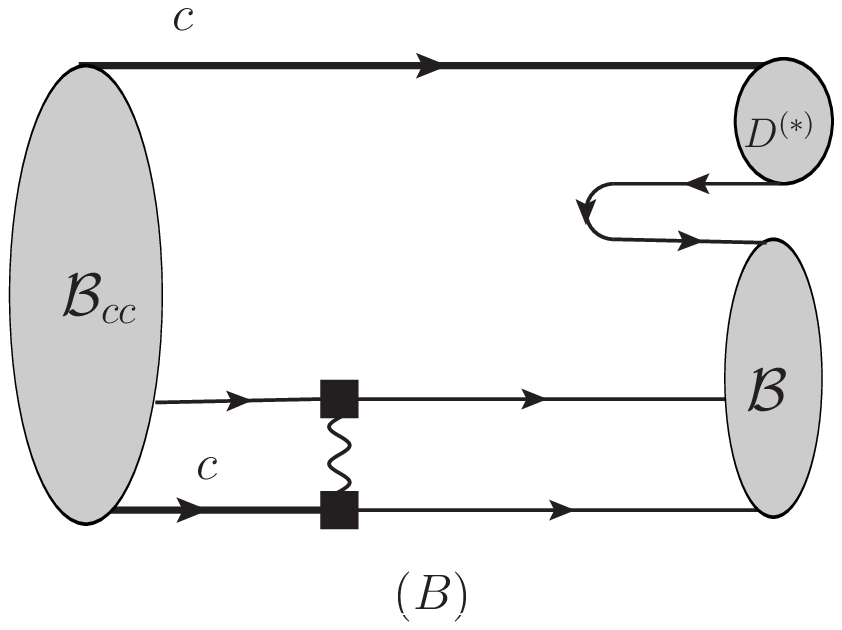}}
\end{minipage}
\end{center}
\caption{Tree level topological diagrams for two body nonleptonic decays of a doubly charm baryons to a baryon and a meson. The wave lines sandwiched by two black squares denote where the four-quark operators are inserted. $T$ denotes external $W$ emission diagram; $C$ represents the internal $W$ emission contribution; $E$ labels $W$ exchange diagrams in which the final state particles share the quarks from the weak vertex; $B$ indicates $W$ exchange mechanism with weakly generating quarks taken all by the final state baryon. Decays calculated in this paper receive no contribution labeled by $B$, although it is listed here.}
\label{fig:topos}
\end{figure}
%++++++++++++++++++++++++++++++++++++++++++++++++++++++++++++++++++++++++++++

The tree level weak interaction contributions to ${\cal B}_{cc} \to {\cal B}_c V$ decays are depicted by Fig. \ref{fig:topos}. Here the diagrams are classified according to the topologies of weak interactions. In each diagram all the strong interactions are included, which means both short distance and long distance contributions are contained. The external $W$ emission contribution is denoted by $T$, and the internal $W$ emission ones are classified into two types represented respectively by $C$ and $C^\prime$. $C$ diagram is the one with both constituent quarks in the light meson taken from the weak vertex, while in $C^\prime$ only the antiquark is generated from the weak vertex, meanwhile the quark participates as a spectator. The $W$ exchange contributions are also divided into two types: one (labeled by $B$) with the weak production quarks all taken by the final state baryon, the other (labeled by $E$) with the two weakly generated quarks shared by the final state particles, furthermore, the $E$ contribution can be divided into two topological diagrams with differently placed weakly decaying $c$ quark. The $B$ diagram, which leads to production of a charm meson, does not contribute in the decays of this work. Such kind of decays will be studied in our future work.

When one considers the short distance contributions, it indicates that $T$ diagram is dominating \cite{Lu:2009cm}. Under the factorization hypothesis it can be calculated with the meson decay constant and the form factors of ${\cal B}_{cc} \to {\cal B}_c$ transition. The $C$ diagram, suppressed by the color factor, can be related to the $T$ diagram by Fierz transformation. Therefore it can also be calculated in the same way. Amplitudes of other diagrams at short distance will be neglected in this paper. For one reason they are expected to be suppressed at least one order \cite{Lu:2009cm}, for another reason long distance dynamics is more important at the scale of charm quark. What's more, there is still not any factorization theory that can deal with baryon decays reliably. It should be emphasised that the insignificance of $C$, $E$ and $B$ contributions at short distance does not mean they are also unimportant at long distance, which can be seen in our results. The long distance contributions, which appear as final-state interactions, are discussed in details in section \ref{ssec:LDcontribution}.

%-------------------------------
\subsection{Short Distance Amplitudes under the Factorization Hypothesis}\label{ssec:SDcontribution}
%-------------------------------
The weak transition of spin-$1/2$ ${\cal B}_{cc}$ to spin-$1/2$ ${\cal B}_{c}$ can be parameterized as
\begin{eqnarray}
\langle {\cal B}_c(p^\prime,s_z^\prime)| (V-A)_\mu |{\cal B}_{cc}(p,s_z)\rangle
=&& \bar u(p^\prime,s^\prime_z)\left[ \gamma_\mu f_1(q^2) + i\sigma_{\mu\nu}\frac{q^\nu}{M} f_2(q^2) +\frac{q^\mu}{M} f_3(q^2) \right] u(p,s_z) \nonumber\\
&&- \bar u(p^\prime,s^\prime_z)\left[ \gamma_\mu g_1(q^2) + i\sigma_{\mu\nu}\frac{q^\nu}{M} g_2(q^2) +\frac{q^\mu}{M} g_3(q^2) \right] \gamma_5 u(p,s_z),\label{eq:ff}
\end{eqnarray}
with $q=p-p^\prime$ and $M$ as the mass of ${\cal B}_{cc}$. The calculation of the form factors $f_i$ and $g_i$, which can be performed under kinds of quark models or sum rules as mentioned in the introduction, is not the task in this work. We will use the recent data obtained under the light-front quark model in Ref. \cite{Wang:2017mqp} as inputs.

The decay constant of a pseudoscalar meson is defined with axial-vector current
\begin{equation}
\langle 0|A_\mu|P(q)\rangle=if_Pq_\mu \, ,
\label{eq:pdc}
\end{equation}
and that of a vector meson is defined with vector current
\begin{equation}
\langle 0|V_\mu|V(q)\rangle=f_Vm_V\epsilon_\mu \, .
\label{eq:vdc}
\end{equation}
Combining Eqs. (\ref{eq:ff}), (\ref{eq:pdc}) and (\ref{eq:vdc}), the short distance factorizable amplitudes of ${\cal B}_{cc}\to{\cal B}_c P$ are given by
\begin{eqnarray}
T_{\mbox{{\tiny SD}}}({\cal B}_{cc}\to{\cal B}_c P)=i\frac{G_F}{\sqrt{2}} V^*_{cq} V_{uD} a_1 f_P
\bar u(p^\prime,s^\prime_z)\left[(M-M^\prime) f_1(m_P^2)+ (M+M^\prime) g_1(m_P^2) \gamma_5 \right] u(p,s_z)\, ,\nonumber\\
C_{\mbox{{\tiny SD}}}({\cal B}_{cc}\to{\cal B}_c P)=i\frac{G_F}{\sqrt{2}} V^*_{cq} V_{uD} a_2 f_P
\bar u(p^\prime,s^\prime_z)\left[(M-M^\prime) f_1(m_P^2)+ (M+M^\prime) g_1(m_P^2) \gamma_5 \right] u(p,s_z)\, .\label{eq:B2BP}
\end{eqnarray}
$a_1=C_2+C_1/N_C$ and $a_2=C_1+C_2/N_C$ are the combinations of Wilson coefficients. In principle the nonfactorizable contributions should also be included in $a_{1,2}$, although they are usually not considered under the factorization hypothesis. Since the decays in this paper happen at the charm scale, $a_1(m_c)$ and $a_2(m_c)$ \cite{Li:2012cfa} are used. $M^\prime$ denotes the mass of ${\cal B}_c$. Terms with $f_3$ and $g_3$ are omitted in Eq. (\ref{eq:B2BP}) because they are suppressed by $m_P^2/M^2$.
The amplitudes of ${\cal B}_{cc}\to{\cal B}_c V$ can be expressed as
\begin{eqnarray}
T_{\mbox{{\tiny SD}}}({\cal B}_{cc}\to{\cal B}_c V)=&&\frac{G_F}{\sqrt{2}} V^*_{cq} V_{uD} a_1 f_V \epsilon^*_\mu
\bar u(p^\prime,s^\prime_z)\left[\left(f_1(m_V^2)-\frac{M+M^\prime}{M}f_2(m_V^2)\right)\gamma^\mu +\frac{2}{M}f_2(m_V^2)p^{\prime\mu} \right. \nonumber\\
&&-\left.\left(g_1(m_V^2)+\frac{M-M^\prime}{M}g_2(m_V^2)\right)\gamma^\mu\gamma_5 -\frac{2}{M}g_2(m_V^2)p^{\prime\mu}\gamma_5\right] u(p,s_z)\, ,\nonumber\\
C_{\mbox{{\tiny SD}}}({\cal B}_{cc}\to{\cal B}_c V)=&&\frac{G_F}{\sqrt{2}} V^*_{cq} V_{uD} a_2 f_V \epsilon^*_\mu
\bar u(p^\prime,s^\prime_z)\left[\left(f_1(m_V^2)-\frac{M+M^\prime}{M}f_2(m_V^2)\right)\gamma^\mu +\frac{2}{M}f_2(m_V^2)p^{\prime\mu}\right. \nonumber\\
&&-\left.\left(g_1(m_V^2)+\frac{M-M^\prime}{M}g_2(m_V^2)\right)\gamma^\mu\gamma_5 -\frac{2}{M}g_2(m_V^2)p^{\prime\mu}\gamma_5\right] u(p,s_z)\, .
\label{eq:B2BV}
\end{eqnarray}

%----------------------------------------------------------------------
\subsection{Long Distance Contributions}\label{ssec:LDcontribution}
%-----------------------------------------------------------------------
The long distance contributions, nonperturbative in nature, are really difficult to be calculated. In this paper we treat them as FSI effects, which is modelled as soft rescattering of two intermediate particles. The FSIs are usually calculated at the hadron level under the one-particle-exchange model, the reasonablity of which is argued in Refs. \cite{FSIs,Cheng:FSIB,Lu:2005mx}.

\begin{figure}[htp]
\begin{center}
\hspace{-0.5cm}
\begin{minipage}{0.25\linewidth}
\centerline{\includegraphics[scale=0.4]{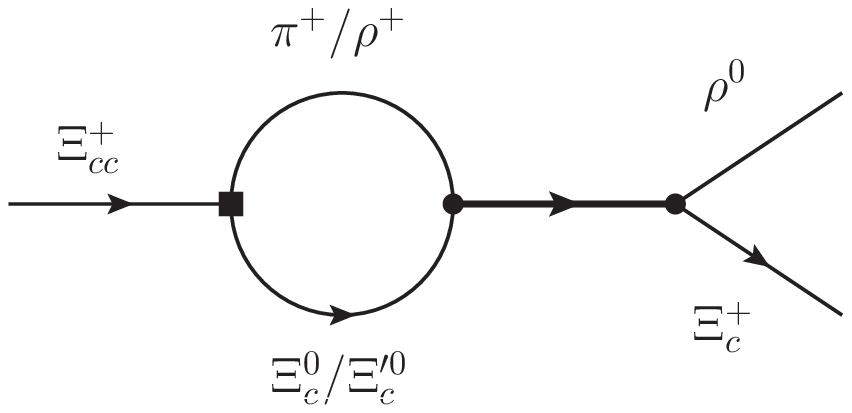}}
(a)
\centerline{\includegraphics[scale=0.4]{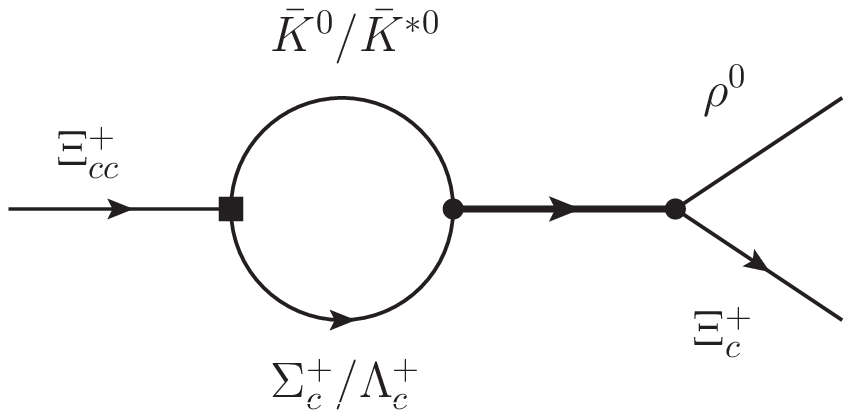}}
(d)
\end{minipage}
\begin{minipage}{0.25\linewidth}
\centerline{\includegraphics[scale=0.45]{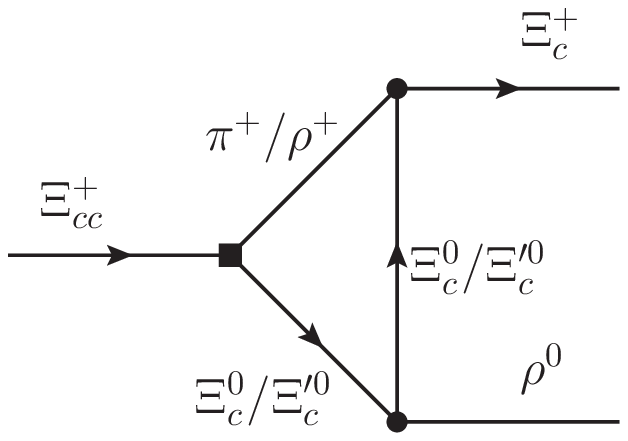}}
(b)
\centerline{\includegraphics[scale=0.45]{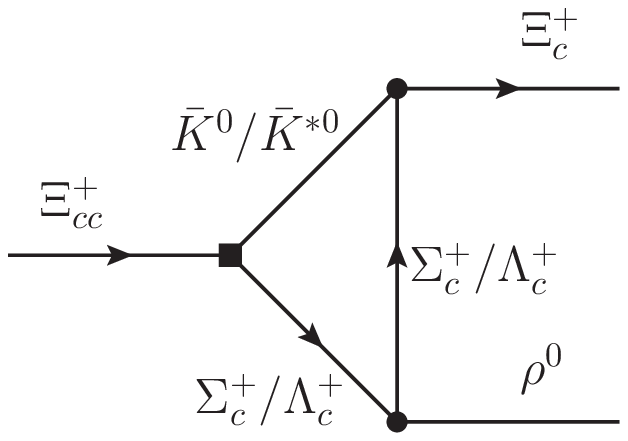}}
(e)
\end{minipage}
\begin{minipage}{0.25\linewidth}
\centerline{\includegraphics[scale=0.45]{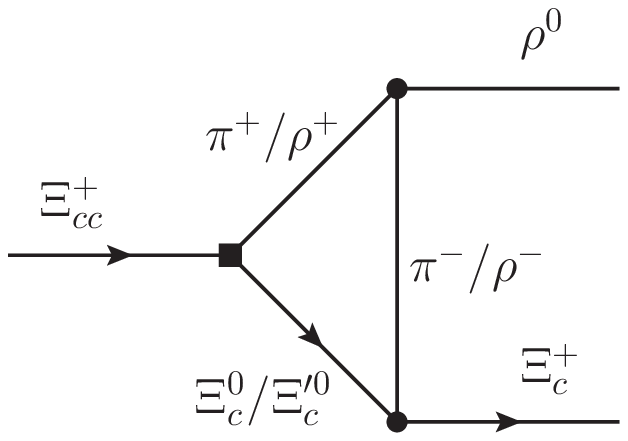}}
(c)
\centerline{\includegraphics[scale=0.45]{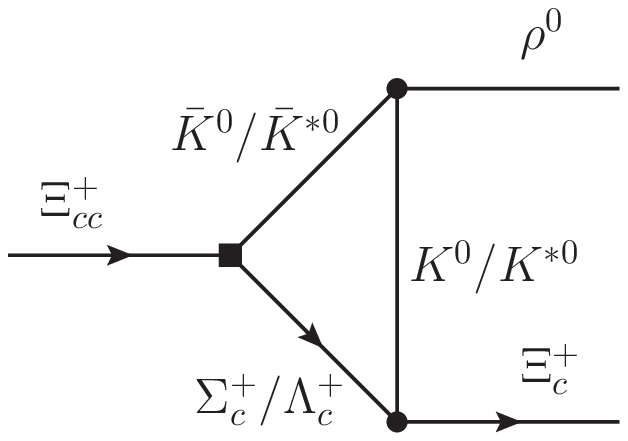}}
(f)
\end{minipage}
\end{center}
\caption{Leading FSI contributions to $\Xi_{cc}^+\to\Xi_c^+\rho^0$ manifested at hadron level. The black squares denote weak vertices and dots represent strong vertices. Diagrams (a), (b) and (c) are induced by the rescattering between $\Xi_c^0/\Xi_c^{\prime 0}$ and $\pi^+/\rho^+$, and diagrams (d), (e) and (f) by the rescattering between $\Sigma_c^+/\Lambda_c^+$ and $\bar K^0/\bar K^{*0}$. Each thick line in diagrams (a) and (d) denotes a resonant structure.}
\label{fig:FSIh}
\end{figure}

We take $\Xi_{cc}^{+}\to\Xi_c^+\rho^0$ as an example to introduce the framework of our calculation. This decay can proceed as $\Xi_{cc}^{+} \to (\Xi_c^0/\Xi_c^{\prime 0})(\pi^+/\rho^+) \to \Xi_c^+\rho^0$ or $\Xi_{cc}^{+} \to (\Sigma_c^+/\Lambda_c^+)(\bar K^0/\bar K^{*0}) \to \Xi_c^+\rho^0$ via exchanging one particle between the intermediate states. Both of the two process are CKM favored, while the intermediate states of the former one are generated by a short distance $T$ diagram and those of the latter one are produced by a short distance $C$ diagram. Utilizing the strong interaction Lagrangian at hadron level, one can draw all the leading diagrams in the meaning of perturbation theory as in Fig. \ref{fig:FSIh}. The Lagrangian employed in this work can be found in Refs. \cite{Yan:1992gz,Casalbuoni:1996pg,Meissner:1987ge,Li:2012bt}. Readers are referred to our previous work \cite{Yu:2017zst} for their specific expressions and we will not list them in this paper any more.

%++++++++++++++++++++++++++++++++++++++++++++++++++++++++++++++++++++++++++++

There are two $s$ channel contributions depicted as Figs. \ref{fig:FSIh}(a) and \ref{fig:FSIh}(d). They are supposed to be suppressed highly by the off-shell effect because of the absence of a candidate resonance particle whose rest energy should be very near to that of $\Xi_{cc}^+$\footnote{It is based on the consideration that $\Xi_{cc}^+$ and $\Xi_{cc}^{++}$ should have nearly the same masses since they are isospin partners.}. If any candidate were to be found in future experiments, its width would play an important role in determining the contribution of this diagram. Since there is no suitable particle and data so far, we just neglect these contributions in our calculation. Besides, there are four $t$ channel contributions. The long distance contributions of $\Xi_{cc}^+\to\Xi_c^+\rho^0$ will be given by these diagrams.

\begin{figure}[htp]
\begin{center}
\centerline{\includegraphics[scale=0.45]{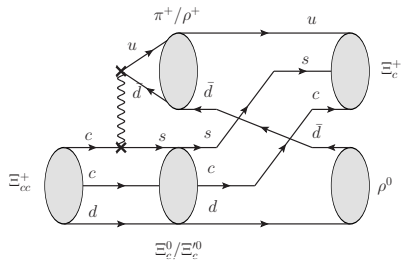}}
\end{center}
\caption{Substructure of Fig. \ref{fig:FSIh}(b) at quark level. The wave line represents a $W^+$ boson.}
\label{fig:FSIq}
\end{figure}

We would like to mention that the topological classification of these contributions can be performed by drawing their substructures at quark level. For example, Fig. \ref{fig:FSIh}(b) is manifested as Fig. \ref{fig:FSIq} at quark level. One can see that the $d$ quark in $\rho^0$ is originally from $\Xi_{cc}^+$, therefore it belongs to the $C^\prime$ contribution.
%++++++++++++++++++++++++++++++++++++++++++++++++++++++++++++++++++++++++++++
\begin{figure}[htp]
\begin{center}
\begin{minipage}{0.25\linewidth}
\centerline{\includegraphics[scale=1]{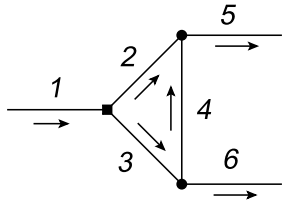}}
\end{minipage}
\end{center}
\caption{Numbers assigned to the lines in a triangle diagram. The arrows define the momentum directions in our calculation.}
\label{fig:number}
\end{figure}
%++++++++++++++++++++++++++++++++++++++++++++++++++++++++++++++++++++++++++++

Now we are at the point of calculating a triangle diagram. In order to simplify the expressions, the particles in a triangle diagram is numbered as in Fig. \ref{fig:number}. In the following paragraphs of this paper, this set of numbers is used when a triangle diagram is calculated. When reading expressions in this work, readers are asked to associate these numbers with particles at corresponding positions. The calculation of FSI effect can be carried on in different ways \cite{FSIs}. In general, the absorptive part of a two-body decay $P1\to P5P6$ can be related to a sum over all possible particle $1$ decay final states $\{q_k\}$, followed by strong $\{q_k\}\to p_5p_6$ rescattering:
\begin{eqnarray}
{\cal A}bs\,M(P1\to P5P6)=\frac{1}{2}\sum_j \left( \Pi_{k=1}^j \int\frac{\mbox{d}^3 q_k}{(2\pi)^3 2E_k} \right)(2\pi)^4
\delta^4(p_5 + p_6 - \sum_{k=1}^j q_k)M(p_1\to\{q_k\})T^*(p_5p_6\to\{q_k\}).
\label{eq:optical}
\end{eqnarray}
In this paper $Pi\,(i=1,2,3,4,5,6)$ represents a particle at position numbered with $i$ in Fig. \ref{fig:number}, and $p_i$ are the corresponding momentum. Basing on the argument that the $2$ body $\rightleftharpoons$ $n$ body rescattering is negligible \cite{FSI:2arguement}, we adopt the scheme in Ref. \cite{Cheng:FSIB} and treat the intermediate particles $P2$ and $P3$ as on-shell. As a result, Eq. (\ref{eq:optical}) is deduced as
\begin{eqnarray}
{\cal A}bs\,M(P1\to P5P6)=\frac{1}{2} \int\frac{\mbox{d}^3 p_2 \mbox{d}^3 p_3}{(2\pi)^6 4E_2 E_3} (2\pi)^4
\delta^4(p_5 + p_6 -p_2 -p_3)M(p_1\to p_2 p_3)T^*(p_5p_6\to p_2 p_3).
\label{eq:optical2}
\end{eqnarray}
The dispersive part can be calculated via the dispersion relation
\begin{equation}
{\cal D}is\,A(m_1^2)=\frac{1}{\pi}\int_s^\infty \frac{{\cal A}bs\,A(s^\prime)}{s^\prime - m_1^2} \mbox{d}s^\prime,
\end{equation}
which suffers from large ambiguities \cite{Cheng:FSIB}, and in this work we assume the absorptive part is dominating and neglect the dispersive part. In future we could come back to this problem if there is plenty of experimental data.

One key phenomenological ingredient is a form factor associated with the exchanged particle. This particle is certainly off-shell, and the form factor is introduced to make the whole framework meaningful in the sense of perturbative calculation. Details are left to the following paragraphs.

%----------------------------------------------------------------------
\subsection{Analytic Expressions}\label{ssec:expressions}
%-----------------------------------------------------------------------
We proceed with the decay $\Xi_{cc}^+\to\Xi_c^+\rho^0$, and present expressions of the amplitudes in this subsection. This decay receives no short distance factorizable contributions. For long distance contributions we need to calculate the diagrams \ref{fig:FSIh}(b), \ref{fig:FSIh}(c), \ref{fig:FSIh}(e) and \ref{fig:FSIh}(f). We introduce the symbol $M_{b/c/e/f}(P2;P3;P4)$ to denote an amplitude. The index $b/c/e/f$ labels the diagram. $P2$, $P3$ and $P4$ represent the particles at positions $2$, $3$ and $4$, respectively. Consequently the absorptive part of Fig. \ref{fig:FSIh}(c) with $P2=\pi^+$ and $P3=P4=\Xi_c^0$ is given by
\begin{eqnarray}
{\cal A}bs\,M_{b}(\pi^+;\Xi_c^0;\Xi_c^0)&=&i\int\frac{|\vec{p_2}|sin\theta d\theta d\varphi}{32\pi^{2}m_{\Xi_{cc}^{+}}}\frac{G_{F}}{\sqrt{2}}V_{cs}^{*}V_{ud}a_{1}f_{\pi}g_{\Xi_{c}^{+}\Xi_{c}^{0}\pi^{+}}
\frac{F^{2}(t,m_{\Xi_{c}^{0}})}{t-m_{\Xi_{c}^{0}}^{2}}\epsilon_{6}^{*\nu}\nonumber\\
&&\times\overline{u}(p_{5},s^{\prime}_{z})\gamma_{5}(\myslash{p_{4}}+m_{\Xi_{c}^{0}})
\left(f_{1\Xi_{c}^{0}\Xi_{c}^{0}\rho^{0}}\gamma_{\nu}
+i\frac{f_{2\Xi_{c}^{0}\Xi_{c}^{0}\rho^{0}}}{2m_{\Xi_{c}^{0}}}
\sigma_{\mu\nu}p_{6}^{\mu}\right)\nonumber\\
&&\times(\myslash{p_{3}}+m_{\Xi_{c}^{0}})
\left[(m_{\Xi_{cc}^{+}}-m_{\Xi_{c}^{0}})f_{1}(m^{2}_{\pi})
+(m_{\Xi_{cc}^{+}}+m_{\Xi_{c}^{0}})g_{1}(m^{2}_{\pi})\gamma_{5}\right]
u(p_{1},s_{z}),
\label{eq:a2PXP}
\end{eqnarray}
where $t=p_4^2$ and $\epsilon_6$ is the polarization vector of $\rho$ in the final state. In Eq. (\ref{eq:a2PXP}) the spin sum of two intermediate $\Xi_c^0$s are performed. The $3$-momentum of final-state baryon is defined at the ``$+$" direction of $z$ axis. $\theta$ and $\phi$ are the polar and azimuthal angles of $\vec{p_3}$ in spherical coordinate system. $g_{\Xi_{c}^{+}\Xi_{c}^{0}\pi^{-}}$, $f_{1\Xi_{c}^{0}\Xi_{c}^{0}\rho^{0}}$ and $f_{2\Xi_{c}^{0}\Xi_{c}^{0}\rho^{0}}$ are strong coupling constants. To account for the off-shell effect, a multiplication of the Breit-Wigner formula and a form factor $F(t,m)$ is associated with the exchanged particle. One can conclude that this diagram contributes little because of highly off-shell effect caused by the large mass of exchanged $\Xi_c^0$. Since it is too small, the term with width of $\Xi_c^0$ in the Breit-Wigner formula is abandoned in Eq. (\ref{eq:a2PXP}). The form factor $F(t,m)$ is parameterized as \cite{Cheng:FSIB}
\begin{equation}
F(t,m)=\left(\frac{\Lambda^2-m^2}{\Lambda^2-t}\right)^n,
\label{eq:Ffactor}
\end{equation}
normalized to $1$ at $t=m^2$. $m$ is the mass of the exchanged particle. The cutoff $\Lambda$ has the form of
\begin{equation}
\Lambda= m + \eta \Lambda_{\rm QCD}
\end{equation}
with $\Lambda_{\rm QCD}=330\,{\rm MeV}$. The phenomenological parameter $\eta$ depends on both exchanged and external particles at the strong vertex. The determination of their values requires huge mount of experimental data, since there are quite a lot of strong vertices in the related decays. For lack of experimental data we use $\eta=1.5$ in this work and range it from $1$ to $2$ for error estimations. Another parameter that needs to be determined is the $n$ in Eq. (\ref{eq:Ffactor}), which is also a phenomenological one extracted from experimental data. We borrow the experience from Ref. \cite{Cheng:FSIB} and set it to $1$.

Similarly, one can get
\begin{eqnarray}
%----------------------------------------------------------------------------------------------------
{\cal A}bs\,M_{b}(\rho^+;\Xi_c^0;\Xi_c^0)&=&-\int\frac{|\vec{p_2}|sin\theta d\theta d\varphi}{32\pi^{2}m_{\Xi_{cc}^{+}}}\frac{G_{F}}{\sqrt{2}}V_{cs}^{*}V_{ud}a_{1}f_{\rho}\frac{F^{2}(t,m_{\Xi_{c}^{0}})}{t-m_{\Xi_{c}^{0}}^{2}}
(-g^{\beta\nu}+\frac{p_{2}^{\beta}p_{2}^{\nu}}{m_{\rho}^{2}})\epsilon_{6}^{*\alpha}\nonumber\\
&&\times \overline{u}(p_{5},s^{\prime}_{z})
\left[f_{1\Xi_{c}^{+}\Xi_{c}^{0}\rho^{+}}\gamma_{\nu}
-i\frac{f_{2\Xi_{c}^{+}\Xi_{c}^{0}\rho^{+}}}{m_{\Xi_{c}^{+}} +m_{\Xi_{c}^{0}}}\sigma_{\mu\nu}p_{2}^{\mu}\right](\myslash{p_{4}}+m_{\Xi_{c}^{0}}) \left[f_{1\Xi_{c}^{0}\Xi_{c}^{0}\rho^{0}}
\gamma_{\alpha}
+i\frac{f_{2\Xi_{c}^{0}\Xi_{c}^{0}\rho^{0}}}{2m_{\Xi_{c}^{0}}}\sigma_{\rho\alpha}p_{6}^{\rho}\right]\nonumber\\
&&\times
(\myslash{p_{3}}+m_{\Xi_{c}^{0}})
\left[\left(f_{1}(m_{\rho}^{2})-\frac{m_{\Xi_{cc}^{+}}+m_{\Xi_{c}^{0}}}{m_{\Xi_{cc}^{+}}}
f_{2}(m_{\rho}^{2})\right)\gamma_{\sigma} +\frac{2}{m_{\Xi_{cc}^{+}}}f_{2}(m_{\rho}^{2})p_{3\sigma}\right.\nonumber\\
&&\left.
-\left(g_{1}(m_{\rho}^{2})+\frac{m_{\Xi_{cc}^{+}}-m_{\Xi_{c}^{0}}}{m_{\Xi_{cc}^{+}}}
g_{2}(m_{\rho}^{2})\right)\gamma_{\sigma}\gamma_{5}\right.
\left.-\frac{2}{m_{\Xi_{cc}^{+}}}g_{2}(m_{\rho}^{2})
p_{3\sigma}\gamma_{5}\right]u(p_{1},s_{z}).
\label{eq:a2VXV}
\end{eqnarray}
Besides the spin sum of two intermediate $\Xi_c^0$s, sum over the polarization states of intermediate $\rho^+$ is performed, too. In the following calculation polarization sum of the intermediate states should always be performed, and we will not remind it one by one any more. It should be stressed that some symbols of strong coupling constants looks similar to those of weak form factors, and one can distinguish them by the feature that strong coupling constants have particle names as indices. The expressions of other amplitudes are listed as follows.
\begin{eqnarray}
{\cal A}bs\,M_{b}(\pi^+;\Xi_c^{0};\Xi_c^{\prime 0})&=&i\int\frac{|\vec{p_2}|sin\theta d\theta d\varphi}{32\pi^{2}m_{\Xi_{cc}^{+}}}\frac{G_{F}}{\sqrt{2}}V_{cs}^{*}V_{ud}a_{1}f_{\pi}g_{\Xi_{c}^{+}\Xi_{c}^{'0}\pi^{+}}
\frac{F^{2}(t,m_{\Xi_{c}^{'0}})}{t-m_{\Xi_{c}^{'0}}^{2}}\epsilon_{6}^{*\nu}\nonumber\\
&&\times\overline{u}(p_{5},s^{\prime}_{z})\gamma_{5}(\myslash{p_{4}}+m_{\Xi_{c}^{'0}})
\left(f_{1\Xi_{c}^{0}\Xi_{c}^{'0}\rho^{0}}\gamma_{\nu}
+i\frac{f_{2\Xi_{c}^{0}\Xi_{c}^{'0}\rho^{0}}}{m_{\Xi_{c}^{0}}+m_{\Xi_{c}^{'0}}}
\sigma_{\mu\nu}p_{6}^{\mu}\right)\nonumber\\
&&\times(\myslash{p_{3}}+m_{\Xi_{c}^{0}})
\left[(m_{\Xi_{cc}^{+}}-m_{\Xi_{c}^{0}})f_{1}(m^{2}_{\pi})
+(m_{\Xi_{cc}^{+}}+m_{\Xi_{c}^{0}})g_{1}(m^{2}_{\pi})\gamma_{5}\right]
u(p_{1},s_{z}),
\end{eqnarray}
\begin{eqnarray}
%----------------------------------------------------------------------------------------------------
{\cal A}bs\,M_{b}(\rho^+;\Xi_c^{0};\Xi_c^{\prime 0})&=&-\int\frac{|\vec{p_2}|sin\theta d\theta d\varphi}{32\pi^{2}m_{\Xi_{cc}^{+}}}\frac{G_{F}}{\sqrt{2}}V_{cs}^{*}V_{ud}a_{1}f_{\rho} \frac{F^{2}(t,m_{\Xi_{c}^{'0}})}{t-m_{\Xi_{c}^{'0}}^{2}}
(-g^{\beta\nu}+\frac{p_{2}^{\beta}p_{2}^{\nu}}{m_{\rho}^{2}})\epsilon_{6}^{*\alpha}\nonumber\\
&&\times\overline{u}(p_{5},s^{\prime}_{z})
\left[f_{1\Xi_{c}^{+}\Xi_{c}^{'0}\rho^{+}}\gamma_{\nu}
-i\frac{f_{2\Xi_{c}^{+}\Xi_{c}^{'0}\rho^{+}}}{m_{\Xi_{c}^{+}} +m_{\Xi_{c}^{'0}}}\sigma_{\mu\nu}p_{2}^{\mu}\right](\myslash{p_{4}}+m_{\Xi_{c}^{'0}})\nonumber\\
&&\times\left[f_{1\Xi_{c}^{0}\Xi_{c}^{'0}\rho^{0}}
\gamma_{\alpha}
 +i\frac{f_{2\Xi_{c}^{0}\Xi_{c}^{'0}\rho^{0}}}{m_{\Xi_{c}^{0}} +m_{\Xi_{c}^{'0}}}\sigma_{\rho\alpha}p_{6}^{\rho}\right]
(\myslash{p_{3}}+m_{\Xi_{c}^{0}})\nonumber\\
&&\times\left[\left(f_{1}(m_{\rho}^{2})-\frac{m_{\Xi_{cc}^{+}}+m_{\Xi_{c}^{0}}}{m_{\Xi_{cc}^{+}}}
f_{2}(m_{\rho}^{2})\right)\gamma_{\sigma}+\frac{2}{m_{\Xi_{cc}^{+}}}f_{2}(m_{\rho}^{2})p_{3\sigma}\right.\nonumber\\
&&\left.
-\left(g_{1}(m_{\rho}^{2})+\frac{m_{\Xi_{cc}^{+}}-m_{\Xi_{c}^{0}}}{m_{\Xi_{cc}^{+}}}
g_{2}(m_{\rho}^{2})\right)\gamma_{\sigma}\gamma_{5}\right.
\left.-\frac{2}{m_{\Xi_{cc}^{+}}}g_{2}(m_{\rho}^{2})
p_{3\sigma}\gamma_{5}\right]u(p_{1},s_{z}),
\end{eqnarray}
\begin{eqnarray}
%----------------------------------------------------------------------------------------------------
{\cal A}bs\,M_{b}(\pi^+;\Xi_c^{\prime 0};\Xi_c^{0})&=&i\int\frac{|\vec{p_2}|sin\theta d\theta d\varphi}{32\pi^{2}m_{\Xi_{cc}^{+}}}\frac{G_{F}}{\sqrt{2}}V_{cs}^{*}V_{ud}a_{1}f_{\pi}g_{\Xi_{c}^{+}\Xi_{c}^{'0}\pi^{+}}
\frac{F^{2}(t,m_{\Xi_{c}^{0}})}{t-m_{\Xi_{c}^{0}}^{2}}\epsilon_{6}^{*\nu}\nonumber\\
&&\times\overline{u}(p_{5},s^{\prime}_{z})\gamma_{5}(\myslash{p_{4}}+m_{\Xi_{c}^{0}})
\left(f_{1\Xi_{c}^{'0}\Xi_{c}^{0}\rho^{0}}\gamma_{\nu}
+i\frac{f_{2\Xi_{c}^{'0}\Xi_{c}^{0}\rho^{0}}}{m_{\Xi_{c}^{'0}}+m_{\Xi_{c}^{0}}}
\sigma_{\mu\nu}p_{6}^{\mu}\right)\nonumber\\
&&\times(\myslash{p_{3}}+m_{\Xi_{c}^{'0}})
\left[(m_{\Xi_{cc}^{+}}-m_{\Xi_{c}^{'0}})f_{1}(m^{2}_{\pi})
+(m_{\Xi_{cc}^{+}}+m_{\Xi_{c}^{'0}})g_{1}(m^{2}_{\pi})\gamma_{5}\right]
u(p_{1},s_{z}),
\end{eqnarray}
\begin{eqnarray}
%----------------------------------------------------------------------------------------------------
{\cal A}bs\,M_{b}(\rho^+;\Xi_c^{\prime 0};\Xi_c^{0})&=&-\int\frac{|\vec{p_2}|sin\theta d\theta d\varphi}{32\pi^{2}m_{\Xi_{cc}^{+}}}\frac{G_{F}}{\sqrt{2}}V_{cs}^{*}V_{ud}a_{1}f_{\rho}
\frac{F^{2}(t,m_{\Xi_{c}^{0}})}{t-m_{\Xi_{c}^{0}}^{2}} (-g^{\beta\nu}+\frac{p_{2}^{\beta}p_{2}^{\nu}}{m_{\rho}^{2}}) \epsilon_{6}^{*\alpha}\nonumber\\ &&\times\overline{u}(p_{5},s^{\prime}_{z})
\left[f_{1\Xi_{c}^{+}\Xi_{c}^{0}\rho^{+}}\gamma_{\nu}
-i\frac{f_{2\Xi_{c}^{+}\Xi_{c}^{0}\rho^{+}}}{m_{\Xi_{c}^{+}} +m_{\Xi_{c}^{0}}}\sigma_{\mu\nu}p_{2}^{\mu}\right](\myslash{p_{4}}+m_{\Xi_{c}^{0}})\nonumber\\
&&\times\left[f_{1\Xi_{c}^{'0}\Xi_{c}^{0}\rho^{0}}
\gamma_{\alpha}
+i\frac{f_{2\Xi_{c}^{'0}\Xi_{c}^{0}\rho^{0}}}{m_{\Xi_{c}^{'0}} +m_{\Xi_{c}^{0}}}\sigma_{\rho\alpha}p_{6}^{\rho}\right]
(\myslash{p_{3}}+m_{\Xi_{c}^{0}})\nonumber\\
&&\times\left[\left(f_{1}(m_{\rho}^{2})-\frac{m_{\Xi_{cc}^{+}}+m_{\Xi_{c}^{'0}}}{m_{\Xi_{cc}^{+}}}
f_{2}(m_{\rho}^{2})\right)\gamma_{\sigma}+\frac{2}{m_{\Xi_{cc}^{+}}}f_{2}(m_{\rho}^{2})p_{3\sigma}\right.\nonumber\\
&&\left.
-\left(g_{1}(m_{\rho}^{2})+\frac{m_{\Xi_{cc}^{+}}-m_{\Xi_{c}^{'0}}}{m_{\Xi_{cc}^{+}}}
g_{2}(m_{\rho}^{2})\right)\gamma_{\sigma}\gamma_{5}\right.
\left.-\frac{2}{m_{\Xi_{cc}^{+}}}g_{2}(m_{\rho}^{2})
p_{3\sigma}\gamma_{5}\right]u(p_{1},s_{z}),
\end{eqnarray}
\begin{eqnarray}
%----------------------------------------------------------------------------------------------------
{\cal A}bs\,M_{b}(\pi^+;\Xi_c^{\prime 0};\Xi_c^{\prime 0})&=&i\int\frac{|\vec{p_2}|sin\theta d\theta d\varphi}{32\pi^{2}m_{\Xi_{cc}^{+}}}\frac{G_{F}}{\sqrt{2}}V_{cs}^{*}V_{ud}a_{1}f_{\pi}g_{\Xi_{c}^{+}\Xi_{c}^{'0}\pi^{+}}
 \frac{F^{2}(t,m_{\Xi_{c}^{'0}})}{t-m_{\Xi_{c}^{'0}}^{2}}\epsilon_{6}^{*\nu}\nonumber\\
&&\times\overline{u}(p_{5},s^{\prime}_{z})\gamma_{5}(\myslash{p_{4}}+m_{\Xi_{c}^{'0}})
\left(f_{1\Xi_{c}^{'0}\Xi_{c}^{'0}\rho^{0}}\gamma_{\nu}
+i\frac{f_{2\Xi_{c}^{'0}\Xi_{c}^{'0}\rho^{0}}}{2m_{\Xi_{c}^{'0}}}
\sigma_{\mu\nu}p_{6}^{\mu}\right)\nonumber\\
&&\times(\myslash{p_{3}}+m_{\Xi_{c}^{'0}})
\left[(m_{\Xi_{cc}^{+}}-m_{\Xi_{c}^{'0}})f_{1}(m^{2}_{\pi})
+(m_{\Xi_{cc}^{+}}+m_{\Xi_{c}^{'0}})g_{1}(m^{2}_{\pi})\gamma_{5}\right]
u(p_{1},s_{z}),
\end{eqnarray}
\begin{eqnarray}
%----------------------------------------------------------------------------------------------------
{\cal A}bs\,M_{b}(\rho^+;\Xi_c^{\prime 0};\Xi_c^{\prime 0})&=&-\int\frac{|\vec{p_2}|sin\theta d\theta d\varphi}{32\pi^{2}m_{\Xi_{cc}^{+}}}\frac{G_{F}}{\sqrt{2}}V_{cs}^{*}V_{ud}a_{1}f_{\rho}
\frac{F^{2}(t,m_{\Xi_{c}^{'0}})}{t-m_{\Xi_{c}^{'0}}^{2}} (-g^{\beta\nu}+\frac{p_{2}^{\beta}p_{2}^{\nu}}{m_{\rho}^{2}}) \epsilon_{6}^{*\alpha}\nonumber\\
&&\times\overline{u}(p_{5},s^{\prime}_{z})
\left[f_{1\Xi_{c}^{+}\Xi_{c}^{'0}\rho^{+}}\gamma_{\nu}
-i\frac{f_{2\Xi_{c}^{+}\Xi_{c}^{'0}\rho^{+}}}{m_{\Xi_{c}^{+}} +m_{\Xi_{c}^{'0}}}\sigma_{\mu\nu}p_{2}^{\mu}\right](\myslash{p_{4}}+m_{\Xi_{c}^{'0}})\nonumber\\
&&\times\left[f_{1\Xi_{c}^{'0}\Xi_{c}^{'0}\rho^{0}}\gamma_{\alpha}
+i\frac{f_{2\Xi_{c}^{'0}\Xi_{c}^{'0}\rho^{0}}}{2m_{\Xi_{c}^{'0}}}\sigma_{\rho\alpha}p_{6}^{\rho}\right]
(\myslash{p_{3}}+m_{\Xi_{c}^{'0}})\nonumber\\
&&\times\left[\left(f_{1}(m_{\rho}^{2})-\frac{m_{\Xi_{cc}^{+}}+m_{\Xi_{c}^{'0}}}{m_{\Xi_{cc}^{+}}}
f_{2}(m_{\rho}^{2})\right)\gamma_{\sigma} +\frac{2}{m_{\Xi_{cc}^{+}}}f_{2}(m_{\rho}^{2})p_{3\sigma}\right.\nonumber\\
&&\left.-\left(g_{1}(m_{\rho}^{2})+\frac{m_{\Xi_{cc}^{+}}-m_{\Xi_{c}^{'0}}}{m_{\Xi_{cc}^{+}}}
g_{2}(m_{\rho}^{2})\right)\gamma_{\sigma}\gamma_{5}\right.
\left.-\frac{2}{m_{\Xi_{cc}^{+}}}g_{2}(m_{\rho}^{2})
p_{3\sigma}\gamma_{5}\right]u(p_{1},s_{z}).
\end{eqnarray}
\begin{eqnarray}
{\cal A}bs\,M_{c}(\pi^+;\Xi_c^0;\pi^-)&=&i\int\frac{|\vec{p_2}|sin\theta d\theta d\varphi}{32\pi^{2}m_{\Xi_{cc}^{+}}}\frac{G_{F}}{\sqrt{2}}V_{cs}^{*}V_{ud}a_{1}f_{\pi^+}
g_{\Xi_{c}^{+}\Xi_c^0 \pi^-}\frac{g_{\rho^{0}\pi^-\pi^+}}{\sqrt{2}}\frac{F^{2}(t,m_{\pi^-})}{t-m_{\pi^-}^{2}}(p_{2}\cdot\epsilon_{5}^{\ast}
-p_{4}\cdot\epsilon_{5}^{\ast})\nonumber\\
&&\times\overline{u}(p_{6},s^{\prime}_{z})\gamma_{5}
(\myslash{p_{3}}+m_{\Xi_c^0})\left[(m_{\Xi_{cc}^{+}}-m_{\Xi_c^0})f_{1}(m^{2}_{\pi^+})
+(m_{\Xi_{cc}^{+}}+m_{\Xi_c^0})g_{1}(m^{2}_{\pi^+})\gamma_{5}\right]
u(p_{1},s_{z}),
\end{eqnarray}
\begin{eqnarray}
{\cal A}bs\,M_{c}(\rho^{+};\Xi_c^0;\rho^{-})&=&\int\frac{|\vec{p_2}|sin\theta d\theta d\varphi}{32\pi^{2}m_{\Xi_{cc}^{+}}}\frac{G_{F}}{\sqrt{2}}V_{cs}^{*}V_{ud}a_{1}f_{\rho^{+}}
\frac{g_{\rho^{0}\rho^{-}\rho^{+}}}{\sqrt{2}} \frac{F^{2}(t,m_{\rho^{-}})}{t-m_{\rho^{-}}^{2}+im_{\rho^{-}}\Gamma_{\rho^{-}}}
\nonumber\\
&&\times\overline{u}(p_{6},s_{z}^\prime)\left(f_{1\Xi_{c}^{+}\Xi_c^0 \rho^{-}}\gamma_{\nu}
+i\frac{f_{2\Xi_{c}^{+}\Xi_c^0 \rho^{-}}}{m_{\Xi_{c}^{+}}+m_{\Xi_c^0}}
\sigma_{\mu\nu}p_{4}^{\mu}\right)(\myslash{p_{3}}+m_{\Xi_c^0})\nonumber\\
&&\times\left[\left(f_{1}(m_{\rho^{+}}^{2})-\frac{m_{\Xi_{cc}^{+}} +m_{\Xi_c^0}}{m_{\Xi_{cc}^{+}}}
f_{2}(m_{\rho^{+}}^{2})\right)\gamma_{\sigma} +\frac{2}{m_{\Xi_{cc}^{+}}}f_{2}(m_{\rho^{+}}^{2})p_{3\sigma}\right.\nonumber\\
&&\left.-\left(g_{1}(m_{\rho^{+}}^{2}) +\frac{m_{\Xi_{cc}^{+}}-m_{\Xi_c^0}}{m_{\Xi_{cc}^{+}}}
g_{2}(m_{\rho^{+}}^{2})\right)\gamma_{\sigma}\gamma_{5} -\frac{2}{m_{\Xi_{cc}^{+}}}g_{2}(m_{\rho^{+}}^{2})
p_{3\sigma}\gamma_{5}\right]u(p_{1},s_{z})\nonumber\\
&&\times[(2p_{5\alpha}-p_{2\alpha})\epsilon_{5\beta}^{\ast}-(p_{5\beta}
+p_{2\beta})\epsilon_{5\alpha}^{\ast}+g_{\alpha\beta}
(2p_{2}\cdot\epsilon_{5}^{\ast})](-g^{\nu\beta}+\frac{p_{4}^{\nu}p_{4}^{\beta}}{m_{\rho^{-}}^{2}})
(-g^{\sigma\alpha}+\frac{p_{2}^{\sigma}p_{2}^{\alpha}}{m_{\rho^{+}}^{2}}),
\end{eqnarray}
\begin{eqnarray}
{\cal A}bs\,M_{c}(\pi^+;\Xi_c^{'0};\pi^-)&=&i\int\frac{|\vec{p_2}|sin\theta d\theta d\varphi}{32\pi^{2}m_{\Xi_{cc}^{+}}}\frac{G_{F}}{\sqrt{2}}V_{cs}^{*}V_{ud}a_{1}f_{\pi^+}
g_{\Xi_{c}^{+}\Xi_c^{'0} \pi^-}\frac{g_{\rho^{0}\pi^-\pi^+}}{\sqrt{2}}\frac{F^{2}(t,m_{\pi^-})}{t-m_{\pi^-}^{2}}(p_{2}\cdot\epsilon_{5}^{\ast}
-p_{4}\cdot\epsilon_{5}^{\ast})\nonumber\\
&&\times\overline{u}(p_{6},s^{\prime}_{z})\gamma_{5}
(\myslash{p_{3}}+m_{\Xi_c^{'0}})\left[(m_{\Xi_{cc}^{+}}-m_{\Xi_c^{'0}})f_{1}(m^{2}_{\pi^+})
+(m_{\Xi_{cc}^{+}}+m_{\Xi_c^{'0}})g_{1}(m^{2}_{\pi^+})\gamma_{5}\right]
u(p_{1},s_{z}),
\end{eqnarray}
\begin{eqnarray}
{\cal A}bs\,M_{c}(\rho^{+};\Xi_c^{'0};\rho^{-})&=&\int\frac{|\vec{p_2}|sin\theta d\theta d\varphi}{32\pi^{2}m_{\Xi_{cc}^{+}}}\frac{G_{F}}{\sqrt{2}}V_{cs}^{*}V_{ud}a_{1}f_{\rho^{+}}
\frac{g_{\rho^{0}\rho^{-}\rho^{+}}}{\sqrt{2}} \frac{F^{2}(t,m_{\rho^{-}})}{t-m_{\rho^{-}}^{2}+im_{\rho^{-}}\Gamma_{\rho^{-}}}
\nonumber\\
&&\times\overline{u}(p_{6},s_{z}^\prime)\left(f_{1\Xi_{c}^{+}\Xi_c^{'0} \rho^{-}}\gamma_{\nu}
+i\frac{f_{2\Xi_{c}^{+}\Xi_c^{'0} \rho^{-}}}{m_{\Xi_{c}^{+}}+m_{\Xi_c^{'0}}}
\sigma_{\mu\nu}p_{4}^{\mu}\right)(\myslash{p_{3}}+m_{\Xi_c^{'0}})\nonumber\\
&&\times\left[\left(f_{1}(m_{\rho^{+}}^{2})-\frac{m_{\Xi_{cc}^{+}} +m_{\Xi_c^{'0}}}{m_{\Xi_{cc}^{+}}}
f_{2}(m_{\rho^{+}}^{2})\right)\gamma_{\sigma} +\frac{2}{m_{\Xi_{cc}^{+}}}f_{2}(m_{\rho^{+}}^{2})p_{3\sigma}\right.\nonumber\\
&&\left.-\left(g_{1}(m_{\rho^{+}}^{2}) +\frac{m_{\Xi_{cc}^{+}}-m_{\Xi_c^{'0}}}{m_{\Xi_{cc}^{+}}}
g_{2}(m_{\rho^{+}}^{2})\right)\gamma_{\sigma}\gamma_{5} -\frac{2}{m_{\Xi_{cc}^{+}}}g_{2}(m_{\rho^{+}}^{2})
p_{3\sigma}\gamma_{5}\right]u(p_{1},s_{z})\nonumber\\
&&\times[(2p_{5\alpha}-p_{2\alpha})\epsilon_{5\beta}^{\ast}-(p_{5\beta}
+p_{2\beta})\epsilon_{5\alpha}^{\ast}+g_{\alpha\beta}
(2p_{2}\cdot\epsilon_{5}^{\ast})](-g^{\nu\beta}+\frac{p_{4}^{\nu}p_{4}^{\beta}}{m_{\rho^{-}}^{2}})
(-g^{\sigma\alpha}+\frac{p_{2}^{\sigma}p_{2}^{\alpha}}{m_{\rho^{+}}^{2}}),
\end{eqnarray}
\begin{eqnarray}
{\cal A}bs\,M_{e}(\overline{K}^{0};\Sigma_c^+;\Lambda_c^+)&=&i\int\frac{|\vec{p_2}|sin\theta d\theta d\varphi}{32\pi^{2}m_{\Xi_{cc}^{+}}}\frac{G_{F}}{\sqrt{2}}V_{cs}^{*}V_{ud}a_{2}f_{\overline{K}^{0}}g_{\Xi_{c}^{+}\Lambda_c^+\overline{K}^{0}}
\frac{F^{2}(t,m_{\Lambda_c^+})}{t-m_{\Lambda_c^+}^{2}}\epsilon_{6}^{*\nu}\nonumber\\
&&\times\overline{u}(p_{5},s^{\prime}_{z})\gamma_{5}(\myslash{p_{4}}+m_{\Lambda_c^+})
\left(f_{1\Sigma_c^+\Lambda_c^+\rho^{0}}\gamma_{\nu}
+i\frac{f_{2\Sigma_c^+\Lambda_c^+\rho^{0}}}{m_{\Sigma_c^+}+m_{\Lambda_c^+}}
\sigma_{\mu\nu}p_{6}^{\mu}\right)\nonumber\\
&&\times(\myslash{p_{3}}+m_{\Lambda_c^+})
\left[(m_{\Xi_{cc}^{+}}-m_{\Lambda_c^+})f_{1}(m^{2}_{\overline{K}^{0}})
+(m_{\Xi_{cc}^{+}}+m_{\Lambda_c^+})g_{1}(m^{2}_{\overline{K}^{0}})\gamma_{5}\right]
u(p_{1},s_{z}),
\end{eqnarray}
\begin{eqnarray}
%----------------------------------------------------------------------------------------------------
{\cal A}bs\,M_{e}(\overline{K}^{*0};\Sigma_c^+;\Lambda_c^+)&=&-\int\frac{|\vec{p_2}|sin\theta d\theta d\varphi}{32\pi^{2}m_{\Xi_{cc}^{+}}}\frac{G_{F}}{\sqrt{2}}V_{cs}^{*}V_{ud}a_{2}f_{\overline{K}^{*0}} \frac{F^{2}(t,m_{\Lambda_c^+})}{t-m_{\Lambda_c^+}^{2}}
(-g^{\beta\nu}+\frac{p_{2}^{\beta}p_{2}^{\nu}}{m_{\overline{K}^{*0}}^{2}})\epsilon_{6}^{*\alpha}\nonumber\\
&&\times\overline{u}(p_{5},s^{\prime}_{z})
\left[f_{1\Xi_{c}^{+}\Lambda_c^+\overline{K}^{*0}}\gamma_{\nu}
-i\frac{f_{2\Xi_{c}^{+}\Lambda_c^+\overline{K}^{*0}}}{m_{\Xi_{c}^{+}} +m_{\Lambda_c^+}}\sigma_{\mu\nu}p_{2}^{\mu}\right](\myslash{p_{4}}+m_{\Lambda_c^+})\nonumber\\
&&\times\left[f_{1\Sigma_c^+\Lambda_c^+\rho^{0}}
\gamma_{\alpha}
 +i\frac{f_{2\Sigma_c^+\Lambda_c^+\rho^{0}}}{m_{\Sigma_c^+}+m_{\Lambda_c^+}}\sigma_{\rho\alpha}p_{6}^{\rho}\right]
(\myslash{p_{3}}+m_{\Sigma_c^+})\nonumber\\
&&\times\left[\left(f_{1}(m_{\overline{K}^{*0}}^{2})-\frac{m_{\Xi_{cc}^{+}}+m_{\Sigma_c^+}}{m_{\Xi_{cc}^{+}}}
f_{2}(m_{\overline{K}^{*0}}^{2})\right)\gamma_{\sigma}+\frac{2}{m_{\Xi_{cc}^{+}}}f_{2}(m_{\overline{K}^{*0}}^{2})p_{3\sigma}\right.\nonumber\\
&&\left.
-\left(g_{1}(m_{\overline{K}^{*0}}^{2})+\frac{m_{\Xi_{cc}^{+}}-m_{\Sigma_c^+}}{m_{\Xi_{cc}^{+}}}
g_{2}(m_{\overline{K}^{*0}}^{2})\right)\gamma_{\sigma}\gamma_{5}\right.
\left.-\frac{2}{m_{\Xi_{cc}^{+}}}g_{2}(m_{\overline{K}^{*0}}^{2})
p_{3\sigma}\gamma_{5}\right]u(p_{1},s_{z}),
\end{eqnarray}
\begin{eqnarray}
{\cal A}bs\,M_{e}(\overline{K}^{0};\Lambda_c^+;\Sigma_c^+)&=&i\int\frac{|\vec{p_2}|sin\theta d\theta d\varphi}{32\pi^{2}m_{\Xi_{cc}^{+}}}\frac{G_{F}}{\sqrt{2}}V_{cs}^{*}V_{ud}a_{2}f_{\overline{K}^{0}}g_{\Xi_{c}^{+}\Sigma_c^+\overline{K}^{0}}
\frac{F^{2}(t,m_{\Sigma_c^+})}{t-m_{\Sigma_c^+}^{2}}\epsilon_{6}^{*\nu}\nonumber\\
&&\times\overline{u}(p_{5},s^{\prime}_{z})\gamma_{5}(\myslash{p_{4}}+m_{\Sigma_c^+})
\left(f_{1\Lambda_c^+\Sigma_c^+\rho^{0}}\gamma_{\nu}
+i\frac{f_{2\Lambda_c^+\Sigma_c^+\rho^{0}}}{m_{\Lambda_c^+}+m_{\Sigma_c^+}}
\sigma_{\mu\nu}p_{6}^{\mu}\right)\nonumber\\
&&\times(\myslash{p_{3}}+m_{\Sigma_c^+})
\left[(m_{\Xi_{cc}^{+}}-m_{\Sigma_c^+})f_{1}(m^{2}_{\overline{K}^{0}})
+(m_{\Xi_{cc}^{+}}+m_{\Sigma_c^+})g_{1}(m^{2}_{\overline{K}^{0}})\gamma_{5}\right]
u(p_{1},s_{z}),
\end{eqnarray}
\begin{eqnarray}
%----------------------------------------------------------------------------------------------------
{\cal A}bs\,M_{e}(\overline{K}^{*0};\Lambda_c^+;\Sigma_c^+)&=&-\int\frac{|\vec{p_2}|sin\theta d\theta d\varphi}{32\pi^{2}m_{\Xi_{cc}^{+}}}\frac{G_{F}}{\sqrt{2}}V_{cs}^{*}V_{ud}a_{2}f_{\overline{K}^{*0}} \frac{F^{2}(t,m_{\Sigma_c^+})}{t-m_{\Sigma_c^+}^{2}}
(-g^{\beta\nu}+\frac{p_{2}^{\beta}p_{2}^{\nu}}{m_{\overline{K}^{*0}}^{2}})\epsilon_{6}^{*\alpha}\nonumber\\
&&\times\overline{u}(p_{5},s^{\prime}_{z})
\left[f_{1\Xi_{c}^{+}\Sigma_c^+\overline{K}^{*0}}\gamma_{\nu}
-i\frac{f_{2\Xi_{c}^{+}\Sigma_c^+\overline{K}^{*0}}}{m_{\Xi_{c}^{+}} +m_{\Sigma_c^+}}\sigma_{\mu\nu}p_{2}^{\mu}\right](\myslash{p_{4}}+m_{\Sigma_c^+})\nonumber\\
&&\times\left[f_{1\Lambda_c^+\Sigma_c^+\rho^{0}}
\gamma_{\alpha}
 +i\frac{f_{2\Lambda_c^+\Sigma_c^+\rho^{0}}}{m_{\Lambda_c^+}+m_{\Sigma_c^+}}\sigma_{\rho\alpha}p_{6}^{\rho}\right]
(\myslash{p_{3}}+m_{\Lambda_c^+})\nonumber\\
&&\times\left[\left(f_{1}(m_{\overline{K}^{*0}}^{2})-\frac{m_{\Xi_{cc}^{+}}+m_{\Lambda_c^+}}{m_{\Xi_{cc}^{+}}}
f_{2}(m_{\overline{K}^{*0}}^{2})\right)\gamma_{\sigma}+\frac{2}{m_{\Xi_{cc}^{+}}}f_{2}(m_{\overline{K}^{*0}}^{2})p_{3\sigma}\right.\nonumber\\
&&\left.
-\left(g_{1}(m_{\overline{K}^{*0}}^{2})+\frac{m_{\Xi_{cc}^{+}}-m_{\Lambda_c^+}}{m_{\Xi_{cc}^{+}}}
g_{2}(m_{\overline{K}^{*0}}^{2})\right)\gamma_{\sigma}\gamma_{5}\right.
\left.-\frac{2}{m_{\Xi_{cc}^{+}}}g_{2}(m_{\overline{K}^{*0}}^{2})
p_{3\sigma}\gamma_{5}\right]u(p_{1},s_{z}),
\end{eqnarray}
\begin{eqnarray}
{\cal A}bs\,M_{f}(\bar K^0;\Lambda_c^+;K^0)&=&i\int\frac{|\vec{p_2}|sin\theta d\theta d\varphi}{32\pi^{2}m_{\Xi_{cc}^{+}}}\frac{G_{F}}{\sqrt{2}}V_{cs}^{*}V_{ud}a_{2}f_{\overline{K}^{0}}
g_{\Xi_{c}^{+}\Lambda_{c}^{+}K^{0}}\frac{g_{\rho^{0}K^{0}\overline{K}^{0}}}{\sqrt{2}}\frac{F^{2}(t,m_{K^{0}})}{t-m_{K^{0}}^{2}}(p_{2}\cdot\epsilon_{5}^{\ast}
-p_{4}\cdot\epsilon_{5}^{\ast})\nonumber\\
&&\times\overline{u}(p_{6},s^{\prime}_{z})\gamma_{5}
(\myslash{p_{3}}+m_{\Lambda_{c}^{+}})\left[(m_{\Xi_{cc}^{+}}-m_{\Lambda_{c}^{+}})f_{1}(m^{2}_{\overline{K}^{0}})
+(m_{\Xi_{cc}^{+}}+m_{\Lambda_{c}^{+}})g_{1}(m^{2}_{\overline{K}^{0}})\gamma_{5}\right]
u(p_{1},s_{z})
\end{eqnarray}
\begin{eqnarray}
{\cal A}bs\,M_{f}(\bar K^{*0};\Lambda_c^+;K^{*0})&=&\int\frac{|\vec{p_2}|sin\theta d\theta d\varphi}{32\pi^{2}m_{\Xi_{cc}^{+}}}\frac{G_{F}}{\sqrt{2}}V_{cs}^{*}V_{ud}a_{2}f_{\overline{K}^{*0}}
\frac{g_{\rho^{0}K^{*0}\overline{K}^{*0}}}{\sqrt{2}} \frac{F^{2}(t,m_{K^{*0}})}{t-m_{K^{*0}}^{2}+im_{K^{*0}}\Gamma_{K^{*0}}}
\nonumber\\
&&\times\overline{u}(p_{6},s_{z}^\prime)\left(f_{1\Xi_{c}^{+}\Lambda_{c}^{+}K^{*0}}\gamma_{\nu}
+i\frac{f_{2\Xi_{c}^{+}\Lambda_{c}^{+}K^{*0}}}{m_{\Xi_{c}^{+}}+m_{\Lambda_{c}^{+}}}
\sigma_{\mu\nu}p_{4}^{\mu}\right)(\myslash{p_{3}}+m_{\Lambda_{c}^{+}})\nonumber\\
&&\times\left[\left(f_{1}(m_{\overline{K}^{*0}}^{2})-\frac{m_{\Xi_{cc}^{+}} +m_{\Lambda_{c}^{+}}}{m_{\Xi_{cc}^{+}}}
f_{2}(m_{\overline{K}^{*0}}^{2})\right)\gamma_{\sigma} +\frac{2}{m_{\Xi_{cc}^{+}}}f_{2}(m_{\overline{K}^{*0}}^{2})p_{3\sigma}\right.\nonumber\\
&&\left.-\left(g_{1}(m_{\overline{K}^{*0}}^{2}) +\frac{m_{\Xi_{cc}^{+}}-m_{\Lambda_{c}^{+}}}{m_{\Xi_{cc}^{+}}}
g_{2}(m_{\overline{K}^{*0}}^{2})\right)\gamma_{\sigma}\gamma_{5} -\frac{2}{m_{\Xi_{cc}^{+}}}g_{2}(m_{\overline{K}^{*0}}^{2})
p_{3\sigma}\gamma_{5}\right]u(p_{1},s_{z})\nonumber\\
&&\times[(2p_{5\alpha}-p_{2\alpha})\epsilon_{5\beta}^{\ast}-(p_{5\beta}
+p_{2\beta})\epsilon_{5\alpha}^{\ast}+g_{\alpha\beta}
(2p_{2}\cdot\epsilon_{5}^{\ast})](-g^{\nu\beta}+\frac{p_{4}^{\nu}p_{4}^{\beta}}{m_{K^{*0}}^{2}})
(-g^{\sigma\alpha}+\frac{p_{2}^{\sigma}p_{2}^{\alpha}}{m_{\overline{K}^{*0}}^{2}}),
\end{eqnarray}
\begin{eqnarray}
{\cal A}bs\,M_{f}(\bar K^{0};\Sigma_c^+;K^{0})&=&i\int\frac{|\vec{p_2}|sin\theta d\theta d\varphi}{32\pi^{2}m_{\Xi_{cc}^{+}}}\frac{G_{F}}{\sqrt{2}}V_{cs}^{*}V_{ud}a_{2}f_{\overline{K}^{0}}
g_{\Xi_{c}^{+}\Sigma_{c}^{+}K^{0}}\frac{g_{\rho^{0}K^{0}\overline{K}^{0}}}{\sqrt{2}}\frac{F^{2}(t,m_{K^{0}})}{t-m_{K^{0}}^{2}}(p_{2}\cdot\epsilon_{5}^{\ast}
-p_{4}\cdot\epsilon_{5}^{\ast})\nonumber\\
&&\times\overline{u}(p_{6},s^{\prime}_{z})\gamma_{5}
(\myslash{p_{3}}+m_{\Sigma_{c}^{+}})\left[(m_{\Xi_{cc}^{+}}-m_{\Sigma_{c}^{+}})f_{1}(m^{2}_{\overline{K}^{0}})
+(m_{\Xi_{cc}^{+}}+m_{\Sigma_{c}^{+}})g_{1}(m^{2}_{\overline{K}^{0}})\gamma_{5}\right]
u(p_{1},s_{z}),
\end{eqnarray}
\begin{eqnarray}
{\cal A}bs\,M_{f}(\bar K^{*0};\Sigma_c^+;K^{*0})&=&\int\frac{|\vec{p_2}|sin\theta d\theta d\varphi}{32\pi^{2}m_{\Xi_{cc}^{+}}}\frac{G_{F}}{\sqrt{2}}V_{cs}^{*}V_{ud}a_{2}f_{\overline{K}^{*0}}
\frac{g_{\rho^{0}K^{*0}\overline{K}^{*0}}}{\sqrt{2}}\frac{F^{2}(t,m_{K^{*0}})}{t-m_{K^{*0}}^{2}+im_{K^{*0}}\Gamma_{K^{*0}}}
\nonumber\\
&&\times\overline{u}(p_{6},s_{z}^\prime)\left(f_{1\Xi_{c}^{+}K^{*0}\Sigma_{c}^{+}}\gamma_{\nu}
+i\frac{f_{2\Xi_{c}^{+}K^{*0}\Sigma_{c}^{+}}}{m_{\Xi_{c}^{+}}+m_{\Sigma_{c}^{+}}}
\sigma_{\mu\nu}p_{4}^{\mu}\right)(\myslash{p_{3}}+m_{\Sigma_{c}^{+}})\nonumber\\
&&\times\left[\left(f_{1}(m_{\overline{K}^{*0}}^{2})-\frac{m_{\Xi_{cc}^{+}}+m_{\Sigma_{c}^{+}}}{m_{\Xi_{cc}^{+}}}
f_{2}(m_{\overline{K}^{*0}}^{2})\right)\gamma_{\sigma} +\frac{2}{m_{\Xi_{cc}^{+}}}f_{2}(m_{\overline{K}^{*0}}^{2})p_{3\sigma}\right.\nonumber\\
&&\left.-\left(g_{1}(m_{\overline{K}^{*0}}^{2})+\frac{m_{\Xi_{cc}^{+}}-m_{\Sigma_{c}^{+}}}{m_{\Xi_{cc}^{+}}}
g_{2}(m_{\overline{K}^{*0}}^{2})\right)\gamma_{\sigma}\gamma_{5}-\frac{2}{m_{\Xi_{cc}^{+}}}g_{2}(m_{\overline{K}^{*0}}^{2})
p_{3\sigma}\gamma_{5}\right]u(p_{1},s_{z})\nonumber\\
&&\times
[(2p_{5\alpha}-p_{2\alpha})\epsilon_{5\beta}^{\ast}-(p_{5\beta}
+p_{2\beta})\epsilon_{5\alpha}^{\ast}+g_{\alpha\beta}
(2p_{2}\cdot\epsilon_{5}^{\ast})](-g^{\nu\beta}+\frac{p_{4}^{\nu}p_{4}^{\beta}}{m_{K^{*0}}^{2}})
(-g^{\sigma\alpha}+\frac{p_{2}^{\sigma}p_{2}^{\alpha}}{m_{\overline{K}^{*0}}^{2}}).
\end{eqnarray}

With all the pieces gathered, the amplitude of $\Xi_{cc}\to\Xi_c^+\rho^0$ is expressed by
\begin{eqnarray}
{\cal A}(\Xi_{cc}\to\Xi_c^+\rho^0)=&&i{\cal A}bs\,\big[M_{b}(\pi^+;\Xi_c^0;\Xi_c^0) + M_{b}(\rho^+;\Xi_c^0;\Xi_c^0)+M_{b}(\pi^+;\Xi_c^{0};\Xi_c^{\prime 0}) + M_{b}(\rho^+;\Xi_c^{0};\Xi_c^{\prime 0}) \nonumber\\
&& +M_{b}(\pi^+;\Xi_c^{\prime 0};\Xi_c^{0})+M_{b}(\rho^+;\Xi_c^{\prime 0};\Xi_c^{0}) +M_{b}(\pi^+;\Xi_c^{\prime 0};\Xi_c^{\prime 0}) +M_{b}(\rho^+;\Xi_c^{\prime 0};\Xi_c^{\prime 0})\nonumber\\
&& + M_{c}(\pi^+;\Xi_c^0;\pi^-) + M_{c}(\rho^{+};\Xi_c^0;\rho^{-}) + M_{c}(\pi^+;\Xi_c^{'0};\pi^-) + M_{c}(\rho^{+};\Xi_c^{'0};\rho^{-}) \nonumber\\
&&+ M_{e}(\overline{K}^{0};\Sigma_c^+;\Lambda_c^+) + M_{e}(\overline{K}^{*0};\Sigma_c^+;\Lambda_c^+) + M_{e}(\overline{K}^{0};\Lambda_c^+;\Sigma_c^+) + M_{e}(\overline{K}^{*0};\Lambda_c^+;\Sigma_c^+)\nonumber\\
&&+M_{f}(\bar K^0;\Lambda_c^+;K^0) +M_{f}(\bar K^{*0};\Lambda_c^+;K^{*0}) + M_{f}(\bar K^{0};\Sigma_c^+;K^{0})+M_{f}(\bar K^{*0};\Sigma_c^+;K^{*0})\big].
\end{eqnarray}

Amplitudes of all the other decays can be calculated in the same way, whose expressions are collected in appendix \ref{app:amps}. Now we are ready to talk about the branching fractions.

%=========================================================
\section{Inputs, Numerical Results, and Discussions}\label{sec:results}
%=========================================================
With amplitudes at hand, a decay width of $P1\to P5P6$ can be calculated at the rest frame of $P1$ by
\begin{eqnarray}
\Gamma(P1\to P5P6)=\frac{\sqrt{(m_1^2-(m_5+m_6)^2)(m_1^2-(m_5-m_6)^2)}}{32\pi m_1^3}\sum_{\rm pol.}|{\cal A}(P1\to P5P6)|^2,
\end{eqnarray}
where the summations are performed over the polarizations of initial and final state particles, and an factor $1/2$ is multiplied to average over the polarizations of the mother particle. To calculate the branching fractions lifetimes of the mother particles are necessary. The recent measurement $\tau_{\Xi_{cc}^{++}}=256\,{\rm fs}$ \cite{Aaij:2018wzf} by the LHCb collaboration is adopted in this work.
%It is argued theoretically that the lifetimes of $\Xi_{cc}^+$ and $\Omega_{cc}$ should be much shorter than that of $\Xi_{cc}^{++}$, which means that their values might be much smaller than $256\,{\rm fs}$.
For lack of experimental data, the decay widths of $\Xi_{cc}^{+}$ and $\Omega_{cc}$ are given instead of the branching fractions.

%++++++++++++++++++++++++++++++++++++++++++++++++++++++++++++++++++++++++++++
\begin{table}[!htbp]
  \centering
 \caption{Decay constants of light pseudoscalar and vector mesons collected from Refs. \cite{PDG,Choi:2015ywa} (in unit of MeV). $f_{\eta_8}$ and $f_{\eta_1}$ are calculated with the formulas in Ref. \cite{Feldmann:1998vh}.}
 \label{tab:decaycons}
  \begin{tabular}{cccccccc}  %c
   \hline\hline %
     $f_{\pi}$   &$f_{K}$    &$f_{\eta_8}$   &$f_{\eta_1}$   &$f_{\rho}$   &$f_{K^*}$   &$f_{\omega}$   &$f_{\phi}$\\
   \hline %
     $130$   &$156$     &$163$          &$152$          &$216$        &$217$       &$195$          &$233$    \\
   \hline\hline
   \end{tabular}
\end{table}
%++++++++++++++++++++++++++++++++++++++++++++++++++++++++++++++++++++++++++++

Besides the inputs already specified, decay constants of light pseudoscalar and vector mesons are collected in Table \ref{tab:decaycons}. The related masses and widths are taken from the Particle Data Group \cite{PDG} and we will not list their values here any longer. $m_{\eta_1}=835.7$ MeV is used as in Ref. \cite{Guo:2015xva} and $m_{\eta_8}=612$ MeV is calculated with the formulas in Ref. \cite{Feldmann:1998sh}. Strong coupling constants are the inputs with the largest amount. Some of them can be found in the literatures, and the unfound ones are calculated with respect to the $SU(3)$ symmetry. The strong coupling constants appeared in our calculation are gathered in the appendix \ref{app:stcouplings}.

In our previous work \cite{Yu:2017zst} the sensitivity of branching ratios to $\eta$ in Eq. (\ref{eq:Ffactor}) has already been investigated. It is found that branching fractions might vary a lot when $\eta$ ranges from $1$ to $2$. However this parameter is purely a phenomenological one, which is determined not by the first principle calculations but experimental data. Since the experimental data is far less enough, we make a rough decision of letting it equal $1.5$ for central values and vary it from $1$ to $2$ for error estimations.

%++++++++++++++++++++++++++++++++++++++++++++++++++++++++++
\begin{figure}[htp]
\begin{center}
\begin{minipage}{0.25\linewidth}
\centerline{\includegraphics[scale=0.5]{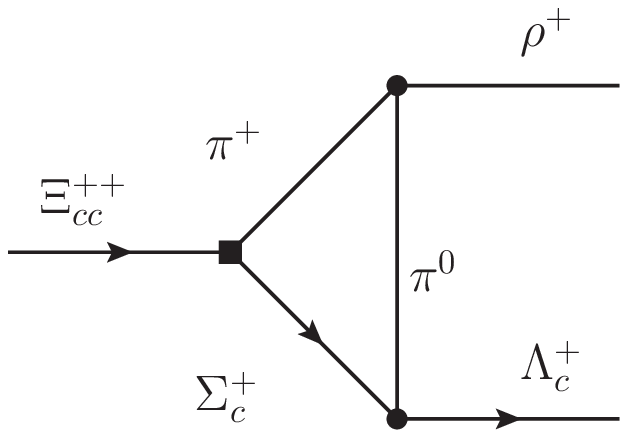}}
\vspace{0.7cm}
(a)
\end{minipage}
\begin{minipage}{0.25\linewidth}
\centerline{\includegraphics[scale=0.3]{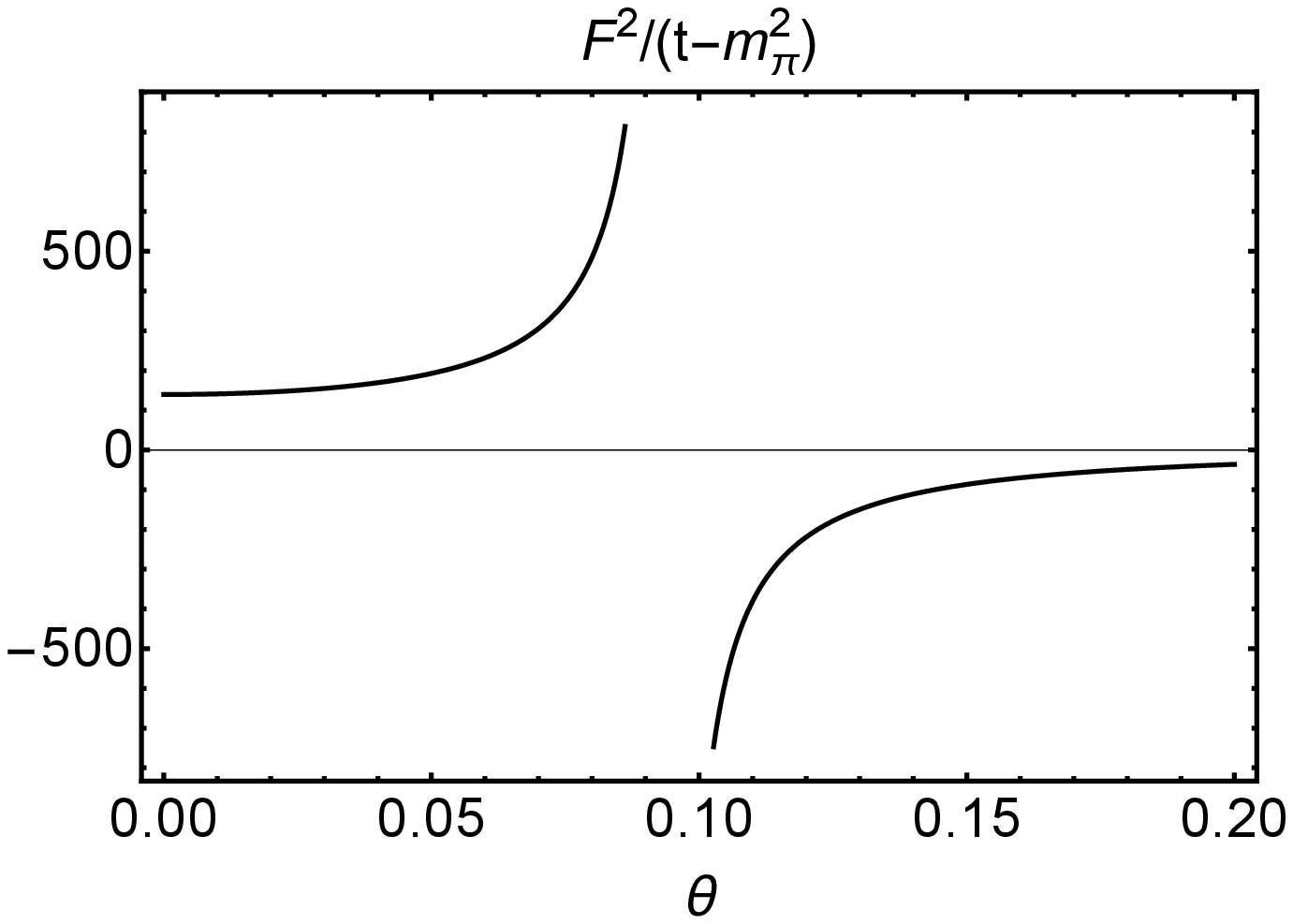}}
(b)
\end{minipage}
\end{center}
\caption{(a): An example with possible contribution of on-shell exchanged particle. (b): Dependence of the Breit-Wigner like factor $F^2(t,m_\pi)/(t-m^2_\pi)$ on $\theta$ in region around the singularity, where $\theta$ is the polar angle of $\Sigma_c^+$'s $3$-momentum obtained at the rest frame of $\Xi_{cc}^{++}$.}
\label{fig:sigularity}
\end{figure}
%++++++++++++++++++++++++++++++++++++++++++++++++++++++++++

It should be mentioned that the exchanged particle is possible to be on-shell in some kinematics region. This case appears when $P3$'s rest energy in Fig. \ref{fig:number} is larger than the rest energy sum of $P4$ and $P6$, and at the same time the rest energy sum of $P4$ and $P2$ is smaller than that of $P5$. It is just the physical constraint under which a $1\,{\rm body}\to 2\,{\rm bodies}$ decay and meanwhile a $2\,{\rm bodies}\to 1\,{\rm body}$ collision take place.  In Fig. \ref{fig:sigularity}(a), for example, one gets $t-m^2_\pi=(1.65 \cos\theta -1.64)\,{\rm GeV}^2$ after some calculation at the rest frame of $\Xi_{cc}^{++}$, where $\theta$ is the polar angle of $\Sigma_c^+$'s $3$-momentum obtained at the rest frame of $\Xi_{cc}^{++}$. It indicates a singularity nearby $\theta=0$, which is depicted in Fig. \ref{fig:sigularity}(b). In this case the abandoned term with pion's lifetime in the Breit-Wigner formula has to be picked up to regular the singularity. However, one needs not to worry about losing control of this singularity, which contributes little because of the cancellation in integration between left and right sides of the pole.

In tables \ref{tab:brsXiccpp}, \ref{tab:brsXiccp} and \ref{tab:brsOmegacc} we collected our results and the corresponding topological contributions to each decay. The short distance contributions are identified with indices ``SD", and the long distance contributions are labelled with the symbols defined in Fig. \ref{fig:topos}.
%
%One may find a new contribution represented by $N$, which has already been mentioned in section \ref{ssec:hamiltonian}. We would like to make the introduction easier by using a specific example, shown in Fig. \ref{fig:Ncontribution}, in which the weak decay is induced at quark level by $c\to u d \bar d$. The $d\bar d$ pair annihilates subsequently, and the weak transition is equivalent to a flavor changing neutral current $c \to u \gamma/g$. The same situation exists in $c \to u s \bar s$ inducing decays. Topologically, it looks the same as $O_{7\gamma}$ or $O_{8g}$ operators in the low energy effective hamiltonian, which are purely loop effects and suppressed highly at short distance in the standard model. However, it is the long distance contributions that are considered here, thus they are kept.
%%++++++++++++++++++++++++++++++++++++++++++++++++++++++++++
%\begin{figure}[htp]
%\begin{center}
%\centerline{\includegraphics[scale=0.5]{NC.eps}}
%\end{center}
%\caption{An example of $N$ contribution depicted at quark level.}
%\label{fig:Ncontribution}
%\end{figure}
%%++++++++++++++++++++++++++++++++++++++++++++++++++++++++++++++++++++++++++++++++++++++++++++++++++++++++++++++++++++++
According to the CKM matrix elements, these decays can also be classified as CKM favored ones induced by $c\to su\bar d$ and labelled with ``CF" in tables \ref{tab:brsXiccpp}, \ref{tab:brsXiccp} and \ref{tab:brsOmegacc}, singly CKM suppressed ones by $c\to du\bar d$ or $c\to su\bar s$ with ``SCS", and doubly CKM suppressed ones by $c\to du\bar s$ with ``DCS".
%++++++++++++++++++++++++++++++++++++++++++++++++++++++++++++++++++++++++++++++++++++++++++++++++++++++++++++++++++++++
\begin{table}[!htbp]
  \centering
 \caption{Our results for branching ratios of $\Xi_{cc}^{++}\to {\cal B}_c V$ and corresponding topological contributions. The ``CF", ``SCS" and ``DCS" represent CKM favored, singly CKM suppressed and doubly CKM suppressed processes, respectively. The errors are estimated by varying $\eta$ from $1$ to $2$, and the central values are given at $\eta=1.5$.}
 \label{tab:brsXiccpp}
  \begin{tabular}{|c|c|c|c|c|c|c|c|}  %c
   \hline %
   Channels   &${\cal BR}(\%)$      &Contributions &CKM    &Channels    &${\cal BR}(\%)$         &Contributions &CKM \\
   \hline %
   $\Xi_{cc}^{++}\rightarrow\Sigma_{c}^{++}\bar K^{*0}$    &$5.40_{-3.66}^{+5.59}$    &$C_{\rm SD},C$ &CF &$\Xi_{cc}^{++}\rightarrow\Xi_{c}^{+}\rho^{+}$    &$15.98_{-3.35}^{+5.33}$    & $T_{\rm SD}, T, C^\prime$ &CF\\
   \hline %
   $\Xi_{cc}^{++}\rightarrow\Xi_{c}^{'+}\rho^{+}$    &$16.54_{-0.72}^{+1.25}$    &$T_{\rm SD},T,C^\prime$  &CF&$\Xi_{cc}^{++}\rightarrow\Sigma_{c}^{+}\rho^{+}$    &$1.05_{-0.06}^{+0.08}$    &$T_{\rm SD},T,C^\prime$  &SCS\\
   \hline %
   $\Xi_{cc}^{++}\rightarrow\Lambda_{c}^{+}\rho^{+}$    &$0.95_{-0.03}^{+0.04}$    &$T_{\rm SD}, T, C^\prime$ &SCS&$\Xi_{cc}^{++}\rightarrow\Sigma_{c}^{++}\rho^{0}$    &$0.45_{-0.31}^{+0.51}$    &$C_{\rm SD},C$ &SCS\\
     \hline %
   $\Xi_{cc}^{++}\rightarrow\Sigma_{c}^{++}\omega$    &$0.14_{-0.09}^{+0.16}$    &$C_{\rm SD},C$   &SCS &$\Xi_{cc}^{++}\rightarrow\Sigma_{c}^{++}\phi$    &$0.09_{-0.06}^{+0.08}$    &$C_{\rm SD},C$ &SCS\\
     \hline %
   $\Xi_{cc}^{++}\rightarrow\Xi_{c}^{+}K^{*+}$    &$0.59_{-0.09}^{+0.16}$    &$T_{\rm SD},T,C^\prime$   &SCS&$\Xi_{cc}^{++}\rightarrow\Xi_{c}^{'+}K^{*+}$    &$0.80_{-0.05}^{+0.10}$    &$T_{\rm SD},T,C^\prime$&SCS\\
     \hline %
   $\Xi_{cc}^{++}\rightarrow\Sigma_{c}^{+}K^{*+}$    &$0.06_{-0.01}^{+0.00}$    &$T_{\rm SD},T,C^\prime$   &DCS&$\Xi_{cc}^{++}\rightarrow\Lambda_{c}^{+}K^{*+}$    &$0.05_{-0.00}^{+0.00}$    &$T_{\rm SD},T,C^\prime$&DCS\\
     \hline %
    $\Xi_{cc}^{++}\rightarrow\Sigma_{c}^{++}K^{*0}$    &$0.02_{-0.01}^{+0.02}$    &$C_{\rm SD},C$ &DCS&      &      && \\
   \hline
   \end{tabular}
\end{table}
%++++++++++++++++++++++++++++++++++++++++++++++++++++++++++++++++++++++++++++++++++++++++++++++++++++++++++++++++++++++

One can see in table \ref{tab:brsXiccpp} that each ${\Xi}_{cc}^{++} \to {\cal B}_c V$ decay receives short distance contribution of $C_{\rm SD}$ or $T_{\rm SD}$. The discovery decay $\Xi_{cc}^{++}\rightarrow\Sigma_{c}^{++}\bar K^{*0}$ is predicted to range from about $1.7\%$ to $11.0\%$, which are slightly different from the values in our previous work \cite{Yu:2017zst}. It is caused by the different inputs used in this paper, such as the life time of $\Xi_{cc}^{++}$, etc. Although they are not chosen as the discovery decays because of the low efficiency of reconstructing $\rho^+$ in the detectors, $\Xi_{cc}^{++}\rightarrow\Xi_{c}^{+}\rho^{+}$ and $\Xi_{cc}^{++}\rightarrow\Xi_{c}^{'+}\rho^{+}$ have the largest branching fractions, whose values are around $16\%$. These two decays as well as the discovery decay are all CKM favored ones. The branching fractions of singly CKM suppressed decays range from order of $10^{-3}$ to $10^{-2}$. The doubly CKM suppressed decays have the smallest branching fractions at the order of $10^{-4}$. Among decays in the same CKM mode, those with $T$ contributions tend to have largest branching fractions. The $C$ type decays are about several times smaller than the $T$ type. The other types of decays are suppressed highly. Under this theoretical framework $T_{\rm SD}$ is absolutely the dominating contribution comparing to the nonfactorizable contributions. As a result one can see that the results with $T_{\rm SD}$ contributions are not sensitive to the variation of $\eta$. However, the picture is totally different for the other types of contributions which increase or decrease rapidly as $\eta$ changes, which indicates very important FSI effects. Similar cases exist in $\Xi_{cc}^+$ and $\Omega_{cc}$ decays.

%++++++++++++++++++++++++++++++++++++++++++++++++++++++++++++++++++++++++++++++++++++++++++++++++++++++++++++++++++++++
\begin{table}[!htbp]
  \centering
 \caption{Our results for decay widths of $\Xi_{cc}^{+}\to {\cal B}_c V$ and corresponding topological contributions. The ``CF", ``SCS" and ``DCS" represent CKM favored, singly CKM suppressed and doubly CKM suppressed processes, respectively. The errors are estimated by varying $\eta$ from $1$ to $2$, and the central values are given at $\eta=1.5$.}
 \label{tab:brsXiccp}
  \begin{tabular}{|c|c|c|c|c|c|c|c|}  %c
   \hline %
   Channels   &$\Gamma/{\rm GeV}$      &Contributions&CKM    &Channels    &$\Gamma/{\rm GeV}$         &Contributions&CKM    \\
   \hline
  $\Xi_{cc}^{+}\rightarrow\Sigma_{c}^{+}\overline{K}^{*0}$    &$(8.42_{-5.74}^{+8.87})*10^{-14}$    &$C_{\rm SD},C,E_2$  &CF&$\Xi_{cc}^{+}\rightarrow\Lambda_{c}^{+}\overline{K}^{*0}$    &$(7.06_{-4.86}^{+7.68})*10^{-14}$    &$C_{\rm SD},C,E_2$ &CF\\
  \hline
  $\Xi_{cc}^{+}\rightarrow\Xi_{c}^{0}\rho^{+}$&$(3.83 _{-0.37 }^{+0.47 })*10^{-13}$&$T_{\rm SD},T,E_1$&CF&$\Xi_{cc}^{+}\rightarrow\Xi_{c}^{'0}\rho^{+}$&$( 4.77_{-0.24 }^{+0.31 })*10^{-13}$&$T_{\rm SD},T,E_1$ &CF\\
  \hline
  $\Xi_{cc}^{+}\rightarrow\Xi_{c}^{+}\rho^{0}$    &$(1.82 _{-1.22 }^{+1.85})*10^{-13}$    &$C^\prime,E_1$ &CF&$\Xi_{cc}^{+}\rightarrow\Xi_{c}^{'+}\rho^{0}$    &$(6.13 _{-4.14 }^{+6.23 })*10^{-14}$    &$C^\prime,E_1$ &CF\\
  \hline
  $\Xi_{cc}^{+}\rightarrow\Xi_{c}^{+}\omega$    &$(1.63 _{-1.14 }^{+1.87 })*10^{-14}$    &$C^\prime,E_1$  &CF&$\Xi_{cc}^{+}\rightarrow\Xi_{c}^{'+}\omega$    &$( 2.47_{-1.70 }^{+2.71 })*10^{-15}$    &$C^\prime,E_1$ &CF\\
  \hline
  $\Xi_{cc}^{+}\rightarrow\Sigma_{c}^{++}K^{*-}$    &$(7.38 _{-5.02 }^{+7.83 })*10^{-16}$    &$E_2$ &CF&$\Xi_{cc}^{+}\rightarrow\Xi_{c}^{+}\phi$    &$(5.12 _{-3.52 }^{+5.59 })*10^{-15}$    &$E_2$&CF\\
  \hline
  $\Xi_{cc}^{+}\rightarrow\Xi_{c}^{'+}\phi$    &$(9.90 _{-6.72 }^{+10.24})*10^{-17}$    &$E_2$ &CF &$\Xi_{cc}^{+}\rightarrow\Omega_{c}^{0}K^{*+}$    &$(2.33 _{-1.54 }^{+2.16 })*10^{-14}$    &$E_1$&CF\\
  \hline
  $\Xi_{cc}^{+}\rightarrow\Sigma_{c}^{+}\rho^{0}$    &$(1.31 _{-0.91 }^{+1.50 })*10^{-14}$    &$C_{\rm SD},C,C^\prime,E_1,E_2$  &SCS&$\Xi_{cc}^{+}\rightarrow\Lambda_{c}^{+}\rho^{0}$    &$(3.36 _{-2.35 }^{+3.95 })*10^{-15}$    &$C_{\rm SD},C,C^\prime,E_1,E_2$ &SCS\\
  \hline
  $\Xi_{cc}^{+}\rightarrow\Sigma_{c}^{+}\omega$    &$(2.21 _{-1.52 }^{+2.37})*10^{-15}$    &$C_{\rm SD},C,C^\prime,E_1,E_2$  &SCS&$\Xi_{cc}^{+}\rightarrow\Lambda_{c}^{+}\omega$    &$(1.01 _{-0.70}^{+1.12})*10^{-15}$    &$C_{\rm SD},C,C^\prime,E_1,E_2$ &SCS\\
  \hline
  $\Xi_{cc}^{+}\rightarrow\Sigma_{c}^{0}\rho^{+}$    &$(6.01 _{-0.43 }^{+0.57 })*10^{-14}$    &$T_{\rm SD},T,E_1$  &SCS&$\Xi_{cc}^{+}\rightarrow\Sigma_{c}^{+}\phi$    &$(1.54 _{-1.01 }^{+1.40 })*10^{-15}$    &$C_{\rm SD},C$  &SCS\\
  \hline
  $\Xi_{cc}^{+}\rightarrow\Lambda_{c}^{+}\phi$    &$(2.61 _{-1.76 }^{+2.67 })*10^{-15}$    &$C_{\rm SD},C$    &SCS&$\Xi_{cc}^{+}\rightarrow\Xi_{c}^{0}K^{*+}$    &$(1.30 _{-0.00 }^{+0.00})*10^{-14}$    &$T_{\rm SD},T,E_1$&SCS\\
  \hline
  $\Xi_{cc}^{+}\rightarrow\Xi_{c}^{'0}K^{*+}$    &$(2.19 _{-0.10}^{+0.19 })*10^{-14}$    &$T_{\rm SD},T,E_1$ &SCS    &$\Xi_{cc}^{+}\rightarrow\Xi_{c}^{+}K^{*0}$    &$(1.00 _{-0.64 }^{+0.86 })*10^{-15}$    &$C^\prime,E_2$ &SCS\\
  \hline
  $\Xi_{cc}^{+}\rightarrow\Xi_{c}^{'+}K^{*0}$    &$(1.53 _{-1.01}^{+1.47 })*10^{-15}$    &$C^\prime,E_2$  &SCS &$\Xi_{cc}^{+}\rightarrow\Sigma_{c}^{++}\rho^{-}$    &$(9.08 _{-6.14 }^{+9.39 })*10^{-17}$    &$E_2$  &SCS\\
  \hline
  $\Xi_{cc}^{+}\rightarrow\Sigma_{c}^{+}K^{*0}$    &$(4.87 _{-3.29 }^{+5.02 })*10^{-16}$    &$C_{\rm SD},C,C^\prime$  &DCS&$\Xi_{cc}^{+}\rightarrow\Lambda_{c}^{+}K^{*0}$    &$(2.54 _{-1.76 }^{+2.86 })*10^{-16}$    &$C_{\rm SD},C,C^\prime$ &DCS\\
  \hline
  $\Xi_{cc}^{+}\rightarrow\Sigma_{c}^{0}K^{*+}$    &$(2.88 _{-0.00 }^{+0.00 })*10^{-15}$    &$T_{\rm SD},T$ &DCS&   &    &    &    \\
  \hline
   \end{tabular}
   \end{table}

%++++++++++++++++++++++++++++++++++++++++++++++++++++++++++++++++++++++++++++++++++++++++++++++++++++++++++++++++++++++

The ${\Xi}_{cc}^{+} \to {\cal B}_c V$ decays, whose results are collected in table \ref{tab:brsXiccp}, have richer dynamics. Besides the contributions in $\Xi_{cc}^{++}$ decays, $E_{1,2}$ contributions come in. Because $\Xi_{cc}^+$ is still not established in experiments, there is no experimental data for its lifetime. Instead of branching fractions we present the decay widths in unit of GeV. Similar to the $\Xi_{cc}^{++}$ decays, the different topological contributions fall in a hierarchy in the same CKM decay mode. The decays with the largest widths are the CKM favored ones with $T$ contributions $\Xi_{cc}^{+}\rightarrow\Xi_{c}^{0}\rho^{+}$ and $\Xi_{cc}^{+}\rightarrow\Xi_{c}^{'0}\rho^{+}$. In a very recent work, the lifetime of $\Xi_{cc}^+$ is calculated to be $45$ fs \cite{Cheng:2018mwu}. If estimated with this value, the branching ratios of these two decays are given as
\begin{eqnarray}
{\cal BR}(\Xi_{cc}^{+}\rightarrow\Xi_{c}^{0}\rho^{+}) &\in& [2.4\%,2.9\%]\, ,\,\,\,\,{\cal BR}(\Xi_{cc}^{+}\rightarrow\Xi_{c}^{'0}\rho^{+}) \in [3.1\%,3.5\%]\,.
\label{eq:Xiccpbrs}
\end{eqnarray}
Although they have large branching fractions, these two decays are not ideal discovery decays of $\Xi_{cc}^+$ because of the low efficiency of reconstructing $\rho^+$. The decay $\Xi_{cc}^+\to\Xi_c^+\rho^0$ has a large branching fraction, too, which is given as\footnote{For estimation of branching fractions, the lifetime of $\Xi_{cc}^+$ is always used as $45$ fs.}
\begin{equation}
{\cal BR}(\Xi_{cc}^+\to\Xi_c^+\rho^0) \in [0.4\%,2.5\%].
\end{equation}
In Ref. \cite{Yu:2017zst} the ratios of the branching fractions are found to be not sensitive to the variation of $\eta$, which can be obtained at a fixed value of $\eta$ and used to pick out the ideal discovery decay. The central values given at $\eta=1.5$ in table \ref{tab:brsXiccp} indicates that ${\cal BR}(\Xi_{cc}^+\to\Xi_c^+\rho^0)$ is expected to have the largest branching fraction except the two CKM favored $T$ decays. Considering $\rho^0$ can be reconstructed by $\pi^+\pi^-$, we think $\Xi_{cc}^+\to\Xi_c^+\pi^+\pi^-$ can be chosen as a candidate for searching $\Xi_{cc}^+$.

When SELEX reported the first observation of $\Xi_{cc}^+$, $\Xi_{cc}^+ \to \Lambda_c K^- \pi^+$ is used for reconstruction \cite{Mattson:2002vu}, which is contributed by $\Xi_{cc}^{+}\rightarrow\Lambda_{c}^{+}\overline{K}^{*0}$ and $\Xi_{cc}^{+}\rightarrow\Sigma_{c}^{++}K^{-}$. The branching fraction of the former one ranges from $0.2\%$ to $1.2\%$ and is only about half of ${\cal BR}(\Xi_{cc}^+\to\Xi_c^+\rho^0)$, while the latter one, which can only occurs through $W$ exchange mechanism, is expected to be highly suppressed.

It is interesting that the following five decays are purely induced by the $W$ exchange mechanism, whose short distance contributions are thought to be highly suppressed:
\begin{eqnarray}
\Xi_{cc}^{+}\rightarrow\Sigma_{c}^{++}K^{*-},\,\, \Xi_{cc}^{+}\rightarrow\Xi_{c}^{+}\phi,\,\, \Xi_{cc}^{+}\rightarrow\Xi_{c}^{'+}\phi,\,\, \Xi_{cc}^{+}\rightarrow\Omega_{c}^{0}K^{*+},\,\, \Xi_{cc}^{+}\rightarrow\Sigma_{c}^{++}\rho^{-}.
\end{eqnarray}
Provided that there is enough data in future experiments, measurement of these decays will help to figure out the role of FSIs in doubly charmed baryon decays. Among these decays the branching fraction of $\Xi_{cc}^{+}\rightarrow\Omega_{c}^{0}K^{*+}$ is possible to be as large as $3.1\mbox{\textperthousand}$ when $\eta=2$, which is hopeful to be observed.

%++++++++++++++++++++++++++++++++++++++++++++++++++++++++++++++++++++++++++++++++++++++++++++++++++++++++++++++++++++++
\begin{table}[!htbp]
  \centering
 \caption{Our results for decay widths of $\Omega_{cc}^+\to {\cal B}_c V$ and corresponding topological contributions. The ``CF", ``SCS" and ``DCS" represent CKM favored, singly CKM suppressed and doubly CKM suppressed processes, respectively. The errors are estimated by varying $\eta$ from $1$ to $2$, and the central values are given at $\eta=1.5$.}
 \label{tab:brsOmegacc}
  \begin{tabular}{|c|c|c|c|c|c|c|c|}
   \hline %
   Channels   &$\Gamma/{\rm GeV}$      &Contributions&CKM    &Channels    &$\Gamma/{\rm GeV}$         &Contributions&CKM    \\
   \hline
   $\Omega_{cc}^{+}\rightarrow\Xi_{c}^{+}\overline{K}^{*0}$    &$(1.38_{-0.95}^{+1.49})*10^{-13}$    &$C_{\rm SD},C,C^\prime$ &CF   &$\Omega_{cc}^{+}\rightarrow\Xi_{c}^{'+}\overline{K}^{*0}$    &$(2.64_{-1.79}^{+2.72})*10^{-13}$    &$C_{\rm SD},C,C^\prime$ &CF\\
   \hline
   $\Omega_{cc}^{+}\rightarrow\Omega_{c}^{0}\rho^{+}$    &$(8.75_{-0.00}^{+0.00})*10^{-13}$    &$T_{\rm SD},T$ &CF&$\Omega_{cc}^{+}\rightarrow\Sigma_{c}^{+}\overline{K}^{*0}$    &$(1.35_{-0.96}^{+1.53})*10^{-15}$    &$C^\prime,E_2$ &SCS\\
   \hline
   $\Omega_{cc}^{+}\rightarrow\Lambda_{c}^{+}\overline{K}^{*0}$    &$(1.00_{-0.70}^{+1.16})*10^{-15}$    &$C^\prime,E_2$ &SCS&$\Omega_{cc}^{+}\rightarrow\Xi_{c}^{+}\rho^{0}$    &$(4.28_{-2.96}^{+4.78})*10^{-14}$    &$C_{\rm SD},C,E_1$ &SCS\\
   \hline
   $\Omega_{cc}^{+}\rightarrow\Xi_{c}^{'+}\rho^{0}$    &$(8.58_{-5.98}^{+9.88})*10^{-15}$    &$C_{\rm SD},C,E_1$ &SCS& $\Omega_{cc}^{+}\rightarrow\Xi_{c}^{+}\omega$    &$(8.22_{-5.77}^{+9.60})*10^{-15}$    &$C_{\rm SD},C,E_1$ &SCS\\
   \hline
   $\Omega_{cc}^{+}\rightarrow\Xi_{c}^{'+}\omega$    &$(6.09_{-4.18}^{+6.52})*10^{-15}$    &$C_{\rm SD},C,E_1$ &SCS&$\Omega_{cc}^{+}\rightarrow\Xi_{c}^{0}\rho^{+}$    &$(2.87_{-0.49}^{+0.69})*10^{-14}$    &$T_{\rm SD},T,E_1$ &SCS\\
   \hline
   $\Omega_{cc}^{+}\rightarrow\Xi_{c}^{'0}\rho^{+}$    &$(2.85_{-0.14}^{+0.19})*10^{-14}$    &$T_{\rm SD},T,E_1$ &SCS&$\Omega_{cc}^{+}\rightarrow\Xi_{c}^{+}\phi$    &$(1.86_{-1.23}^{+1.74})*10^{-15}$    &$C_{\rm SD},C,C^\prime,E_2$&SCS\\
   \hline
   $\Omega_{cc}^{+}\rightarrow\Xi_{c}^{'+}\phi$    &$(9.45_{-6.23}^{+8.84})*10^{-15}$    &$C_{\rm SD},C,C^\prime,E_2$&SCS&$\Omega_{cc}^{+}\rightarrow\Omega_{c}^{0}K^{*+}$    &$(4.18_{-0.01}^{+0.02})*10^{-14}$    &$T_{\rm SD},T,E_1$ &SCS\\
   \hline
   $\Omega_{cc}^{+}\rightarrow\Sigma_{c}^{++}K^{*-}$    &$(1.63_{-1.16}^{+2.06})*10^{-17}$    &$E_2$&SCS&$\Omega_{cc}^{+}\rightarrow\Xi_{c}^{'+}K^{*0}$    &$(4.77_{-3.26}^{+5.07})*10^{-16}$    &$C_{\rm SD},C,E_2$ &DCS\\
   \hline
   $\Omega_{cc}^{+}\rightarrow\Sigma_{c}^{+}\phi$    &$(8.45_{-5.81}^{+9.15})*10^{-17}$    &$C^\prime$ &DCS& $\Omega_{cc}^{+}\rightarrow\Lambda_{c}^{+}\phi$    &$(4.25_{-2.93}^{+4.64})*10^{-17}$    &$C^\prime$ &DCS\\
   \hline
   $\Omega_{cc}^{+}\rightarrow\Xi_{c}^{0}K^{*+}$    &$(1.00_{-0.00}^{+0.01})*10^{-15}$    &$T_{\rm SD},T,E_1$&DCS&$\Omega_{cc}^{+}\rightarrow\Xi_{c}^{\prime 0}K^{*+}$    &$(1.12_{-0.77}^{+1.22})*10^{-16}$    &$T_{\rm SD},T,E_1$&DCS\\
   \hline
   $\Omega_{cc}^{+}\rightarrow\Xi_{c}^{+}K^{*0}$    &$(6.24_{-4.23}^{+6.49})*10^{-16}$    &$C_{\rm SD},C,E_2$ &DCS&$\Omega_{cc}^{+}\rightarrow\Sigma_{c}^{++}\rho^{-}$    &$(1.20_{-0.84}^{+1.38})*10^{-17}$    &$E_2$&DCS\\
   \hline
   $\Omega_{cc}^{+}\rightarrow\Sigma_{c}^{+}\rho^{0}$    &$(4.27_{-2.98}^{+4.87})*10^{-17}$    &$E_1,E_2$ &DCS&$\Omega_{cc}^{+}\rightarrow\Lambda_{c}^{+}\rho^{0}$    &$(4.10_{-2.87}^{+4.74})*10^{-17}$    &$E_1,E_2$&DCS\\
   \hline
   $\Omega_{cc}^{+}\rightarrow\Sigma_{c}^{+}\omega$    &$(1.74_{-1.23}^{+2.05})*10^{-17}$    &$E_1,E_2$ &DCS&$\Omega_{cc}^{+}\rightarrow\Lambda_{c}^{+}\omega$    &$(1.76_{-1.27}^{+2.23})*10^{-17}$    &$E_1,E_2$&DCS \\
   \hline
   $\Omega_{cc}^{+}\rightarrow\Sigma_{c}^{0}\rho^{+}$    &$(1.39_{-0.97}^{+1.62})*10^{-16}$    &$E_1$ &DCS&   &    &    & \\
   \hline
\end{tabular}
\end{table}
 %++++++++++++++++++++++++++++++++++++++++++++++++++++++++++++++++++++++++++++++++++++++++++++++++++++++++++++++++++++++

There are only three CKM favored $\Omega_{cc}^+\to {\cal B}_c V$ decays, all of which have large decay widths. In Ref. \cite{Cheng:2018mwu}, the life time of $\Omega_{cc}^+$ is evaluated with large ambiguity as $75\sim 180$ fs. Estimated with $\tau_{\Omega_{cc}}=75$ fs, branching fractions of the three CKM favored decays are given as
\begin{eqnarray}
{\cal BR}(\Omega_{cc}^{+}\rightarrow\Xi_{c}^{+}\overline{K}^{*0})\in [0.5\%,3.3\%]  ,\,\,\,
{\cal BR}(\Omega_{cc}^{+}\rightarrow\Xi_{c}^{'+}\overline{K}^{*0})\in [1.0\%,6.1\%]  ,\,\,\,
{\cal BR}(\Omega_{cc}^{+}\rightarrow\Omega_{c}^{0}\rho^{+}) \approx 10.0\%  .
\end{eqnarray}
Considering the reconstruction efficiency, $\Omega_{cc}^{+}\rightarrow\Omega_{c}^{0}\rho^{+}$ is not an ideal discovery decay for $\Omega_{cc}^+$, although it has a large branching fraction. Among these decays $\Omega_{cc}^{+}\rightarrow\Xi_{c}^{+}\overline{K}^{*0}$ can be a candidate of searching $\Omega_{cc}^+$ with $\bar K^{*0}$ reconstructed by $K^-\pi^+$.

Besides some decays induced purely by the $W$ exchange mechanism, there are also two pure $C^\prime$ type decays $\Omega_{cc}^{+}\rightarrow\Sigma_{c}^{+}\phi$ and $\Omega_{cc}^{+}\rightarrow\Lambda_{c}^{+}\phi$. Research on these decays will help to understand the importance of this type of contributions. However, they have tiny branching fractions at the order of $10^{-5}$ or $10^{-6}$ because of the doubly CKM suppression, which hints that it may not be easy to  observe them in experiments.

%=========================================================
\section{Summary}\label{sec:summary}
%=========================================================
Inspired by the discovery of $\Xi_{cc}^{++}$, recently doubly heavy baryons are being studied intensively. In purpose of studying the weak decays of heavy baryons and presenting some research channels for unfound doubly charm baryons, we investigate the two body nonleptonic decays of a spin-$1/2$ doubly charm baryon (${\cal B}_{cc}$) to a spin-$1/2$ singly charm baryon (${\cal B}_{c}$) and a light vector meson ($V$). These decays happen at quark level through the weak decays of a charm quark. It is usually thought that the scale of the charm quark is not large enough to activate the perturbative theory. Therefore the long distance contributions are expected to play an important role in these decays. In calculation we employ a phenomenological model used in Ref. \cite{Yu:2017zst}. Under this framework the short distance contributions are calculated under the factorization hypothesis and the long distance contributions are modelled as the final-state interactions (FSIs) which are calculated with the one-particle-exchange model.

The contributions in doubly charm baryon decays induced by the charged current of weak interactions can be classified into six types labelled by $T$, $C$, $C^\prime$, $E_{1,2}$, and $B$. Among the same CKM decay mode the $T$ topological diagrams are found to have the largest contribution, and $C$ diagrams contributes two or several times smaller than $T$. The other topological diagrams are relatively suppressed. It is found that the short distance factorizable contribution $T_{\rm SD}$ is dominating. As a result branching fractions with $T_{\rm SD}$ are not so sensitive to $\eta$, an phenomenological parameter in the form factor associated with the exchanged particle in FSIs to account for the off-shell effects and to render the whole calculation meaningful in the perturbative sense. However, branching fractions with $C_{\rm SD}$ or purely nonfactorizable contributions vary rapidly as $\eta$ changes.

The branching fractions of $\Xi_{cc}^{++} \to {\cal B}_{c} V$ decays are presented with $\tau_{\Xi_{cc}^{++}}=256\,{\rm fs}$. The three CKM favored decays $\Xi_{cc}^{++}\rightarrow\Sigma_{c}^{++}\bar K^{*0}$, $\Xi_{cc}^{++}\rightarrow\Xi_{c}^{+}\rho^{+}$ and $\Xi_{cc}^{++}\rightarrow\Xi_{c}^{'+}\rho^{+}$ have large branching fractions of $5\%$ to $20\%$. The branching fractions of singly CKM suppressed decays $\Xi_{cc}^{++}\rightarrow\Sigma_{c}^{+}\rho^{+}$ and $\Xi_{cc}^{++}\rightarrow\Lambda_{c}^{+}\rho^{+}$ are also possible to reach the percent level owing to the $T$ contributions.

The decay widths of $\Xi_{cc}^{+} \to {\cal B}_{c} V$ and $\Omega_{cc}^{+} \to {\cal B}_{c} V$ are given instead of branching fractions for lack of experimental data for $\tau_{\Xi_{cc}^{+}}$ and $\tau_{\Omega_{cc}^{+}}$. Calculated with $\tau_{\Xi_{cc}^+}=45\,{\rm fs}$, $\Xi_{cc}^{+}\rightarrow\Xi_{c}^{0}\rho^{+}$ and $\Xi_{cc}^{+}\rightarrow\Xi_{c}^{'0}\rho^{+}$ are evaluated to have the largest branching fraction of about $3\%$. However, they are not ideal candidates of discovery decays because of the low reconstruction efficiency of $\rho^+$ in experiments. Except these two decays, the $\Xi_{cc}^+\to\Xi_c^+\rho^0$ is expected to have the largest branching fraction, which is evaluated to range from $0.4\%$ to $2.5\%$. Considering $\rho^0$ can be reconstructed by the easily detected $\pi^+\pi^-$, $\Xi_{cc}^+\to\Xi_c^+\pi^+\pi^-$ can be proposed as the candidate for searching $\Xi_{cc}^+$.

In $\Omega_{cc}^{+} \to {\cal B}_{c} V$ mode the larges branching fractions belong to the three CKM favored decays, whose values are estimated to be ${\cal BR}(\Omega_{cc}^{+}\rightarrow\Xi_{c}^{+}\overline{K}^{*0})\in [0.5\%,3.3\%]$, ${\cal BR}(\Omega_{cc}^{+}\rightarrow\Xi_{c}^{'+}\overline{K}^{*0})\in [1.0\%,6.1\%]$  and ${\cal BR}(\Omega_{cc}^{+}\rightarrow\Omega_{c}^{0}\rho^{+}) \approx 10.0\%$ with $\tau_{\Omega_{cc}^+}=75\,{\rm fs}$. Taking the detection efficiency into consideration, we think $\Omega_{cc}^{+}\rightarrow\Xi_{c}^{+}K^-\pi^+$ is a favorable mode to search $\Omega_{cc}^+$.

Some pure $W$ exchange decays are founded, which are highly suppressed at short distance. These decays are thought to be activated almost by the long distance effects. Observation on these decays in the future experiments can help to understand the role of FSIs in charm baryon decays.

%====================================================================
\section{Acknowledgement}
%====================================================================
This work was supported in part by the National Natural Science Foundation of China under the Grant Nos. 11505098, 11765012 and 11791240174, and the plan of \emph{Young Creative Talents} under \emph{the Talent of Prairie} project of the Inner Mongolia. We thank Z.-R. Huang, H.-Y. Jiang, Y. Li, X. Liu, Y.-L. Shen, Z.-G. Wang, and F.-S. Yu for helpful discussions. We thank W. Wang and Z.-X. Zhao for reading the manuscript and giving their advices for modifications. We also thank the referee very much for the suggestions of discovery channels.
%====================================================================
\appendix
%====================================================================
\section{Expressions of Amplitudes}
\label{app:amps}
%====================================================================
The expressions of amplitudes for all the ${\cal B}_{cc}\to {\cal B}_{c}V$ decays are collected in the section. In order to make the expressions simpler, we define function ${\cal M}(P1,P2,P3,P4,P5,P6)$ as the absorptive part of a triangle diagram shown in Fig. \ref{fig:number}. The isospin factors of external particles are already included. The absorptive part in Eq. (\ref{eq:a2PXP}) is related to this function as
{\scriptsize
\begin{equation}
{\cal A}bs\,M_{a2}(\pi^+;\Xi_c^0;\pi^-)={\cal M}(\Xi_{cc}^{+}, \pi^+, \Xi_c^0, \pi^-, \rho^0,\Xi_c^{+}).
\end{equation}
}
Amplitudes of all ${\cal B}_{cc} \to {\cal B}_c V$ decays are given as follows with the help of this function.
{\scriptsize
\begin{eqnarray}
{\cal A}(\Xi_{cc}^{++} \to \Sigma_c^{++} \bar K^{*0})
&=&C_{\rm SD}(\Xi_{cc}^{++} \to \Sigma_c^{++} \bar K^{*0})+ i [
    {\cal M}(\Xi_{cc}^{++}, \pi^+, \Xi_c^+, K^-, \bar K^{*0},\Sigma_c^{++})
  + {\cal M}(\Xi_{cc}^{++}, \rho^+,\Xi_c^+, K^{*-}, \bar K^{*0},\Sigma_c^{++}) \nonumber\\
&&+ {\cal M}(\Xi_{cc}^{++}, \pi^+, \Xi_c^{\prime +}, K^-, \bar K^{*0},\Sigma_c^{++})
  + {\cal M}(\Xi_{cc}^{++}, \rho^+,\Xi_c^{\prime +}, K^{*-}, \bar K^{*0},\Sigma_c^{++})
  + {\cal M}(\Xi_{cc}^{++}, \pi^+, \Xi_c^{ +}, \Lambda_c^+,\Sigma_c^{++}, \bar K^{*0})\nonumber\\
&&+ {\cal M}(\Xi_{cc}^{++}, \rho^+,\Xi_c^{ +}, \Lambda_c^+,\Sigma_c^{++}, \bar K^{*0})
  + {\cal M}(\Xi_{cc}^{++}, \pi^+, \Xi_c^{ +}, \Sigma_c^+,\Sigma_c^{++}, \bar K^{*0})
  + {\cal M}(\Xi_{cc}^{++}, \rho^+,\Xi_c^{ +}, \Sigma_c^+,\Sigma_c^{++}, \bar K^{*0})\nonumber\\
&&+ {\cal M}(\Xi_{cc}^{++}, \pi^+, \Xi_c^{\prime +}, \Lambda_c^+,\Sigma_c^{++}, \bar K^{*0})
  + {\cal M}(\Xi_{cc}^{++}, \rho^+,\Xi_c^{\prime +}, \Lambda_c^+,\Sigma_c^{++}, \bar K^{*0})
  + {\cal M}(\Xi_{cc}^{++}, \pi^+, \Xi_c^{\prime +},\Sigma_c^+,\Sigma_c^{++},\bar K^{*0})\nonumber\\
&&+ {\cal M}(\Xi_{cc}^{++}, \rho^+,\Xi_c^{\prime +}, \Sigma_c^+,\Sigma_c^{++}, \bar K^{*0})],
\end{eqnarray}
\begin{eqnarray}
{\cal A}(\Xi_{cc}^{++} \to \Xi_c^+\rho^+)
&=&T_{\rm SD}(\Xi_{cc}^{++} \to \Xi_c^{+} \rho^+)+ i [
    {\cal M}(\Xi_{cc}^{++},  \pi^+,     \Xi_c^{ +},     \pi^0 ,     \rho^+,      \Xi_c^{+})
  + {\cal M}(\Xi_{cc}^{++},  \rho^+,    \Xi_c^{ +},     \rho^0 ,    \rho^+,      \Xi_c^{+})\nonumber\\
&&+ {\cal M}(\Xi_{cc}^{++},  \pi^+,     \Xi_c^{'+},     \pi^0 ,     \rho^+,      \Xi_c^{+})
  + {\cal M}(\Xi_{cc}^{++},  \rho^+,    \Xi_c^{'+},     \rho^0 ,    \rho^+,      \Xi_c^{+})
  + {\cal M}(\Xi_{cc}^{++},  \pi^+,     \Xi_c^{ +},     \Xi_c^{0},    \Xi_c^{+},   \rho^+)\nonumber\\
&&+ {\cal M}(\Xi_{cc}^{++},  \rho^+,    \Xi_c^{ +},     \Xi_c^{0},    \Xi_c^{+},   \rho^+)
  + {\cal M}(\Xi_{cc}^{++},  \pi^+,     \Xi_c^{ +},     \Xi_c^{'0},   \Xi_c^{+},   \rho^+)
  + {\cal M}(\Xi_{cc}^{++},  \rho^+,    \Xi_c^{ +},     \Xi_c^{'0},   \Xi_c^{+},   \rho^+)\nonumber\\
&&+ {\cal M}(\Xi_{cc}^{++},  \pi^+,     \Xi_c^{'+},     \Xi_c^{ 0},   \Xi_c^{+},   \rho^+)
  + {\cal M}(\Xi_{cc}^{++},  \rho^+,    \Xi_c^{'+},     \Xi_c^{ 0},   \Xi_c^{+},   \rho^+)
  + {\cal M}(\Xi_{cc}^{++},  \pi^+,      \Xi_c^{'+},     \Xi_c^{'0},   \Xi_c^{+},   \rho^+)\nonumber\\
&&+ {\cal M}(\Xi_{cc}^{++},  \rho^+,     \Xi_c^{'+},     \Xi_c^{'0},   \Xi_c^{+},   \rho^+)
  + {\cal M}(\Xi_{cc}^{++},  \bar K^0,   \Sigma_c^{++},   K^+,        \rho^+,      \Xi_c^{+})
  + {\cal M}(\Xi_{cc}^{++},  \bar K^{*0}, \Sigma_c^{++},   K^{*+},      \rho^+,      \Xi_c^{+})\nonumber\\
&&+ {\cal M}(\Xi_{cc}^{++},  \bar K^0,    \Sigma_c^{++},  \Lambda_c^+,  \Xi_c^{+},   \rho^+)
  + {\cal M}(\Xi_{cc}^{++},  \bar K^{*0}, \Sigma_c^{++},  \Lambda_c^+,  \Xi_c^{+},   \rho^+)
  + {\cal M}(\Xi_{cc}^{++},  \bar K^0,    \Sigma_c^{++},  \Sigma_c^{+},  \Xi_c^{+},   \rho^+)\nonumber\\
&&+ {\cal M}(\Xi_{cc}^{++},  \bar K^{*0}, \Sigma_c^{++},  \Sigma_c^{+},  \Xi_c^{+},   \rho^+)],
\end{eqnarray}
\begin{eqnarray}
{\cal A}(\Xi_{cc}^{++} \to \Xi_c^{'+} \rho^+)
&=&T_{\rm SD}(\Xi_{cc}^{++} \to \Xi_c^{'+} \rho^+)+ i [
    {\cal M}(\Xi_{cc}^{++},  \pi^+,     \Xi_c^{ +},     \pi^0 ,     \rho^+,      \Xi_c^{'+})
  + {\cal M}(\Xi_{cc}^{++},  \rho^+,    \Xi_c^{ +},     \rho^0 ,    \rho^+,      \Xi_c^{'+})\nonumber\\
&&+ {\cal M}(\Xi_{cc}^{++},  \pi^+,     \Xi_c^{'+},     \pi^0 ,     \rho^+,      \Xi_c^{'+})
  + {\cal M}(\Xi_{cc}^{++},  \rho^+,    \Xi_c^{'+},     \rho^0 ,    \rho^+,      \Xi_c^{'+})
  + {\cal M}(\Xi_{cc}^{++},  \pi^+,     \Xi_c^{ +},     \Xi_c^{0},    \Xi_c^{'+},   \rho^+)\nonumber\\
&&+ {\cal M}(\Xi_{cc}^{++},  \rho^+,    \Xi_c^{ +},     \Xi_c^{0},    \Xi_c^{'+},   \rho^+)
  + {\cal M}(\Xi_{cc}^{++},  \pi^+,     \Xi_c^{ +},     \Xi_c^{'0},   \Xi_c^{'+},   \rho^+)
  + {\cal M}(\Xi_{cc}^{++},  \rho^+,    \Xi_c^{ +},     \Xi_c^{'0},   \Xi_c^{'+},   \rho^+)\nonumber\\
&&+ {\cal M}(\Xi_{cc}^{++},  \pi^+,     \Xi_c^{'+},     \Xi_c^{ 0},   \Xi_c^{'+},   \rho^+)
  + {\cal M}(\Xi_{cc}^{++},  \rho^+,    \Xi_c^{'+},     \Xi_c^{ 0},   \Xi_c^{'+},   \rho^+)
  + {\cal M}(\Xi_{cc}^{++},  \pi^+,      \Xi_c^{'+},     \Xi_c^{'0},   \Xi_c^{'+},   \rho^+)\nonumber\\
&&+ {\cal M}(\Xi_{cc}^{++},  \rho^+,     \Xi_c^{'+},     \Xi_c^{'0},   \Xi_c^{'+},   \rho^+)
  + {\cal M}(\Xi_{cc}^{++},  \bar K^0,   \Sigma_c^{++},   K^+,         \rho^+,      \Xi_c^{'+})
  + {\cal M}(\Xi_{cc}^{++},  \bar K^{*0}, \Sigma_c^{++},  K^{*+},      \rho^+,      \Xi_c^{'+})\nonumber\\
&&+ {\cal M}(\Xi_{cc}^{++},  \bar K^0,    \Sigma_c^{++},  \Lambda_c^+,  \Xi_c^{'+},   \rho^+)
  + {\cal M}(\Xi_{cc}^{++},  \bar K^{*0}, \Sigma_c^{++},  \Lambda_c^+,  \Xi_c^{'+},   \rho^+)
  + {\cal M}(\Xi_{cc}^{++},  \bar K^0,    \Sigma_c^{++},  \Sigma_c^{+},  \Xi_c^{'+},   \rho^+)\nonumber\\
&&+ {\cal M}(\Xi_{cc}^{++},  \bar K^{*0}, \Sigma_c^{++},  \Sigma_c^{+},  \Xi_c^{'+},   \rho^+)],
\end{eqnarray}
\begin{eqnarray}
{\cal A}(\Xi_{cc}^{++} \to \Sigma_c^{+} \rho^+)
&=&T_{\rm SD}(\Xi_{cc}^{++} \to \Sigma_c^{+} \rho^+)+ i [
    {\cal M}(\Xi_{cc}^{++},  \pi^+,     \Lambda_c^+,     \pi^0 ,       \rho^+,      \Sigma_c^{+})
  + {\cal M}(\Xi_{cc}^{++},  \rho^+,    \Lambda_c^+,     \rho^0 ,      \rho^+,      \Sigma_c^{+})\nonumber\\
&&+ {\cal M}(\Xi_{cc}^{++},  \pi^+,     \Lambda_c^+,     \Sigma_c^{0}, \Sigma_c^{+},   \rho^+)
  + {\cal M}(\Xi_{cc}^{++},  \rho^+,    \Lambda_c^+,     \Sigma_c^{0}, \Sigma_c^{+},   \rho^+)
  + {\cal M}(\Xi_{cc}^{++},  \pi^+,     \Sigma_c^+,      \Sigma_c^{0}, \Sigma_c^{+},   \rho^+)\nonumber\\
&&+ {\cal M}(\Xi_{cc}^{++},  \rho^+,    \Sigma_c^{+},    \Sigma_c^{0}, \Sigma_c^{+},   \rho^+)
  + {\cal M}(\Xi_{cc}^{++},  \pi^0 ,    \Sigma_c^{++},   \pi^+,        \rho^+,      \Sigma_c^{+})
  + {\cal M}(\Xi_{cc}^{++},  \rho^0 ,   \Sigma_c^{++},   \rho^+,       \rho^+,      \Sigma_c^{+})\nonumber\\
&&+ {\cal M}(\Xi_{cc}^{++},  \pi^0 ,    \Sigma_c^{++},  \Lambda_c^+,  \Sigma_c^{+},   \rho^+)
  + {\cal M}(\Xi_{cc}^{++},  \rho^0 ,   \Sigma_c^{++},  \Lambda_c^+,  \Sigma_c^{+},   \rho^+)
  + {\cal M}(\Xi_{cc}^{++},  \eta_8 ,     \Sigma_c^{++},  \Sigma_c^+,   \Sigma_c^{+},   \rho^+)\nonumber\\
&&+ {\cal M}(\Xi_{cc}^{++},  \omega ,   \Sigma_c^{++},  \Sigma_c^+,   \Sigma_c^{+},   \rho^+)
  + {\cal M}(\Xi_{cc}^{++},   K^+,        \Xi_c^+,       \bar K^0,     \rho^+,      \Sigma_c^{+})
  + {\cal M}(\Xi_{cc}^{++},   K^{*+},     \Xi_c^+,       \bar K^{*0},  \rho^+,      \Sigma_c^{+})\nonumber\\
&&+ {\cal M}(\Xi_{cc}^{++},   K^{ +},     \Xi_c^{'+},    \bar K^  0 ,  \rho^+,      \Sigma_c^{+})
  + {\cal M}(\Xi_{cc}^{++},   K^{*+},     \Xi_c^{'+},    \bar K^{*0},  \rho^+,      \Sigma_c^{+})
  + {\cal M}(\Xi_{cc}^{++},   K^{ +},     \Xi_c^{ +},    \Xi_c^{ 0},   \Sigma_c^{+},   \rho^+)\nonumber\\
&&+ {\cal M}(\Xi_{cc}^{++},   K^{*+},     \Xi_c^{ +},    \Xi_c^{ 0},   \Sigma_c^{+},   \rho^+)
  + {\cal M}(\Xi_{cc}^{++},   K^+,        \Xi_c^{ +},    \Xi_c^{'0},   \Sigma_c^{+},   \rho^+)
  + {\cal M}(\Xi_{cc}^{++},   K^{*+},     \Xi_c^{ +},    \Xi_c^{'0},   \Sigma_c^{+},   \rho^+)\nonumber\\
&&+ {\cal M}(\Xi_{cc}^{++},   K^{ +},     \Xi_c^{'+},    \Xi_c^{ 0},   \Sigma_c^{+},   \rho^+)
  + {\cal M}(\Xi_{cc}^{++},   K^{*+},     \Xi_c^{'+},    \Xi_c^{ 0},   \Sigma_c^{+},   \rho^+)
  + {\cal M}(\Xi_{cc}^{++},   K^{ +},     \Xi_c^{'+},    \Xi_c^{'0},   \Sigma_c^{+},   \rho^+)\nonumber\\
&&+ {\cal M}(\Xi_{cc}^{++},   K^{*+},     \Xi_c^{'+},    \Xi_c^{'0},   \Sigma_c^{+},   \rho^+)
  + {\cal M}(\Xi_{cc}^{++},  \eta_1 ,     \Sigma_c^{++},  \Sigma_c^+,   \Sigma_c^{+},   \rho^+)],
\end{eqnarray}
\begin{eqnarray}
{\cal A}(\Xi_{cc}^{++} \to \Lambda_c^{+} \rho^+)
&=&T_{\rm SD}(\Xi_{cc}^{++} \to \Lambda_c^{+} \rho^+)+ i [
    {\cal M}(\Xi_{cc}^{++},  \pi^+,     \Sigma_c^+,      \pi^0 ,     \rho^+,     \Lambda_c^{+})
  + {\cal M}(\Xi_{cc}^{++},  \rho^+,    \Sigma_c^+,      \rho^0 ,    \rho^+,     \Lambda_c^{+})\nonumber\\
&&+ {\cal M}(\Xi_{cc}^{++},  \pi^+,     \Lambda_c^+,     \Sigma_c^{0}, \Lambda_c^{+},   \rho^+)
  + {\cal M}(\Xi_{cc}^{++},  \rho^+,    \Lambda_c^+,     \Sigma_c^{0}, \Lambda_c^{+},   \rho^+)
  + {\cal M}(\Xi_{cc}^{++},  \pi^+,     \Sigma_c^+,      \Sigma_c^{0}, \Lambda_c^{+},   \rho^+)\nonumber\\
&&+ {\cal M}(\Xi_{cc}^{++},  \rho^+,    \Sigma_c^+,      \Sigma_c^{0}, \Lambda_c^{+},   \rho^+)
  + {\cal M}(\Xi_{cc}^{++},  \pi^0 ,  \Sigma_c^{++},   \pi^+,        \rho^+,     \Lambda_c^{+})
  + {\cal M}(\Xi_{cc}^{++},  \rho^0 , \Sigma_c^{++},   \rho^+,       \rho^+,     \Lambda_c^{+})\nonumber\\
&&+ {\cal M}(\Xi_{cc}^{++},  \pi^0 ,  \Sigma_c^{++},   \Sigma_c^{+}, \Lambda_c^{+},   \rho^+)
  + {\cal M}(\Xi_{cc}^{++},  \rho^0 , \Sigma_c^{++},   \Sigma_c^{+}, \Lambda_c^{+},   \rho^+)
  + {\cal M}(\Xi_{cc}^{++},  \eta_8 ,   \Sigma_c^{++},   \Lambda_c^{+},\Lambda_c^{+},   \rho^+)\nonumber\\
&&+ {\cal M}(\Xi_{cc}^{++},  \omega , \Sigma_c^{++},   \Lambda_c^{+},\Lambda_c^{+},   \rho^+)
  + {\cal M}(\Xi_{cc}^{++},   K^+,      \Xi_c^{+},       \bar K^0,     \rho^+,     \Lambda_c^{+})
  + {\cal M}(\Xi_{cc}^{++},   K^{*+},   \Xi_c^{+},       \bar K^{*0},  \rho^+,     \Lambda_c^{+})\nonumber\\
&&+ {\cal M}(\Xi_{cc}^{++},   K^+,      \Xi_c^{'+},      \bar K^0,     \rho^+,     \Lambda_c^{+})
  + {\cal M}(\Xi_{cc}^{++},   K^{*+},   \Xi_c^{'+},      \bar K^{*0},  \rho^+,     \Lambda_c^{+})
  + {\cal M}(\Xi_{cc}^{++},   K^+,      \Xi_c^{ +},      \Xi_c^{0},    \Lambda_c^{+},   \rho^+)\nonumber\\
&&+ {\cal M}(\Xi_{cc}^{++},   K^{*+},   \Xi_c^{ +},      \Xi_c^{0},    \Lambda_c^{+},   \rho^+)
  + {\cal M}(\Xi_{cc}^{++},   K^+,      \Xi_c^{ +},      \Xi_c^{'0},   \Lambda_c^{+},   \rho^+)
  + {\cal M}(\Xi_{cc}^{++},   K^{*+},   \Xi_c^{ +},      \Xi_c^{'0},   \Lambda_c^{+},   \rho^+)\nonumber\\
&&+ {\cal M}(\Xi_{cc}^{++},   K^+,      \Xi_c^{'+},      \Xi_c^{ 0},   \Lambda_c^{+},   \rho^+)
  + {\cal M}(\Xi_{cc}^{++},   K^{*+},   \Xi_c^{'+},      \Xi_c^{ 0},   \Lambda_c^{+},   \rho^+)
  + {\cal M}(\Xi_{cc}^{++},   K^+,      \Xi_c^{'+},      \Xi_c^{'0},   \Lambda_c^{+},   \rho^+)\nonumber\\
&&+ {\cal M}(\Xi_{cc}^{++},   K^{*+},   \Xi_c^{'+},      \Xi_c^{'0},   \Lambda_c^{+},   \rho^+)
  + {\cal M}(\Xi_{cc}^{++},  \eta_1 ,   \Sigma_c^{++},   \Lambda_c^{+},\Lambda_c^{+},   \rho^+)],
\end{eqnarray}
\begin{eqnarray}
{\cal A}(\Xi_{cc}^{++} \to \Sigma_c^{++} \rho^{0})
&=&C_{\rm SD}(\Xi_{cc}^{++} \to \Sigma_c^{++} \rho^{0+} )+ i [
    {\cal M}(\Xi_{cc}^{++},  \pi^+,     \Lambda_c^+,   \pi^-,        \rho^0 ,    \Sigma_c^{++})
  + {\cal M}(\Xi_{cc}^{++},  \rho^+,    \Lambda_c^+,   \rho^-,       \rho^0 ,    \Sigma_c^{++})\nonumber\\
&&+ {\cal M}(\Xi_{cc}^{++},  \pi^+,     \Sigma_c^{+},  \pi^-,        \rho^0 ,    \Sigma_c^{++})
  + {\cal M}(\Xi_{cc}^{++},  \rho^+,    \Sigma_c^{+},  \rho^-,       \rho^0 ,    \Sigma_c^{++})
  + {\cal M}(\Xi_{cc}^{++},  \pi^+,     \Lambda_c^{+}, \Sigma_c^{+}, \Sigma_c^{++},\rho^0 )\nonumber\\
&&+ {\cal M}(\Xi_{cc}^{++},  \rho^+,    \Lambda_c^{+}, \Sigma_c^{+}, \Sigma_c^{++},\rho^0 )
  + {\cal M}(\Xi_{cc}^{++},  \pi^+,     \Sigma_c^{+} , \Lambda_c^{+},\Sigma_c^{++},\rho^0 )
  + {\cal M}(\Xi_{cc}^{++},  \rho^+,    \Sigma_c^{+} , \Lambda_c^{+},\Sigma_c^{++},\rho^0 ) \nonumber\\
&&+ {\cal M}(\Xi_{cc}^{++},   K^+,      \Xi_c^{+},      K^-,         \rho^0 ,    \Sigma_c^{++})
  + {\cal M}(\Xi_{cc}^{++},   K^{*+},   \Xi_c^{+},      K^{*-},      \rho^0 ,    \Sigma_c^{++})
  + {\cal M}(\Xi_{cc}^{++},   K^+,      \Xi_c^{'+},     K^-,         \rho^0 ,    \Sigma_c^{++})\nonumber\\
&&+ {\cal M}(\Xi_{cc}^{++},   K^{*+},   \Xi_c^{'+},     K^{*-},      \rho^0 ,    \Sigma_c^{++})
  + {\cal M}(\Xi_{cc}^{++},   K^+ ,     \Xi_c^{+},     \Xi_c^{+},    \Sigma_c^{++},   \rho^0 )
  + {\cal M}(\Xi_{cc}^{++},   K^{*+},   \Xi_c^{+},     \Xi_c^{+},    \Sigma_c^{++},   \rho^0 )\nonumber\\
&&+ {\cal M}(\Xi_{cc}^{++},   K^+ ,     \Xi_c^{'+},    \Xi_c^{+},    \Sigma_c^{++},   \rho^0 )
  + {\cal M}(\Xi_{cc}^{++},   K^{*+} ,  \Xi_c^{'+},    \Xi_c^{+},    \Sigma_c^{++},   \rho^0 )
  + {\cal M}(\Xi_{cc}^{++},   K^+ ,     \Xi_c^{+},     \Xi_c^{'+},   \Sigma_c^{++},   \rho^0 )\nonumber\\
&&+ {\cal M}(\Xi_{cc}^{++},   K^{*+},   \Xi_c^{+},     \Xi_c^{'+},   \Sigma_c^{++},   \rho^0 )
  + {\cal M}(\Xi_{cc}^{++},   K^+ ,     \Xi_c^{'+},    \Xi_c^{'+},   \Sigma_c^{++},   \rho^0 )
  + {\cal M}(\Xi_{cc}^{++},   K^{*+} ,  \Xi_c^{'+},    \Xi_c^{'+},   \Sigma_c^{++},   \rho^0 )],
\end{eqnarray}
\begin{eqnarray}
{\cal A}(\Xi_{cc}^{++} \to \Sigma_c^{++} \omega)
&=&C_{\rm SD}(\Xi_{cc}^{++} \to \Sigma_c^{++} \omega)  + i [
    {\cal M}(\Xi_{cc}^{++},  \pi^+,     \Lambda_c^+,    \Lambda_c^+, \Sigma_c^{++},  \omega )
  + {\cal M}(\Xi_{cc}^{++},  \rho^+,    \Lambda_c^+,    \Lambda_c^+, \Sigma_c^{++},  \omega )\nonumber\\
&&+ {\cal M}(\Xi_{cc}^{++},  \pi^+,     \Sigma_c^+,     \Sigma_c^+,  \Sigma_c^{++},  \omega )
  + {\cal M}(\Xi_{cc}^{++},  \rho^+,    \Sigma_c^+,     \Sigma_c^+,  \Sigma_c^{++},  \omega )
  + {\cal M}(\Xi_{cc}^{++},   K^+,      \Xi_c^+,          K^-,       \omega ,    \Sigma_c^{++})\nonumber\\
&&+ {\cal M}(\Xi_{cc}^{++},   K^{*+},   \Xi_c^+,          K^{*-},    \omega ,    \Sigma_c^{++})
  + {\cal M}(\Xi_{cc}^{++},   K^+,      \Xi_c^{'+},       K^-,       \omega ,    \Sigma_c^{++})
  + {\cal M}(\Xi_{cc}^{++},   K^{*+},   \Xi_c^{'+},       K^{*-},    \omega ,    \Sigma_c^{++})\nonumber\\
&&+ {\cal M}(\Xi_{cc}^{++},   K^+,      \Xi_c^+ ,       \Xi_c{0},    \Sigma_c^{++},  \omega )
  + {\cal M}(\Xi_{cc}^{++},   K^{*+},   \Xi_c^+,        \Xi_c{0},    \Sigma_c^{++},  \omega )
  + {\cal M}(\Xi_{cc}^{++},   K^+,      \Xi_c^{'+},     \Xi_c{0},    \Sigma_c^{++},  \omega )\nonumber\\
&&+ {\cal M}(\Xi_{cc}^{++},   K^{*+},   \Xi_c^{'+},     \Xi_c{0},    \Sigma_c^{++},  \omega )
  + {\cal M}(\Xi_{cc}^{++},   K^+,      \Xi_c^+,        \Xi_c{'0},   \Sigma_c^{++},  \omega )
  + {\cal M}(\Xi_{cc}^{++},   K^{*+},   \Xi_c^+,        \Xi_c{'0},   \Sigma_c^{++},  \omega )\nonumber\\
&&+ {\cal M}(\Xi_{cc}^{++},   K^+,      \Xi_c^{'+},     \Xi_c{'0},   \Sigma_c^{++},  \omega )
  + {\cal M}(\Xi_{cc}^{++},   K^{*+},   \Xi_c^{'+},     \Xi_c{'0},   \Sigma_c^{++},  \omega )],
\end{eqnarray}
\begin{eqnarray}
{\cal A}(\Xi_{cc}^{++} \to \Sigma_c^{++} \phi)
&=&C_{\rm SD}(\Xi_{cc}^{++} \to \Sigma_c^{++} \phi )+ i [
    {\cal M}(\Xi_{cc}^{++},   K^+,      \Xi_c^+,         K^-,        \phi,         \Sigma_c^{++})
  + {\cal M}(\Xi_{cc}^{++},   K^{*+},   \Xi_c^+,         K^{*-},     \phi,         \Sigma_c^{++})\nonumber\\
&&+ {\cal M}(\Xi_{cc}^{++},   K^+,      \Xi_c^{'+},      K^-,        \phi,         \Sigma_c^{++})
  + {\cal M}(\Xi_{cc}^{++},   K^{*+},   \Xi_c^{'+},      K^{*-},     \phi,         \Sigma_c^{++})
  + {\cal M}(\Xi_{cc}^{++},   K^+,      \Xi_c^+,        \Xi_c^+ ,    \Sigma_c^{++},    \phi)\nonumber\\
&&+ {\cal M}(\Xi_{cc}^{++},   K^{*+},   \Xi_c^+,        \Xi_c^+ ,    \Sigma_c^{++},    \phi)
  + {\cal M}(\Xi_{cc}^{++},   K^+,      \Xi_c^+,        \Xi_c^{'+},  \Sigma_c^{++},    \phi)
  + {\cal M}(\Xi_{cc}^{++},   K^{*+},   \Xi_c^+,        \Xi_c^{'+},  \Sigma_c^{++},    \phi)\nonumber\\
&&+ {\cal M}(\Xi_{cc}^{++},   K^+,      \Xi_c^{'+},     \Xi_c^+ ,    \Sigma_c^{++},    \phi)
  + {\cal M}(\Xi_{cc}^{++},   K^{*+},   \Xi_c^{'+},     \Xi_c^+ ,    \Sigma_c^{++},    \phi)
  + {\cal M}(\Xi_{cc}^{++},   K^+,      \Xi_c^{'+},     \Xi_c^{'+},  \Sigma_c^{++},    \phi)\nonumber\\
&&+ {\cal M}(\Xi_{cc}^{++},   K^{*+},   \Xi_c^{'+},     \Xi_c^{'+},  \Sigma_c^{++},    \phi)],
\end{eqnarray}
\begin{eqnarray}
{\cal A}(\Xi_{cc}^{++} \to \Xi_c^+ K^{*+})
&=&T_{\rm SD}(\Xi_{cc}^{++} \to \Xi_c^+ K^{*+}) + i [
    {\cal M}(\Xi_{cc}^{++},   K^+,      \Xi_c^+,        \pi^0 ,     K^{*+},       \Xi_c^+)
  + {\cal M}(\Xi_{cc}^{++},   K^{*+},   \Xi_c^+,        \rho^0 ,    K^{*+},       \Xi_c^+)\nonumber\\
&&+ {\cal M}(\Xi_{cc}^{++},   K^+,      \Xi_c^+,        \eta_8 ,      K^{*+},       \Xi_c^+)
  + {\cal M}(\Xi_{cc}^{++},   K^{*+},   \Xi_c^+,        \omega ,    K^{*+},       \Xi_c^+)
  + {\cal M}(\Xi_{cc}^{++},   K^+,      \Xi_c^{'+},     \pi^0 ,     K^{*+},       \Xi_c^+)\nonumber\\
&&+ {\cal M}(\Xi_{cc}^{++},   K^{*+},   \Xi_c^{'+},     \rho^0 ,    K^{*+},       \Xi_c^+)
  + {\cal M}(\Xi_{cc}^{++},   K^+,      \Xi_c^{'+},     \eta_8 ,      K^{*+},       \Xi_c^+)
  + {\cal M}(\Xi_{cc}^{++},   K^{*+},   \Xi_c^{'+},     \omega ,    K^{*+},       \Xi_c^+)\nonumber\\
&&+ {\cal M}(\Xi_{cc}^{++},   K^+,      \Xi_c^+,        \Omega_c,     \Xi_c^+,       K^{*+})
  + {\cal M}(\Xi_{cc}^{++},   K^{*+},   \Xi_c^+,        \Omega_c,     \Xi_c^+,       K^{*+})
  + {\cal M}(\Xi_{cc}^{++},   K^+,      \Xi_c^{'+},     \Omega_c,     \Xi_c^+,       K^{*+})\nonumber\\
&&+ {\cal M}(\Xi_{cc}^{++},   K^{*+},   \Xi_c^{'+},     \Omega_c,     \Xi_c^+,       K^{*+})
  + {\cal M}(\Xi_{cc}^{++},   \eta_8 ,    \Xi_c^{++},       K^+,        K^{*+},       \Xi_c^+)
  + {\cal M}(\Xi_{cc}^{++},   \phi,     \Xi_c^{++},       K^{*+},     K^{*+},       \Xi_c^+)\nonumber\\
&&+ {\cal M}(\Xi_{cc}^{++},   \eta_8,    \Xi_c^{++},     \Xi_c^+,      \Xi_c^+,       K^{*+})
  + {\cal M}(\Xi_{cc}^{++},   \phi,     \Xi_c^{++},     \Xi_c^+,      \Xi_c^+,       K^{*+})
  + {\cal M}(\Xi_{cc}^{++},   \eta_8 ,    \Xi_c^{++},     \Xi_c^{'+},   \Xi_c^+,       K^{*+})\nonumber\\
&&+ {\cal M}(\Xi_{cc}^{++},   \phi,     \Xi_c^{++},     \Xi_c^{'+},   \Xi_c^+,       K^{*+})
  + {\cal M}(\Xi_{cc}^{++},   \pi^+,    \Lambda_c^+,      K^{ 0},     K^{*+},       \Xi_c^+)
  + {\cal M}(\Xi_{cc}^{++},   \rho^+,   \Lambda_c^+,      K^{*0},     K^{*+},       \Xi_c^+)\nonumber\\
&&+ {\cal M}(\Xi_{cc}^{++},   \pi^+,    \Sigma_c^+,       K^{ 0},     K^{*+},       \Xi_c^+)
  + {\cal M}(\Xi_{cc}^{++},   \rho^+,   \Sigma_c^+,       K^{*0},     K^{*+},       \Xi_c^+)
  + {\cal M}(\Xi_{cc}^{++},   \pi^+,    \Lambda_c^+,    \Xi_c^{ 0},   \Xi_c^+,       K^{*+})\nonumber\\
&&+ {\cal M}(\Xi_{cc}^{++},   \rho^+,   \Lambda_c^+,    \Xi_c^{ 0},   \Xi_c^+,       K^{*+})
  + {\cal M}(\Xi_{cc}^{++},   \pi^+,    \Sigma_c^+,     \Xi_c^{ 0},   \Xi_c^+,       K^{*+})
  + {\cal M}(\Xi_{cc}^{++},   \rho^+,   \Sigma_c^+,     \Xi_c^{ 0},   \Xi_c^+,       K^{*+})\nonumber\\
&&+ {\cal M}(\Xi_{cc}^{++},   \pi^+,    \Lambda_c^+,    \Xi_c^{'0},   \Xi_c^+,       K^{*+})
  + {\cal M}(\Xi_{cc}^{++},   \rho^+,   \Lambda_c^+,    \Xi_c^{'0},   \Xi_c^+,       K^{*+})
  + {\cal M}(\Xi_{cc}^{++},   \pi^+,    \Sigma_c^+,     \Xi_c^{'0},   \Xi_c^+,       K^{*+})\nonumber\\
&&+ {\cal M}(\Xi_{cc}^{++},   \rho^+,   \Sigma_c^+,     \Xi_c^{'0},   \Xi_c^+,       K^{*+})
  + {\cal M}(\Xi_{cc}^{++},    K^+,     \Xi_c^+,        \eta_8 ,        K^{*+},       \Xi_c^+)
  + {\cal M}(\Xi_{cc}^{++},    K^{*+},  \Xi_c^+,        \phi,         K^{*+},       \Xi_c^+)\nonumber\\
&&+ {\cal M}(\Xi_{cc}^{++},    K^+,     \Xi_c^{'+},     \eta_8 ,        K^{*+},       \Xi_c^+)
  + {\cal M}(\Xi_{cc}^{++},    K^{*+},  \Xi_c^{'+},     \phi,         K^{*+},       \Xi_c^+)
  + {\cal M}(\Xi_{cc}^{++},   \eta_1,    \Sigma_c^{++},     \Xi_c^+,      \Xi_c^+,       K^{*+})],
\end{eqnarray}
\begin{eqnarray}
{\cal A}(\Xi_{cc}^{++} \to \Xi_c^{'+} K^{*+})
&=&T_{\rm SD}(\Xi_{cc}^{++} \to \Xi_c^{'+} K^{*+}) + i [
    {\cal M}(\Xi_{cc}^{++},   K^+,      \Xi_c^+,        \pi^0   ,     K^{*+},       \Xi_c^{'+})
  + {\cal M}(\Xi_{cc}^{++},   K^{*+},   \Xi_c^+,        \rho^0   ,    K^{*+},       \Xi_c^{'+})\nonumber\\
&&+ {\cal M}(\Xi_{cc}^{++},   K^+,      \Xi_c^+,        \eta_8   ,      K^{*+},       \Xi_c^{'+})
  + {\cal M}(\Xi_{cc}^{++},   K^{*+},   \Xi_c^+,        \omega   ,    K^{*+},       \Xi_c^{'+})
  + {\cal M}(\Xi_{cc}^{++},   K^+,      \Xi_c^{'+},     \pi^0   ,     K^{*+},       \Xi_c^{'+})\nonumber\\
&&+ {\cal M}(\Xi_{cc}^{++},   K^{*+},   \Xi_c^{'+},     \rho^0   ,    K^{*+},       \Xi_c^{'+})
  + {\cal M}(\Xi_{cc}^{++},   K^+,      \Xi_c^{'+},     \eta_8   ,      K^{*+},       \Xi_c^{'+})
  + {\cal M}(\Xi_{cc}^{++},   K^{*+},   \Xi_c^{'+},     \omega   ,    K^{*+},       \Xi_c^{'+})\nonumber\\
&&+ {\cal M}(\Xi_{cc}^{++},   K^+,      \Xi_c^+,        \Omega_c,     \Xi_c^{'+},       K^{*+})
  + {\cal M}(\Xi_{cc}^{++},   K^{*+},   \Xi_c^+,        \Omega_c,     \Xi_c^{'+},       K^{*+})
  + {\cal M}(\Xi_{cc}^{++},   K^+,      \Xi_c^{'+},     \Omega_c,     \Xi_c^{'+},       K^{*+})\nonumber\\
&&+ {\cal M}(\Xi_{cc}^{++},   K^{*+},   \Xi_c^{'+},     \Omega_c,     \Xi_c^{'+},       K^{*+})
  + {\cal M}(\Xi_{cc}^{++},   \eta_8   ,  \Xi_c^{++},       K^+,        K^{*+},       \Xi_c^{'+})
  + {\cal M}(\Xi_{cc}^{++},   \phi,     \Xi_c^{++},       K^{*+},     K^{*+},       \Xi_c^{'+})\nonumber\\
&&+ {\cal M}(\Xi_{cc}^{++},   \eta_8   ,  \Xi_c^{++},     \Xi_c^+,      \Xi_c^{'+},       K^{*+})
  + {\cal M}(\Xi_{cc}^{++},   \phi,     \Xi_c^{++},     \Xi_c^+,      \Xi_c^{'+},       K^{*+})
  + {\cal M}(\Xi_{cc}^{++},   \eta_8   ,  \Xi_c^{++},     \Xi_c^{'+},   \Xi_c^{'+},       K^{*+})\nonumber\\
&&+ {\cal M}(\Xi_{cc}^{++},   \phi,     \Xi_c^{++},     \Xi_c^{'+},   \Xi_c^{'+},       K^{*+})
  + {\cal M}(\Xi_{cc}^{++},   \pi^+,    \Lambda_c^+,      K^{ 0},     K^{*+},       \Xi_c^{'+})
  + {\cal M}(\Xi_{cc}^{++},   \rho^+,   \Lambda_c^+,      K^{*0},     K^{*+},       \Xi_c^{'+})\nonumber\\
&&+ {\cal M}(\Xi_{cc}^{++},   \pi^+,    \Sigma_c^+,       K^{ 0},     K^{*+},       \Xi_c^{'+})
  + {\cal M}(\Xi_{cc}^{++},   \rho^+,   \Sigma_c^+,       K^{*0},     K^{*+},       \Xi_c^{'+})
  + {\cal M}(\Xi_{cc}^{++},   \pi^+,    \Lambda_c^+,    \Xi_c^{ 0},   \Xi_c^{'+},       K^{*+})\nonumber\\
&&+ {\cal M}(\Xi_{cc}^{++},   \rho^+,   \Lambda_c^+,    \Xi_c^{ 0},   \Xi_c^{'+},       K^{*+})
  + {\cal M}(\Xi_{cc}^{++},   \pi^+,    \Sigma_c^+,     \Xi_c^{ 0},   \Xi_c^{'+},       K^{*+})
  + {\cal M}(\Xi_{cc}^{++},   \rho^+,   \Sigma_c^+,     \Xi_c^{ 0},   \Xi_c^{'+},       K^{*+})\nonumber\\
&&+ {\cal M}(\Xi_{cc}^{++},   \pi^+,    \Lambda_c^+,    \Xi_c^{'0},   \Xi_c^{'+},       K^{*+})
  + {\cal M}(\Xi_{cc}^{++},   \rho^+,   \Lambda_c^+,    \Xi_c^{'0},   \Xi_c^{'+},       K^{*+})
  + {\cal M}(\Xi_{cc}^{++},   \pi^+,    \Sigma_c^+,     \Xi_c^{'0},   \Xi_c^{'+},       K^{*+})\nonumber\\
&&+ {\cal M}(\Xi_{cc}^{++},   \rho^+,   \Sigma_c^+,     \Xi_c^{'0},   \Xi_c^{'+},       K^{*+})
  + {\cal M}(\Xi_{cc}^{++},    K^+,     \Xi_c^+,        \eta_8   ,      K^{*+},       \Xi_c^{'+})
  + {\cal M}(\Xi_{cc}^{++},    K^{*+},  \Xi_c^+,        \phi,         K^{*+},       \Xi_c^{'+})\nonumber\\
&&+ {\cal M}(\Xi_{cc}^{++},    K^+,     \Xi_c^{'+},     \eta_8   ,      K^{*+},       \Xi_c^{'+})
  + {\cal M}(\Xi_{cc}^{++},    K^{*+},  \Xi_c^{'+},     \phi,         K^{*+},       \Xi_c^{'+})
  + {\cal M}(\Xi_{cc}^{++},   \eta_1   ,  \Sigma_c^{++},     \Xi_c^{'+},   \Xi_c^{'+},       K^{*+})],
\end{eqnarray}
\begin{eqnarray}
{\cal A}(\Xi_{cc}^{++} \to \Sigma_c^{ +} K^{*+})
&=&T_{\rm SD}(\Xi_{cc}^{++} \to \Xi_c^{'+} K^{*+}) + i [
    {\cal M}(\Xi_{cc}^{++},   K^+,      \Lambda_c^+,     \pi^0   ,     K^{*+},      \Sigma_c^{+})
  + {\cal M}(\Xi_{cc}^{++},   K^{*+},   \Lambda_c^+,     \rho^0   ,    K^{*+},      \Sigma_c^{+})\nonumber\\
&&+ {\cal M}(\Xi_{cc}^{++},   K^+,      \Sigma_c^+,      \eta_8   ,      K^{*+},      \Sigma_c^{+})
  + {\cal M}(\Xi_{cc}^{++},   K^{*+},   \Sigma_c^+,      \omega   ,    K^{*+},      \Sigma_c^{+})
  + {\cal M}(\Xi_{cc}^{++},   K^+,      \Lambda_c^+,     \Xi_c^{ 0},   \Sigma_c^{+},   K^{*+})\nonumber\\
&&+ {\cal M}(\Xi_{cc}^{++},   K^{*+},   \Lambda_c^+,     \Xi_c^{ 0},   \Sigma_c^{+},   K^{*+})
  + {\cal M}(\Xi_{cc}^{++},   K^+,      \Lambda_c^+,     \Xi_c^{'0},   \Sigma_c^{+},   K^{*+})
  + {\cal M}(\Xi_{cc}^{++},   K^{*+},   \Lambda_c^+,     \Xi_c^{'0},   \Sigma_c^{+},   K^{*+}) \nonumber\\
&&+ {\cal M}(\Xi_{cc}^{++},   K^+,      \Sigma_c^+,      \Xi_c^{ 0},   \Sigma_c^{+},   K^{*+})
  + {\cal M}(\Xi_{cc}^{++},   K^{*+},   \Sigma_c^+,      \Xi_c^{ 0},   \Sigma_c^{+},   K^{*+})
  + {\cal M}(\Xi_{cc}^{++},   K^+,      \Sigma_c^+,      \Xi_c^{'0},   \Sigma_c^{+},   K^{*+})\nonumber\\
&&+ {\cal M}(\Xi_{cc}^{++},   K^{*+},   \Sigma_c^+,      \Xi_c^{'0},   \Sigma_c^{+},   K^{*+})
  + {\cal M}(\Xi_{cc}^{++},   K^0,      \Sigma_c^{++},   \pi^+,        K^{*+},      \Sigma_c^{+})
  + {\cal M}(\Xi_{cc}^{++},   K^{*0},   \Sigma_c^{++},   \rho^+,       K^{*+},      \Sigma_c^{+})\nonumber\\
&&+ {\cal M}(\Xi_{cc}^{++},   K^0,      \Sigma_c^{++},   \Xi_c^{ +},   \Sigma_c^{+},   K^{*+})
  + {\cal M}(\Xi_{cc}^{++},   K^{*0},   \Sigma_c^{++},   \Xi_c^{ +},   \Sigma_c^{+},   K^{*+})
  + {\cal M}(\Xi_{cc}^{++},   K^0,      \Sigma_c^{++},   \Xi_c^{'+},   \Sigma_c^{+},   K^{*+})\nonumber\\
&&+ {\cal M}(\Xi_{cc}^{++},   K^{*0},   \Sigma_c^{++},   \Xi_c^{'+},   \Sigma_c^{+},   K^{*+})],
\end{eqnarray}
\begin{eqnarray}
{\cal A}(\Xi_{cc}^{++} \to \Lambda_c^{ +} K^{*+})
&=&T_{\rm SD}(\Xi_{cc}^{++} \to \Xi_c^{'+} K^{*+}) + i [
    {\cal M}(\Xi_{cc}^{++},   K^+,      \Lambda_c^+,     \eta_8   ,      K^{*+},     \Lambda_c^{+})
  + {\cal M}(\Xi_{cc}^{++},   K^{*+},   \Lambda_c^+,     \omega   ,    K^{*+},     \Lambda_c^{+})\nonumber\\
&&+ {\cal M}(\Xi_{cc}^{++},   K^+,      \Sigma_c^+,      \pi^0   ,     K^{*+},     \Lambda_c^{+})
  + {\cal M}(\Xi_{cc}^{++},   K^{*+},   \Sigma_c^+,      \rho^0   ,    K^{*+},     \Lambda_c^{+})
  + {\cal M}(\Xi_{cc}^{++},   K^+,      \Lambda_c^+,     \Xi_c^{ 0},   \Lambda_c^{+},   K^{*+})\nonumber\\
&&+ {\cal M}(\Xi_{cc}^{++},   K^{*+},   \Lambda_c^+,     \Xi_c^{ 0},   \Lambda_c^{+},   K^{*+})
  + {\cal M}(\Xi_{cc}^{++},   K^+,      \Lambda_c^+,     \Xi_c^{'0},   \Lambda_c^{+},   K^{*+})
  + {\cal M}(\Xi_{cc}^{++},   K^{*+},   \Lambda_c^+,     \Xi_c^{'0},   \Lambda_c^{+},   K^{*+})\nonumber\\
&&+ {\cal M}(\Xi_{cc}^{++},   K^+,      \Sigma_c^+,      \Xi_c^{ 0},   \Lambda_c^{+},   K^{*+})
  + {\cal M}(\Xi_{cc}^{++},   K^{*+},   \Sigma_c^+,      \Xi_c^{ 0},   \Lambda_c^{+},   K^{*+})
  + {\cal M}(\Xi_{cc}^{++},   K^+,      \Sigma_c^+,      \Xi_c^{'0},   \Lambda_c^{+},   K^{*+})\nonumber\\
&&+ {\cal M}(\Xi_{cc}^{++},   K^{*+},   \Sigma_c^+,      \Xi_c^{'0},   \Lambda_c^{+},   K^{*+})
  + {\cal M}(\Xi_{cc}^{++},   K^0,      \Sigma_c^{++},   \pi^+,        K^{*+},     \Lambda_c^{+})
  + {\cal M}(\Xi_{cc}^{++},   K^{*0},   \Sigma_c^{++},   \rho^+,       K^{*+},     \Lambda_c^{+})\nonumber\\
&&+ {\cal M}(\Xi_{cc}^{++},   K^0,      \Sigma_c^{++},   \Xi_c^{ +},   \Lambda_c^{+},   K^{*+})
  + {\cal M}(\Xi_{cc}^{++},   K^{*0},   \Sigma_c^{++},   \Xi_c^{ +},   \Lambda_c^{+},   K^{*+})
  + {\cal M}(\Xi_{cc}^{++},   K^0,      \Sigma_c^{++},   \Xi_c^{'+},   \Lambda_c^{+},   K^{*+})\nonumber\\
&&+ {\cal M}(\Xi_{cc}^{++},   K^{*0},   \Sigma_c^{++},   \Xi_c^{'+},   \Lambda_c^{+},   K^{*+})],
\end{eqnarray}
\begin{eqnarray}
{\cal A}(\Xi_{cc}^{++} \to \Sigma_c^{++} K^{*0})
&=&C_{\rm SD}(\Xi_{cc}^{++} \to \Sigma_c^{++} K^{*0}) + i [
    {\cal M}(\Xi_{cc}^{++},   K^{ +},   \Lambda_c^+,     \pi^- ,        K^{*0},    \Sigma_c^{++})
  + {\cal M}(\Xi_{cc}^{++},   K^{*+},   \Lambda_c^+,     \rho^- ,       K^{*0},    \Sigma_c^{++})\nonumber\\
&&+ {\cal M}(\Xi_{cc}^{++},   K^{ +},   \Sigma_c^{+},    \pi^- ,        K^{*0},    \Sigma_c^{++})
  + {\cal M}(\Xi_{cc}^{++},   K^{*+},   \Sigma_c^{+},    \rho^- ,       K^{*0},    \Sigma_c^{++})
  + {\cal M}(\Xi_{cc}^{++},   K^{ +},   \Lambda_c^{+},   \Xi_c^{ +},   \Sigma_c^{++},   K^{*0})\nonumber\\
&&+ {\cal M}(\Xi_{cc}^{++},   K^{*+},   \Lambda_c^{+},   \Xi_c^{ +},   \Sigma_c^{++},   K^{*0})
  + {\cal M}(\Xi_{cc}^{++},   K^{ +},   \Lambda_c^{+},   \Xi_c^{'+},   \Sigma_c^{++},   K^{*0})
  + {\cal M}(\Xi_{cc}^{++},   K^{*+},   \Lambda_c^{+},   \Xi_c^{'+},   \Sigma_c^{++},   K^{*0})\nonumber\\
&&+ {\cal M}(\Xi_{cc}^{++},   K^{ +},   \Sigma_c^{+},    \Xi_c^{ +},   \Sigma_c^{++},   K^{*0})
  + {\cal M}(\Xi_{cc}^{++},   K^{*+},   \Sigma_c^{+},    \Xi_c^{ +},   \Sigma_c^{++},   K^{*0})
  + {\cal M}(\Xi_{cc}^{++},   K^{ +},   \Sigma_c^{+},    \Xi_c^{'+},   \Sigma_c^{++},   K^{*0})\nonumber\\
&&+ {\cal M}(\Xi_{cc}^{++},   K^{*+},   \Sigma_c^{+},    \Xi_c^{'+},   \Sigma_c^{++},   K^{*0})],
\end{eqnarray}
\begin{eqnarray}
{\cal A}(\Xi_{cc}^{+} \to \Sigma_c^{ +}\bar K^{*0})
&=&C_{\rm SD}(\Xi_{cc}^{+} \to \Sigma_c^{ +} \bar K^{*0}) + i [
    {\cal M}(\Xi_{cc}^{+},   \pi^{+},    \Xi_c^{ 0},      K^{ -},    \bar K^{*0},   \Sigma_c^{+})
  + {\cal M}(\Xi_{cc}^{+},   \rho^{+},   \Xi_c^{ 0},      K^{*-},    \bar K^{*0},   \Sigma_c^{+})\nonumber\\
&&+ {\cal M}(\Xi_{cc}^{+},   \pi^{+},    \Xi_c^{'0},      K^{ -},    \bar K^{*0},   \Sigma_c^{+})
  + {\cal M}(\Xi_{cc}^{+},   \rho^{+},   \Xi_c^{'0},      K^{*-},    \bar K^{*0},   \Sigma_c^{+})
  + {\cal M}(\Xi_{cc}^{+},   \pi^{+},    \Xi_c^{ 0},   \Sigma_c^{0}, \Sigma_c^{+},  \bar K^{*0})\nonumber\\
&&+ {\cal M}(\Xi_{cc}^{+},   \rho^{+},   \Xi_c^{ 0},   \Sigma_c^{0}, \Sigma_c^{+},  \bar K^{*0})
  + {\cal M}(\Xi_{cc}^{+},   \pi^{+},    \Xi_c^{'0},   \Sigma_c^{0}, \Sigma_c^{+},  \bar K^{*0})
  + {\cal M}(\Xi_{cc}^{+},   \rho^{+},   \Xi_c^{'0},   \Sigma_c^{0}, \Sigma_c^{+},  \bar K^{*0})\nonumber\\
&&+ {\cal M}(\Xi_{cc}^{+}, \bar K^{ 0}, \Lambda_c^{+}, \pi^0   ,     \bar K^{*0},   \Sigma_c^{+})
  + {\cal M}(\Xi_{cc}^{+}, \bar K^{*0}, \Lambda_c^{+}, \rho^0   ,    \bar K^{*0},   \Sigma_c^{+})
  + {\cal M}(\Xi_{cc}^{+}, \bar K^{ 0}, \Sigma_c^{+},  \eta_8   ,      \bar K^{*0},   \Sigma_c^{+})\nonumber\\
&&+ {\cal M}(\Xi_{cc}^{+}, \bar K^{*0}, \Sigma_c^{+},  \omega   ,    \bar K^{*0},   \Sigma_c^{+})],
\end{eqnarray}
\begin{eqnarray}
{\cal A}(\Xi_{cc}^{+} \to \Lambda_c^{ +}\bar K^{*0})
&=&C_{\rm SD}(\Xi_{cc}^{+} \to \Lambda_c^{ +} \bar K^{*0}) + i [
    {\cal M}(\Xi_{cc}^{+},   \pi^{+},    \Xi_c^{ 0},      K^{ -},    \bar K^{*0},  \Lambda_c^{+})
  + {\cal M}(\Xi_{cc}^{+},   \rho^{+},   \Xi_c^{ 0},      K^{*-},    \bar K^{*0},  \Lambda_c^{+})\nonumber\\
&&+ {\cal M}(\Xi_{cc}^{+},   \pi^{+},    \Xi_c^{'0},      K^{ -},    \bar K^{*0},  \Lambda_c^{+})
  + {\cal M}(\Xi_{cc}^{+},   \rho^{+},   \Xi_c^{'0},      K^{*-},    \bar K^{*0},  \Lambda_c^{+})
  + {\cal M}(\Xi_{cc}^{+},   \pi^{+},    \Xi_c^{ 0},   \Sigma_c^{0}, \Lambda_c^{+},  \bar K^{*0})\nonumber\\
&&+ {\cal M}(\Xi_{cc}^{+},   \rho^{+},   \Xi_c^{ 0},   \Sigma_c^{0}, \Lambda_c^{+},  \bar K^{*0})
  + {\cal M}(\Xi_{cc}^{+},   \pi^{+},    \Xi_c^{'0},   \Sigma_c^{0}, \Lambda_c^{+},  \bar K^{*0})
  + {\cal M}(\Xi_{cc}^{+},   \rho^{+},   \Xi_c^{'0},   \Sigma_c^{0}, \Lambda_c^{+},  \bar K^{*0})\nonumber\\
&&+ {\cal M}(\Xi_{cc}^{+}, \bar K^{ 0}, \Lambda_c^{+}, \eta_8   ,      \bar K^{*0},  \Lambda_c^{+})
  + {\cal M}(\Xi_{cc}^{+}, \bar K^{*0}, \Lambda_c^{+}, \omega   ,    \bar K^{*0},  \Lambda_c^{+})
  + {\cal M}(\Xi_{cc}^{+}, \bar K^{ 0}, \Sigma_c^{+},  \pi^0   ,     \bar K^{*0},  \Lambda_c^{+})\nonumber\\
&&+ {\cal M}(\Xi_{cc}^{+}, \bar K^{*0}, \Sigma_c^{+},  \rho^0   ,    \bar K^{*0},  \Lambda_c^{+})],
\end{eqnarray}
\begin{eqnarray}
{\cal A}(\Xi_{cc}^{+} \to \Xi_c^{ 0}\rho^+)
&=&T_{\rm SD}(\Xi_{cc}^{+} \to \Xi_c^{ 0} \rho^+) + i [
    {\cal M}(\Xi_{cc}^{+},   \bar K^{ 0},   \Lambda_c^{+},     K^{ +},     \rho^+,    \Xi_c^{ 0})
  + {\cal M}(\Xi_{cc}^{+},   \bar K^{*0},   \Lambda_c^{+},     K^{*+},     \rho^+,    \Xi_c^{ 0})\nonumber\\
&&+ {\cal M}(\Xi_{cc}^{+},   \bar K^{ 0},   \Sigma_c^{+},      K^{ +},     \rho^+,    \Xi_c^{ 0})
  + {\cal M}(\Xi_{cc}^{+},   \bar K^{*0},   \Sigma_c^{+},      K^{*+},     \rho^+,    \Xi_c^{ 0})
  + {\cal M}(\Xi_{cc}^{+},   \bar K^{ 0},   \Lambda_c^{+}, \Sigma_c^{0},   \Xi_c^{ 0},   \rho^+)\nonumber\\
&&+ {\cal M}(\Xi_{cc}^{+},   \bar K^{*0},   \Lambda_c^{+}, \Sigma_c^{0},   \Xi_c^{ 0},   \rho^+)
  + {\cal M}(\Xi_{cc}^{+},   \bar K^{ 0},   \Sigma_c^{+},  \Sigma_c^{0},   \Xi_c^{ 0},   \rho^+)
  + {\cal M}(\Xi_{cc}^{+},   \bar K^{*0},   \Sigma_c^{+},  \Sigma_c^{0},   \Xi_c^{ 0},   \rho^+)\nonumber\\
&&+ {\cal M}(\Xi_{cc}^{+},      \pi^+,      \Xi_c^{ 0},    \pi^0   ,       \rho^+,    \Xi_c^{ 0})
  + {\cal M}(\Xi_{cc}^{+},      \rho^+,     \Xi_c^{ 0},    \rho^0   ,      \rho^+,    \Xi_c^{ 0})
  + {\cal M}(\Xi_{cc}^{+},      \pi^+,      \Xi_c^{'0},    \pi^0   ,       \rho^+,    \Xi_c^{ 0})\nonumber\\
&&+ {\cal M}(\Xi_{cc}^{+},      \rho^+,     \Xi_c^{'0},    \rho^0   ,      \rho^+,    \Xi_c^{ 0})],
\end{eqnarray}
\begin{eqnarray}
{\cal A}(\Xi_{cc}^{+} \to \Xi_c^{'0}\rho^+)
&=&T_{\rm SD}(\Xi_{cc}^{+} \to \Xi_c^{'0} \rho^+) + i [
    {\cal M}(\Xi_{cc}^{+},   \bar K^{ 0},   \Lambda_c^{+},     K^{ +},     \rho^+,    \Xi_c^{'0})
  + {\cal M}(\Xi_{cc}^{+},   \bar K^{*0},   \Lambda_c^{+},     K^{*+},     \rho^+,    \Xi_c^{' 0})\nonumber\\
&&+ {\cal M}(\Xi_{cc}^{+},   \bar K^{ 0},   \Sigma_c^{+},      K^{ +},     \rho^+,    \Xi_c^{'0})
  + {\cal M}(\Xi_{cc}^{+},   \bar K^{*0},   \Sigma_c^{+},      K^{*+},     \rho^+,    \Xi_c^{'0})
  + {\cal M}(\Xi_{cc}^{+},   \bar K^{ 0},   \Lambda_c^{+}, \Sigma_c^{0},   \Xi_c^{'0},   \rho^+)\nonumber\\
&&+ {\cal M}(\Xi_{cc}^{+},   \bar K^{*0},   \Lambda_c^{+}, \Sigma_c^{0},   \Xi_c^{'0},   \rho^+)
  + {\cal M}(\Xi_{cc}^{+},   \bar K^{ 0},   \Sigma_c^{+},  \Sigma_c^{0},   \Xi_c^{'0},   \rho^+)
  + {\cal M}(\Xi_{cc}^{+},   \bar K^{*0},   \Sigma_c^{+},  \Sigma_c^{0},   \Xi_c^{'0},   \rho^+)\nonumber\\
&&+ {\cal M}(\Xi_{cc}^{+},      \pi^+,      \Xi_c^{ 0},    \pi^0   ,       \rho^+,    \Xi_c^{'0})
  + {\cal M}(\Xi_{cc}^{+},      \rho^+,     \Xi_c^{ 0},    \rho^0   ,      \rho^+,    \Xi_c^{'0})
  + {\cal M}(\Xi_{cc}^{+},      \pi^+,      \Xi_c^{'0},    \pi^0   ,       \rho^+,    \Xi_c^{'0})\nonumber\\
&&+ {\cal M}(\Xi_{cc}^{+},      \rho^+,     \Xi_c^{'0},    \rho^0   ,      \rho^+,    \Xi_c^{' 0})],
\end{eqnarray}
\begin{eqnarray}
{\cal A}(\Xi_{cc}^{+} \to \Xi_c^{ +}\rho^0)
&=&i [
    {\cal M}(\Xi_{cc}^{+},      \pi^+,       \Xi_c^{ 0},      \pi^{ -},      \rho^0   , \Xi_c^{ +})
  + {\cal M}(\Xi_{cc}^{+},      \rho^+,      \Xi_c^{ 0},      \rho^{-},      \rho^0   , \Xi_c^{ +})
  + {\cal M}(\Xi_{cc}^{+},      \pi^+,       \Xi_c^{'0},      \pi^{ -},      \rho^0   , \Xi_c^{ +})\nonumber\\
&&+ {\cal M}(\Xi_{cc}^{+},      \rho^+,      \Xi_c^{'0},      \rho^{-},      \rho^0   , \Xi_c^{ +})
  + {\cal M}(\Xi_{cc}^{+},      \pi^+,       \Xi_c^{ 0},    \Xi_c^{ 0},    \Xi_c^{ +}, \rho^0 )
  + {\cal M}(\Xi_{cc}^{+},      \rho^+,      \Xi_c^{ 0},    \Xi_c^{ 0},    \Xi_c^{ +}, \rho^0 )\nonumber\\
&&+ {\cal M}(\Xi_{cc}^{+},      \pi^+,       \Xi_c^{ 0},    \Xi_c^{'0},    \Xi_c^{ +}, \rho^0 )
  + {\cal M}(\Xi_{cc}^{+},      \rho^+,      \Xi_c^{ 0},    \Xi_c^{'0},    \Xi_c^{ +}, \rho^0   )
  + {\cal M}(\Xi_{cc}^{+},      \pi^+,       \Xi_c^{'0},    \Xi_c^{ 0},    \Xi_c^{ +}, \rho^0   )\nonumber\\
&&+ {\cal M}(\Xi_{cc}^{+},      \rho^+,      \Xi_c^{'0},    \Xi_c^{ 0},    \Xi_c^{ +}, \rho^0   )
  + {\cal M}(\Xi_{cc}^{+},      \pi^+,       \Xi_c^{'0},    \Xi_c^{'0},    \Xi_c^{ +}, \rho^0   )
  + {\cal M}(\Xi_{cc}^{+},      \rho^+,      \Xi_c^{'0},    \Xi_c^{'0},    \Xi_c^{ +}, \rho^0   )\nonumber\\
&&+ {\cal M}(\Xi_{cc}^{+},   \bar K^{ 0},    \Xi_c^{ 0},      K^{ 0},      \rho^0   , \Xi_c^{ +})
  + {\cal M}(\Xi_{cc}^{+},   \bar K^{*0},    \Xi_c^{ 0},      K^{*0},      \rho^0   , \Xi_c^{ +})
  + {\cal M}(\Xi_{cc}^{+},   \bar K^{ 0},    \Xi_c^{'0},      K^{ 0},      \rho^0   , \Xi_c^{ +})\nonumber\\
&&+ {\cal M}(\Xi_{cc}^{+},   \bar K^{*0},    \Xi_c^{'0},      K^{*0},      \rho^0   , \Xi_c^{ +})
  + {\cal M}(\Xi_{cc}^{+},   \bar K^{ 0},    \Lambda_c^+,   \Sigma_c^+,    \Xi_c^{ +}, \rho^0 )
  + {\cal M}(\Xi_{cc}^{+},   \bar K^{*0},    \Lambda_c^+,   \Sigma_c^+,    \Xi_c^{ +}, \rho^0 )\nonumber\\
&&+ {\cal M}(\Xi_{cc}^{+},   \bar K^{ 0},    \Sigma_c^+,    \Lambda_c^+,   \Xi_c^{ +}, \rho^0 )
  + {\cal M}(\Xi_{cc}^{+},   \bar K^{*0},    \Sigma_c^+,    \Lambda_c^+,   \Xi_c^{ +}, \rho^0 )],
\end{eqnarray}
\begin{eqnarray}
{\cal A}(\Xi_{cc}^{+} \to \Xi_c^{'+}\rho^0)
&=&i [
    {\cal M}(\Xi_{cc}^{+},      \pi^+,       \Xi_c^{ 0},      \pi^{ -},      \rho^0   , \Xi_c^{'+})
  + {\cal M}(\Xi_{cc}^{+},      \rho^+,      \Xi_c^{ 0},       \rho^{-},      \rho^0   , \Xi_c^{'+})
  + {\cal M}(\Xi_{cc}^{+},      \pi^+,       \Xi_c^{'0},     \pi^{ -},      \rho^0   , \Xi_c^{'+})\nonumber\\
&&+ {\cal M}(\Xi_{cc}^{+},      \rho^+,      \Xi_c^{'0},       \rho^{-},      \rho^0   , \Xi_c^{'+})
  + {\cal M}(\Xi_{cc}^{+},      \pi^+,       \Xi_c^{ 0},    \Xi_c^{ 0},    \Xi_c^{'+}, \rho^0 )
  + {\cal M}(\Xi_{cc}^{+},      \rho^+,      \Xi_c^{ 0},    \Xi_c^{ 0},    \Xi_c^{'+}, \rho^0 )\nonumber\\
&&+ {\cal M}(\Xi_{cc}^{+},      \pi^+,       \Xi_c^{ 0},    \Xi_c^{'0},    \Xi_c^{'+}, \rho^0 )
  + {\cal M}(\Xi_{cc}^{+},      \rho^+,      \Xi_c^{ 0},    \Xi_c^{'0},    \Xi_c^{'+}, \rho^0 )
  + {\cal M}(\Xi_{cc}^{+},      \pi^+,       \Xi_c^{'0},    \Xi_c^{ 0},    \Xi_c^{'+}, \rho^0 )\nonumber\\
&&+ {\cal M}(\Xi_{cc}^{+},      \rho^+,      \Xi_c^{'0},    \Xi_c^{ 0},    \Xi_c^{'+}, \rho^0 )
  + {\cal M}(\Xi_{cc}^{+},      \pi^+,       \Xi_c^{'0},    \Xi_c^{'0},    \Xi_c^{'+}, \rho^0 )
  + {\cal M}(\Xi_{cc}^{+},      \rho^+,      \Xi_c^{'0},    \Xi_c^{'0},    \Xi_c^{'+}, \rho^0(d))\nonumber\\
&&+ {\cal M}(\Xi_{cc}^{+},   \bar K^{ 0},    \Xi_c^{ 0},      K^{ 0},      \rho^0   , \Xi_c^{'+})
  + {\cal M}(\Xi_{cc}^{+},   \bar K^{*0},    \Xi_c^{ 0},      K^{*0},      \rho^0   , \Xi_c^{'+})
  + {\cal M}(\Xi_{cc}^{+},   \bar K^{ 0},    \Xi_c^{'0},      K^{ 0},      \rho^0   , \Xi_c^{'+})\nonumber\\
&&+ {\cal M}(\Xi_{cc}^{+},   \bar K^{*0},    \Xi_c^{'0},      K^{*0},      \rho^0   , \Xi_c^{'+})
  + {\cal M}(\Xi_{cc}^{+},   \bar K^{ 0},    \Lambda_c^+,   \Sigma_c^+,    \Xi_c^{'+}, \rho^0 )
  + {\cal M}(\Xi_{cc}^{+},   \bar K^{*0},    \Lambda_c^+,   \Sigma_c^+,    \Xi_c^{'+}, \rho^0 )\nonumber\\
&&+ {\cal M}(\Xi_{cc}^{+},   \bar K^{ 0},    \Sigma_c^+,    \Lambda_c^+,   \Xi_c^{'+}, \rho^0 )
  + {\cal M}(\Xi_{cc}^{+},   \bar K^{*0},    \Sigma_c^+,    \Lambda_c^+,   \Xi_c^{'+}, \rho^0 )],
\end{eqnarray}
\begin{eqnarray}
{\cal A}(\Xi_{cc}^{+} \to \Xi_c^{+}\omega)
&=&i [
    {\cal M}(\Xi_{cc}^{+},      \pi^+,       \Xi_c^{ 0},    \Xi_c^{ 0},  \Xi_c^{+}, \omega  )
  + {\cal M}(\Xi_{cc}^{+},      \rho^+,      \Xi_c^{ 0},    \Xi_c^{ 0},  \Xi_c^{+}, \omega )
  + {\cal M}(\Xi_{cc}^{+},      \pi^+,       \Xi_c^{ 0},    \Xi_c^{'0},  \Xi_c^{+}, \omega )\nonumber\\
&&+ {\cal M}(\Xi_{cc}^{+},      \rho^+,      \Xi_c^{'0},    \Xi_c^{'0},  \Xi_c^{+}, \omega )
  + {\cal M}(\Xi_{cc}^{+},      \pi^+,       \Xi_c^{'0},    \Xi_c^{ 0},  \Xi_c^{+}, \omega )
  + {\cal M}(\Xi_{cc}^{+},      \rho^+,      \Xi_c^{'0},    \Xi_c^{ 0},  \Xi_c^{+}, \omega )\nonumber\\
&&+ {\cal M}(\Xi_{cc}^{+},      \pi^+,       \Xi_c^{'0},    \Xi_c^{'0},  \Xi_c^{+}, \omega )
  + {\cal M}(\Xi_{cc}^{+},      \rho^+,      \Xi_c^{'0},    \Xi_c^{'0},  \Xi_c^{+}, \omega )
  + {\cal M}(\Xi_{cc}^{+},    \bar K^{ 0},   \Lambda_c^+,     K^{ 0},    \omega   , \Xi_c^{+})\nonumber\\
&&+ {\cal M}(\Xi_{cc}^{+},    \bar K^{*0},   \Lambda_c^+,     K^{*0} ,   \omega   , \Xi_c^{+})
  + {\cal M}(\Xi_{cc}^{+},    \bar K^{ 0},   \Sigma_c^+,      K^{ 0},    \omega   , \Xi_c^{+})
  + {\cal M}(\Xi_{cc}^{+},    \bar K^{*0},   \Sigma_c^+,      K^{*0} ,   \omega   , \Xi_c^{+})\nonumber\\
&&+ {\cal M}(\Xi_{cc}^{+},    \bar K^{ 0},   \Lambda_c^+,   \Lambda_c^+, \Xi_c^{+}, \omega )
  + {\cal M}(\Xi_{cc}^{+},    \bar K^{*0},   \Lambda_c^+,   \Lambda_c^+, \Xi_c^{+}, \omega )
  + {\cal M}(\Xi_{cc}^{+},    \bar K^{ 0},   \Sigma_c^+,    \Sigma_c^+,  \Xi_c^{+}, \omega )\nonumber\\
&&+ {\cal M}(\Xi_{cc}^{+},    \bar K^{*0},   \Sigma_c^+,    \Sigma_c^+,  \Xi_c^{+}, \omega )],
\end{eqnarray}
\begin{eqnarray}
{\cal A}(\Xi_{cc}^{+} \to \Xi_c^{'+}\omega)
&=&i [
    {\cal M}(\Xi_{cc}^{+},      \pi^+,       \Xi_c^{ 0},    \Xi_c^{ 0},  \Xi_c^{'+}, \omega   )
  + {\cal M}(\Xi_{cc}^{+},      \rho^+,      \Xi_c^{ 0},    \Xi_c^{ 0},  \Xi_c^{'+}, \omega )
  + {\cal M}(\Xi_{cc}^{+},      \pi^+,       \Xi_c^{ 0},    \Xi_c^{'0},  \Xi_c^{'+}, \omega )\nonumber\\
&&+ {\cal M}(\Xi_{cc}^{+},      \rho^+,      \Xi_c^{'0},    \Xi_c^{'0},  \Xi_c^{'+}, \omega )
  + {\cal M}(\Xi_{cc}^{+},      \pi^+,       \Xi_c^{'0},    \Xi_c^{ 0},  \Xi_c^{'+}, \omega )
  + {\cal M}(\Xi_{cc}^{+},      \rho^+,      \Xi_c^{'0},    \Xi_c^{ 0},  \Xi_c^{'+}, \omega )\nonumber\\
&&+ {\cal M}(\Xi_{cc}^{+},      \pi^+,       \Xi_c^{'0},    \Xi_c^{'0},  \Xi_c^{'+}, \omega )
  + {\cal M}(\Xi_{cc}^{+},      \rho^+,      \Xi_c^{'0},    \Xi_c^{'0},  \Xi_c^{'+}, \omega )
  + {\cal M}(\Xi_{cc}^{+},    \bar K^{ 0},   \Lambda_c^+,     K^{ 0},    \omega   , \Xi_c^{'+})\nonumber\\
&&+ {\cal M}(\Xi_{cc}^{+},    \bar K^{*0},   \Lambda_c^+,     K^{*0} ,   \omega   , \Xi_c^{'+})
  + {\cal M}(\Xi_{cc}^{+},    \bar K^{ 0},   \Sigma_c^+,      K^{ 0},    \omega   , \Xi_c^{'+})
  + {\cal M}(\Xi_{cc}^{+},    \bar K^{*0},   \Sigma_c^+,      K^{*0} ,   \omega   , \Xi_c^{'+})\nonumber\\
&&+ {\cal M}(\Xi_{cc}^{+},    \bar K^{ 0},   \Lambda_c^+,   \Lambda_c^+, \Xi_c^{'+}, \omega )
  + {\cal M}(\Xi_{cc}^{+},    \bar K^{*0},   \Lambda_c^+,   \Lambda_c^+, \Xi_c^{'+}, \omega )
  + {\cal M}(\Xi_{cc}^{+},    \bar K^{ 0},   \Sigma_c^+,    \Sigma_c^+,  \Xi_c^{'+}, \omega )\nonumber\\
&&+ {\cal M}(\Xi_{cc}^{+},    \bar K^{*0},   \Sigma_c^+,    \Sigma_c^+,  \Xi_c^{+'}, \omega )],
\end{eqnarray}
\begin{eqnarray}
{\cal A}(\Xi_{cc}^{+} \to \Sigma_{c}^{+}\rho^{0})
&=&i [
    {\cal M}(\Xi_{cc}^{+},      \pi^{+},       \Sigma_{c}^{0},      \pi^{-} ,    \rho^{0}   ,   \Sigma_{c}^{+})
  + {\cal M}(\Xi_{cc}^{+},      \rho^{+},      \Sigma_{c}^{0},      \rho^{-} ,   \rho^{0}   ,   \Sigma_{c}^{+})
  + {\cal M}(\Xi_{cc}^{+},      \pi^{+},       \Sigma_{c}^{0},     \Sigma_{c}^{0}, \Sigma_{c}^{+},  \rho^{0} )\nonumber\\
&&+ {\cal M}(\Xi_{cc}^{+},      \rho^{+},      \Sigma_{c}^{0},     \Sigma_{c}^{0}, \Sigma_{c}^{+},  \rho^{0} )
  + {\cal M}(\Xi_{cc}^{+},      \eta_8 ,          \Lambda_{c}^{+},     \Sigma_{c}^{+}, \Sigma_{c}^{+},  \rho^{0} )
  + {\cal M}(\Xi_{cc}^{+},      \omega ,        \Lambda_{c}^{+},     \Sigma_{c}^{+}, \Sigma_{c}^{+},  \rho^{0} )\nonumber\\
&&+ {\cal M}(\Xi_{cc}^{+},      \pi^{0} ,       \Sigma_{c}^{+},     \Lambda_{c}^{+}, \Sigma_{c}^{+},  \rho^{0} )
  + {\cal M}(\Xi_{cc}^{+},      \rho^{0} ,      \Sigma_{c}^{+},     \Lambda_{c}^{+}, \Sigma_{c}^{+},  \rho^{0} )
  + {\cal M}(\Xi_{cc}^{+},      K^{+},       \Xi_{c}^{0},     \Xi_{c}^{0}, \Sigma_{c}^{+},  \rho^{0} )\nonumber\\
&&+ {\cal M}(\Xi_{cc}^{+},      K^{*+},      \Xi_{c}^{0},     \Xi_{c}^{0}, \Sigma_{c}^{+},  \rho^{0} )
  + {\cal M}(\Xi_{cc}^{+},      K^{+},       \Xi_{c}^{'0},     \Xi_{c}^{0}, \Sigma_{c}^{+},  \rho^{0} )
  + {\cal M}(\Xi_{cc}^{+},      K^{*+},      \Xi_{c}^{'0},     \Xi_{c}^{0}, \Sigma_{c}^{+},  \rho^{0} )\nonumber\\
&&+ {\cal M}(\Xi_{cc}^{+},      K^{+},       \Xi_{c}^{0},     \Xi_{c}^{'0}, \Sigma_{c}^{+},  \rho^{0} )
  + {\cal M}(\Xi_{cc}^{+},      K^{*+},      \Xi_{c}^{0},     \Xi_{c}^{'0}, \Sigma_{c}^{+},  \rho^{0} )
  + {\cal M}(\Xi_{cc}^{+},      K^{+},       \Xi_{c}^{'0},     \Xi_{c}^{'0}, \Sigma_{c}^{+},  \rho^{0} )\nonumber\\
&&+ {\cal M}(\Xi_{cc}^{+},      K^{*+},      \Xi_{c}^{'0},     \Xi_{c}^{'0}, \Sigma_{c}^{+},  \rho^{0} )
  + {\cal M}(\Xi_{cc}^{+},      K^{+},       \Xi_{c}^{0},      K^{-} ,    \rho^{0}   ,   \Sigma_{c}^{+})
  + {\cal M}(\Xi_{cc}^{+},      K^{*+},      \Xi_{c}^{0},      K^{*-} ,   \rho^{0}   ,   \Sigma_{c}^{+})\nonumber\\
&&+ {\cal M}(\Xi_{cc}^{+},      K^{+},       \Xi_{c}^{'0},      K^{-} ,    \rho^{0}   ,   \Sigma_{c}^{+})
  + {\cal M}(\Xi_{cc}^{+},      K^{*+},      \Xi_{c}^{'0},      K^{*-} ,   \rho^{0}   ,   \Sigma_{c}^{+})
  + {\cal M}(\Xi_{cc}^{+},      \eta_1 ,          \Lambda_{c}^{+},     \Sigma_{c}^{+}, \Sigma_{c}^{+},  \rho^{0} )],
\end{eqnarray}
\begin{eqnarray}
{\cal A}(\Xi_{cc}^{+} \to \Lambda_{c}^{+}\rho^{0})
&=&i [
    {\cal M}(\Xi_{cc}^{+},      \pi^{+},       \Sigma_{c}^{0},      \pi^{-} ,    \rho^{0}   ,   \Lambda_{c}^{+})
  + {\cal M}(\Xi_{cc}^{+},      \rho^{+},      \Sigma_{c}^{0},      \rho^{-} ,   \rho^{0}   ,   \Lambda_{c}^{+})
  + {\cal M}(\Xi_{cc}^{+},      \pi^{+},       \Sigma_{c}^{0},     \Sigma_{c}^{0}, \Lambda_{c}^{+},  \rho^{0} )\nonumber\\
&&+ {\cal M}(\Xi_{cc}^{+},      \rho^{+},      \Sigma_{c}^{0},     \Sigma_{c}^{0}, \Lambda_{c}^{+},  \rho^{0} )
  + {\cal M}(\Xi_{cc}^{+},      \pi^{0} ,       \Lambda_{c}^{+},     \Sigma_{c}^{+}, \Lambda_{c}^{+},  \rho^{0} )
  + {\cal M}(\Xi_{cc}^{+},      \rho^{0} ,      \Lambda_{c}^{+},     \Sigma_{c}^{+}, \Lambda_{c}^{+},  \rho^{0} )\nonumber\\
&&+ {\cal M}(\Xi_{cc}^{+},      \eta_8 ,          \Sigma_{c}^{+},     \Lambda_{c}^{+}, \Lambda_{c}^{+},  \rho^{0} )
  + {\cal M}(\Xi_{cc}^{+},      \omega ,        \Sigma_{c}^{+},     \Lambda_{c}^{+}, \Lambda_{c}^{+},  \rho^{0} )
  + {\cal M}(\Xi_{cc}^{+},      K^{+},       \Xi_{c}^{0},     \Xi_{c}^{0}, \Lambda_{c}^{+},  \rho^{0} )\nonumber\\
&&+ {\cal M}(\Xi_{cc}^{+},      K^{*+},      \Xi_{c}^{0},     \Xi_{c}^{0}, \Lambda_{c}^{+},  \rho^{0} )
  + {\cal M}(\Xi_{cc}^{+},      K^{+},       \Xi_{c}^{'0},     \Xi_{c}^{0}, \Lambda_{c}^{+},  \rho^{0} )
  + {\cal M}(\Xi_{cc}^{+},      K^{*+},      \Xi_{c}^{'0},     \Xi_{c}^{0}, \Lambda_{c}^{+},  \rho^{0} )\nonumber\\
&&+ {\cal M}(\Xi_{cc}^{+},      K^{+},       \Xi_{c}^{0},     \Xi_{c}^{'0}, \Lambda_{c}^{+},  \rho^{0} )
  + {\cal M}(\Xi_{cc}^{+},      K^{*+},      \Xi_{c}^{0},     \Xi_{c}^{'0}, \Lambda_{c}^{+},  \rho^{0} )
  + {\cal M}(\Xi_{cc}^{+},      K^{+},       \Xi_{c}^{'0},     \Xi_{c}^{'0}, \Lambda_{c}^{+},  \rho^{0} )\nonumber\\
&&+ {\cal M}(\Xi_{cc}^{+},      K^{*+},      \Xi_{c}^{'0},     \Xi_{c}^{'0}, \Lambda_{c}^{+},  \rho^{0} )
  + {\cal M}(\Xi_{cc}^{+},      K^{+},       \Xi_{c}^{0},      K^{-} ,    \rho^{0}   ,   \Lambda_{c}^{+})
  + {\cal M}(\Xi_{cc}^{+},      K^{*+},      \Xi_{c}^{0},      K^{*-} ,   \rho^{0}   ,   \Lambda_{c}^{+})\nonumber\\
&&+ {\cal M}(\Xi_{cc}^{+},      K^{+},       \Xi_{c}^{'0},      K^{-} ,    \rho^{0} ,   \Lambda_{c}^{+})
  + {\cal M}(\Xi_{cc}^{+},      K^{*+},      \Xi_{c}^{'0},      K^{*-} ,   \rho^{0} ,   \Lambda_{c}^{+})
  + {\cal M}(\Xi_{cc}^{+},      \eta_1 ,          \Sigma_{c}^{+},     \Lambda_{c}^{+}, \Lambda_{c}^{+},  \rho^{0} )],
\end{eqnarray}

\begin{eqnarray}
{\cal A}(\Xi_{cc}^{+} \to \Sigma_{c}^{+}\omega)
&=&i [
    {\cal M}(\Xi_{cc}^{+},      \pi^{+},       \Sigma_{c}^{0},     \Sigma_{c}^{0}, \Sigma_{c}^{+},  \omega )
  + {\cal M}(\Xi_{cc}^{+},      \rho^{+},      \Sigma_{c}^{0},     \Sigma_{c}^{0}, \Sigma_{c}^{+},  \omega )
  + {\cal M}(\Xi_{cc}^{+},      \pi^{0} ,       \Lambda_{c}^{+},     \Lambda_{c}^{+}, \Sigma_{c}^{+},  \omega )\nonumber\\
&&+ {\cal M}(\Xi_{cc}^{+},      \rho^{0} ,      \Lambda_{c}^{+},     \Lambda_{c}^{+}, \Sigma_{c}^{+},  \omega )
  + {\cal M}(\Xi_{cc}^{+},      \eta_8 ,          \Sigma_{c}^{+},     \Sigma_{c}^{+}, \Sigma_{c}^{+},  \omega )
  + {\cal M}(\Xi_{cc}^{+},      \omega ,        \Sigma_{c}^{+},     \Sigma_{c}^{+}, \Sigma_{c}^{+},  \omega )\nonumber\\
&&+ {\cal M}(\Xi_{cc}^{+},      K^{+},       \Xi_{c}^{0},     \Xi_{c}^{0}, \Sigma_{c}^{+},  \omega )
  + {\cal M}(\Xi_{cc}^{+},      K^{*+},      \Xi_{c}^{0},     \Xi_{c}^{0}, \Sigma_{c}^{+},  \omega )
  + {\cal M}(\Xi_{cc}^{+},      K^{+},       \Xi_{c}^{'0},     \Xi_{c}^{0}, \Sigma_{c}^{+},  \omega )\nonumber\\
&&+ {\cal M}(\Xi_{cc}^{+},      K^{*+},      \Xi_{c}^{'0},     \Xi_{c}^{0}, \Sigma_{c}^{+},  \omega )
  + {\cal M}(\Xi_{cc}^{+},      K^{+},       \Xi_{c}^{0},     \Xi_{c}^{'0}, \Sigma_{c}^{+},  \omega )
  + {\cal M}(\Xi_{cc}^{+},      K^{*+},      \Xi_{c}^{0},     \Xi_{c}^{'0}, \Sigma_{c}^{+},  \omega )\nonumber\\
&&+ {\cal M}(\Xi_{cc}^{+},      K^{+},       \Xi_{c}^{'0},     \Xi_{c}^{'0}, \Sigma_{c}^{+},  \omega )
  + {\cal M}(\Xi_{cc}^{+},      K^{*+},      \Xi_{c}^{'0},     \Xi_{c}^{'0}, \Sigma_{c}^{+},  \omega )
  + {\cal M}(\Xi_{cc}^{+},      K^{+},       \Xi_{c}^{0},      K^{-} ,    \omega  ,   \Sigma_{c}^{+})\nonumber\\
&&+ {\cal M}(\Xi_{cc}^{+},      K^{*+},      \Xi_{c}^{0},      K^{*-} ,   \omega  ,   \Sigma_{c}^{+})
  + {\cal M}(\Xi_{cc}^{+},      K^{+},       \Xi_{c}^{'0},      K^{-} ,    \omega  ,   \Sigma_{c}^{+})
  + {\cal M}(\Xi_{cc}^{+},      K^{*+},      \Xi_{c}^{'0},      K^{*-} ,   \omega  ,   \Sigma_{c}^{+})\nonumber\\
&&+ {\cal M}(\Xi_{cc}^{+},      \eta_1 ,          \Sigma_{c}^{+},     \Sigma_{c}^{+}, \Sigma_{c}^{+},  \omega )],
\end{eqnarray}
\begin{eqnarray}
{\cal A}(\Xi_{cc}^{+} \to \Lambda_{c}^{+}\omega)
&=&i [
    {\cal M}(\Xi_{cc}^{+},      \pi^{+},       \Sigma_{c}^{0},     \Sigma_{c}^{0}, \Lambda_{c}^{+},  \omega )
  + {\cal M}(\Xi_{cc}^{+},      \rho^{+},      \Sigma_{c}^{0},     \Sigma_{c}^{0}, \Lambda_{c}^{+},  \omega )
  + {\cal M}(\Xi_{cc}^{+},      \eta_8 ,          \Lambda_{c}^{+},     \Lambda_{c}^{+}, \Lambda_{c}^{+},  \omega )\nonumber\\
&&+ {\cal M}(\Xi_{cc}^{+},      \omega ,        \Lambda_{c}^{+},     \Lambda_{c}^{+}, \Lambda_{c}^{+},  \omega )
  + {\cal M}(\Xi_{cc}^{+},      \pi^{0} ,       \Sigma_{c}^{+},     \Sigma_{c}^{+}, \Lambda_{c}^{+}, \omega )
  + {\cal M}(\Xi_{cc}^{+},      \rho^{0} ,      \Sigma_{c}^{+},     \Sigma_{c}^{+}, \Lambda_{c}^{+},  \omega )\nonumber\\
&&+ {\cal M}(\Xi_{cc}^{+},      K^{+},       \Xi_{c}^{0},     \Xi_{c}^{0}, \Lambda_{c}^{+},  \omega )
  + {\cal M}(\Xi_{cc}^{+},      K^{*+},      \Xi_{c}^{0},     \Xi_{c}^{0}, \Lambda_{c}^{+},  \omega )
  + {\cal M}(\Xi_{cc}^{+},      K^{+},       \Xi_{c}^{'0},     \Xi_{c}^{0}, \Lambda_{c}^{+},  \omega )\nonumber\\
&&+ {\cal M}(\Xi_{cc}^{+},      K^{*+},      \Xi_{c}^{'0},     \Xi_{c}^{0}, \Lambda_{c}^{+},  \omega )
  + {\cal M}(\Xi_{cc}^{+},      K^{+},       \Xi_{c}^{0},     \Xi_{c}^{'0}, \Lambda_{c}^{+},  \omega  )
  + {\cal M}(\Xi_{cc}^{+},      K^{*+},      \Xi_{c}^{0},     \Xi_{c}^{'0}, \Lambda_{c}^{+},  \omega )\nonumber\\
&&+ {\cal M}(\Xi_{cc}^{+},      K^{+},       \Xi_{c}^{'0},     \Xi_{c}^{'0}, \Lambda_{c}^{+},  \omega )
  + {\cal M}(\Xi_{cc}^{+},      K^{*+},      \Xi_{c}^{'0},     \Xi_{c}^{'0}, \Lambda_{c}^{+},  \omega )
  + {\cal M}(\Xi_{cc}^{+},      K^{+},       \Xi_{c}^{0},      K^{-} ,    \omega  ,   \Lambda_{c}^{+})\nonumber\\
&&+ {\cal M}(\Xi_{cc}^{+},      K^{*+},      \Xi_{c}^{0},      K^{*-} ,   \omega  ,   \Lambda_{c}^{+})
  + {\cal M}(\Xi_{cc}^{+},      K^{+},       \Xi_{c}^{'0},      K^{-} ,   \omega  ,   \Lambda_{c}^{+})
  + {\cal M}(\Xi_{cc}^{+},      K^{*+},      \Xi_{c}^{'0},      K^{*-} ,  \omega  ,   \Lambda_{c}^{+})\nonumber\\
&&+ {\cal M}(\Xi_{cc}^{+},      \eta_1 ,          \Lambda_{c}^{+},     \Lambda_{c}^{+}, \Lambda_{c}^{+},  \omega )],
\end{eqnarray}

\begin{eqnarray}
{\cal A}(\Xi_{cc}^{+} \to \Sigma_{c}^{0}\rho^{+})
&=&T_{\rm SD}(\Xi_{cc}^{+} \to \Sigma_{c}^{+}\rho^{+})+ i [
    {\cal M}(\Xi_{cc}^{+}, \pi^{0}   ,    \Lambda_{c}^{+},   \pi^{+},   \rho^{+},\Sigma_{c}^{0})
  + {\cal M}(\Xi_{cc}^{+}, \rho^{0}   ,   \Lambda_{c}^{+},   \rho^{+},  \rho^{+},\Sigma_{c}^{0})\nonumber\\
&&+ {\cal M}(\Xi_{cc}^{+}, \pi^{0}   ,    \Sigma_{c}^{+},    \pi^{+},   \rho^{+},\Sigma_{c}^{0})
  + {\cal M}(\Xi_{cc}^{+}, \rho^{0}   ,   \Sigma_{c}^{+},    \rho^{+},  \rho^{+},\Sigma_{c}^{0})
  + {\cal M}(\Xi_{cc}^{+}, \pi^{0}   ,    \Lambda_{c}^{+},   \Sigma_{c}^{0},    \Sigma_{c}^{0},\rho^{+})\nonumber\\
&&+ {\cal M}(\Xi_{cc}^{+}, \rho^{0}   ,   \Lambda_{c}^{+},   \Sigma_{c}^{0},    \Sigma_{c}^{0},\rho^{+})
  + {\cal M}(\Xi_{cc}^{+}, \pi^{0}   ,    \Sigma_{c}^{+},    \Sigma_{c}^{0},    \Sigma_{c}^{0},\rho^{+})
  + {\cal M}(\Xi_{cc}^{+}, \rho^{0}   ,   \Sigma_{c}^{+},    \Sigma_{c}^{0},    \Sigma_{c}^{0},\rho^{+})\nonumber\\
&&+ {\cal M}(\Xi_{cc}^{+}, \eta_8   ,       \Lambda_{c}^{+},   \Sigma_{c}^{0},    \Sigma_{c}^{0},\rho^{+})
  + {\cal M}(\Xi_{cc}^{+}, \omega   ,     \Lambda_{c}^{+},   \Sigma_{c}^{0},    \Sigma_{c}^{0},\rho^{+})
  + {\cal M}(\Xi_{cc}^{+}, \eta_8   ,       \Sigma_{c}^{+},    \Sigma_{c}^{0},   \Sigma_{c}^{0},\rho^{+})\nonumber\\
&&+ {\cal M}(\Xi_{cc}^{+}, \omega   ,     \Sigma_{c}^{+},    \Sigma_{c}^{0},   \Sigma_{c}^{0},\rho^{+})
  + {\cal M}(\Xi_{cc}^{+}, \pi^{+},       \Sigma_{c}^{0},    \pi^{0}   ,   \rho^{+},\Sigma_{c}^{0})
  + {\cal M}(\Xi_{cc}^{+}, \rho^{+},      \Sigma_{c}^{0},    \rho^{0}   ,  \rho^{+},\Sigma_{c}^{0})\nonumber\\
&&+ {\cal M}(\Xi_{cc}^{+},  K^{+},    \Xi_{c}^{0},     \overline{K}^{0},   \rho^{+},\Sigma_{c}^{0})
  + {\cal M}(\Xi_{cc}^{+},  K^{+},    \Xi_{c}^{0},     \overline{K}^{*0},  \rho^{+},\Sigma_{c}^{0})
  + {\cal M}(\Xi_{cc}^{+},  K^{+},    \Xi_{c}^{'0},    \overline{K}^{0},   \rho^{+},\Sigma_{c}^{0})\nonumber\\
&&+ {\cal M}(\Xi_{cc}^{+},  K^{+},    \Xi_{c}^{'0},    \overline{K}^{*0},  \rho^{+},\Sigma_{c}^{0})
  + {\cal M}(\Xi_{cc}^{+}, \eta_1   ,       \Lambda_{c}^{+},   \Sigma_{c}^{0},    \Sigma_{c}^{0},\rho^{+})
  + {\cal M}(\Xi_{cc}^{+}, \eta_1   ,       \Sigma_{c}^{+},    \Sigma_{c}^{0},   \Sigma_{c}^{0},\rho^{+})],
\end{eqnarray}
\begin{eqnarray}
{\cal A}(\Xi_{cc}^{+} \to \Sigma_{c}^{+}\phi)
&=&C_{\rm SD}(\Xi_{cc}^{+} \to \Sigma_{c}^{+}\phi)+ i [
    {\cal M}(\Xi_{cc}^{+}, K^{+},  \Xi_{c}^{0},   K^{-},   \phi,\Sigma_{c}^{+})
  + {\cal M}(\Xi_{cc}^{+}, K^{*+}, \Xi_{c}^{0},   K^{*-},  \phi,\Sigma_{c}^{+})\nonumber\\
&&+ {\cal M}(\Xi_{cc}^{+}, K^{+},  \Xi_{c}^{'0},  K^{-},   \phi,\Sigma_{c}^{+})
  + {\cal M}(\Xi_{cc}^{+}, K^{*+},  \Xi_{c}^{'0}, K^{*-},   \phi,\Sigma_{c}^{+})
  + {\cal M}(\Xi_{cc}^{+}, K^{+},   \Xi_{c}^{0},  \Xi_{c}^{0},    \Sigma_{c}^{+},\phi)\nonumber\\
&&+ {\cal M}(\Xi_{cc}^{+}, K^{*+},  \Xi_{c}^{0},  \Xi_{c}^{0},    \Sigma_{c}^{+},\phi)
  + {\cal M}(\Xi_{cc}^{+}, K^{+},   \Xi_{c}^{0},  \Xi_{c}^{'0},    \Sigma_{c}^{+},\phi)
  + {\cal M}(\Xi_{cc}^{+}, K^{*+},   \Xi_{c}^{0},  \Xi_{c}^{'0},    \Sigma_{c}^{+},\phi)\nonumber\\
&&+ {\cal M}(\Xi_{cc}^{+}, K^{+},    \Xi_{c}^{'0},  \Xi_{c}^{0},    \Sigma_{c}^{+},\phi)
  + {\cal M}(\Xi_{cc}^{+}, K^{*+},   \Xi_{c}^{'0},   \Xi_{c}^{0},    \Sigma_{c}^{+},\phi)
  + {\cal M}(\Xi_{cc}^{+}, K^{+},     \Xi_{c}^{'0},    \Xi_{c}^{'0},    \Sigma_{c}^{+},\phi)\nonumber\\
&&+ {\cal M}(\Xi_{cc}^{+}, K^{*+},    \Xi_{c}^{'0},    \Xi_{c}^{'0},    \Sigma_{c}^{+},\phi)
  ],
\end{eqnarray}
\begin{eqnarray}
{\cal A}(\Xi_{cc}^{+} \to \Lambda_{c}^{+}\phi)
&=&C_{\rm SD}(\Xi_{cc}^{+} \to \Lambda_{c}^{+}\phi)+ i [
    {\cal M}(\Xi_{cc}^{+}, K^{+},  \Xi_{c}^{0},   K^{-},   \phi,\Lambda_{c}^{+})
  + {\cal M}(\Xi_{cc}^{+}, K^{*+}, \Xi_{c}^{0},   K^{*-},  \phi,\Lambda_{c}^{+})\nonumber\\
&&+ {\cal M}(\Xi_{cc}^{+}, K^{+},  \Xi_{c}^{'0},  K^{-},   \phi,\Lambda_{c}^{+})
  + {\cal M}(\Xi_{cc}^{+}, K^{*+},  \Xi_{c}^{'0}, K^{*-},   \phi,\Lambda_{c}^{+})
  + {\cal M}(\Xi_{cc}^{+}, K^{+},   \Xi_{c}^{0},  \Xi_{c}^{0},    \Lambda_{c}^{+},\phi)\nonumber\\
&&+ {\cal M}(\Xi_{cc}^{+}, K^{*+},  \Xi_{c}^{0},  \Xi_{c}^{0},    \Lambda_{c}^{+},\phi)
  + {\cal M}(\Xi_{cc}^{+}, K^{+},   \Xi_{c}^{0},  \Xi_{c}^{'0},    \Lambda_{c}^{+},\phi)
  + {\cal M}(\Xi_{cc}^{+}, K^{*+},   \Xi_{c}^{0},  \Xi_{c}^{'0},    \Lambda_{c}^{+},\phi)\nonumber\\
&&+ {\cal M}(\Xi_{cc}^{+}, K^{+},    \Xi_{c}^{'0},  \Xi_{c}^{0},    \Lambda_{c}^{+},\phi)
  + {\cal M}(\Xi_{cc}^{+}, K^{*+},   \Xi_{c}^{'0},   \Xi_{c}^{0},    \Lambda_{c}^{+},\phi)
  + {\cal M}(\Xi_{cc}^{+}, K^{+},     \Xi_{c}^{'0},    \Xi_{c}^{'0},    \Lambda_{c}^{+},\phi)\nonumber\\
&&+ {\cal M}(\Xi_{cc}^{+}, K^{*+},    \Xi_{c}^{'0},    \Xi_{c}^{'0},    \Lambda_{c}^{+},\phi)
  ],
\end{eqnarray}
\begin{eqnarray}
{\cal A}(\Xi_{cc}^{+} \to \Xi_{c}^{0}K^{*+})
&=&T_{\rm SD}(\Xi_{cc}^{+} \to \Xi_{c}^{0}K^{*+})+ i [
    {\cal M}(\Xi_{cc}^{+}, \eta_8  ,   \Lambda_{c}^{+},   K^{+},    K^{*+},\Xi_{c}^{0})
  + {\cal M}(\Xi_{cc}^{+}, \phi,     \Lambda_{c}^{+},   K^{*+},   K^{*+},\Xi_{c}^{0})\nonumber\\
&&+ {\cal M}(\Xi_{cc}^{+}, \eta_8  ,   \Sigma_{c}^{+},   K^{+},    K^{*+},\Xi_{c}^{0})
  + {\cal M}(\Xi_{cc}^{+}, \phi,     \Sigma_{c}^{+},   K^{*+},   K^{*+},\Xi_{c}^{0})
  + {\cal M}(\Xi_{cc}^{+}, \eta_8   ,  \Lambda_{c}^{+},  \Xi_{c}^{0},    \Xi_{c}^{0},K^{*0})\nonumber\\
&&+ {\cal M}(\Xi_{cc}^{+}, \phi,     \Lambda_{c}^{+},  \Xi_{c}^{0},    \Xi_{c}^{0},K^{*0})
  + {\cal M}(\Xi_{cc}^{+}, \eta_8   ,  \Lambda_{c}^{+},  \Xi_{c}^{'0},    \Xi_{c}^{0},K^{*0})
  + {\cal M}(\Xi_{cc}^{+}, \phi,     \Lambda_{c}^{+},  \Xi_{c}^{'0},    \Xi_{c}^{0},K^{*0})\nonumber\\
&&+ {\cal M}(\Xi_{cc}^{+}, \eta_8   ,  \Sigma_{c}^{+},  \Xi_{c}^{0},    \Xi_{c}^{0},K^{*0})
  + {\cal M}(\Xi_{cc}^{+}, \phi,     \Sigma_{c}^{+},  \Xi_{c}^{0},    \Xi_{c}^{0},K^{*0})
  + {\cal M}(\Xi_{cc}^{+}, \eta_8   ,  \Sigma_{c}^{+},  \Xi_{c}^{'0},    \Xi_{c}^{0},K^{*0})\nonumber\\
&&+ {\cal M}(\Xi_{cc}^{+}, \phi,     \Sigma_{c}^{+},  \Xi_{c}^{'0},    \Xi_{c}^{0},K^{*0})
  + {\cal M}(\Xi_{cc}^{+}, \pi^{+},  \Sigma_{c}^{0},   K^{0},    K^{*+},\Xi_{c}^{0})
  + {\cal M}(\Xi_{cc}^{+}, \rho^{+},  \Sigma_{c}^{0},   K^{*0},   K^{*+},\Xi_{c}^{0})\nonumber\\
&&+ {\cal M}(\Xi_{cc}^{+}, \pi^{0}   ,  \Lambda_{c}^{+},  \Xi_{c}^{0},    \Xi_{c}^{0},K^{*0})
  + {\cal M}(\Xi_{cc}^{+}, \rho^{0}   ,  \Lambda_{c}^{+},  \Xi_{c}^{0},    \Xi_{c}^{0},K^{*0})
  + {\cal M}(\Xi_{cc}^{+}, \pi^{0}   ,   \Sigma_{c}^{+},   \Xi_{c}^{0},    \Xi_{c}^{0},K^{*0})\nonumber\\
&&+ {\cal M}(\Xi_{cc}^{+}, \rho^{0}  ,   \Sigma_{c}^{+},  \Xi_{c}^{0},    \Xi_{c}^{0},K^{*0})
  + {\cal M}(\Xi_{cc}^{+}, \pi^{0}    ,  \Lambda_{c}^{+},  \Xi_{c}^{'0},    \Xi_{c}^{0},K^{*0})
  + {\cal M}(\Xi_{cc}^{+}, \rho^{0}   ,  \Lambda_{c}^{+},  \Xi_{c}^{'0},    \Xi_{c}^{0},K^{*0})\nonumber\\
&&+ {\cal M}(\Xi_{cc}^{+}, \pi^{0}   ,   \Sigma_{c}^{+},   \Xi_{c}^{'0},    \Xi_{c}^{0},K^{*0})
  + {\cal M}(\Xi_{cc}^{+}, \rho^{0}  ,   \Sigma_{c}^{+},  \Xi_{c}^{'0},    \Xi_{c}^{0},K^{*0})
  + {\cal M}(\Xi_{cc}^{+}, \eta_8      ,  \Lambda_{c}^{+},  \Xi_{c}^{0},    \Xi_{c}^{0},K^{*0})\nonumber\\
&&+ {\cal M}(\Xi_{cc}^{+}, \omega    ,  \Lambda_{c}^{+},  \Xi_{c}^{0},    \Xi_{c}^{0},K^{*0})
  + {\cal M}(\Xi_{cc}^{+}, \eta_8    ,   \Sigma_{c}^{+},   \Xi_{c}^{0},    \Xi_{c}^{0},K^{*0})
  + {\cal M}(\Xi_{cc}^{+}, \omega    ,   \Sigma_{c}^{+},  \Xi_{c}^{0},    \Xi_{c}^{0},K^{*0})\nonumber\\
&&+ {\cal M}(\Xi_{cc}^{+}, \eta_8    ,  \Lambda_{c}^{+},  \Xi_{c}^{'0},    \Xi_{c}^{0},K^{*0})
  + {\cal M}(\Xi_{cc}^{+}, \omega   ,  \Lambda_{c}^{+},  \Xi_{c}^{'0},    \Xi_{c}^{0},K^{*0})
  + {\cal M}(\Xi_{cc}^{+}, \eta_8   ,   \Sigma_{c}^{+},   \Xi_{c}^{'0},    \Xi_{c}^{0},K^{*0})\nonumber\\
&&+ {\cal M}(\Xi_{cc}^{+}, \omega   ,   \Sigma_{c}^{+},  \Xi_{c}^{'0},    \Xi_{c}^{0},K^{*0})
  + {\cal M}(\Xi_{cc}^{+}, K^{+},  \Xi_{c}^{0},   \eta_8   ,    K^{*+},\Xi_{c}^{0})
  + {\cal M}(\Xi_{cc}^{+}, K^{+},  \Xi_{c}^{0},  \phi,        K^{*+},\Xi_{c}^{0})\nonumber\\
&&+ {\cal M}(\Xi_{cc}^{+}, K^{+},  \Xi_{c}^{'0},  \eta_8   ,    K^{*+},\Xi_{c}^{0})
  + {\cal M}(\Xi_{cc}^{+}, K^{+},  \Xi_{c}^{'0},   \phi,      K^{*+},\Xi_{c}^{0})
  + {\cal M}(\Xi_{cc}^{+}, \eta_1   ,  \Lambda_{c}^{+},  \Xi_{c}^{0},    \Xi_{c}^{0},K^{*0})\nonumber\\
&&+ {\cal M}(\Xi_{cc}^{+}, \eta_1   ,  \Sigma_{c}^{+},  \Xi_{c}^{0},    \Xi_{c}^{0},K^{*0})],
\end{eqnarray}
\begin{eqnarray}
{\cal A}(\Xi_{cc}^{+} \to \Xi_{c}^{'0}K^{*+})
&=&T_{\rm SD}(\Xi_{cc}^{+} \to \Xi_{c}^{'0}K^{*+})+ i [
    {\cal M}(\Xi_{cc}^{+}, \eta_8   ,  \Lambda_{c}^{+},   K^{+},    K^{*+},\Xi_{c}^{'0})
  + {\cal M}(\Xi_{cc}^{+}, \phi,     \Lambda_{c}^{+},   K^{*+},   K^{*+},\Xi_{c}^{'0})\nonumber\\
&&+ {\cal M}(\Xi_{cc}^{+}, \eta_8  ,  \Sigma_{c}^{+},   K^{+},    K^{*+},\Xi_{c}^{'0})
  + {\cal M}(\Xi_{cc}^{+}, \phi,     \Sigma_{c}^{+},   K^{*+},   K^{*+},\Xi_{c}^{'0})
  + {\cal M}(\Xi_{cc}^{+}, \eta_8  ,  \Lambda_{c}^{+},  \Xi_{c}^{0},    \Xi_{c}^{'0},K^{*0})\nonumber\\
&&+ {\cal M}(\Xi_{cc}^{+}, \phi,     \Lambda_{c}^{+},  \Xi_{c}^{0},    \Xi_{c}^{'0},K^{*0})
  + {\cal M}(\Xi_{cc}^{+}, \eta_8 ,  \Lambda_{c}^{+},  \Xi_{c}^{'0},    \Xi_{c}^{'0},K^{*0})
  + {\cal M}(\Xi_{cc}^{+}, \phi,     \Lambda_{c}^{+},  \Xi_{c}^{'0},    \Xi_{c}^{'0},K^{*0})\nonumber\\
&&+ {\cal M}(\Xi_{cc}^{+}, \eta_8  ,  \Sigma_{c}^{+},  \Xi_{c}^{0},    \Xi_{c}^{'0},K^{*0})
  + {\cal M}(\Xi_{cc}^{+}, \phi,     \Sigma_{c}^{+},  \Xi_{c}^{0},    \Xi_{c}^{'0},K^{*0})
  + {\cal M}(\Xi_{cc}^{+}, \eta_8  ,  \Sigma_{c}^{+},  \Xi_{c}^{'0},    \Xi_{c}^{'0},K^{*0})\nonumber\\
&&+ {\cal M}(\Xi_{cc}^{+}, \phi,     \Sigma_{c}^{+},  \Xi_{c}^{'0},    \Xi_{c}^{'0},K^{*0})
  + {\cal M}(\Xi_{cc}^{+}, \pi^{+},  \Sigma_{c}^{0},   K^{0},    K^{*+},\Xi_{c}^{'0})
  + {\cal M}(\Xi_{cc}^{+}, \rho^{+},  \Sigma_{c}^{0},   K^{*0},   K^{*+},\Xi_{c}^{'0})\nonumber\\
&&+ {\cal M}(\Xi_{cc}^{+}, \pi^{0} ,  \Lambda_{c}^{+},  \Xi_{c}^{0},    \Xi_{c}^{'0},K^{*0})
  + {\cal M}(\Xi_{cc}^{+}, \rho^{0} ,  \Lambda_{c}^{+},  \Xi_{c}^{0},    \Xi_{c}^{'0},K^{*0})
  + {\cal M}(\Xi_{cc}^{+}, \pi^{0} ,   \Sigma_{c}^{+},   \Xi_{c}^{0},    \Xi_{c}^{'0},K^{*0})\nonumber\\
&&+ {\cal M}(\Xi_{cc}^{+}, \rho^{0} ,   \Sigma_{c}^{+},  \Xi_{c}^{0},    \Xi_{c}^{'0},K^{*0})
  + {\cal M}(\Xi_{cc}^{+}, \pi^{0} ,  \Lambda_{c}^{+},  \Xi_{c}^{'0},    \Xi_{c}^{'0},K^{*0})
  + {\cal M}(\Xi_{cc}^{+}, \rho^{0} ,  \Lambda_{c}^{+},  \Xi_{c}^{'0},    \Xi_{c}^{'0},K^{*0})\nonumber\\
&&+ {\cal M}(\Xi_{cc}^{+}, \pi^{0} ,   \Sigma_{c}^{+},   \Xi_{c}^{'0},    \Xi_{c}^{'0},K^{*0})
  + {\cal M}(\Xi_{cc}^{+}, \rho^{0} ,   \Sigma_{c}^{+},  \Xi_{c}^{'0},    \Xi_{c}^{'0},K^{*0})
  + {\cal M}(\Xi_{cc}^{+}, \eta_8 ,  \Lambda_{c}^{+},  \Xi_{c}^{0},    \Xi_{c}^{'0},K^{*0})\nonumber\\
&&+ {\cal M}(\Xi_{cc}^{+}, \omega ,  \Lambda_{c}^{+},  \Xi_{c}^{0},    \Xi_{c}^{'0},K^{*0})
  + {\cal M}(\Xi_{cc}^{+}, \eta_8 ,   \Sigma_{c}^{+},   \Xi_{c}^{0},    \Xi_{c}^{'0},K^{*0})
  + {\cal M}(\Xi_{cc}^{+}, \omega ,   \Sigma_{c}^{+},  \Xi_{c}^{0},    \Xi_{c}^{'0},K^{*0})\nonumber\\
&&+ {\cal M}(\Xi_{cc}^{+}, \eta_8 ,  \Lambda_{c}^{+},  \Xi_{c}^{'0},    \Xi_{c}^{'0},K^{*0})
  + {\cal M}(\Xi_{cc}^{+}, \omega ,  \Lambda_{c}^{+},  \Xi_{c}^{'0},    \Xi_{c}^{'0},K^{*0})
  + {\cal M}(\Xi_{cc}^{+}, \eta_8 ,   \Sigma_{c}^{+},   \Xi_{c}^{'0},    \Xi_{c}^{'0},K^{*0})\nonumber\\
&&+ {\cal M}(\Xi_{cc}^{+}, \omega ,   \Sigma_{c}^{+},  \Xi_{c}^{'0},    \Xi_{c}^{'0},K^{*0})
  + {\cal M}(\Xi_{cc}^{+}, K^{+},  \Xi_{c}^{0},   \eta_8 ,    K^{*+},\Xi_{c}^{'0})
  + {\cal M}(\Xi_{cc}^{+}, K^{+},  \Xi_{c}^{0},  \phi,        K^{*+},\Xi_{c}^{'0})\nonumber\\
&&+ {\cal M}(\Xi_{cc}^{+}, K^{+},  \Xi_{c}^{'0},  \eta_8 ,    K^{*+},\Xi_{c}^{'0})
  + {\cal M}(\Xi_{cc}^{+}, K^{+},  \Xi_{c}^{'0},   \phi,      K^{*+},\Xi_{c}^{'0})
  + {\cal M}(\Xi_{cc}^{+}, \eta_1 ,  \Lambda_{c}^{+},  \Xi_{c}^{'0},    \Xi_{c}^{'0},K^{*0})\nonumber\\
&&+ {\cal M}(\Xi_{cc}^{+}, \eta_1  ,  \Sigma_{c}^{+},  \Xi_{c}^{'0},    \Xi_{c}^{'0},K^{*0})],
\end{eqnarray}
\begin{eqnarray}
{\cal A}(\Xi_{cc}^{+} \to \Xi_{c}^{+}K^{*0})
&=&i [
    {\cal M}(\Xi_{cc}^{+}, K^{+},  \Xi_{c}^{0},   \pi^{-},    K^{*0},\Xi_{c}^{+})
  + {\cal M}(\Xi_{cc}^{+}, K^{*+}, \Xi_{c}^{0},   \rho^{-},    K^{*0},\Xi_{c}^{+})
  + {\cal M}(\Xi_{cc}^{+}, K^{+},  \Xi_{c}^{'0},  \pi^{-},    K^{*0},\Xi_{c}^{+})\nonumber\\
&&+ {\cal M}(\Xi_{cc}^{+}, K^{*+}, \Xi_{c}^{'0},  \rho^{-},   K^{*0},\Xi_{c}^{+})
  + {\cal M}(\Xi_{cc}^{+}, K^{+},  \Xi_{c}^{0},   \Omega_{c}^{0},    \Xi_{c}^{+},K^{*0})
  + {\cal M}(\Xi_{cc}^{+}, K^{*+}, \Xi_{c}^{0},   \Omega_{c}^{0},    \Xi_{c}^{+},K^{*0})\nonumber\\
&&+ {\cal M}(\Xi_{cc}^{+}, K^{+},  \Xi_{c}^{'0},  \Omega_{c}^{0},    \Xi_{c}^{+},K^{*0})
  + {\cal M}(\Xi_{cc}^{+}, K^{*+}, \Xi_{c}^{'0},  \Omega_{c}^{0},    \Xi_{c}^{+},K^{*0})
  + {\cal M}(\Xi_{cc}^{+}, \eta_8,  \Lambda_{c}^{+},   K^{0},    K^{*0},\Xi_{c}^{+})\nonumber\\
&&+ {\cal M}(\Xi_{cc}^{+}, \phi,     \Lambda_{c}^{+},   K^{*0},   K^{*0},\Xi_{c}^{+})
  + {\cal M}(\Xi_{cc}^{+}, \eta_8,  \Sigma_{c}^{+},    K^{0},    K^{*0},\Xi_{c}^{+})
  + {\cal M}(\Xi_{cc}^{+}, \phi,     \Sigma_{c}^{+},    K^{*0},   K^{*+},\Xi_{c}^{+})\nonumber\\
&&+ {\cal M}(\Xi_{cc}^{+}, \eta_8,  \Lambda_{c}^{+},   \Xi_{c}^{+},    \Xi_{c}^{+},K^{*0})
  + {\cal M}(\Xi_{cc}^{+}, \phi,     \Lambda_{c}^{+},   \Xi_{c}^{+},    \Xi_{c}^{+},K^{*0})
  + {\cal M}(\Xi_{cc}^{+}, \eta_8,  \Lambda_{c}^{+},   \Xi_{c}^{'+},    \Xi_{c}^{+},K^{*0})\nonumber\\
&&+ {\cal M}(\Xi_{cc}^{+}, \phi,     \Lambda_{c}^{+},   \Xi_{c}^{'+},    \Xi_{c}^{+},K^{*0})
  + {\cal M}(\Xi_{cc}^{+}, \eta_8,  \Sigma_{c}^{+},   \Xi_{c}^{+},    \Xi_{c}^{+},K^{*0})
  + {\cal M}(\Xi_{cc}^{+}, \phi,     \Sigma_{c}^{+},   \Xi_{c}^{+},    \Xi_{c}^{+},K^{*0})\nonumber\\
&&+ {\cal M}(\Xi_{cc}^{+}, \eta_8,  \Sigma_{c}^{+},   \Xi_{c}^{'+},    \Xi_{c}^{+},K^{*0})
  + {\cal M}(\Xi_{cc}^{+}, \phi,     \Sigma_{c}^{+},   \Xi_{c}^{'+},    \Xi_{c}^{+},K^{*0})
  + {\cal M}(\Xi_{cc}^{+}, \pi^{0}   ,  \Lambda_{c}^{+},   K^{0},    K^{*0},\Xi_{c}^{+})\nonumber\\
&&+ {\cal M}(\Xi_{cc}^{+}, \rho^{0}   ,  \Lambda_{c}^{+},   K^{*0},   K^{*0},\Xi_{c}^{+})
  + {\cal M}(\Xi_{cc}^{+}, \pi^{0}   ,  \Sigma_{c}^{+},   K^{0},    K^{*0},\Xi_{c}^{+})
  + {\cal M}(\Xi_{cc}^{+}, \rho^{0}   ,  \Sigma_{c}^{+},   K^{*0},   K^{*0},\Xi_{c}^{+})\nonumber\\
&&+ {\cal M}(\Xi_{cc}^{+}, \eta_8     ,   \Lambda_{c}^{+},    K^{0},    K^{*0},\Xi_{c}^{+})
  + {\cal M}(\Xi_{cc}^{+}, \omega    ,  \Lambda_{c}^{+},   K^{*0},   K^{*0},\Xi_{c}^{+})
  + {\cal M}(\Xi_{cc}^{+}, \eta_8    ,    \Sigma_{c}^{+},    K^{0},    K^{*0},\Xi_{c}^{+})\nonumber\\
&&+ {\cal M}(\Xi_{cc}^{+}, \omega    ,  \Sigma_{c}^{+},   K^{*0},   K^{*0},\Xi_{c}^{+})
  + {\cal M}(\Xi_{cc}^{+}, \pi^{+},  \Sigma_{c}^{0},   \Xi_{c}^{0},    \Xi_{c}^{+},K^{*0})
  + {\cal M}(\Xi_{cc}^{+}, \rho^{+}, \Sigma_{c}^{0},   \Xi_{c}^{0},    \Xi_{c}^{+},K^{*0})\nonumber\\
&&+ {\cal M}(\Xi_{cc}^{+}, \pi^{+},  \Sigma_{c}^{0},  \Xi_{c}^{'0},    \Xi_{c}^{+},K^{*0})
  + {\cal M}(\Xi_{cc}^{+}, \rho^{+}, \Sigma_{c}^{0},  \Xi_{c}^{'0},    \Xi_{c}^{+},K^{*0})
  + {\cal M}(\Xi_{cc}^{+}, \eta_1,  \Lambda_{c}^{+},   \Xi_{c}^{+},    \Xi_{c}^{+},K^{*0})\nonumber\\
&&+ {\cal M}(\Xi_{cc}^{+}, \eta_1 ,  \Sigma_{c}^{+},   \Xi_{c}^{+},    \Xi_{c}^{+},K^{*0})
  ],
\end{eqnarray}
\begin{eqnarray}
{\cal A}(\Xi_{cc}^{+} \to \Xi_{c}^{'+}K^{*0})
&=&i [
    {\cal M}(\Xi_{cc}^{+}, K^{+},  \Xi_{c}^{0},   \pi^{-},    K^{*0},\Xi_{c}^{'+})
  + {\cal M}(\Xi_{cc}^{+}, K^{*+}, \Xi_{c}^{0},   \rho^{-},    K^{*0},\Xi_{c}^{'+})
  + {\cal M}(\Xi_{cc}^{+}, K^{+},  \Xi_{c}^{'0},  \pi^{-},    K^{*0},\Xi_{c}^{'+})\nonumber\\
&&+ {\cal M}(\Xi_{cc}^{+}, K^{*+}, \Xi_{c}^{'0},  \rho^{-},   K^{*0},\Xi_{c}^{'+})
  + {\cal M}(\Xi_{cc}^{+}, K^{+},  \Xi_{c}^{0},   \Omega_{c}^{0},    \Xi_{c}^{'+},K^{*0})
  + {\cal M}(\Xi_{cc}^{+}, K^{*+}, \Xi_{c}^{0},   \Omega_{c}^{0},    \Xi_{c}^{'+},K^{*0})\nonumber\\
&&+ {\cal M}(\Xi_{cc}^{+}, K^{+},  \Xi_{c}^{'0},  \Omega_{c}^{0},    \Xi_{c}^{'+},K^{*0})
  + {\cal M}(\Xi_{cc}^{+}, K^{*+}, \Xi_{c}^{'0},  \Omega_{c}^{0},    \Xi_{c}^{'+},K^{*0})
  + {\cal M}(\Xi_{cc}^{+}, \eta_8,  \Lambda_{c}^{+},   K^{0},    K^{*0},\Xi_{c}^{'+})\nonumber\\
&&+ {\cal M}(\Xi_{cc}^{+}, \phi,     \Lambda_{c}^{+},   K^{*0},   K^{*0},\Xi_{c}^{'+})
  + {\cal M}(\Xi_{cc}^{+}, \eta_8,  \Sigma_{c}^{+},    K^{0},    K^{*0},\Xi_{c}^{'+})
  + {\cal M}(\Xi_{cc}^{+}, \phi,     \Sigma_{c}^{+},    K^{*0},   K^{*+},\Xi_{c}^{'+})\nonumber\\
&&+ {\cal M}(\Xi_{cc}^{+}, \eta_8,  \Lambda_{c}^{+},   \Xi_{c}^{+},    \Xi_{c}^{'+},K^{*0})
  + {\cal M}(\Xi_{cc}^{+}, \phi,     \Lambda_{c}^{+},   \Xi_{c}^{+},    \Xi_{c}^{'+},K^{*0})
  + {\cal M}(\Xi_{cc}^{+}, \eta_8,  \Lambda_{c}^{+},   \Xi_{c}^{'+},    \Xi_{c}^{'+},K^{*0})\nonumber\\
&&+ {\cal M}(\Xi_{cc}^{+}, \phi,     \Lambda_{c}^{+},   \Xi_{c}^{'+},    \Xi_{c}^{'+},K^{*0})
  + {\cal M}(\Xi_{cc}^{+}, \eta_8,  \Sigma_{c}^{+},   \Xi_{c}^{+},    \Xi_{c}^{'+},K^{*0})
  + {\cal M}(\Xi_{cc}^{+}, \phi,     \Sigma_{c}^{+},   \Xi_{c}^{+},    \Xi_{c}^{'+},K^{*0})\nonumber\\
&&+ {\cal M}(\Xi_{cc}^{+}, \eta_8,  \Sigma_{c}^{+},   \Xi_{c}^{'+},    \Xi_{c}^{'+},K^{*0})
  + {\cal M}(\Xi_{cc}^{+}, \phi,     \Sigma_{c}^{+},   \Xi_{c}^{'+},    \Xi_{c}^{'+},K^{*0})
  + {\cal M}(\Xi_{cc}^{+}, \pi^{0}   ,  \Lambda_{c}^{+},   K^{0},    K^{*0},\Xi_{c}^{'+})\nonumber\\
&&+ {\cal M}(\Xi_{cc}^{+}, \rho^{0}   ,  \Lambda_{c}^{+},   K^{*0},   K^{*0},\Xi_{c}^{'+})
  + {\cal M}(\Xi_{cc}^{+}, \pi^{0}   ,  \Sigma_{c}^{+},   K^{0},    K^{*0},\Xi_{c}^{'+})
  + {\cal M}(\Xi_{cc}^{+}, \rho^{0}   ,  \Sigma_{c}^{+},   K^{*0},   K^{*0},\Xi_{c}^{'+})\nonumber\\
&&+ {\cal M}(\Xi_{cc}^{+}, \eta_8,   \Lambda_{c}^{+},    K^{0},    K^{*0},\Xi_{c}^{'+})
  + {\cal M}(\Xi_{cc}^{+}, \omega    ,  \Lambda_{c}^{+},   K^{*0},   K^{*0},\Xi_{c}^{'+})
  + {\cal M}(\Xi_{cc}^{+}, \eta_8,    \Sigma_{c}^{+},    K^{0},    K^{*0},\Xi_{c}^{'+})\nonumber\\
&&+ {\cal M}(\Xi_{cc}^{+}, \omega    ,  \Sigma_{c}^{+},   K^{*0},   K^{*0},\Xi_{c}^{'+})
  + {\cal M}(\Xi_{cc}^{+}, \pi^{+},  \Sigma_{c}^{0},   \Xi_{c}^{0},    \Xi_{c}^{'+},K^{*0})
  + {\cal M}(\Xi_{cc}^{+}, \rho^{+}, \Sigma_{c}^{0},   \Xi_{c}^{0},    \Xi_{c}^{'+},K^{*0})\nonumber\\
&&+ {\cal M}(\Xi_{cc}^{+}, \pi^{+},  \Sigma_{c}^{0},  \Xi_{c}^{'0},    \Xi_{c}^{'+},K^{*0})
  + {\cal M}(\Xi_{cc}^{+}, \rho^{+}, \Sigma_{c}^{0},  \Xi_{c}^{'0},    \Xi_{c}^{'+},K^{*0})
  + {\cal M}(\Xi_{cc}^{+}, \eta_1,  \Lambda_{c}^{+},   \Xi_{c}^{'+},    \Xi_{c}^{'+},K^{*0})\nonumber\\
&&+ {\cal M}(\Xi_{cc}^{+}, \eta_1,  \Sigma_{c}^{+},   \Xi_{c}^{'+},    \Xi_{c}^{'+},K^{*0})],
\end{eqnarray}
\begin{eqnarray}
{\cal A}(\Xi_{cc}^{+} \to \Sigma_{c}^{0}K^{*+})
&=&T_{\rm SD}(\Xi_{cc}^{+} \to \Sigma_{c}^{0}K^{*+})+ i [
   {\cal M}(\Xi_{cc}^{+}, K^{0}, \Lambda_{c}^{+},   \pi^{+},    K^{*+},\Sigma_{c}^{0})
  + {\cal M}(\Xi_{cc}^{+}, K^{*0}, \Lambda_{c}^{+}, \rho^{+},   K^{*+},\Sigma_{c}^{0})\nonumber\\
&&+{\cal M}(\Xi_{cc}^{+}, K^{0}, \Lambda_{c}^{+},    \pi^{+},   K^{*+},\Sigma_{c}^{0})
  + {\cal M}(\Xi_{cc}^{+}, K^{*0}, \Lambda_{c}^{+},  \rho^{+},  K^{*+},\Sigma_{c}^{0})
  +{\cal M}(\Xi_{cc}^{+}, K^{0},   \Lambda_{c}^{+},  \Xi_{c}^{0},   \Sigma_{c}^{0},K^{*+})\nonumber\\
&&+ {\cal M}(\Xi_{cc}^{+}, K^{*0}, \Lambda_{c}^{+},  \Xi_{c}^{0},   \Sigma_{c}^{0},K^{*+})
  +{\cal M}(\Xi_{cc}^{+}, K^{0},   \Lambda_{c}^{+},  \Xi_{c}^{'0},   \Sigma_{c}^{0},K^{*+})
  + {\cal M}(\Xi_{cc}^{+}, K^{*0}, \Lambda_{c}^{+},  \Xi_{c}^{'0},  \Sigma_{c}^{0},K^{*+})\nonumber\\
&&+{\cal M}(\Xi_{cc}^{+}, K^{0},   \Sigma_{c}^{+},  \Xi_{c}^{0},   \Sigma_{c}^{0},K^{*+})
  + {\cal M}(\Xi_{cc}^{+}, K^{*0}, \Sigma_{c}^{+},  \Xi_{c}^{0},  \Sigma_{c}^{0},K^{*+})
  +{\cal M}(\Xi_{cc}^{+}, K^{0},   \Sigma_{c}^{+},  \Xi_{c}^{'0},   \Sigma_{c}^{0},K^{*+})\nonumber\\
&&+ {\cal M}(\Xi_{cc}^{+}, K^{*0}, \Sigma_{c}^{+},  \Xi_{c}^{'0},  \Sigma_{c}^{0},K^{*+})
  ],
\end{eqnarray}
\begin{eqnarray}
{\cal A}(\Xi_{cc}^{+} \to \Sigma_{c}^{+}K^{*0})
&=&C_{\rm SD}(\Xi_{cc}^{+} \to \Sigma_{c}^{+}K^{*0})+ i [
    {\cal M}(\Xi_{cc}^{+}, K^{+}, \Sigma_{c}^{0},  \pi^{-}, K^{*0},\Sigma_{c}^{+})
  + {\cal M}(\Xi_{cc}^{+}, K^{*+}, \Sigma_{c}^{0}, \rho^{-}, K^{*0},\Sigma_{c}^{+})\nonumber\\
&&+ {\cal M}(\Xi_{cc}^{+}, K^{+},  \Sigma_{c}^{0},  \Xi_{c}^{0}, \Sigma_{c}^{+},K^{*0})
  + {\cal M}(\Xi_{cc}^{+}, K^{*+}, \Sigma_{c}^{0},  \Xi_{c}^{0}, \Sigma_{c}^{+},K^{*0})
  + {\cal M}(\Xi_{cc}^{+}, K^{+},   \Sigma_{c}^{0},  \Xi_{c}^{'0}, \Sigma_{c}^{+},K^{*0})\nonumber\\
&&+ {\cal M}(\Xi_{cc}^{+}, K^{*+},   \Sigma_{c}^{0},  \Xi_{c}^{'0}, \Sigma_{c}^{+},K^{*0})
  +{\cal M}(\Xi_{cc}^{+}, K^{0}, \Lambda_{c}^{+},  \pi^{0}    ,   K^{*0},\Sigma_{c}^{+})
  + {\cal M}(\Xi_{cc}^{+}, K^{*0}, \Lambda_{c}^{+}, \rho^{0}    , K^{*0},\Sigma_{c}^{+})\nonumber\\
&&+{\cal M}(\Xi_{cc}^{+}, K^{0}, \Sigma_{c}^{+},  \eta_8     ,      K^{*0},\Sigma_{c}^{+})
  + {\cal M}(\Xi_{cc}^{+}, K^{*0}, \Sigma_{c}^{+}, \omega     ,   K^{*0},\Sigma_{c}^{+})
  +{\cal M}(\Xi_{cc}^{+}, K^{0},   \Lambda_{c}^{+},  \Xi_{c}^{+},   \Sigma_{c}^{+},K^{*0})\nonumber\\
&&+ {\cal M}(\Xi_{cc}^{+}, K^{*0}, \Lambda_{c}^{+},  \Xi_{c}^{+},   \Sigma_{c}^{+},K^{*0})
  +{\cal M}(\Xi_{cc}^{+}, K^{0},   \Lambda_{c}^{+},  \Xi_{c}^{'+},   \Sigma_{c}^{+},K^{*0})
  + {\cal M}(\Xi_{cc}^{+}, K^{*0}, \Lambda_{c}^{+},  \Xi_{c}^{'+},  \Sigma_{c}^{+},K^{*0})\nonumber\\
&&+{\cal M}(\Xi_{cc}^{+}, K^{0},   \Sigma_{c}^{+},  \Xi_{c}^{+},   \Sigma_{c}^{+},K^{*0})
  + {\cal M}(\Xi_{cc}^{+}, K^{*0}, \Sigma_{c}^{+},  \Xi_{c}^{+},  \Sigma_{c}^{+},K^{*0})
  +{\cal M}(\Xi_{cc}^{+}, K^{0},   \Sigma_{c}^{+},  \Xi_{c}^{'+},   \Sigma_{c}^{+},K^{*0})\nonumber\\
&&+ {\cal M}(\Xi_{cc}^{+}, K^{*0}, \Sigma_{c}^{+},  \Xi_{c}^{'+},  \Sigma_{c}^{+},K^{*0})],
\end{eqnarray}
\begin{eqnarray}
{\cal A}(\Xi_{cc}^{+} \to \Lambda_{c}^{+}K^{*0})
&=&C_{\rm SD}(\Xi_{cc}^{+} \to \Lambda_{c}^{+}K^{*0})+ i [
    {\cal M}(\Xi_{cc}^{+}, K^{+}, \Sigma_{c}^{0},  \pi^{-}, K^{*0},\Lambda_{c}^{+})
  + {\cal M}(\Xi_{cc}^{+}, K^{*+}, \Sigma_{c}^{0}, \rho^{-}, K^{*0},\Lambda_{c}^{+})\nonumber\\
&&+ {\cal M}(\Xi_{cc}^{+}, K^{+},  \Sigma_{c}^{0},  \Xi_{c}^{0}, \Lambda_{c}^{+},K^{*0})
  + {\cal M}(\Xi_{cc}^{+}, K^{*+}, \Sigma_{c}^{0},  \Xi_{c}^{0}, \Lambda_{c}^{+},K^{*0})
  + {\cal M}(\Xi_{cc}^{+}, K^{+},   \Sigma_{c}^{0},  \Xi_{c}^{'0}, \Lambda_{c}^{+},K^{*0})\nonumber\\
&&+ {\cal M}(\Xi_{cc}^{+}, K^{*+},   \Sigma_{c}^{0},  \Xi_{c}^{'0}, \Lambda_{c}^{+},K^{*0})
  +{\cal M}(\Xi_{cc}^{+}, K^{0}, \Lambda_{c}^{+},  \eta_8    ,      K^{*0},\Lambda_{c}^{+})
  + {\cal M}(\Xi_{cc}^{+}, K^{*0}, \Lambda_{c}^{+}, \omega    ,   K^{*0},\Lambda_{c}^{+})\nonumber\\
&&+{\cal M}(\Xi_{cc}^{+}, K^{0}, \Sigma_{c}^{+},  \pi^{0}     ,   K^{*0},\Lambda_{c}^{+})
  + {\cal M}(\Xi_{cc}^{+}, K^{*0}, \Sigma_{c}^{+}, \rho^{0}     , K^{*0},\Lambda_{c}^{+})
  +{\cal M}(\Xi_{cc}^{+}, K^{0},   \Lambda_{c}^{+},  \Xi_{c}^{+},   \Lambda_{c}^{+},K^{*0})\nonumber\\
&&+ {\cal M}(\Xi_{cc}^{+}, K^{*0}, \Lambda_{c}^{+},  \Xi_{c}^{+},  \Lambda_{c}^{+},K^{*0})
  +{\cal M}(\Xi_{cc}^{+}, K^{0},   \Lambda_{c}^{+},  \Xi_{c}^{'+},   \Lambda_{c}^{+},K^{*0})
  + {\cal M}(\Xi_{cc}^{+}, K^{*0}, \Lambda_{c}^{+},  \Xi_{c}^{'+},  \Lambda_{c}^{+},K^{*0})\nonumber\\
&&+{\cal M}(\Xi_{cc}^{+}, K^{0},   \Sigma_{c}^{+},  \Xi_{c}^{+},   \Lambda_{c}^{+},K^{*0})
  + {\cal M}(\Xi_{cc}^{+}, K^{*0}, \Sigma_{c}^{+},  \Xi_{c}^{+},  \Lambda_{c}^{+},K^{*0})
  +{\cal M}(\Xi_{cc}^{+}, K^{0},   \Sigma_{c}^{+},  \Xi_{c}^{'+},   \Lambda_{c}^{+},K^{*0})\nonumber\\
&&+ {\cal M}(\Xi_{cc}^{+}, K^{*0}, \Sigma_{c}^{+},  \Xi_{c}^{'+},  \Lambda_{c}^{+},K^{*0})],
\end{eqnarray}
\begin{eqnarray}
{\cal A}(\Omega_{cc}^{+} \to \Xi_{c}^{+}\overline{K}^{*0})
&=&C_{\rm SD}(\Omega_{cc}^{+} \to \Xi_{c}^{+}\overline{K}^{*0})+ i [
    {\cal M}(\Omega_{cc}^{+}, \pi^{+}, \Omega_{c}^{0},  K^{-}, \overline{K}^{*0},\Xi_{c}^{+})
  + {\cal M}(\Omega_{cc}^{+}, \rho^{+}, \Omega_{c}^{0}, K^{*-}, \overline{K}^{*0},\Xi_{c}^{+})\nonumber\\
&&+ {\cal M}(\Omega_{cc}^{+}, \pi^{+},  \Omega_{c}^{0},  \Xi_{c}^{0}, \Xi_{c}^{+},\overline{K}^{*0})
  + {\cal M}(\Omega_{cc}^{+}, \rho^{+}, \Omega_{c}^{0},  \Xi_{c}^{0}, \Xi_{c}^{+},\overline{K}^{*0})
  + {\cal M}(\Omega_{cc}^{+}, \pi^{+},   \Omega_{c}^{0},  \Xi_{c}^{'0}, \Xi_{c}^{+},\overline{K}^{*0})\nonumber\\
&&+ {\cal M}(\Omega_{cc}^{+}, \rho^{+},   \Omega_{c}^{0},  \Xi_{c}^{'0}, \Xi_{c}^{+},\overline{K}^{*0})
  +{\cal M}(\Omega_{cc}^{+}, \overline{K}^{0},  \Xi_{c}^{+}, \eta_8, \overline{K}^{*0},\Xi_{c}^{+})
  + {\cal M}(\Omega_{cc}^{+}, \overline{K}^{*0}, \Xi_{c}^{+}, \phi, \overline{K}^{*0},\Xi_{c}^{+})\nonumber\\
&&+{\cal M}(\Omega_{cc}^{+}, \overline{K}^{0},  \Xi_{c}^{'+}, \eta_8, \overline{K}^{*0},\Xi_{c}^{+})
  + {\cal M}(\Omega_{cc}^{+}, \overline{K}^{*0}, \Xi_{c}^{'+}, \phi, \overline{K}^{*0},\Xi_{c}^{+})
  +{\cal M}(\Omega_{cc}^{+}, \overline{K}^{0},  \Xi_{c}^{+}, \Lambda_{c}^{+}, \Xi_{c}^{+},\overline{K}^{*0})\nonumber\\
&&+ {\cal M}(\Omega_{cc}^{+}, \overline{K}^{*0}, \Xi_{c}^{+}, \Lambda_{c}^{+},   \Xi_{c}^{+},\overline{K}^{*0})
  +{\cal M}(\Omega_{cc}^{+}, \overline{K}^{0},  \Xi_{c}^{+}, \Sigma_{c}^{+}, \Xi_{c}^{+},\overline{K}^{*0})
  + {\cal M}(\Omega_{cc}^{+}, \overline{K}^{*0}, \Xi_{c}^{+}, \Sigma_{c}^{+},   \Xi_{c}^{+},\overline{K}^{*0})\nonumber\\
&&+{\cal M}(\Omega_{cc}^{+}, \overline{K}^{0},  \Xi_{c}^{'+}, \Lambda_{c}^{+}, \Xi_{c}^{+},\overline{K}^{*0})
  + {\cal M}(\Omega_{cc}^{+}, \overline{K}^{*0}, \Xi_{c}^{'+}, \Lambda_{c}^{+},   \Xi_{c}^{+},\overline{K}^{*0})
  +{\cal M}(\Omega_{cc}^{+}, \overline{K}^{0},  \Xi_{c}^{'+}, \Sigma_{c}^{+}, \Xi_{c}^{+},\overline{K}^{*0})\nonumber\\
&&+ {\cal M}(\Omega_{cc}^{+}, \overline{K}^{*0}, \Xi_{c}^{'+}, \Sigma_{c}^{+},   \Xi_{c}^{+},\overline{K}^{*0})
],
\end{eqnarray}
\begin{eqnarray}
{\cal A}(\Omega_{cc}^{+} \to \Xi_{c}^{'+}\overline{K}^{*0})
&=&C_{\rm SD}(\Omega_{cc}^{+} \to \Xi_{c}^{'+}\overline{K}^{*0})+ i [
    {\cal M}(\Omega_{cc}^{+}, \pi^{+}, \Omega_{c}^{0},  K^{-}, \overline{K}^{*0},\Xi_{c}^{'+})
  + {\cal M}(\Omega_{cc}^{+}, \rho^{+}, \Omega_{c}^{0}, K^{*-}, \overline{K}^{*0},\Xi_{c}^{'+})\nonumber\\
&&+ {\cal M}(\Omega_{cc}^{+}, \pi^{+},  \Omega_{c}^{0},  \Xi_{c}^{0}, \Xi_{c}^{'+},\overline{K}^{*0})
  + {\cal M}(\Omega_{cc}^{+}, \rho^{+}, \Omega_{c}^{0},  \Xi_{c}^{0}, \Xi_{c}^{'+},\overline{K}^{*0})
  + {\cal M}(\Omega_{cc}^{+}, \pi^{+},   \Omega_{c}^{0},  \Xi_{c}^{'0}, \Xi_{c}^{'+},\overline{K}^{*0})\nonumber\\
&&+ {\cal M}(\Omega_{cc}^{+}, \rho^{+},   \Omega_{c}^{0},  \Xi_{c}^{'0}, \Xi_{c}^{'+},\overline{K}^{*0})
  +{\cal M}(\Omega_{cc}^{+}, \overline{K}^{0},  \Xi_{c}^{+}, \eta_8, \overline{K}^{*0},\Xi_{c}^{'+})
  + {\cal M}(\Omega_{cc}^{+}, \overline{K}^{*0}, \Xi_{c}^{+}, \phi, \overline{K}^{*0},\Xi_{c}^{'+})\nonumber\\
&&+{\cal M}(\Omega_{cc}^{+}, \overline{K}^{0},  \Xi_{c}^{'+}, \eta_8, \overline{K}^{*0},\Xi_{c}^{'+})
  + {\cal M}(\Omega_{cc}^{+}, \overline{K}^{*0}, \Xi_{c}^{'+}, \phi, \overline{K}^{*0},\Xi_{c}^{'+})
  +{\cal M}(\Omega_{cc}^{+}, \overline{K}^{0},  \Xi_{c}^{+}, \Lambda_{c}^{+}, \Xi_{c}^{'+},\overline{K}^{*0})\nonumber\\
&&+ {\cal M}(\Omega_{cc}^{+}, \overline{K}^{*0}, \Xi_{c}^{+}, \Lambda_{c}^{+},   \Xi_{c}^{'+},\overline{K}^{*0})
  +{\cal M}(\Omega_{cc}^{+}, \overline{K}^{0},  \Xi_{c}^{+}, \Sigma_{c}^{+}, \Xi_{c}^{'+},\overline{K}^{*0})
  + {\cal M}(\Omega_{cc}^{+}, \overline{K}^{*0}, \Xi_{c}^{+}, \Sigma_{c}^{+},   \Xi_{c}^{'+},\overline{K}^{*0})\nonumber\\
&&+{\cal M}(\Omega_{cc}^{+}, \overline{K}^{0},  \Xi_{c}^{'+}, \Lambda_{c}^{+}, \Xi_{c}^{'+},\overline{K}^{*0})
  + {\cal M}(\Omega_{cc}^{+}, \overline{K}^{*0}, \Xi_{c}^{'+}, \Lambda_{c}^{+},   \Xi_{c}^{'+},\overline{K}^{*0})
  +{\cal M}(\Omega_{cc}^{+}, \overline{K}^{0},  \Xi_{c}^{'+}, \Sigma_{c}^{+}, \Xi_{c}^{'+},\overline{K}^{*0})\nonumber\\
&&+ {\cal M}(\Omega_{cc}^{+}, \overline{K}^{*0}, \Xi_{c}^{'+}, \Sigma_{c}^{+},   \Xi_{c}^{'+},\overline{K}^{*0})
],
\end{eqnarray}
\begin{eqnarray}
{\cal A}(\Omega_{cc}^{+} \to \Omega_{c}^{0}\rho^{+})
&=&T_{\rm SD}(\Omega_{cc}^{+} \to \Omega_{c}^{0}\rho^{+})+ i [
    {\cal M}(\Omega_{cc}^{+}, \overline{K}^{0}, \Xi_{c}^{+},  K^{+}, \rho^{+},\Omega_{c}^{0})
  + {\cal M}(\Omega_{cc}^{+}, \overline{K}^{*0}, \Xi_{c}^{+}, K^{*+},\rho^{+},\Omega_{c}^{0})\nonumber\\
&&+ {\cal M}(\Omega_{cc}^{+}, \overline{K}^{0},  \Xi_{c}^{'+}, K^{+},\rho^{+},\Omega_{c}^{0})
  + {\cal M}(\Omega_{cc}^{+}, \overline{K}^{*0}, \Xi_{c}^{'+}, K^{*+},\rho^{+},\Omega_{c}^{0})
  + {\cal M}(\Omega_{cc}^{+}, \overline{K}^{0},  \Xi_{c}^{+},  \Xi_{c}^{0}, \Omega_{c}^{0},\rho^{+})\nonumber\\
&&+ {\cal M}(\Omega_{cc}^{+}, \overline{K}^{*0},  \Xi_{c}^{+},  \Xi_{c}^{0},  \Omega_{c}^{0},\rho^{+})
  + {\cal M}(\Omega_{cc}^{+}, \overline{K}^{0},   \Xi_{c}^{+},  \Xi_{c}^{'0}, \Omega_{c}^{0},\rho^{+})
  + {\cal M}(\Omega_{cc}^{+}, \overline{K}^{*0},   \Xi_{c}^{+},  \Xi_{c}^{'0}, \Omega_{c}^{0},\rho^{+})\nonumber\\
&&+ {\cal M}(\Omega_{cc}^{+}, \overline{K}^{0},  \Xi_{c}^{'+},  \Xi_{c}^{0}, \Omega_{c}^{0},\rho^{+})
  + {\cal M}(\Omega_{cc}^{+}, \overline{K}^{*0},  \Xi_{c}^{'+},  \Xi_{c}^{0},  \Omega_{c}^{0},\rho^{+})
  + {\cal M}(\Omega_{cc}^{+}, \overline{K}^{0},   \Xi_{c}^{'+},  \Xi_{c}^{'0}, \Omega_{c}^{0},\rho^{+})\nonumber\\
&&+ {\cal M}(\Omega_{cc}^{+}, \overline{K}^{*0},   \Xi_{c}^{'+},  \Xi_{c}^{'0}, \Omega_{c}^{0},\rho^{+})
  ],
\end{eqnarray}
\begin{eqnarray}
{\cal A}(\Omega_{cc}^{+} \to \Sigma_{c}^{+}\overline{K}^{*0})
&=&i [
    {\cal M}(\Omega_{cc}^{+}, \pi^{+}, \Xi_{c}^{0},    K^{-}, \overline{K}^{*0},\Sigma_{c}^{+})
  + {\cal M}(\Omega_{cc}^{+}, \rho^{+}, \Xi_{c}^{0},   K^{*-},\overline{K}^{*0},\Sigma_{c}^{+})
  + {\cal M}(\Omega_{cc}^{+}, \pi^{+},  \Xi_{c}^{'0},  K^{-}, \overline{K}^{*0},\Sigma_{c}^{+})\nonumber\\
&&+ {\cal M}(\Omega_{cc}^{+}, \rho^{+},  \Xi_{c}^{'0}, K^{*-},\overline{K}^{*0},\Sigma_{c}^{+})
  + {\cal M}(\Omega_{cc}^{+},  \pi^{+},  \Xi_{c}^{0},   \Sigma_{c}^{0}, \Sigma_{c}^{+},\overline{K}^{*0})
  + {\cal M}(\Omega_{cc}^{+}, \rho^{+},  \Xi_{c}^{0},   \Sigma_{c}^{0},  \Sigma_{c}^{+},\overline{K}^{*0})\nonumber\\
&&+ {\cal M}(\Omega_{cc}^{+},  \pi^{+},   \Xi_{c}^{'0},  \Sigma_{c}^{0}, \Sigma_{c}^{+},\overline{K}^{*0})
  + {\cal M}(\Omega_{cc}^{+}, \rho^{+},   \Xi_{c}^{'0},  \Sigma_{c}^{0},  \Sigma_{c}^{+},\overline{K}^{*0})
  + {\cal M}(\Omega_{cc}^{+}, \pi^{0}   , \Xi_{c}^{+},    \overline{K}^{0}, \overline{K}^{*0},\Sigma_{c}^{+})\nonumber\\
&&+ {\cal M}(\Omega_{cc}^{+}, \rho^{0}  , \Xi_{c}^{+},   \overline{K}^{*0},\overline{K}^{*0},\Sigma_{c}^{+})
  + {\cal M}(\Omega_{cc}^{+}, \eta_8,     \Xi_{c}^{+},   \overline{K}^{0}, \overline{K}^{*0},\Sigma_{c}^{+})
  + {\cal M}(\Omega_{cc}^{+}, \omega   ,   \Xi_{c}^{+},   \overline{K}^{*0},\overline{K}^{*0},\Sigma_{c}^{+})\nonumber\\
&&+ {\cal M}(\Omega_{cc}^{+}, \pi^{0}    , \Xi_{c}^{'+},    \overline{K}^{0}, \overline{K}^{*0},\Sigma_{c}^{+})
  + {\cal M}(\Omega_{cc}^{+}, \rho^{0}   , \Xi_{c}^{'+},   \overline{K}^{*0},\overline{K}^{*0},\Sigma_{c}^{+})
  + {\cal M}(\Omega_{cc}^{+}, \eta_8,     \Xi_{c}^{'+},   \overline{K}^{0}, \overline{K}^{*0},\Sigma_{c}^{+})\nonumber\\
&&+ {\cal M}(\Omega_{cc}^{+}, \omega   ,   \Xi_{c}^{'+},   \overline{K}^{*0},\overline{K}^{*0},\Sigma_{c}^{+})
  + {\cal M}(\Omega_{cc}^{+}, \pi^{0}    , \Xi_{c}^{+},    \Lambda_{c}^{+}, \Sigma_{c}^{+},\overline{K}^{*0})
  + {\cal M}(\Omega_{cc}^{+}, \rho^{0}   , \Xi_{c}^{+},   \Lambda_{c}^{+},\Sigma_{c}^{+},\overline{K}^{*0})\nonumber\\
&&+ {\cal M}(\Omega_{cc}^{+}, \eta_8,     \Xi_{c}^{+},   \Sigma_{c}^{+}, \Sigma_{c}^{+},\overline{K}^{*0})
  + {\cal M}(\Omega_{cc}^{+}, \omega    ,   \Xi_{c}^{+},   \Sigma_{c}^{+},\Sigma_{c}^{+},\overline{K}^{*0})
  + {\cal M}(\Omega_{cc}^{+}, \pi^{0}   , \Xi_{c}^{'+},    \Lambda_{c}^{+}, \Sigma_{c}^{+},\overline{K}^{*0})\nonumber\\
&&+ {\cal M}(\Omega_{cc}^{+}, \rho^{0}   , \Xi_{c}^{'+},   \Lambda_{c}^{+},\Sigma_{c}^{+},\overline{K}^{*0})
  + {\cal M}(\Omega_{cc}^{+}, \eta_8,     \Xi_{c}^{'+},   \Sigma_{c}^{+}, \Sigma_{c}^{+},\overline{K}^{*0})
  + {\cal M}(\Omega_{cc}^{+}, \omega   ,   \Xi_{c}^{'+},   \Sigma_{c}^{+},\Sigma_{c}^{+},\overline{K}^{*0})\nonumber\\
&&+ {\cal M}(\Omega_{cc}^{+}, \eta_8, \Xi_{c}^{+},    \overline{K}^{0}, \overline{K}^{*0},\Sigma_{c}^{+})
  + {\cal M}(\Omega_{cc}^{+}, \phi,    \Xi_{c}^{+},     \overline{K}^{*0},\overline{K}^{*0},\Sigma_{c}^{+})
  + {\cal M}(\Omega_{cc}^{+}, \eta_8,  \Xi_{c}^{'+},   \overline{K}^{0}, \overline{K}^{*0},\Sigma_{c}^{+})\nonumber\\
&&+ {\cal M}(\Omega_{cc}^{+}, \phi,     \Xi_{c}^{'+},   \overline{K}^{*0},\overline{K}^{*0},\Sigma_{c}^{+})
  + {\cal M}(\Omega_{cc}^{+}, K^{+},  \Omega_{c}^{0},    \Xi_{c}^{0},  \Sigma_{c}^{+},\overline{K}^{*0})
  + {\cal M}(\Omega_{cc}^{+}, K^{*+},  \Omega_{c}^{0},   \Xi_{c}^{0},  \Sigma_{c}^{+},\overline{K}^{*0})\nonumber\\
&&+ {\cal M}(\Omega_{cc}^{+}, K^{+},   \Omega_{c}^{0},   \Xi_{c}^{'0},  \Sigma_{c}^{+},\overline{K}^{*0})
  + {\cal M}(\Omega_{cc}^{+}, K^{*+},  \Omega_{c}^{0},   \Xi_{c}^{'0},  \Sigma_{c}^{+},\overline{K}^{*0})
  + {\cal M}(\Omega_{cc}^{+}, \eta_1,     \Xi_{c}^{+},   \Sigma_{c}^{+}, \Sigma_{c}^{+},\overline{K}^{*0})\nonumber\\
&&+ {\cal M}(\Omega_{cc}^{+}, \eta_1,     \Xi_{c}^{'+},   \Sigma_{c}^{+}, \Sigma_{c}^{+},\overline{K}^{*0})],
\end{eqnarray}
\begin{eqnarray}
{\cal A}(\Omega_{cc}^{+} \to \Lambda_{c}^{+}\overline{K}^{*0})
&=&i [
    {\cal M}(\Omega_{cc}^{+}, \pi^{+}, \Xi_{c}^{0},    K^{-}, \overline{K}^{*0},\Lambda_{c}^{+})
  + {\cal M}(\Omega_{cc}^{+}, \rho^{+}, \Xi_{c}^{0},   K^{*-},\overline{K}^{*0},\Lambda_{c}^{+})
  + {\cal M}(\Omega_{cc}^{+}, \pi^{+},  \Xi_{c}^{'0},  K^{-}, \overline{K}^{*0},\Lambda_{c}^{+})\nonumber\\
&&+ {\cal M}(\Omega_{cc}^{+}, \rho^{+},  \Xi_{c}^{'0}, K^{*-},\overline{K}^{*0},\Lambda_{c}^{+})
  + {\cal M}(\Omega_{cc}^{+},  \pi^{+},  \Xi_{c}^{0},   \Sigma_{c}^{0}, \Lambda_{c}^{+},\overline{K}^{*0})
  + {\cal M}(\Omega_{cc}^{+}, \rho^{+},  \Xi_{c}^{0},   \Sigma_{c}^{0},  \Lambda_{c}^{+},\overline{K}^{*0})\nonumber\\
&&+ {\cal M}(\Omega_{cc}^{+},  \pi^{+},   \Xi_{c}^{'0},  \Sigma_{c}^{0}, \Lambda_{c}^{+},\overline{K}^{*0})
  + {\cal M}(\Omega_{cc}^{+}, \rho^{+},   \Xi_{c}^{'0},  \Sigma_{c}^{0},  \Lambda_{c}^{+},\overline{K}^{*0})
  + {\cal M}(\Omega_{cc}^{+}, \pi^{0}   , \Xi_{c}^{+},    \overline{K}^{0}, \overline{K}^{*0},\Lambda_{c}^{+})\nonumber\\
&&+ {\cal M}(\Omega_{cc}^{+}, \rho^{0}   , \Xi_{c}^{+},   \overline{K}^{*0},\overline{K}^{*0},\Lambda_{c}^{+})
  + {\cal M}(\Omega_{cc}^{+}, \eta_8,     \Xi_{c}^{+},   \overline{K}^{0}, \overline{K}^{*0},\Lambda_{c}^{+})
  + {\cal M}(\Omega_{cc}^{+}, \omega   ,   \Xi_{c}^{+},   \overline{K}^{*0},\overline{K}^{*0},\Lambda_{c}^{+})\nonumber\\
&&+ {\cal M}(\Omega_{cc}^{+}, \pi^{0}   , \Xi_{c}^{'+},    \overline{K}^{0}, \overline{K}^{*0},\Lambda_{c}^{+})
  + {\cal M}(\Omega_{cc}^{+}, \rho^{0}   , \Xi_{c}^{'+},   \overline{K}^{*0},\overline{K}^{*0},\Lambda_{c}^{+})
  + {\cal M}(\Omega_{cc}^{+}, \eta_8,     \Xi_{c}^{'+},   \overline{K}^{0}, \overline{K}^{*0},\Lambda_{c}^{+})\nonumber\\
&&+ {\cal M}(\Omega_{cc}^{+}, \omega   ,   \Xi_{c}^{'+},   \overline{K}^{*0},\overline{K}^{*0},\Lambda_{c}^{+})
  + {\cal M}(\Omega_{cc}^{+}, \eta_8,     \Xi_{c}^{+},   \Lambda_{c}^{+}, \Lambda_{c}^{+},\overline{K}^{*0})
  + {\cal M}(\Omega_{cc}^{+}, \omega   ,   \Xi_{c}^{+},   \Lambda_{c}^{+},\Lambda_{c}^{+},\overline{K}^{*0})\nonumber\\
&&+ {\cal M}(\Omega_{cc}^{+}, \pi^{0}    , \Xi_{c}^{+},    \Sigma_{c}^{+}, \Lambda_{c}^{+},\overline{K}^{*0})
  + {\cal M}(\Omega_{cc}^{+}, \rho^{0}   , \Xi_{c}^{+},   \Sigma_{c}^{+},\Lambda_{c}^{+},\overline{K}^{*0})
  + {\cal M}(\Omega_{cc}^{+}, \eta_8,     \Xi_{c}^{'+},   \Lambda_{c}^{+}, \Lambda_{c}^{+},\overline{K}^{*0})\nonumber\\
&&+ {\cal M}(\Omega_{cc}^{+}, \omega   ,   \Xi_{c}^{'+},   \Lambda_{c}^{+},\Lambda_{c}^{+},\overline{K}^{*0})
  + {\cal M}(\Omega_{cc}^{+}, \pi^{0}   , \Xi_{c}^{'+},    \Sigma_{c}^{+}, \Lambda_{c}^{+},\overline{K}^{*0})
  + {\cal M}(\Omega_{cc}^{+}, \rho^{0}   , \Xi_{c}^{'+},   \Sigma_{c}^{+},\Lambda_{c}^{+},\overline{K}^{*0})\nonumber\\
&&+ {\cal M}(\Omega_{cc}^{+}, \eta_8, \Xi_{c}^{+},    \overline{K}^{0}, \overline{K}^{*0},\Lambda_{c}^{+})
  + {\cal M}(\Omega_{cc}^{+}, \phi,    \Xi_{c}^{+},     \overline{K}^{*0},\overline{K}^{*0},\Lambda_{c}^{+})
  + {\cal M}(\Omega_{cc}^{+}, \eta_8,  \Xi_{c}^{'+},   \overline{K}^{0}, \overline{K}^{*0},\Lambda_{c}^{+})\nonumber\\
&&+ {\cal M}(\Omega_{cc}^{+}, \phi,     \Xi_{c}^{'+},   \overline{K}^{*0},\overline{K}^{*0},\Lambda_{c}^{+})
  + {\cal M}(\Omega_{cc}^{+}, K^{+},  \Omega_{c}^{0},    \Xi_{c}^{0},  \Lambda_{c}^{+},\overline{K}^{*0})
  + {\cal M}(\Omega_{cc}^{+}, K^{*+},  \Omega_{c}^{0},   \Xi_{c}^{0},  \Lambda_{c}^{+},\overline{K}^{*0})\nonumber\\
&&+ {\cal M}(\Omega_{cc}^{+}, K^{+},   \Omega_{c}^{0},   \Xi_{c}^{'0},  \Lambda_{c}^{+},\overline{K}^{*0})
  + {\cal M}(\Omega_{cc}^{+}, K^{*+},  \Omega_{c}^{0},   \Xi_{c}^{'0},  \Lambda_{c}^{+},\overline{K}^{*0})
  + {\cal M}(\Omega_{cc}^{+}, \eta_1,     \Xi_{c}^{+},   \Lambda_{c}^{+}, \Lambda_{c}^{+},\overline{K}^{*0})\nonumber\\
&&+ {\cal M}(\Omega_{cc}^{+}, \eta_1,     \Xi_{c}^{'+},   \Lambda_{c}^{+}, \Lambda_{c}^{+},\overline{K}^{*0})
  ],
\end{eqnarray}
\begin{eqnarray}
{\cal A}(\Omega_{cc}^{+} \to \Xi_{c}^{+}\rho^{0})
&=&C_{\rm SD}(\Omega_{cc}^{+} \to \Xi_{c}^{+}\rho^{0} )+ i [
    {\cal M}(\Omega_{cc}^{+}, \pi^{+}, \Xi_{c}^{0},  \pi^{-}, \rho^{0} ,\Xi_{c}^{+})
  + {\cal M}(\Omega_{cc}^{+}, \rho^{+}, \Xi_{c}^{0}, \rho^{-},\rho^{0} ,\Xi_{c}^{+})\nonumber\\
&&+ {\cal M}(\Omega_{cc}^{+}, \pi^{+},  \Xi_{c}^{'0},  \pi^{-}, \rho^{0} ,\Xi_{c}^{+})
  + {\cal M}(\Omega_{cc}^{+}, \rho^{+},  \Xi_{c}^{'0}, \rho^{-},\rho^{0} ,\Xi_{c}^{+})
  + {\cal M}(\Omega_{cc}^{+},  \pi^{+},  \Xi_{c}^{0},   \Xi_{c}^{0}, \Xi_{c}^{+},\rho^{0} )\nonumber\\
&&+ {\cal M}(\Omega_{cc}^{+}, \rho^{+},  \Xi_{c}^{0},   \Xi_{c}^{0},  \Xi_{c}^{+},\rho^{0} )
  + {\cal M}(\Omega_{cc}^{+},  \pi^{+},   \Xi_{c}^{0},   \Xi_{c}^{'0}, \Xi_{c}^{+},\rho^{0} )
  + {\cal M}(\Omega_{cc}^{+}, \rho^{+},   \Xi_{c}^{0},   \Xi_{c}^{'0},  \Xi_{c}^{+},\rho^{0} )\nonumber\\
&&+ {\cal M}(\Omega_{cc}^{+},  \pi^{+},  \Xi_{c}^{'0},   \Xi_{c}^{0}, \Xi_{c}^{+},\rho^{0} )
  + {\cal M}(\Omega_{cc}^{+}, \rho^{+},  \Xi_{c}^{'0},   \Xi_{c}^{0},  \Xi_{c}^{+},\rho^{0} )
  + {\cal M}(\Omega_{cc}^{+},  \pi^{+},   \Xi_{c}^{'0},   \Xi_{c}^{'0}, \Xi_{c}^{+},\rho^{0} )\nonumber\\
&&+ {\cal M}(\Omega_{cc}^{+}, \rho^{+},   \Xi_{c}^{'0},   \Xi_{c}^{'0},  \Xi_{c}^{+},\rho^{0} )
  + {\cal M}(\Omega_{cc}^{+},  \eta_8,  \Xi_{c}^{+},   \Xi_{c}^{+}, \Xi_{c}^{+},\rho^{0} )
  + {\cal M}(\Omega_{cc}^{+},  \phi,     \Xi_{c}^{+},   \Xi_{c}^{+},  \Xi_{c}^{+},\rho^{0} )\nonumber\\
&&+ {\cal M}(\Omega_{cc}^{+},  \eta_8,   \Xi_{c}^{'+},   \Xi_{c}^{+}, \Xi_{c}^{+},\rho^{0} )
  + {\cal M}(\Omega_{cc}^{+},  \phi,      \Xi_{c}^{'+},   \Xi_{c}^{+},  \Xi_{c}^{+},\rho^{0} )
  + {\cal M}(\Omega_{cc}^{+},  \eta_8,   \Xi_{c}^{+},   \Xi_{c}^{'+}, \Xi_{c}^{+},\rho^{0} )\nonumber\\
&&+ {\cal M}(\Omega_{cc}^{+},  \phi,      \Xi_{c}^{+},   \Xi_{c}^{'+},  \Xi_{c}^{+},\rho^{0} )
  + {\cal M}(\Omega_{cc}^{+},  \eta_8,   \Xi_{c}^{'+},   \Xi_{c}^{'+}, \Xi_{c}^{+},\rho^{0} )
  + {\cal M}(\Omega_{cc}^{+},  \phi,      \Xi_{c}^{'+},   \Xi_{c}^{'+},  \Xi_{c}^{+},\rho^{0} )\nonumber\\
&&+ {\cal M}(\Omega_{cc}^{+}, K^{+},  \Omega_{c}^{0},  K^{-}, \rho^{0} ,\Xi_{c}^{+})
  + {\cal M}(\Omega_{cc}^{+}, K^{*+},  \Omega_{c}^{0}, K^{*-},\rho^{0} ,\Xi_{c}^{+})
  + {\cal M}(\Omega_{cc}^{+},  \eta_1,  \Xi_{c}^{+},   \Xi_{c}^{+}, \Xi_{c}^{+},\rho^{0} )\nonumber\\
&&+ {\cal M}(\Omega_{cc}^{+},  \eta_1,   \Xi_{c}^{'+},   \Xi_{c}^{+}, \Xi_{c}^{+},\rho^{0} )
  ],
\end{eqnarray}
\begin{eqnarray}
{\cal A}(\Omega_{cc}^{+} \to \Xi_{c}^{'+}\rho^{0})
&=&C_{\rm SD}(\Omega_{cc}^{+} \to \Xi_{c}^{'+}\rho^{0} )+ i [
    {\cal M}(\Omega_{cc}^{+}, \pi^{+}, \Xi_{c}^{0},  \pi^{-}, \rho^{0} ,\Xi_{c}^{'+})
  + {\cal M}(\Omega_{cc}^{+}, \rho^{+}, \Xi_{c}^{0}, \rho^{-},\rho^{0} ,\Xi_{c}^{'+})\nonumber\\
&&+ {\cal M}(\Omega_{cc}^{+}, \pi^{+},  \Xi_{c}^{'0},  \pi^{-}, \rho^{0} ,\Xi_{c}^{'+})
  + {\cal M}(\Omega_{cc}^{+}, \rho^{+},  \Xi_{c}^{'0}, \rho^{-},\rho^{0} ,\Xi_{c}^{'+})
  + {\cal M}(\Omega_{cc}^{+},  \pi^{+},  \Xi_{c}^{0},   \Xi_{c}^{0}, \Xi_{c}^{'+},\rho^{0} )\nonumber\\
&&+ {\cal M}(\Omega_{cc}^{+}, \rho^{+},  \Xi_{c}^{0},   \Xi_{c}^{0},  \Xi_{c}^{'+},\rho^{0} )
  + {\cal M}(\Omega_{cc}^{+},  \pi^{+},   \Xi_{c}^{0},   \Xi_{c}^{'0}, \Xi_{c}^{'+},\rho^{0} )
  + {\cal M}(\Omega_{cc}^{+}, \rho^{+},   \Xi_{c}^{0},   \Xi_{c}^{'0},  \Xi_{c}^{'+},\rho^{0} )\nonumber\\
&&+ {\cal M}(\Omega_{cc}^{+},  \pi^{+},  \Xi_{c}^{'0},   \Xi_{c}^{0}, \Xi_{c}^{'+},\rho^{0} )
  + {\cal M}(\Omega_{cc}^{+}, \rho^{+},  \Xi_{c}^{'0},   \Xi_{c}^{0},  \Xi_{c}^{'+},\rho^{0} )
  + {\cal M}(\Omega_{cc}^{+},  \pi^{+},   \Xi_{c}^{'0},   \Xi_{c}^{'0}, \Xi_{c}^{'+},\rho^{0} )\nonumber\\
&&+ {\cal M}(\Omega_{cc}^{+}, \rho^{+},   \Xi_{c}^{'0},   \Xi_{c}^{'0},  \Xi_{c}^{'+},\rho^{0} )
  + {\cal M}(\Omega_{cc}^{+},  \eta_8,  \Xi_{c}^{+},   \Xi_{c}^{+}, \Xi_{c}^{'+},\rho^{0} )
  + {\cal M}(\Omega_{cc}^{+},  \phi,     \Xi_{c}^{+},   \Xi_{c}^{+},  \Xi_{c}^{'+},\rho^{0} )\nonumber\\
&&+ {\cal M}(\Omega_{cc}^{+},  \eta_8,   \Xi_{c}^{'+},   \Xi_{c}^{+}, \Xi_{c}^{'+},\rho^{0} )
  + {\cal M}(\Omega_{cc}^{+},  \phi,      \Xi_{c}^{'+},   \Xi_{c}^{+},  \Xi_{c}^{'+},\rho^{0} )
  + {\cal M}(\Omega_{cc}^{+},  \eta_8,   \Xi_{c}^{+},   \Xi_{c}^{'+}, \Xi_{c}^{'+},\rho^{0} )\nonumber\\
&&+ {\cal M}(\Omega_{cc}^{+},  \phi,      \Xi_{c}^{+},   \Xi_{c}^{'+},  \Xi_{c}^{'+},\rho^{0} )
  + {\cal M}(\Omega_{cc}^{+},  \eta_8,   \Xi_{c}^{'+},   \Xi_{c}^{'+}, \Xi_{c}^{'+},\rho^{0} )
  + {\cal M}(\Omega_{cc}^{+},  \phi,      \Xi_{c}^{'+},   \Xi_{c}^{'+},  \Xi_{c}^{'+},\rho^{0} )\nonumber\\
&&+ {\cal M}(\Omega_{cc}^{+}, K^{+},  \Omega_{c}^{0},  K^{-}, \rho^{0} ,\Xi_{c}^{'+})
  + {\cal M}(\Omega_{cc}^{+}, K^{*+},  \Omega_{c}^{0}, K^{*-},\rho^{0} ,\Xi_{c}^{'+})
  + {\cal M}(\Omega_{cc}^{+},  \eta_1,   \Xi_{c}^{+},   \Xi_{c}^{'+}, \Xi_{c}^{'+},\rho^{0} )\nonumber\\
&&+ {\cal M}(\Omega_{cc}^{+},  \eta_1,   \Xi_{c}^{'+},   \Xi_{c}^{'+}, \Xi_{c}^{'+},\rho^{0} )
  ],
\end{eqnarray}
\begin{eqnarray}
{\cal A}(\Omega_{cc}^{+} \to \Xi_{c}^{+}\omega)
&=&C_{\rm SD}(\Omega_{cc}^{+} \to \Xi_{c}^{+}\omega )+ i [
    {\cal M}(\Omega_{cc}^{+},  \pi^{+},  \Xi_{c}^{0},   \Xi_{c}^{0}, \Xi_{c}^{+},\omega )
  + {\cal M}(\Omega_{cc}^{+}, \rho^{+},  \Xi_{c}^{0},   \Xi_{c}^{0},  \Xi_{c}^{+},\omega )\nonumber\\
&&+ {\cal M}(\Omega_{cc}^{+},  \pi^{+},   \Xi_{c}^{0},   \Xi_{c}^{'0}, \Xi_{c}^{+},\omega )
  + {\cal M}(\Omega_{cc}^{+}, \rho^{+},   \Xi_{c}^{0},   \Xi_{c}^{'0},  \Xi_{c}^{+},\omega )
  + {\cal M}(\Omega_{cc}^{+},  \pi^{+},  \Xi_{c}^{'0},   \Xi_{c}^{0}, \Xi_{c}^{+},\omega )\nonumber\\
&&+ {\cal M}(\Omega_{cc}^{+}, \rho^{+},  \Xi_{c}^{'0},   \Xi_{c}^{0},  \Xi_{c}^{+},\omega )
  + {\cal M}(\Omega_{cc}^{+},  \pi^{+},   \Xi_{c}^{'0},   \Xi_{c}^{'0}, \Xi_{c}^{+},\omega )
  + {\cal M}(\Omega_{cc}^{+}, \rho^{+},   \Xi_{c}^{'0},   \Xi_{c}^{'0},  \Xi_{c}^{+},\omega )\nonumber\\
&&+ {\cal M}(\Omega_{cc}^{+},  \eta_8,  \Xi_{c}^{+},   \Xi_{c}^{+}, \Xi_{c}^{+},\omega )
  + {\cal M}(\Omega_{cc}^{+},  \phi,     \Xi_{c}^{+},   \Xi_{c}^{+},  \Xi_{c}^{+},\omega )
  + {\cal M}(\Omega_{cc}^{+},  \eta_8,   \Xi_{c}^{'+},   \Xi_{c}^{+}, \Xi_{c}^{+},\omega )\nonumber\\
&&+ {\cal M}(\Omega_{cc}^{+},  \phi,      \Xi_{c}^{'+},   \Xi_{c}^{+},  \Xi_{c}^{+},\omega )
  + {\cal M}(\Omega_{cc}^{+},  \eta_8,   \Xi_{c}^{+},   \Xi_{c}^{'+}, \Xi_{c}^{+},\omega )
  + {\cal M}(\Omega_{cc}^{+},  \phi,      \Xi_{c}^{+},   \Xi_{c}^{'+},  \Xi_{c}^{+},\omega )\nonumber\\
&&+ {\cal M}(\Omega_{cc}^{+},  \eta_8,   \Xi_{c}^{'+},   \Xi_{c}^{'+}, \Xi_{c}^{+},\omega )
  + {\cal M}(\Omega_{cc}^{+},  \phi,      \Xi_{c}^{'+},   \Xi_{c}^{'+},  \Xi_{c}^{+},\omega )
  + {\cal M}(\Omega_{cc}^{+}, K^{+},  \Omega_{c}^{0},  K^{-}, \omega ,\Xi_{c}^{+})\nonumber\\
&&+ {\cal M}(\Omega_{cc}^{+}, K^{*+},  \Omega_{c}^{0}, K^{*-},\omega ,\Xi_{c}^{+})
  + {\cal M}(\Omega_{cc}^{+},  \eta_1,  \Xi_{c}^{+},   \Xi_{c}^{+}, \Xi_{c}^{+},\omega )
  + {\cal M}(\Omega_{cc}^{+},  \eta_1,   \Xi_{c}^{'+},   \Xi_{c}^{+}, \Xi_{c}^{+},\omega )
  ],
\end{eqnarray}
\begin{eqnarray}
{\cal A}(\Omega_{cc}^{+} \to \Xi_{c}^{'+}\omega)
&=&C_{\rm SD}(\Omega_{cc}^{+} \to \Xi_{c}^{'+}\omega )+ i [
    {\cal M}(\Omega_{cc}^{+},  \pi^{+},  \Xi_{c}^{0},   \Xi_{c}^{0}, \Xi_{c}^{'+},\omega )
  + {\cal M}(\Omega_{cc}^{+}, \rho^{+},  \Xi_{c}^{0},   \Xi_{c}^{0},  \Xi_{c}^{'+},\omega )\nonumber\\
&&+ {\cal M}(\Omega_{cc}^{+},  \pi^{+},   \Xi_{c}^{0},   \Xi_{c}^{'0}, \Xi_{c}^{'+},\omega )
  + {\cal M}(\Omega_{cc}^{+}, \rho^{+},   \Xi_{c}^{0},   \Xi_{c}^{'0},  \Xi_{c}^{'+},\omega )
  + {\cal M}(\Omega_{cc}^{+},  \pi^{+},  \Xi_{c}^{'0},   \Xi_{c}^{0}, \Xi_{c}^{'+},\omega )\nonumber\\
&&+ {\cal M}(\Omega_{cc}^{+}, \rho^{+},  \Xi_{c}^{'0},   \Xi_{c}^{0},  \Xi_{c}^{'+},\omega )
  + {\cal M}(\Omega_{cc}^{+},  \pi^{+},   \Xi_{c}^{'0},   \Xi_{c}^{'0}, \Xi_{c}^{'+},\omega )
  + {\cal M}(\Omega_{cc}^{+}, \rho^{+},   \Xi_{c}^{'0},   \Xi_{c}^{'0},  \Xi_{c}^{'+},\omega )\nonumber\\
&&+ {\cal M}(\Omega_{cc}^{+},  \eta_8,  \Xi_{c}^{+},   \Xi_{c}^{+}, \Xi_{c}^{'+},\omega )
  + {\cal M}(\Omega_{cc}^{+},  \phi,     \Xi_{c}^{+},   \Xi_{c}^{+},  \Xi_{c}^{'+},\omega )
  + {\cal M}(\Omega_{cc}^{+},  \eta_8,   \Xi_{c}^{'+},   \Xi_{c}^{+}, \Xi_{c}^{'+},\omega )\nonumber\\
&&+ {\cal M}(\Omega_{cc}^{+},  \phi,      \Xi_{c}^{'+},   \Xi_{c}^{+},  \Xi_{c}^{'+},\omega )
  + {\cal M}(\Omega_{cc}^{+},  \eta_8,   \Xi_{c}^{+},   \Xi_{c}^{'+}, \Xi_{c}^{'+},\omega )
  + {\cal M}(\Omega_{cc}^{+},  \phi,      \Xi_{c}^{+},   \Xi_{c}^{'+},  \Xi_{c}^{'+},\omega )\nonumber\\
&&+ {\cal M}(\Omega_{cc}^{+},  \eta_8,   \Xi_{c}^{'+},   \Xi_{c}^{'+}, \Xi_{c}^{'+},\omega )
  + {\cal M}(\Omega_{cc}^{+},  \phi,      \Xi_{c}^{'+},   \Xi_{c}^{'+},  \Xi_{c}^{'+},\omega )
  + {\cal M}(\Omega_{cc}^{+}, K^{+},  \Omega_{c}^{0},  K^{-}, \omega ,\Xi_{c}^{'+})\nonumber\\
&&+ {\cal M}(\Omega_{cc}^{+}, K^{*+},  \Omega_{c}^{0}, K^{*-},\omega ,\Xi_{c}^{'+})
  + {\cal M}(\Omega_{cc}^{+},  \eta_1,   \Xi_{c}^{+},   \Xi_{c}^{'+}, \Xi_{c}^{'+},\omega )
  + {\cal M}(\Omega_{cc}^{+},  \eta_1,   \Xi_{c}^{'+},   \Xi_{c}^{'+}, \Xi_{c}^{'+},\omega )
  ],
\end{eqnarray}
\begin{eqnarray}
{\cal A}(\Omega_{cc}^{+} \to \Xi_{c}^{0}\rho^{+})
&=&T_{\rm SD}(\Omega_{cc}^{+} \to \Xi_{c}^{0}\rho^{+})+ i [
    {\cal M}(\Omega_{cc}^{+},  \pi^{0} , \Xi_{c}^{+},  \pi^{+}, \rho^{+},\Xi_{c}^{0})
  + {\cal M}(\Omega_{cc}^{+},  \rho^{0} , \Xi_{c}^{+}, \rho^{+},\rho^{+},\Xi_{c}^{0})\nonumber\\
&&+{\cal M}(\Omega_{cc}^{+},  \pi^{0} ,   \Xi_{c}^{'+},  \pi^{+}, \rho^{+},\Xi_{c}^{0})
  + {\cal M}(\Omega_{cc}^{+},  \rho^{0} , \Xi_{c}^{'+}, \rho^{+},\rho^{+},\Xi_{c}^{0})
  +{\cal M}(\Omega_{cc}^{+},  \pi^{0} ,   \Xi_{c}^{+},   \Xi_{c}^{0},\Xi_{c}^{0},\rho^{+})\nonumber\\
&&+ {\cal M}(\Omega_{cc}^{+},  \rho^{0} , \Xi_{c}^{+},   \Xi_{c}^{0},\Xi_{c}^{0},\rho^{+})
  +{\cal M}(\Omega_{cc}^{+},  \eta_8,      \Xi_{c}^{+},   \Xi_{c}^{0},\Xi_{c}^{0},\rho^{+})
  + {\cal M}(\Omega_{cc}^{+}, \omega ,    \Xi_{c}^{+},   \Xi_{c}^{0},\Xi_{c}^{0},\rho^{+})\nonumber\\
&&+{\cal M}(\Omega_{cc}^{+},  \pi^{0} ,   \Xi_{c}^{+},   \Xi_{c}^{'0},\Xi_{c}^{0},\rho^{+})
  + {\cal M}(\Omega_{cc}^{+},  \rho^{0} , \Xi_{c}^{+},   \Xi_{c}^{'0},\Xi_{c}^{0},\rho^{+})
  +{\cal M}(\Omega_{cc}^{+},  \eta_8,      \Xi_{c}^{+},   \Xi_{c}^{'0},\Xi_{c}^{0},\rho^{+})\nonumber\\
&&+ {\cal M}(\Omega_{cc}^{+}, \omega ,    \Xi_{c}^{+},   \Xi_{c}^{'0},\Xi_{c}^{0},\rho^{+})
  +{\cal M}(\Omega_{cc}^{+},  \pi^{0} ,   \Xi_{c}^{'+},   \Xi_{c}^{0},\Xi_{c}^{0},\rho^{+})
  + {\cal M}(\Omega_{cc}^{+},  \rho^{0} , \Xi_{c}^{'+},   \Xi_{c}^{0},\Xi_{c}^{0},\rho^{+})\nonumber\\
&&+{\cal M}(\Omega_{cc}^{+},  \eta_8,      \Xi_{c}^{'+},   \Xi_{c}^{0},\Xi_{c}^{0},\rho^{+})
  + {\cal M}(\Omega_{cc}^{+}, \omega ,    \Xi_{c}^{'+},   \Xi_{c}^{0},\Xi_{c}^{0},\rho^{+})
  +{\cal M}(\Omega_{cc}^{+},  \pi^{0} ,   \Xi_{c}^{'+},   \Xi_{c}^{'0},\Xi_{c}^{0},\rho^{+})\nonumber\\
&&+ {\cal M}(\Omega_{cc}^{+},  \rho^{0}  , \Xi_{c}^{'+},   \Xi_{c}^{'0},\Xi_{c}^{0},\rho^{+})
  + {\cal M}(\Omega_{cc}^{+},  \eta_8,      \Xi_{c}^{'+},   \Xi_{c}^{'0},\Xi_{c}^{0},\rho^{+})
  + {\cal M}(\Omega_{cc}^{+}, \omega ,    \Xi_{c}^{'+},   \Xi_{c}^{'0},\Xi_{c}^{0},\rho^{+})\nonumber\\
&&+ {\cal M}(\Omega_{cc}^{+}, \pi^{+}, \Xi_{c}^{0},  \pi^{0} , \rho^{+},\Xi_{c}^{0})
  + {\cal M}(\Omega_{cc}^{+}, \rho^{+}, \Xi_{c}^{0}, \rho^{0} ,\rho^{+},\Xi_{c}^{0})
  + {\cal M}(\Omega_{cc}^{+}, \pi^{+},  \Xi_{c}^{'0}, \pi^{0} , \rho^{+},\Xi_{c}^{0})\nonumber\\
&&+ {\cal M}(\Omega_{cc}^{+}, \rho^{+}, \Xi_{c}^{'0}, \rho^{0} ,\rho^{+},\Xi_{c}^{0})
  +{\cal M}(\Omega_{cc}^{+},  \eta_8,   \Xi_{c}^{+},   \Xi_{c}^{0},\Xi_{c}^{0},\rho^{+})
  + {\cal M}(\Omega_{cc}^{+},  \phi,     \Xi_{c}^{+},   \Xi_{c}^{0},\Xi_{c}^{0},\rho^{+})\nonumber\\
&&+{\cal M}(\Omega_{cc}^{+},   \eta_8,  \Xi_{c}^{'+},   \Xi_{c}^{0},\Xi_{c}^{0},\rho^{+})
  + {\cal M}(\Omega_{cc}^{+},  \phi,     \Xi_{c}^{'+},   \Xi_{c}^{0},\Xi_{c}^{0},\rho^{+})
  +{\cal M}(\Omega_{cc}^{+},  \eta_8,   \Xi_{c}^{+},   \Xi_{c}^{'0},\Xi_{c}^{0},\rho^{+})\nonumber\\
&&+ {\cal M}(\Omega_{cc}^{+},  \phi,     \Xi_{c}^{+},   \Xi_{c}^{'0},\Xi_{c}^{0},\rho^{+})
  +{\cal M}(\Omega_{cc}^{+},   \eta_8,  \Xi_{c}^{'+},   \Xi_{c}^{'0},\Xi_{c}^{0},\rho^{+})
  + {\cal M}(\Omega_{cc}^{+},  \phi,     \Xi_{c}^{'+},   \Xi_{c}^{'0},\Xi_{c}^{0},\rho^{+})\nonumber\\
&&+ {\cal M}(\Omega_{cc}^{+}, K^{+}, \Omega_{c}^{0},  \overline{K}^{0}, \rho^{+},\Xi_{c}^{0})
  + {\cal M}(\Omega_{cc}^{+}, K^{*+}, \Omega_{c}^{0}, \overline{K}^{*0},\rho^{+},\Xi_{c}^{0})
  +{\cal M}(\Omega_{cc}^{+},  \eta_1,      \Xi_{c}^{+},   \Xi_{c}^{0},\Xi_{c}^{0},\rho^{+})\nonumber\\
&&+{\cal M}(\Omega_{cc}^{+},  \eta_1,      \Xi_{c}^{'+},   \Xi_{c}^{0},\Xi_{c}^{0},\rho^{+})],
\end{eqnarray}
\begin{eqnarray}
{\cal A}(\Omega_{cc}^{+} \to \Xi_{c}^{'0}\rho^{+})
&=&T_{\rm SD}(\Omega_{cc}^{+} \to \Xi_{c}^{'0}\rho^{+})+ i [
    {\cal M}(\Omega_{cc}^{+},  \pi^{0} , \Xi_{c}^{+},  \pi^{+}, \rho^{+},\Xi_{c}^{'0})
  + {\cal M}(\Omega_{cc}^{+},  \rho^{0} , \Xi_{c}^{+}, \rho^{+},\rho^{+},\Xi_{c}^{'0})\nonumber\\
&&+{\cal M}(\Omega_{cc}^{+},  \pi^{0} ,   \Xi_{c}^{'+},  \pi^{+}, \rho^{+},\Xi_{c}^{'0})
  + {\cal M}(\Omega_{cc}^{+},  \rho^{0} , \Xi_{c}^{'+}, \rho^{+},\rho^{+},\Xi_{c}^{'0})
  +{\cal M}(\Omega_{cc}^{+},  \pi^{0} ,   \Xi_{c}^{+},   \Xi_{c}^{0},\Xi_{c}^{'0},\rho^{+})\nonumber\\
&&+ {\cal M}(\Omega_{cc}^{+},  \rho^{0} , \Xi_{c}^{+},   \Xi_{c}^{0},\Xi_{c}^{'0},\rho^{+})
  +{\cal M}(\Omega_{cc}^{+},  \eta_8,      \Xi_{c}^{+},   \Xi_{c}^{0},\Xi_{c}^{'0},\rho^{+})
  + {\cal M}(\Omega_{cc}^{+}, \omega ,    \Xi_{c}^{+},   \Xi_{c}^{0},\Xi_{c}^{'0},\rho^{+})\nonumber\\
&&+{\cal M}(\Omega_{cc}^{+},  \pi^{0} ,   \Xi_{c}^{+},   \Xi_{c}^{'0},\Xi_{c}^{'0},\rho^{+})
  + {\cal M}(\Omega_{cc}^{+},  \rho^{0} , \Xi_{c}^{+},   \Xi_{c}^{'0},\Xi_{c}^{'0},\rho^{+})
  +{\cal M}(\Omega_{cc}^{+},  \eta_8,      \Xi_{c}^{+},   \Xi_{c}^{'0},\Xi_{c}^{'0},\rho^{+})\nonumber\\
&&+ {\cal M}(\Omega_{cc}^{+}, \omega ,    \Xi_{c}^{+},   \Xi_{c}^{'0},\Xi_{c}^{'0},\rho^{+})
  +{\cal M}(\Omega_{cc}^{+},  \pi^{0} ,   \Xi_{c}^{'+},   \Xi_{c}^{0},\Xi_{c}^{'0},\rho^{+})
  + {\cal M}(\Omega_{cc}^{+},  \rho^{0} , \Xi_{c}^{'+},   \Xi_{c}^{0},\Xi_{c}^{'0},\rho^{+})\nonumber\\
&&+{\cal M}(\Omega_{cc}^{+},  \eta_8,      \Xi_{c}^{'+},   \Xi_{c}^{0},\Xi_{c}^{'0},\rho^{+})
  + {\cal M}(\Omega_{cc}^{+}, \omega ,    \Xi_{c}^{'+},   \Xi_{c}^{0},\Xi_{c}^{'0},\rho^{+})
  +{\cal M}(\Omega_{cc}^{+},  \pi^{0} ,   \Xi_{c}^{'+},   \Xi_{c}^{'0},\Xi_{c}^{'0},\rho^{+})\nonumber\\
&&+ {\cal M}(\Omega_{cc}^{+},  \rho^{0} , \Xi_{c}^{'+},   \Xi_{c}^{'0},\Xi_{c}^{'0},\rho^{+})
  + {\cal M}(\Omega_{cc}^{+},  \eta_8,      \Xi_{c}^{'+},   \Xi_{c}^{'0},\Xi_{c}^{'0},\rho^{+})
  + {\cal M}(\Omega_{cc}^{+}, \omega ,    \Xi_{c}^{'+},   \Xi_{c}^{'0},\Xi_{c}^{'0},\rho^{+})\nonumber\\
&&+ {\cal M}(\Omega_{cc}^{+}, \pi^{+}, \Xi_{c}^{0},  \pi^{0} , \rho^{+},\Xi_{c}^{'0})
  + {\cal M}(\Omega_{cc}^{+}, \rho^{+}, \Xi_{c}^{0}, \rho^{0} ,\rho^{+},\Xi_{c}^{'0})
  + {\cal M}(\Omega_{cc}^{+}, \pi^{+},  \Xi_{c}^{'0}, \pi^{0} , \rho^{+},\Xi_{c}^{'0})\nonumber\\
&&+ {\cal M}(\Omega_{cc}^{+}, \rho^{+}, \Xi_{c}^{'0}, \rho^{0} ,\rho^{+},\Xi_{c}^{'0})
  +{\cal M}(\Omega_{cc}^{+},  \eta_8,   \Xi_{c}^{+},   \Xi_{c}^{0},\Xi_{c}^{'0},\rho^{+})
  + {\cal M}(\Omega_{cc}^{+},  \phi,     \Xi_{c}^{+},   \Xi_{c}^{0},\Xi_{c}^{'0},\rho^{+})\nonumber\\
&&+{\cal M}(\Omega_{cc}^{+},   \eta_8,  \Xi_{c}^{'+},   \Xi_{c}^{0},\Xi_{c}^{'0},\rho^{+})
  + {\cal M}(\Omega_{cc}^{+},  \phi,     \Xi_{c}^{'+},   \Xi_{c}^{0},\Xi_{c}^{'0},\rho^{+})
  +{\cal M}(\Omega_{cc}^{+},  \eta_8,   \Xi_{c}^{+},   \Xi_{c}^{'0},\Xi_{c}^{'0},\rho^{+})\nonumber\\
&&+ {\cal M}(\Omega_{cc}^{+},  \phi,     \Xi_{c}^{+},   \Xi_{c}^{'0},\Xi_{c}^{'0},\rho^{+})
  +{\cal M}(\Omega_{cc}^{+},   \eta_8,  \Xi_{c}^{'+},   \Xi_{c}^{'0},\Xi_{c}^{'0},\rho^{+})
  + {\cal M}(\Omega_{cc}^{+},  \phi,     \Xi_{c}^{'+},   \Xi_{c}^{'0},\Xi_{c}^{'0},\rho^{+})\nonumber\\
&&+ {\cal M}(\Omega_{cc}^{+}, K^{+}, \Omega_{c}^{0},  \overline{K}^{0}, \rho^{+},\Xi_{c}^{'0})
  + {\cal M}(\Omega_{cc}^{+}, K^{*+}, \Omega_{c}^{0}, \overline{K}^{*0},\rho^{+},\Xi_{c}^{'0})
  +{\cal M}(\Omega_{cc}^{+},  \eta_1,      \Xi_{c}^{+},   \Xi_{c}^{'0},\Xi_{c}^{'0},\rho^{+})\nonumber\\
&&+ {\cal M}(\Omega_{cc}^{+},  \eta_1,      \Xi_{c}^{'+},   \Xi_{c}^{'0},\Xi_{c}^{'0},\rho^{+})],
\end{eqnarray}
\begin{eqnarray}
{\cal A}(\Omega_{cc}^{+} \to \Xi_{c}^{+}\phi)
&=&C_{\rm SD}(\Omega_{cc}^{+} \to \Xi_{c}^{+}\phi)+ i [
    {\cal M}(\Omega_{cc}^{+},   K^{+}, \Omega_{c}^{0},     K^{-}, \phi,\Xi_{c}^{+})
  + {\cal M}(\Omega_{cc}^{+},   K^{*+}, \Omega_{c}^{0},    K^{*-},\phi,\Xi_{c}^{+})\nonumber\\
&&+ {\cal M}(\Omega_{cc}^{+},   K^{+},  \Omega_{c}^{0},    \Omega_{c}^{0},\Xi_{c}^{+},\phi)
  + {\cal M}(\Omega_{cc}^{+},   K^{*+}, \Omega_{c}^{0},    \Omega_{c}^{0},\Xi_{c}^{+},\phi)
  + {\cal M}(\Omega_{cc}^{+},   \eta_8,  \Xi_{c}^{+}, \Xi_{c}^{+},\Xi_{c}^{+},\phi)\nonumber\\
&&+ {\cal M}(\Omega_{cc}^{+},   \phi,     \Xi_{c}^{+}, \Xi_{c}^{+},\Xi_{c}^{+},\phi)
  + {\cal M}(\Omega_{cc}^{+},   \eta_8,  \Xi_{c}^{+}, \Xi_{c}^{'+},\Xi_{c}^{+},\phi)
  + {\cal M}(\Omega_{cc}^{+},   \phi,     \Xi_{c}^{+}, \Xi_{c}^{'+},\Xi_{c}^{+},\phi)\nonumber\\
&&+ {\cal M}(\Omega_{cc}^{+},   \eta_8,  \Xi_{c}^{'+}, \Xi_{c}^{+},\Xi_{c}^{+},\phi)
  + {\cal M}(\Omega_{cc}^{+},   \phi,     \Xi_{c}^{'+}, \Xi_{c}^{+},\Xi_{c}^{+},\phi)
  + {\cal M}(\Omega_{cc}^{+},   \eta_8,  \Xi_{c}^{'+}, \Xi_{c}^{'+},\Xi_{c}^{+},\phi)\nonumber\\
&&+ {\cal M}(\Omega_{cc}^{+},   \phi,     \Xi_{c}^{'+}, \Xi_{c}^{'+},\Xi_{c}^{+},\phi)
  + {\cal M}(\Omega_{cc}^{+},   \pi^{+},  \Xi_{c}^{0}, \Xi_{c}^{0},\Xi_{c}^{+},\phi)
  + {\cal M}(\Omega_{cc}^{+},  \rho^{+},  \Xi_{c}^{0}, \Xi_{c}^{0},\Xi_{c}^{+},\phi) \nonumber\\
&&+ {\cal M}(\Omega_{cc}^{+},   \pi^{+},  \Xi_{c}^{'0}, \Xi_{c}^{0},\Xi_{c}^{+},\phi)
  + {\cal M}(\Omega_{cc}^{+},  \rho^{+},  \Xi_{c}^{'0}, \Xi_{c}^{0},\Xi_{c}^{+},\phi)
  + {\cal M}(\Omega_{cc}^{+},   \pi^{+},  \Xi_{c}^{0}, \Xi_{c}^{'0},\Xi_{c}^{+},\phi)  \nonumber\\
&&+ {\cal M}(\Omega_{cc}^{+},  \rho^{+},  \Xi_{c}^{0}, \Xi_{c}^{'0},\Xi_{c}^{+},\phi)
  + {\cal M}(\Omega_{cc}^{+},   \pi^{+},  \Xi_{c}^{'0}, \Xi_{c}^{'0},\Xi_{c}^{+},\phi)
  + {\cal M}(\Omega_{cc}^{+},  \rho^{+},  \Xi_{c}^{'0}, \Xi_{c}^{'0},\Xi_{c}^{+},\phi) \nonumber\\
&&+ {\cal M}(\Omega_{cc}^{+},   \eta_1,  \Xi_{c}^{+}, \Xi_{c}^{+},\Xi_{c}^{+},\phi)
  + {\cal M}(\Omega_{cc}^{+},   \eta_1,  \Xi_{c}^{'+}, \Xi_{c}^{+},\Xi_{c}^{+},\phi)],
\end{eqnarray}
\begin{eqnarray}
{\cal A}(\Omega_{cc}^{+} \to \Xi_{c}^{'+}\phi)
&=&C_{\rm SD}(\Omega_{cc}^{+} \to \Xi_{c}^{'+}\phi)+ i [
    {\cal M}(\Omega_{cc}^{+},  K^{+}, \Omega_{c}^{0},     K^{-}, \phi,\Xi_{c}^{'+})
  + {\cal M}(\Omega_{cc}^{+},  K^{*+}, \Omega_{c}^{0},    K^{*-},\phi,\Xi_{c}^{'+})\nonumber\\
&&+{\cal M}(\Omega_{cc}^{+},   K^{+},  \Omega_{c}^{0},    \Omega_{c}^{0},\Xi_{c}^{'+},\phi)
  + {\cal M}(\Omega_{cc}^{+},  K^{*+}, \Omega_{c}^{0},    \Omega_{c}^{0},\Xi_{c}^{'+},\phi)
  +{\cal M}(\Omega_{cc}^{+},   \eta_8,  \Xi_{c}^{+}, \Xi_{c}^{+},\Xi_{c}^{'+},\phi)\nonumber\\
&&+ {\cal M}(\Omega_{cc}^{+},  \phi,     \Xi_{c}^{+}, \Xi_{c}^{+},\Xi_{c}^{'+},\phi)
  +{\cal M}(\Omega_{cc}^{+},   \eta_8,  \Xi_{c}^{+}, \Xi_{c}^{'+},\Xi_{c}^{'+},\phi)
  + {\cal M}(\Omega_{cc}^{+},  \phi,     \Xi_{c}^{+}, \Xi_{c}^{'+},\Xi_{c}^{'+},\phi)\nonumber\\
&&+{\cal M}(\Omega_{cc}^{+},   \eta_8,  \Xi_{c}^{'+}, \Xi_{c}^{+},\Xi_{c}^{'+},\phi)
  + {\cal M}(\Omega_{cc}^{+},  \phi,     \Xi_{c}^{'+}, \Xi_{c}^{+},\Xi_{c}^{'+},\phi)
  +{\cal M}(\Omega_{cc}^{+},   \eta_8,  \Xi_{c}^{'+}, \Xi_{c}^{'+},\Xi_{c}^{'+},\phi)\nonumber\\
&&+ {\cal M}(\Omega_{cc}^{+},  \phi,     \Xi_{c}^{'+}, \Xi_{c}^{'+},\Xi_{c}^{'+},\phi)
  +{\cal M}(\Omega_{cc}^{+},   \pi^{+},  \Xi_{c}^{0}, \Xi_{c}^{0},\Xi_{c}^{'+},\phi)
  + {\cal M}(\Omega_{cc}^{+},  \rho^{+},  \Xi_{c}^{0}, \Xi_{c}^{0},\Xi_{c}^{'+},\phi)\nonumber\\
&&+{\cal M}(\Omega_{cc}^{+},   \pi^{+},  \Xi_{c}^{'0}, \Xi_{c}^{0},\Xi_{c}^{'+},\phi)
  + {\cal M}(\Omega_{cc}^{+},  \rho^{+},  \Xi_{c}^{'0}, \Xi_{c}^{0},\Xi_{c}^{'+},\phi)
  +{\cal M}(\Omega_{cc}^{+},   \pi^{+},  \Xi_{c}^{0}, \Xi_{c}^{'0},\Xi_{c}^{'+},\phi)\nonumber\\
&&+{\cal M}(\Omega_{cc}^{+},  \rho^{+},  \Xi_{c}^{0}, \Xi_{c}^{'0},\Xi_{c}^{'+},\phi)
  +{\cal M}(\Omega_{cc}^{+},   \pi^{+},  \Xi_{c}^{'0}, \Xi_{c}^{'0},\Xi_{c}^{'+},\phi)
  + {\cal M}(\Omega_{cc}^{+},  \rho^{+},  \Xi_{c}^{'0}, \Xi_{c}^{'0},\Xi_{c}^{'+},\phi)\nonumber\\
&&+{\cal M}(\Omega_{cc}^{+},   \eta_1,  \Xi_{c}^{+}, \Xi_{c}^{'+},\Xi_{c}^{'+},\phi)
  +{\cal M}(\Omega_{cc}^{+},   \eta_1,  \Xi_{c}^{'+}, \Xi_{c}^{'+},\Xi_{c}^{'+},\phi) ],
\end{eqnarray}
\begin{eqnarray}
{\cal A}(\Omega_{cc}^{+} \to \Omega_{c}^{0}K^{*+})
&=&T_{\rm SD}(\Omega_{cc}^{+} \to \Omega_{c}^{0}K^{*+})+ i [
    {\cal M}(\Omega_{cc}^{+},  \eta_8, \Xi_{c}^{+},     K^{+},     K^{*+},     \Omega_{c}^{0})
  + {\cal M}(\Omega_{cc}^{+},  \phi,    \Xi_{c}^{+},     K^{*+},    K^{*+},     \Omega_{c}^{0})\nonumber\\
&&+ {\cal M}(\Omega_{cc}^{+},  \eta_8, \Xi_{c}^{'+},    K^{+},    K^{*+},     \Omega_{c}^{0})
  + {\cal M}(\Omega_{cc}^{+},  \phi,    \Xi_{c}^{'+},    K^{*+},   K^{*+},     \Omega_{c}^{0})
  +{\cal M}(\Omega_{cc}^{+},   \eta_8, \Xi_{c}^{+},     \Omega_{c}^{0},\Omega_{c}^{0},K^{*+})\nonumber\\
&&+ {\cal M}(\Omega_{cc}^{+},  \phi,    \Xi_{c}^{+},     \Omega_{c}^{0},\Omega_{c}^{0},K^{*+})
  + {\cal M}(\Omega_{cc}^{+},  \eta_8, \Xi_{c}^{'+},     \Omega_{c}^{0},\Omega_{c}^{0},K^{*+})
  + {\cal M}(\Omega_{cc}^{+},  \phi,    \Xi_{c}^{'+},     \Omega_{c}^{0},\Omega_{c}^{0},K^{*+})\nonumber\\
&&+{\cal M}(\Omega_{cc}^{+},  \pi^{+},     \Xi_{c}^{0},      K^{0},    K^{*+},     \Omega_{c}^{0})
  + {\cal M}(\Omega_{cc}^{+},  \rho^{+},    \Xi_{c}^{0},     K^{*0},   K^{*+},     \Omega_{c}^{0})
  + {\cal M}(\Omega_{cc}^{+},  \pi^{+},     \Xi_{c}^{'0},    K^{0},    K^{*+},     \Omega_{c}^{0})\nonumber\\
&&+ {\cal M}(\Omega_{cc}^{+},  \rho^{+},    \Xi_{c}^{'0},    K^{*0},   K^{*+},     \Omega_{c}^{0})
  +{\cal M}(\Omega_{cc}^{+},   \eta_8, \Xi_{c}^{+},     \Omega_{c}^{0},\Omega_{c}^{0},K^{*+})
  + {\cal M}(\Omega_{cc}^{+},  \phi,    \Xi_{c}^{+},     \Omega_{c}^{0},\Omega_{c}^{0},K^{*+})\nonumber\\
&&+ {\cal M}(\Omega_{cc}^{+},  \eta_8, \Xi_{c}^{'+},     \Omega_{c}^{0},\Omega_{c}^{0},K^{*+})
  + {\cal M}(\Omega_{cc}^{+},  \phi,    \Xi_{c}^{'+},     \Omega_{c}^{0},\Omega_{c}^{0},K^{*+})
  + {\cal M}(\Omega_{cc}^{+},  K^{+},    \Omega_{c}^{0},     \eta_8, K^{*+},     \Omega_{c}^{0})\nonumber\\
&&+ {\cal M}(\Omega_{cc}^{+},  K^{*+},    \Omega_{c}^{0},    \phi,    K^{*+},     \Omega_{c}^{0})
  + {\cal M}(\Omega_{cc}^{+},   \eta_1, \Xi_{c}^{+},     \Omega_{c}^{0},\Omega_{c}^{0},K^{*+})
  + {\cal M}(\Omega_{cc}^{+},  \eta_1, \Xi_{c}^{'+},     \Omega_{c}^{0},\Omega_{c}^{0},K^{*+})\nonumber\\
],
\end{eqnarray}
\begin{eqnarray}
{\cal A}(\Omega_{cc}^{+} \to \Xi_{c}^{+}K^{*0})
&=&C_{\rm SD}(\Omega_{cc}^{+} \to \Xi_{c}^{+}K^{*0})+ i [
    {\cal M}(\Omega_{cc}^{+},  K^{+},     \Xi_{c}^{0},     \pi^{-},     K^{*0},     \Xi_{c}^{+})
  + {\cal M}(\Omega_{cc}^{+},  K^{*+},    \Xi_{c}^{0},     \rho^{-},    K^{*0},     \Xi_{c}^{+})\nonumber\\
&&+ {\cal M}(\Omega_{cc}^{+},  K^{+},     \Xi_{c}^{'0},     \pi^{-},     K^{*0},     \Xi_{c}^{+})
  + {\cal M}(\Omega_{cc}^{+},  K^{*+},    \Xi_{c}^{'0},     \rho^{-},    K^{*0},     \Xi_{c}^{+})
  +{\cal M}(\Omega_{cc}^{+},   K^{+},     \Xi_{c}^{0},     \Omega_{c}^{0},\Xi_{c}^{+},K^{*0})\nonumber\\
&&+ {\cal M}(\Omega_{cc}^{+},  K^{*+},    \Xi_{c}^{0},     \Omega_{c}^{0},\Xi_{c}^{+},K^{*0})
  + {\cal M}(\Omega_{cc}^{+},  K^{+},     \Xi_{c}^{'0},     \Omega_{c}^{0},\Xi_{c}^{+},K^{*0})
  + {\cal M}(\Omega_{cc}^{+},  K^{*+},    \Xi_{c}^{'0},     \Omega_{c}^{0},\Xi_{c}^{+},K^{*0})\nonumber\\
&&+{\cal M}(\Omega_{cc}^{+},   K^{0},     \Xi_{c}^{+},     \eta_8,     K^{*0},     \Xi_{c}^{+})
  + {\cal M}(\Omega_{cc}^{+},  K^{*0},    \Xi_{c}^{+},     \phi,       K^{*0},     \Xi_{c}^{+})
  + {\cal M}(\Omega_{cc}^{+},  K^{0},     \Xi_{c}^{'+},     \eta_8,   K^{*0},     \Xi_{c}^{+})\nonumber\\
&&+ {\cal M}(\Omega_{cc}^{+},  K^{*0},    \Xi_{c}^{'+},     \phi,      K^{*0},     \Xi_{c}^{+})],
\end{eqnarray}
\begin{eqnarray}
{\cal A}(\Omega_{cc}^{+} \to \Xi_{c}^{'+}K^{*0})
&=&C_{\rm SD}(\Omega_{cc}^{+} \to \Xi_{c}^{'+}K^{*0})+ i [
    {\cal M}(\Omega_{cc}^{+},  K^{+},     \Xi_{c}^{0},     \pi^{-},     K^{*0},     \Xi_{c}^{'+})
  + {\cal M}(\Omega_{cc}^{+},  K^{*+},    \Xi_{c}^{0},     \rho^{-},    K^{*0},     \Xi_{c}^{'+})\nonumber\\
&&+ {\cal M}(\Omega_{cc}^{+},  K^{+},     \Xi_{c}^{'0},     \pi^{-},     K^{*0},     \Xi_{c}^{'+})
  + {\cal M}(\Omega_{cc}^{+},  K^{*+},    \Xi_{c}^{'0},     \rho^{-},    K^{*0},     \Xi_{c}^{'+})
  +{\cal M}(\Omega_{cc}^{+},   K^{+},     \Xi_{c}^{0},     \Omega_{c}^{0},\Xi_{c}^{'+},K^{*0})\nonumber\\
&&+ {\cal M}(\Omega_{cc}^{+},  K^{*+},    \Xi_{c}^{0},     \Omega_{c}^{0},\Xi_{c}^{'+},K^{*0})
  + {\cal M}(\Omega_{cc}^{+},  K^{+},     \Xi_{c}^{'0},     \Omega_{c}^{0},\Xi_{c}^{'+},K^{*0})
  + {\cal M}(\Omega_{cc}^{+},  K^{*+},    \Xi_{c}^{'0},     \Omega_{c}^{0},\Xi_{c}^{'+},K^{*0})\nonumber\\
&&+{\cal M}(\Omega_{cc}^{+},   K^{0},     \Xi_{c}^{+},     \eta_8,     K^{*0},     \Xi_{c}^{'+})
  + {\cal M}(\Omega_{cc}^{+},  K^{*0},    \Xi_{c}^{+},     \phi,       K^{*0},     \Xi_{c}^{'+})
  + {\cal M}(\Omega_{cc}^{+},  K^{0},     \Xi_{c}^{'+},     \eta_8,   K^{*0},     \Xi_{c}^{'+})\nonumber\\
&&+ {\cal M}(\Omega_{cc}^{+},  K^{*0},    \Xi_{c}^{'+},     \phi,      K^{*0},     \Xi_{c}^{'+})],
\end{eqnarray}
\begin{eqnarray}
{\cal A}(\Omega_{cc}^{+} \to \Sigma_{c}^{+}\phi)
&=&i [
    {\cal M}(\Omega_{cc}^{+},  K^{+},     \Xi_{c}^{0},     K^{-},     \phi,     \Sigma_{c}^{+})
  + {\cal M}(\Omega_{cc}^{+},  K^{*+},    \Xi_{c}^{0},     K^{*-},    \phi,     \Sigma_{c}^{+})
  + {\cal M}(\Omega_{cc}^{+},  K^{+},     \Xi_{c}^{'0},     K^{-},     \phi,     \Sigma_{c}^{+})\nonumber\\
&&+ {\cal M}(\Omega_{cc}^{+},  K^{*+},    \Xi_{c}^{'0},     K^{*-},    \phi,     \Sigma_{c}^{+})
  + {\cal M}(\Omega_{cc}^{+},  K^{+},     \Xi_{c}^{0},    \Xi_{c}^{0},\Sigma_{c}^{+},\phi)
  + {\cal M}(\Omega_{cc}^{+},  K^{*+},    \Xi_{c}^{0},     \Xi_{c}^{0},\Sigma_{c}^{+},\phi)\nonumber\\
&&+ {\cal M}(\Omega_{cc}^{+},  K^{+},     \Xi_{c}^{'0},    \Xi_{c}^{0},\Sigma_{c}^{+},\phi)
  + {\cal M}(\Omega_{cc}^{+},  K^{*+},    \Xi_{c}^{'0},    \Xi_{c}^{0},\Sigma_{c}^{+},\phi)
  + {\cal M}(\Omega_{cc}^{+},  K^{+},     \Xi_{c}^{0},    \Xi_{c}^{'0},\Sigma_{c}^{+},\phi)\nonumber\\
&&+ {\cal M}(\Omega_{cc}^{+},  K^{*+},    \Xi_{c}^{0},     \Xi_{c}^{'0},\Sigma_{c}^{+},\phi)
  + {\cal M}(\Omega_{cc}^{+},  K^{+},     \Xi_{c}^{'0},    \Xi_{c}^{'0},\Sigma_{c}^{+},\phi)
  + {\cal M}(\Omega_{cc}^{+},  K^{*+},    \Xi_{c}^{'0},    \Xi_{c}^{'0},\Sigma_{c}^{+},\phi)\nonumber\\
&&+{\cal M}(\Omega_{cc}^{+},   K^{0},     \Xi_{c}^{+},     \bar K^{ 0},     \phi,     \Sigma_{c}^{+})
  + {\cal M}(\Omega_{cc}^{+},  K^{*0},    \Xi_{c}^{+},    \bar K^{*0},        \phi,     \Sigma_{c}^{+})
  + {\cal M}(\Omega_{cc}^{+},  K^{0},     \Xi_{c}^{'+},    \bar K^{ 0},     \phi,     \Sigma_{c}^{+})\nonumber\\
&&+ {\cal M}(\Omega_{cc}^{+},  K^{*0},    \Xi_{c}^{'+},    \bar K^{*0},        \phi,     \Sigma_{c}^{+})
  + {\cal M}(\Omega_{cc}^{+},  K^{0},     \Xi_{c}^{+},    \Xi_{c}^{+},\Sigma_{c}^{+},\phi)
  + {\cal M}(\Omega_{cc}^{+},  K^{*0},    \Xi_{c}^{+},    \Xi_{c}^{+},\Sigma_{c}^{+},\phi)\nonumber\\
&&+ {\cal M}(\Omega_{cc}^{+},  K^{0},     \Xi_{c}^{+},    \Xi_{c}^{'+},\Sigma_{c}^{+},\phi)
  + {\cal M}(\Omega_{cc}^{+},  K^{*0},    \Xi_{c}^{+},    \Xi_{c}^{'+},\Sigma_{c}^{+},\phi)
  + {\cal M}(\Omega_{cc}^{+},  K^{0},     \Xi_{c}^{'+},    \Xi_{c}^{+},\Sigma_{c}^{+},\phi)\nonumber\\
&&+ {\cal M}(\Omega_{cc}^{+},  K^{*0},    \Xi_{c}^{'+},    \Xi_{c}^{+},\Sigma_{c}^{+},\phi)
  + {\cal M}(\Omega_{cc}^{+},  K^{0},     \Xi_{c}^{'+},    \Xi_{c}^{'+},\Sigma_{c}^{+},\phi)
  + {\cal M}(\Omega_{cc}^{+},  K^{*0},    \Xi_{c}^{'+},    \Xi_{c}^{'+},\Sigma_{c}^{+},\phi)],
\end{eqnarray}
\begin{eqnarray}
{\cal A}(\Omega_{cc}^{+} \to \Lambda_{c}^{+}\phi)
&=&i [
    {\cal M}(\Omega_{cc}^{+},  K^{+},     \Xi_{c}^{0},     K^{-},     \phi,     \Lambda_{c}^{+})
  + {\cal M}(\Omega_{cc}^{+},  K^{*+},    \Xi_{c}^{0},     K^{*-},    \phi,     \Lambda_{c}^{+})
  + {\cal M}(\Omega_{cc}^{+},  K^{+},     \Xi_{c}^{'0},     K^{-},     \phi,     \Lambda_{c}^{+})\nonumber\\
&&+ {\cal M}(\Omega_{cc}^{+},  K^{*+},    \Xi_{c}^{'0},     K^{*-},    \phi,     \Lambda_{c}^{+})
  + {\cal M}(\Omega_{cc}^{+},  K^{+},     \Xi_{c}^{0},    \Xi_{c}^{0},\Lambda_{c}^{+},\phi)
  + {\cal M}(\Omega_{cc}^{+},  K^{*+},    \Xi_{c}^{0},     \Xi_{c}^{0},\Lambda_{c}^{+},\phi)\nonumber\\
&&+ {\cal M}(\Omega_{cc}^{+},  K^{+},     \Xi_{c}^{'0},    \Xi_{c}^{0},\Lambda_{c}^{+},\phi)
  + {\cal M}(\Omega_{cc}^{+},  K^{*+},    \Xi_{c}^{'0},    \Xi_{c}^{0},\Lambda_{c}^{+},\phi)
  + {\cal M}(\Omega_{cc}^{+},  K^{+},     \Xi_{c}^{0},    \Xi_{c}^{'0},\Lambda_{c}^{+},\phi)\nonumber\\
&&+ {\cal M}(\Omega_{cc}^{+},  K^{*+},    \Xi_{c}^{0},     \Xi_{c}^{'0},\Lambda_{c}^{+},\phi)
  + {\cal M}(\Omega_{cc}^{+},  K^{+},     \Xi_{c}^{'0},    \Xi_{c}^{'0},\Lambda_{c}^{+},\phi)
  + {\cal M}(\Omega_{cc}^{+},  K^{*+},    \Xi_{c}^{'0},    \Xi_{c}^{'0},\Lambda_{c}^{+},\phi)\nonumber\\
&&+{\cal M}(\Omega_{cc}^{+},   K^{0},     \Xi_{c}^{+},     \bar K^0,     \phi,     \Lambda_{c}^{+})
  + {\cal M}(\Omega_{cc}^{+},  K^{*0},    \Xi_{c}^{+},    \bar K^{*0},        \phi,     \Lambda_{c}^{+})
  + {\cal M}(\Omega_{cc}^{+},  K^{0},     \Xi_{c}^{'+},    \bar K^0,     \phi,     \Lambda_{c}^{+})\nonumber\\
&&+ {\cal M}(\Omega_{cc}^{+},  K^{*0},    \Xi_{c}^{'+},    \bar K^{*0},        \phi,     \Lambda_{c}^{+})
  + {\cal M}(\Omega_{cc}^{+},  K^{0},     \Xi_{c}^{+},    \Xi_{c}^{+},\Lambda_{c}^{+},\phi)
  + {\cal M}(\Omega_{cc}^{+},  K^{*0},    \Xi_{c}^{+},    \Xi_{c}^{+},\Lambda_{c}^{+},\phi)\nonumber\\
&&+ {\cal M}(\Omega_{cc}^{+},  K^{0},     \Xi_{c}^{+},    \Xi_{c}^{'+},\Lambda_{c}^{+},\phi)
  + {\cal M}(\Omega_{cc}^{+},  K^{*0},    \Xi_{c}^{+},    \Xi_{c}^{'+},\Lambda_{c}^{+},\phi)
  + {\cal M}(\Omega_{cc}^{+},  K^{0},     \Xi_{c}^{'+},    \Xi_{c}^{+},\Lambda_{c}^{+},\phi)\nonumber\\
&&+ {\cal M}(\Omega_{cc}^{+},  K^{*0},    \Xi_{c}^{'+},    \Xi_{c}^{+},\Lambda_{c}^{+},\phi)
  + {\cal M}(\Omega_{cc}^{+},  K^{0},     \Xi_{c}^{'+},    \Xi_{c}^{'+},\Lambda_{c}^{+},\phi)
  + {\cal M}(\Omega_{cc}^{+},  K^{*0},    \Xi_{c}^{'+},    \Xi_{c}^{'+},\Lambda_{c}^{+},\phi)],
\end{eqnarray}
\begin{eqnarray}
{\cal A}(\Omega_{cc}^{+} \to \Xi_{c}^{0}K^{*+})
&=&T_{\rm SD}(\Omega_{cc}^{+} \to \Xi_{c}^{0}K^{*+})+ i [
    {\cal M}(\Omega_{cc}^{+},  K^{0},     \Xi_{c}^{+},     \pi^{+},     K^{*+},     \Xi_{c}^{0})
  + {\cal M}(\Omega_{cc}^{+},  K^{*0},    \Xi_{c}^{+},     \rho^{+},    K^{*+},     \Xi_{c}^{0})\nonumber\\
&&+ {\cal M}(\Omega_{cc}^{+},  K^{0},     \Xi_{c}^{'+},     \pi^{+},     K^{*+},     \Xi_{c}^{0})
  + {\cal M}(\Omega_{cc}^{+},  K^{*0},    \Xi_{c}^{'+},     \rho^{+},    K^{*+},     \Xi_{c}^{0})
  +{\cal M}(\Omega_{cc}^{+},   K^{0},     \Xi_{c}^{+},     \Omega_{c}^{0},K^{*+},     \Xi_{c}^{0})\nonumber\\
&&+ {\cal M}(\Omega_{cc}^{+},  K^{*0},    \Xi_{c}^{+},     \Omega_{c}^{0},K^{*+},     \Xi_{c}^{0})
  + {\cal M}(\Omega_{cc}^{+},  K^{0},     \Xi_{c}^{'+},     \Omega_{c}^{0},K^{*+},     \Xi_{c}^{0})
  + {\cal M}(\Omega_{cc}^{+},  K^{*0},    \Xi_{c}^{'+},     \Omega_{c}^{0},K^{*+},     \Xi_{c}^{0})\nonumber\\
&&+{\cal M}(\Omega_{cc}^{+},   K^{+},     \Xi_{c}^{0},     \eta_8,     K^{*+},     \Xi_{c}^{0})
  + {\cal M}(\Omega_{cc}^{+},  K^{*+},    \Xi_{c}^{0},     \phi,       K^{*+},     \Xi_{c}^{0})
  + {\cal M}(\Omega_{cc}^{+},  K^{+},     \Xi_{c}^{'0},     \eta_8,   K^{*+},     \Xi_{c}^{0})\nonumber\\
&&+ {\cal M}(\Omega_{cc}^{+},  K^{*+},    \Xi_{c}^{'0},     \phi,      K^{*+},     \Xi_{c}^{0})],
\end{eqnarray}
\begin{eqnarray}
{\cal A}(\Omega_{cc}^{+} \to \Xi_{c}^{'0}K^{*+})
&=&T_{\rm SD}(\Omega_{cc}^{+} \to \Xi_{c}^{'0}K^{*+})+ i [
    {\cal M}(\Omega_{cc}^{+},  K^{0},     \Xi_{c}^{+},     \pi^{+},     K^{*+},     \Xi_{c}^{'0})
  + {\cal M}(\Omega_{cc}^{+},  K^{*0},    \Xi_{c}^{+},     \rho^{+},    K^{*+},     \Xi_{c}^{'0})\nonumber\\
&&+ {\cal M}(\Omega_{cc}^{+},  K^{0},     \Xi_{c}^{'+},     \pi^{+},     K^{*+},     \Xi_{c}^{'0})
  + {\cal M}(\Omega_{cc}^{+},  K^{*0},    \Xi_{c}^{'+},     \rho^{+},    K^{*+},     \Xi_{c}^{'0})
  +{\cal M}(\Omega_{cc}^{+},   K^{0},     \Xi_{c}^{+},     \Omega_{c}^{0},K^{*+},     \Xi_{c}^{'0})\nonumber\\
&&+ {\cal M}(\Omega_{cc}^{+},  K^{*0},    \Xi_{c}^{+},     \Omega_{c}^{0},K^{*+},     \Xi_{c}^{'0})
  + {\cal M}(\Omega_{cc}^{+},  K^{0},     \Xi_{c}^{'+},     \Omega_{c}^{0},K^{*+},     \Xi_{c}^{'0})
  + {\cal M}(\Omega_{cc}^{+},  K^{*0},    \Xi_{c}^{'+},     \Omega_{c}^{0},K^{*+},     \Xi_{c}^{'0})\nonumber\\
&&+{\cal M}(\Omega_{cc}^{+},   K^{+},     \Xi_{c}^{0},     \eta_8,     K^{*+},     \Xi_{c}^{'0})
  + {\cal M}(\Omega_{cc}^{+},  K^{*+},    \Xi_{c}^{0},     \phi,       K^{*+},     \Xi_{c}^{'0})
  + {\cal M}(\Omega_{cc}^{+},  K^{+},     \Xi_{c}^{'0},     \eta_8,   K^{*+},     \Xi_{c}^{'0})\nonumber\\
&&+ {\cal M}(\Omega_{cc}^{+},  K^{*+},    \Xi_{c}^{'0},     \phi,      K^{*+},     \Xi_{c}^{'0})],
\end{eqnarray}
\begin{eqnarray}
{\cal A}(\Xi_{cc}^{+} \to \Sigma_c^{++}K^{*-})
&=&i [
    {\cal M}(\Xi_{cc}^{+}, \overline{K}^{0}, \Lambda_{c}^{+},\pi^{-},K^{*-},\Sigma_c^{++})
  + {\cal M}(\Xi_{cc}^{+}, \overline{K}^{*0},\Lambda_{c}^{+},\rho^{-}, K^{*-},\Sigma_c^{++})
  + {\cal M}(\Xi_{cc}^{+}, \overline{K}^{0}, \Sigma_{c}^{+}, \pi^{-}, K^{*-},\Sigma_c^{++}) \nonumber\\
&&+ {\cal M}(\Xi_{cc}^{+}, \overline{K}^{*0},\Sigma_{c}^{+}, \rho^{-},K^{*-},\Sigma_c^{++})
  + {\cal M}(\Xi_{cc}^{+}, \pi^{+}, \Xi_{c}^{0}, \Lambda_{c}^{+},\Sigma_c^{++},K^{*-})
  + {\cal M}(\Xi_{cc}^{+}, \rho^{+},\Xi_{c}^{0}, \Lambda_{c}^{+},\Sigma_c^{++},K^{*-})\nonumber\\
&&+ {\cal M}(\Xi_{cc}^{+}, \pi^{+}, \Xi_{c}^{'0}, \Lambda_{c}^{+},\Sigma_c^{++}, K^{*-})
  + {\cal M}(\Xi_{cc}^{+}, \rho^{+},\Xi_{c}^{'0}, \Lambda_{c}^{+},\Sigma_c^{++}, K^{*-})
  + {\cal M}(\Xi_{cc}^{+}, \pi^{+}, \Xi_{c}^{0}, \Sigma_{c}^{+},\Sigma_c^{++},K^{*-})\nonumber\\
&&+ {\cal M}(\Xi_{cc}^{+}, \rho^{+},\Xi_{c}^{0}, \Sigma_{c}^{+},\Sigma_c^{++},K^{*-})
  + {\cal M}(\Xi_{cc}^{+}, \pi^{+}, \Xi_{c}^{'0},\Sigma_{c}^{+},\Sigma_c^{++},K^{*-})
  + {\cal M}(\Xi_{cc}^{+}, \rho^{+},\Xi_{c}^{'0},\Sigma_{c}^{+},\Sigma_c^{++},K^{*-})],
\end{eqnarray}
\begin{eqnarray}
{\cal A}(\Xi_{cc}^{+} \to \Xi_c^{+}\phi)
&=&i [
    {\cal M}(\Xi_{cc}^{+}, \overline{K}^{0}, \Lambda_{c}^{+}, K^{0},\phi,\Xi_c^{+})
  + {\cal M}(\Xi_{cc}^{+}, \overline{K}^{*0},\Lambda_{c}^{+}, K^{*0}, \phi,\Xi_c^{+})
  + {\cal M}(\Xi_{cc}^{+}, \overline{K}^{0}, \Sigma_{c}^{+}, K^{0}, \phi,\Xi_c^{+}) \nonumber\\
&&+ {\cal M}(\Xi_{cc}^{+}, \overline{K}^{*0},\Sigma_{c}^{+}, K^{*0},\phi,\Xi_c^{+})
  + {\cal M}(\Xi_{cc}^{+}, \pi^{+}, \Xi_{c}^{0}, \Xi_{c}^{0},\Xi_c^{+},\phi)
  + {\cal M}(\Xi_{cc}^{+}, \rho^{+},\Xi_{c}^{0}, \Xi_{c}^{0},\Xi_c^{+},\phi)\nonumber\\
&&+ {\cal M}(\Xi_{cc}^{+}, \pi^{+}, \Xi_{c}^{'0}, \Xi_{c}^{0},\Xi_c^{+}, \phi)
  + {\cal M}(\Xi_{cc}^{+}, \rho^{+},\Xi_{c}^{'0}, \Xi_{c}^{0},\Xi_c^{+}, \phi)
  + {\cal M}(\Xi_{cc}^{+}, \pi^{+}, \Xi_{c}^{0}, \Xi_{c}^{'0},\Xi_c^{+},\phi)\nonumber\\
&&+ {\cal M}(\Xi_{cc}^{+}, \rho^{+},\Xi_{c}^{0}, \Xi_{c}^{'0},\Xi_c^{+}, \phi)
  + {\cal M}(\Xi_{cc}^{+}, \pi^{+}, \Xi_{c}^{'0},\Xi_{c}^{'0},\Xi_c^{+}, \phi)
  + {\cal M}(\Xi_{cc}^{+}, \rho^{+},\Xi_{c}^{'0},\Xi_{c}^{'0},\Xi_c^{+}, \phi)],
\end{eqnarray}
\begin{eqnarray}
{\cal A}(\Xi_{cc}^{+} \to \Xi_c^{'+}\phi)
&=&i [
    {\cal M}(\Xi_{cc}^{+}, \overline{K}^{0}, \Lambda_{c}^{+}, K^{0},\phi,\Xi_c^{'+})
  + {\cal M}(\Xi_{cc}^{+}, \overline{K}^{*0},\Lambda_{c}^{+}, K^{*0}, \phi,\Xi_c^{'+})
  + {\cal M}(\Xi_{cc}^{+}, \overline{K}^{0}, \Sigma_{c}^{+}, K^{0}, \phi,\Xi_c^{'+}) \nonumber\\
&&+ {\cal M}(\Xi_{cc}^{+}, \overline{K}^{*0},\Sigma_{c}^{+}, K^{*0},\phi,\Xi_c^{'+})
  + {\cal M}(\Xi_{cc}^{+}, \pi^{+}, \Xi_{c}^{0}, \Xi_{c}^{0},\Xi_c^{'+},\phi)
  + {\cal M}(\Xi_{cc}^{+}, \rho^{+},\Xi_{c}^{0}, \Xi_{c}^{0},\Xi_c^{'+},\phi)\nonumber\\
&&+ {\cal M}(\Xi_{cc}^{+}, \pi^{+}, \Xi_{c}^{'0}, \Xi_{c}^{0},\Xi_c^{'+}, \phi)
  + {\cal M}(\Xi_{cc}^{+}, \rho^{+},\Xi_{c}^{'0}, \Xi_{c}^{0},\Xi_c^{'+}, \phi)
  + {\cal M}(\Xi_{cc}^{+}, \pi^{+}, \Xi_{c}^{0}, \Xi_{c}^{'0},\Xi_c^{'+},\phi)\nonumber\\
&&+ {\cal M}(\Xi_{cc}^{+}, \rho^{+},\Xi_{c}^{0}, \Xi_{c}^{'0},\Xi_c^{'+}, \phi)
  + {\cal M}(\Xi_{cc}^{+}, \pi^{+}, \Xi_{c}^{'0},\Xi_{c}^{'0},\Xi_c^{'+}, \phi)
  + {\cal M}(\Xi_{cc}^{+}, \rho^{+},\Xi_{c}^{'0},\Xi_{c}^{'0},\Xi_c^{'+}, \phi)],
\end{eqnarray}
\begin{eqnarray}
{\cal A}(\Xi_{cc}^{+} \to \Omega_c^{0} K^{*+})
&=&i [
    {\cal M}(\Xi_{cc}^{+}, \pi^{+}, \Xi_{c}^{0}, K^{0}, K^{*+},\Omega_c^{0})
  + {\cal M}(\Xi_{cc}^{+}, \rho^{+},\Xi_{c}^{0}, K^{*0}, K^{*+},\Omega_c^{0})
  + {\cal M}(\Xi_{cc}^{+}, \pi^{+}, \Xi_{c}^{'0}, K^{0}, K^{*+},\Omega_c^{0}) \nonumber\\
&&+ {\cal M}(\Xi_{cc}^{+}, \rho^{+},\Xi_{c}^{'0}, K^{*0}, K^{*+},\Omega_c^{0})
  + {\cal M}(\Xi_{cc}^{+}, \overline{K}^{0}, \Lambda_c^{+}, \Xi_{c}^{0},\Omega_c^{0}, K^{*+})
  + {\cal M}(\Xi_{cc}^{+}, \overline{K}^{*0},\Lambda_c^{+}, \Xi_{c}^{0},\Omega_c^{0}, K^{*+})\nonumber\\
&&+ {\cal M}(\Xi_{cc}^{+}, \overline{K}^{0}, \Sigma_c^{+}, \Xi_{c}^{0},\Omega_c^{0}, K^{*+})
  + {\cal M}(\Xi_{cc}^{+}, \overline{K}^{*0},\Sigma_c^{+}, \Xi_{c}^{0},\Omega_c^{0}, K^{*+})
  + {\cal M}(\Xi_{cc}^{+}, \overline{K}^{0}, \Lambda_c^{+}, \Xi_{c}^{'0},\Omega_c^{0},K^{*+})\nonumber\\
&&+ {\cal M}(\Xi_{cc}^{+}, \overline{K}^{*0},\Lambda_c^{+}, \Xi_{c}^{'0},\Omega_c^{0}, K^{*+})
  + {\cal M}(\Xi_{cc}^{+}, \overline{K}^{0}, \Sigma_c^{+},\Xi_{c}^{'0},\Omega_c^{0},K^{*+})
  + {\cal M}(\Xi_{cc}^{+}, \overline{K}^{*0},\Sigma_c^{+},\Xi_{c}^{'0},\Omega_c^{0}, K^{*+})],
\end{eqnarray}
\begin{eqnarray}
{\cal A}(\Xi_{cc}^{+} \to \Sigma_c^{++} \rho^{-})
&=&i [
    {\cal M}(\Xi_{cc}^{+}, \pi^{0} , \Lambda_{c}^{+}, \pi^{-}, \rho^{-},\Sigma_c^{++})
  + {\cal M}(\Xi_{cc}^{+}, \rho^{0} ,\Lambda_{c}^{+},\rho^{-}, \rho^{-},\Sigma_c^{++})
  + {\cal M}(\Xi_{cc}^{+}, \pi^{0} , \Sigma_{c}^{+}, \pi^{-}, \rho^{-},\Sigma_c^{++})\nonumber\\
&&+ {\cal M}(\Xi_{cc}^{+}, \rho^{0} ,\Sigma_{c}^{+},\rho^{-}, \rho^{-},\Sigma_c^{++})
  + {\cal M}(\Xi_{cc}^{+}, \pi^{+},\Sigma_{c}^{0},\Lambda_{c}^{+}, \Sigma_c^{++},\rho^{-})
  + {\cal M}(\Xi_{cc}^{+}, \rho^{+},\Sigma_{c}^{0},\Lambda_{c}^{+}, \Sigma_c^{++},\rho^{-})\nonumber\\
&&+ {\cal M}(\Xi_{cc}^{+}, \pi^{+},\Sigma_{c}^{0},\Sigma_{c}^{+}, \Sigma_c^{++},\rho^{-})
  + {\cal M}(\Xi_{cc}^{+}, \rho^{+},\Sigma_{c}^{0},\Sigma_{c}^{+}, \Sigma_c^{++},\rho^{-})
  + {\cal M}(\Xi_{cc}^{+}, K^{+},\Xi_{c}^{0},\Xi_{c}^{+}, \Sigma_c^{++},\rho^{-})\nonumber\\
&&+ {\cal M}(\Xi_{cc}^{+}, K^{*+},\Xi_{c}^{0},\Xi_{c}^{+}, \Sigma_c^{++},\rho^{-})
  + {\cal M}(\Xi_{cc}^{+}, K^{+},\Xi_{c}^{'0},\Xi_{c}^{+}, \Sigma_c^{++},\rho^{-})
  + {\cal M}(\Xi_{cc}^{+}, K^{*+},\Xi_{c}^{'0},\Xi_{c}^{+}, \Sigma_c^{++},\rho^{-})\nonumber\\
&&+ {\cal M}(\Xi_{cc}^{+}, K^{+},\Xi_{c}^{0},\Xi_{c}^{'+}, \Sigma_c^{++},\rho^{-})
  + {\cal M}(\Xi_{cc}^{+}, K^{*+},\Xi_{c}^{0},\Xi_{c}^{'+}, \Sigma_c^{++},\rho^{-})
  + {\cal M}(\Xi_{cc}^{+}, K^{+},\Xi_{c}^{'0},\Xi_{c}^{'+}, \Sigma_c^{++},\rho^{-})\nonumber\\
&&+ {\cal M}(\Xi_{cc}^{+}, K^{*+},\Xi_{c}^{'0},\Xi_{c}^{'+}, \Sigma_c^{++},\rho^{-}) ],
\end{eqnarray}
\begin{eqnarray}
{\cal A}(\Omega_{cc}^{+} \to \Sigma_c^{++} K^{*-})
&=&i [
    {\cal M}(\Omega_{cc}^{+}, \eta_8, \Xi_{c}^{+}, K^{-}, K^{*-},\Sigma_c^{++})
  + {\cal M}(\Omega_{cc}^{+}, \phi, \Xi_{c}^{+}, K^{*-}, K^{*-},\Sigma_c^{++})
  + {\cal M}(\Omega_{cc}^{+}, \eta_8, \Xi_{c}^{'+}, K^{-}, K^{*-},\Sigma_c^{++})\nonumber\\
&&+ {\cal M}(\Omega_{cc}^{+}, \phi, \Xi_{c}^{'+},K^{*-}, K^{*-},\Sigma_c^{++})
  + {\cal M}(\Omega_{cc}^{+}, K^{+}, \Omega_{c}^{0}, \Xi_{c}^{+}, \Sigma_c^{++},K^{*-})
  + {\cal M}(\Omega_{cc}^{+}, K^{*+}, \Omega_{c}^{0}, \Xi_{c}^{+}, \Sigma_c^{++},K^{*-})\nonumber\\
&&+ {\cal M}(\Omega_{cc}^{+}, K^{+}, \Omega_{c}^{0}, \Xi_{c}^{'+}, \Sigma_c^{++},K^{*-})
  + {\cal M}(\Omega_{cc}^{+}, K^{*+}, \Omega_{c}^{0}, \Xi_{c}^{'+}, \Sigma_c^{++},K^{*-})
  + {\cal M}(\Omega_{cc}^{+}, \pi^{+}, \Xi_{c}^{0}, \Lambda_{c}^{+}, \Sigma_c^{++},K^{*-})\nonumber\\
&&+ {\cal M}(\Omega_{cc}^{+}, \rho^{+}, \Xi_{c}^{0}, \Lambda_{c}^{+}, \Sigma_c^{++},K^{*-})
  + {\cal M}(\Omega_{cc}^{+}, \pi^{+}, \Xi_{c}^{'0}, \Lambda_{c}^{+}, \Sigma_c^{++},K^{*-})
  + {\cal M}(\Omega_{cc}^{+}, \rho^{+}, \Xi_{c}^{'0}, \Lambda_{c}^{+}, \Sigma_c^{++},K^{*-})\nonumber\\
&&+ {\cal M}(\Omega_{cc}^{+}, \pi^{+}, \Xi_{c}^{0}, \Sigma_{c}^{+}, \Sigma_c^{++},K^{*-})
  + {\cal M}(\Omega_{cc}^{+}, \rho^{+}, \Xi_{c}^{0}, \Sigma_{c}^{+}, \Sigma_c^{++},K^{*-})
  + {\cal M}(\Omega_{cc}^{+}, \pi^{+}, \Xi_{c}^{'0}, \Sigma_{c}^{+}, \Sigma_c^{++},K^{*-})\nonumber\\
&&+ {\cal M}(\Omega_{cc}^{+}, \rho^{+}, \Xi_{c}^{'0}, \Sigma_{c}^{+}, \Sigma_c^{++},K^{*-})],
\end{eqnarray}
\begin{eqnarray}
{\cal A}(\Omega_{cc}^{+} \to \Sigma_c^{++} \rho^{-})
&=&i [
    {\cal M}(\Omega_{cc}^{+}, K^{0}, \Xi_{c}^{+}, K^{-}, \rho^{-},\Sigma_c^{++})
  + {\cal M}(\Omega_{cc}^{+}, K^{*0}, \Xi_{c}^{+}, K^{*-}, \rho^{-},\Sigma_c^{++})
  + {\cal M}(\Omega_{cc}^{+}, K^{0}, \Xi_{c}^{'+}, K^{-}, \rho^{-},\Sigma_c^{++})\nonumber\\
&&+ {\cal M}(\Omega_{cc}^{+}, K^{*0}, \Xi_{c}^{'+},K^{*-}, \rho^{-},\Sigma_c^{++})
  + {\cal M}(\Omega_{cc}^{+}, K^{+}, \Xi_{c}^{0}, \Xi_{c}^{+}, \Sigma_c^{++},\rho^{-})
  + {\cal M}(\Omega_{cc}^{+}, K^{*+}, \Xi_{c}^{0}, \Xi_{c}^{+}, \Sigma_c^{++},\rho^{-})\nonumber\\
&&+ {\cal M}(\Omega_{cc}^{+}, K^{+}, \Xi_{c}^{0}, \Xi_{c}^{'+}, \Sigma_c^{++},\rho^{-})
  + {\cal M}(\Omega_{cc}^{+}, K^{*+}, \Xi_{c}^{0}, \Xi_{c}^{'+}, \Sigma_c^{++},\rho^{-})
  + {\cal M}(\Omega_{cc}^{+}, K^{+}, \Xi_{c}^{'0}, \Xi_{c}^{+}, \Sigma_c^{++},\rho^{-})\nonumber\\
&&+ {\cal M}(\Omega_{cc}^{+}, K^{*+}, \Xi_{c}^{'0}, \Xi_{c}^{+}, \Sigma_c^{++},\rho^{-})
  + {\cal M}(\Omega_{cc}^{+}, K^{+}, \Xi_{c}^{'0}, \Xi_{c}^{'+}, \Sigma_c^{++},\rho^{-})
  + {\cal M}(\Omega_{cc}^{+}, K^{*+}, \Xi_{c}^{'0}, \Xi_{c}^{'+}, \Sigma_c^{++},\rho^{-})],
\end{eqnarray}
\begin{eqnarray}
{\cal A}(\Omega_{cc}^{+} \to \Sigma_c^{+}\rho^{0})
&=&i [
    {\cal M}(\Omega_{cc}^{+}, K^{0}, \Xi_{c}^{+}, \bar K^{*0}, \rho^{0} ,\Sigma_c^{+})
  + {\cal M}(\Omega_{cc}^{+}, K^{*0}, \Xi_{c}^{+}, \bar K^{*0}, \rho^{0} ,\Sigma_c^{+})
  + {\cal M}(\Omega_{cc}^{+}, K^{0}, \Xi_{c}^{'+}, \bar K^{*0}, \rho^{0} ,\Sigma_c^{+})\nonumber\\
&&+ {\cal M}(\Omega_{cc}^{+}, K^{*0}, \Xi_{c}^{'+}, \bar K^{*0}, \rho^{0} ,\Sigma_c^{+})
  + {\cal M}(\Omega_{cc}^{+}, K^{0}, \Xi_{c}^{+}, \Xi_{c}^{+}, \Sigma_c^{+},\rho^{0} )
  + {\cal M}(\Omega_{cc}^{+}, K^{*0}, \Xi_{c}^{+}, \Xi_{c}^{+}, \Sigma_c^{+},\rho^{0} )\nonumber\\
&&+{\cal M}(\Omega_{cc}^{+}, K^{0}, \Xi_{c}^{+}, \Xi_{c}^{'+}, \Sigma_c^{+},\rho^{0} )
  +{\cal M}(\Omega_{cc}^{+}, K^{*0}, \Xi_{c}^{+}, \Xi_{c}^{'+}, \Sigma_c^{+},\rho^{0} )
  + {\cal M}(\Omega_{cc}^{+}, K^{0}, \Xi_{c}^{'+}, \Xi_{c}^{+}, \Sigma_c^{+},\rho^{0} )\nonumber\\
&&+ {\cal M}(\Omega_{cc}^{+}, K^{*0}, \Xi_{c}^{'+}, \Xi_{c}^{+}, \Sigma_c^{+},\rho^{0} )
  +{\cal M}(\Omega_{cc}^{+}, K^{0}, \Xi_{c}^{'+}, \Xi_{c}^{'+}, \Sigma_c^{+},\rho^{0} )
  +{\cal M}(\Omega_{cc}^{+}, K^{*0}, \Xi_{c}^{'+}, \Xi_{c}^{'+}, \Sigma_c^{+},\rho^{0} )\nonumber\\
&&+ {\cal M}(\Omega_{cc}^{+}, K^{+}, \Xi_{c}^{0}, K^{-}, \rho^{0} , \Sigma_c^{+})
  + {\cal M}(\Omega_{cc}^{+}, K^{*+}, \Xi_{c}^{0}, K^{*-}, \rho^{0} , \Sigma_c^{+})
  + {\cal M}(\Omega_{cc}^{+}, K^{+}, \Xi_{c}^{'0}, K^{-}, \rho^{0} , \Sigma_c^{+})\nonumber\\
&&+ {\cal M}(\Omega_{cc}^{+}, K^{*+}, \Xi_{c}^{'0}, K^{*-}, \rho^{0} , \Sigma_c^{+})
  + {\cal M}(\Omega_{cc}^{+}, K^{+}, \Xi_{c}^{0}, \Xi_{c}^{0}, \Sigma_c^{+}, \rho^{0} )
  + {\cal M}(\Omega_{cc}^{+}, K^{*+}, \Xi_{c}^{0}, \Xi_{c}^{0}, \Sigma_c^{+}, \rho^{0} )\nonumber\\
&&+ {\cal M}(\Omega_{cc}^{+}, K^{+}, \Xi_{c}^{0}, \Xi_{c}^{'0}, \Sigma_c^{+}, \rho^{0} )
  + {\cal M}(\Omega_{cc}^{+}, K^{*+}, \Xi_{c}^{0}, \Xi_{c}^{'0}, \Sigma_c^{+}, \rho^{0} )
  + {\cal M}(\Omega_{cc}^{+}, K^{+}, \Xi_{c}^{'0}, \Xi_{c}^{0}, \Sigma_c^{+}, \rho^{0} )\nonumber\\
&&+ {\cal M}(\Omega_{cc}^{+}, K^{*+}, \Xi_{c}^{'0}, \Xi_{c}^{0}, \Sigma_c^{+}, \rho^{0} )
  + {\cal M}(\Omega_{cc}^{+}, K^{+}, \Xi_{c}^{'0}, \Xi_{c}^{'0}, \Sigma_c^{+}, \rho^{0} )
  + {\cal M}(\Omega_{cc}^{+}, K^{*+}, \Xi_{c}^{'0}, \Xi_{c}^{'0}, \Sigma_c^{+}, \rho^{0} )
  ],
\end{eqnarray}
\begin{eqnarray}
{\cal A}(\Omega_{cc}^{+} \to \Lambda_c^{+}\rho^{0})
&=&i [
    {\cal M}(\Omega_{cc}^{+}, K^{0}, \Xi_{c}^{+}, \bar K^{*0}, \rho^{0} ,\Lambda_c^{+})
  + {\cal M}(\Omega_{cc}^{+}, K^{*0}, \Xi_{c}^{+}, \bar K^{*0}, \rho^{0} ,\Lambda_c^{+})
  + {\cal M}(\Omega_{cc}^{+}, K^{0}, \Xi_{c}^{'+}, \bar K^{*0}, \rho^{0} ,\Lambda_c^{+})\nonumber\\
&&+ {\cal M}(\Omega_{cc}^{+}, K^{*0}, \Xi_{c}^{'+}, \bar K^{*0}, \rho^{0} ,\Lambda_c^{+})
  + {\cal M}(\Omega_{cc}^{+}, K^{0}, \Xi_{c}^{+}, \Xi_{c}^{+}, \Lambda_c^{+},\rho^{0} )
  + {\cal M}(\Omega_{cc}^{+}, K^{*0}, \Xi_{c}^{+}, \Xi_{c}^{+}, \Lambda_c^{+},\rho^{0} )\nonumber\\
&&+{\cal M}(\Omega_{cc}^{+}, K^{0}, \Xi_{c}^{+}, \Xi_{c}^{'+}, \Lambda_c^{+},\rho^{0} )
  +{\cal M}(\Omega_{cc}^{+}, K^{*0}, \Xi_{c}^{+}, \Xi_{c}^{'+}, \Lambda_c^{+},\rho^{0} )
  + {\cal M}(\Omega_{cc}^{+}, K^{0}, \Xi_{c}^{'+}, \Xi_{c}^{+}, \Lambda_c^{+},\rho^{0} )\nonumber\\
&&+ {\cal M}(\Omega_{cc}^{+}, K^{*0}, \Xi_{c}^{'+}, \Xi_{c}^{+}, \Lambda_c^{+},\rho^{0} )
  +{\cal M}(\Omega_{cc}^{+}, K^{0}, \Xi_{c}^{'+}, \Xi_{c}^{'+}, \Lambda_c^{+},\rho^{0} )
  +{\cal M}(\Omega_{cc}^{+}, K^{*0}, \Xi_{c}^{'+}, \Xi_{c}^{'+}, \Lambda_c^{+},\rho^{0} )\nonumber\\
&&+ {\cal M}(\Omega_{cc}^{+}, K^{+}, \Xi_{c}^{0}, K^{-}, \rho^{0} , \Lambda_c^{+})
  + {\cal M}(\Omega_{cc}^{+}, K^{*+}, \Xi_{c}^{0}, K^{*-}, \rho^{0} , \Lambda_c^{+})
  + {\cal M}(\Omega_{cc}^{+}, K^{+}, \Xi_{c}^{'0}, K^{-}, \rho^{0} , \Lambda_c^{+})\nonumber\\
&&+ {\cal M}(\Omega_{cc}^{+}, K^{*+}, \Xi_{c}^{'0}, K^{*-}, \rho^{0} , \Lambda_c^{+} )
  + {\cal M}(\Omega_{cc}^{+}, K^{+}, \Xi_{c}^{0}, \Xi_{c}^{0}, \Lambda_c^{+}, \rho^{0} )
  + {\cal M}(\Omega_{cc}^{+}, K^{*+}, \Xi_{c}^{0}, \Xi_{c}^{0}, \Lambda_c^{+}, \rho^{0} )\nonumber\\
&&+ {\cal M}(\Omega_{cc}^{+}, K^{+}, \Xi_{c}^{0}, \Xi_{c}^{'0}, \Lambda_c^{+}, \rho^{0} )
  + {\cal M}(\Omega_{cc}^{+}, K^{*+}, \Xi_{c}^{0}, \Xi_{c}^{'0}, \Lambda_c^{+}, \rho^{0} )
  + {\cal M}(\Omega_{cc}^{+}, K^{+}, \Xi_{c}^{'0}, \Xi_{c}^{0}, \Lambda_c^{+}, \rho^{0} )\nonumber\\
&&+ {\cal M}(\Omega_{cc}^{+}, K^{*+}, \Xi_{c}^{'0}, \Xi_{c}^{0}, \Lambda_c^{+}, \rho^{0} )
  + {\cal M}(\Omega_{cc}^{+}, K^{+}, \Xi_{c}^{'0}, \Xi_{c}^{'0}, \Lambda_c^{+}, \rho^{0} )
  + {\cal M}(\Omega_{cc}^{+}, K^{*+}, \Xi_{c}^{'0}, \Xi_{c}^{'0}, \Lambda_c^{+}, \rho^{0} )
  ],
\end{eqnarray}
\begin{eqnarray}
{\cal A}(\Omega_{cc}^{+} \to \Sigma_c^{+}\omega)
&=&i [
    {\cal M}(\Omega_{cc}^{+}, K^{0}, \Xi_{c}^{+}, \bar K^{*0}, \omega ,\Sigma_c^{+})
  + {\cal M}(\Omega_{cc}^{+}, K^{*0}, \Xi_{c}^{+}, \bar K^{*0}, \omega ,\Sigma_c^{+})
  + {\cal M}(\Omega_{cc}^{+}, K^{0}, \Xi_{c}^{'+}, \bar K^{*0}, \omega ,\Sigma_c^{+})\nonumber\\
&&+ {\cal M}(\Omega_{cc}^{+}, K^{*0}, \Xi_{c}^{'+}, \bar K^{*0}, \omega ,\Sigma_c^{+})
  + {\cal M}(\Omega_{cc}^{+}, K^{0}, \Xi_{c}^{+}, \Xi_{c}^{+}, \Sigma_c^{+},\omega )
  + {\cal M}(\Omega_{cc}^{+}, K^{*0}, \Xi_{c}^{+}, \Xi_{c}^{+}, \Sigma_c^{+},\omega )\nonumber\\
&&+{\cal M}(\Omega_{cc}^{+}, K^{0}, \Xi_{c}^{+}, \Xi_{c}^{'+}, \Sigma_c^{+},\omega )
  +{\cal M}(\Omega_{cc}^{+}, K^{*0}, \Xi_{c}^{+}, \Xi_{c}^{'+}, \Sigma_c^{+},\omega )
  + {\cal M}(\Omega_{cc}^{+}, K^{0}, \Xi_{c}^{'+}, \Xi_{c}^{+}, \Sigma_c^{+},\omega )\nonumber\\
&&+ {\cal M}(\Omega_{cc}^{+}, K^{*0}, \Xi_{c}^{'+}, \Xi_{c}^{+}, \Sigma_c^{+},\omega )
  +{\cal M}(\Omega_{cc}^{+}, K^{0}, \Xi_{c}^{'+}, \Xi_{c}^{'+}, \Sigma_c^{+},\omega )
  +{\cal M}(\Omega_{cc}^{+}, K^{*0}, \Xi_{c}^{'+}, \Xi_{c}^{'+}, \Sigma_c^{+},\omega )\nonumber\\
&&+ {\cal M}(\Omega_{cc}^{+}, K^{+}, \Xi_{c}^{0}, K^{-}, \omega , \Sigma_c^{+})
  + {\cal M}(\Omega_{cc}^{+}, K^{*+}, \Xi_{c}^{0}, K^{*-}, \omega , \Sigma_c^{+})
  + {\cal M}(\Omega_{cc}^{+}, K^{+}, \Xi_{c}^{'0}, K^{-}, \omega , \Sigma_c^{+})\nonumber\\
&&+ {\cal M}(\Omega_{cc}^{+}, K^{*+}, \Xi_{c}^{'0}, K^{*-}, \omega , \Sigma_c^{+})
  + {\cal M}(\Omega_{cc}^{+}, K^{+}, \Xi_{c}^{0}, \Xi_{c}^{0}, \Sigma_c^{+}, \omega )
  + {\cal M}(\Omega_{cc}^{+}, K^{*+}, \Xi_{c}^{0}, \Xi_{c}^{0}, \Sigma_c^{+}, \omega )\nonumber\\
&&+ {\cal M}(\Omega_{cc}^{+}, K^{+}, \Xi_{c}^{0}, \Xi_{c}^{'0}, \Sigma_c^{+}, \omega )
  + {\cal M}(\Omega_{cc}^{+}, K^{*+}, \Xi_{c}^{0}, \Xi_{c}^{'0}, \Sigma_c^{+}, \omega )
  + {\cal M}(\Omega_{cc}^{+}, K^{+}, \Xi_{c}^{'0}, \Xi_{c}^{0}, \Sigma_c^{+}, \omega )\nonumber\\
&&+ {\cal M}(\Omega_{cc}^{+}, K^{*+}, \Xi_{c}^{'0}, \Xi_{c}^{0}, \Sigma_c^{+}, \omega )
  + {\cal M}(\Omega_{cc}^{+}, K^{+}, \Xi_{c}^{'0}, \Xi_{c}^{'0}, \Sigma_c^{+}, \omega )
  + {\cal M}(\Omega_{cc}^{+}, K^{*+}, \Xi_{c}^{'0}, \Xi_{c}^{'0}, \Sigma_c^{+}, \omega )
  ],
\end{eqnarray}
\begin{eqnarray}
{\cal A}(\Omega_{cc}^{+} \to \Lambda_c^{+}\omega)
&=&i [
    {\cal M}(\Omega_{cc}^{+}, K^{0}, \Xi_{c}^{+}, \bar K^{*0}, \omega ,\Lambda_c^{+})
  + {\cal M}(\Omega_{cc}^{+}, K^{*0}, \Xi_{c}^{+}, \bar K^{*0}, \omega ,\Lambda_c^{+})
  + {\cal M}(\Omega_{cc}^{+}, K^{0}, \Xi_{c}^{'+}, \bar K^{*0}, \omega ,\Lambda_c^{+})\nonumber\\
&&+ {\cal M}(\Omega_{cc}^{+}, K^{*0}, \Xi_{c}^{'+}, \bar K^{*0}, \omega ,\Lambda_c^{+})
  + {\cal M}(\Omega_{cc}^{+}, K^{0}, \Xi_{c}^{+}, \Xi_{c}^{+}, \Lambda_c^{+},\omega )
  + {\cal M}(\Omega_{cc}^{+}, K^{*0}, \Xi_{c}^{+}, \Xi_{c}^{+}, \Lambda_c^{+},\omega )\nonumber\\
&&+{\cal M}(\Omega_{cc}^{+}, K^{0}, \Xi_{c}^{+}, \Xi_{c}^{'+}, \Lambda_c^{+},\omega )
  +{\cal M}(\Omega_{cc}^{+}, K^{*0}, \Xi_{c}^{+}, \Xi_{c}^{'+}, \Lambda_c^{+},\omega )
  + {\cal M}(\Omega_{cc}^{+}, K^{0}, \Xi_{c}^{'+}, \Xi_{c}^{+}, \Lambda_c^{+},\omega )\nonumber\\
&&+ {\cal M}(\Omega_{cc}^{+}, K^{*0}, \Xi_{c}^{'+}, \Xi_{c}^{+}, \Lambda_c^{+},\omega )
  +{\cal M}(\Omega_{cc}^{+}, K^{0}, \Xi_{c}^{'+}, \Xi_{c}^{'+}, \Lambda_c^{+},\omega )
  +{\cal M}(\Omega_{cc}^{+}, K^{*0}, \Xi_{c}^{'+}, \Xi_{c}^{'+}, \Lambda_c^{+},\omega )\nonumber\\
&&+ {\cal M}(\Omega_{cc}^{+}, K^{+}, \Xi_{c}^{0}, K^{-}, \omega , \Lambda_c^{+})
  + {\cal M}(\Omega_{cc}^{+}, K^{*+}, \Xi_{c}^{0}, K^{*-}, \omega , \Lambda_c^{+})
  + {\cal M}(\Omega_{cc}^{+}, K^{+}, \Xi_{c}^{'0}, K^{-}, \omega , \Lambda_c^{+})\nonumber\\
&&+ {\cal M}(\Omega_{cc}^{+}, K^{*+}, \Xi_{c}^{'0}, K^{*-}, \omega , \Lambda_c^{+})
  + {\cal M}(\Omega_{cc}^{+}, K^{+}, \Xi_{c}^{0}, \Xi_{c}^{0}, \Lambda_c^{+}, \omega )
  + {\cal M}(\Omega_{cc}^{+}, K^{*+}, \Xi_{c}^{0}, \Xi_{c}^{0}, \Lambda_c^{+}, \omega )\nonumber\\
&&+ {\cal M}(\Omega_{cc}^{+}, K^{+}, \Xi_{c}^{0}, \Xi_{c}^{'0}, \Lambda_c^{+}, \omega )
  + {\cal M}(\Omega_{cc}^{+}, K^{*+}, \Xi_{c}^{0}, \Xi_{c}^{'0}, \Lambda_c^{+}, \omega )
  + {\cal M}(\Omega_{cc}^{+}, K^{+}, \Xi_{c}^{'0}, \Xi_{c}^{0}, \Lambda_c^{+}, \omega )\nonumber\\
&&+ {\cal M}(\Omega_{cc}^{+}, K^{*+}, \Xi_{c}^{'0}, \Xi_{c}^{0}, \Lambda_c^{+}, \omega )
  + {\cal M}(\Omega_{cc}^{+}, K^{+}, \Xi_{c}^{'0}, \Xi_{c}^{'0}, \Lambda_c^{+}, \omega )
  + {\cal M}(\Omega_{cc}^{+}, K^{*+}, \Xi_{c}^{'0}, \Xi_{c}^{'0}, \Lambda_c^{+}, \omega )
  ],
\end{eqnarray}
\begin{eqnarray}
{\cal A}(\Omega_{cc}^{+} \to \Sigma_c^{0} \rho^{+})
&=&i [
    {\cal M}(\Omega_{cc}^{+}, K^{+}, \Xi_{c}^{0}, K^{0}, \rho^{+},\Sigma_c^{0})
  + {\cal M}(\Omega_{cc}^{+}, K^{*+}, \Xi_{c}^{0}, \bar K^{*0}, \rho^{+},\Sigma_c^{0})
  + {\cal M}(\Omega_{cc}^{+}, K^{+}, \Xi_{c}^{'0}, K^{0}, \rho^{+},\Sigma_c^{0})\nonumber\\
&&+ {\cal M}(\Omega_{cc}^{+}, K^{*+}, \Xi_{c}^{'0}, \bar K^{*0}, \rho^{+},\Sigma_c^{0})
  + {\cal M}(\Omega_{cc}^{+}, K^{0}, \Xi_{c}^{+}, \Xi_{c}^{0}, \Sigma_c^{0},\rho^{+})
  + {\cal M}(\Omega_{cc}^{+}, K^{*0}, \Xi_{c}^{+}, \Xi_{c}^{0}, \Sigma_c^{0},\rho^{+})\nonumber\\
&&+ {\cal M}(\Omega_{cc}^{+}, K^{0}, \Xi_{c}^{+}, \Xi_{c}^{'0}, \Sigma_c^{0},\rho^{+})
  + {\cal M}(\Omega_{cc}^{+}, K^{*0}, \Xi_{c}^{+}, \Xi_{c}^{'0}, \Sigma_c^{0},\rho^{+})
  + {\cal M}(\Omega_{cc}^{+}, K^{0}, \Xi_{c}^{'+}, \Xi_{c}^{0}, \Sigma_c^{0},\rho^{+})\nonumber\\
&&+ {\cal M}(\Omega_{cc}^{+}, K^{*0}, \Xi_{c}^{'+}, \Xi_{c}^{0}, \Sigma_c^{0},\rho^{+})
  + {\cal M}(\Omega_{cc}^{+}, K^{0}, \Xi_{c}^{'+}, \Xi_{c}^{'0}, \Sigma_c^{0},\rho^{+})
  + {\cal M}(\Omega_{cc}^{+}, K^{*0}, \Xi_{c}^{'+}, \Xi_{c}^{'0}, \Sigma_c^{0},\rho^{+})].
\end{eqnarray}

}

%====================================================================
\section{Strong Coupling Constants}
\label{app:stcouplings}
%====================================================================
In this section we list all the strong coupling constants used in our calculation. Some of these values are taken from Refs. \cite{Cheng:FSIB,Aliev:2010yx,Aliev:2010nh,Khodjamirian:2011jp,Azizi:2014bua,Yu:2016pyo,Azizi:2015tya,Ballon-Bayona:2017bwk}. For those that can not be found directly in the literatures, we calculate them under the assumption of $SU(3)$ symmetry. In calculation we perform $SU(2)$ transformation prior to $SU(3)$ ones.

According to which $SU(3)$ multiplets  do the particles belong to, the vertices in this paper can be divided into $8$ types: $VPP$, $VVV$, ${\cal B}_{c3}{\cal B}_{c3}P$, ${\cal B}_{c6}{\cal B}_{c6}P$, ${\cal B}_{c6}{\cal B}_{c3}P$,  ${\cal B}_{c3}{\cal B}_{c3}V$, ${\cal B}_{c6}{\cal B}_{c6}V$, and ${\cal B}_{c6}{\cal B}_{c3}V$. $P$ denotes a light pseudoscalar meson, and $V$ represents a light vector meson. The singly charm baryons can be classified into two $SU(3)$ multiplets: a triplet labeled by ${\cal B}_{c3}$ and a sextet by ${\cal B}_{c6}$. With these label-definitions one can know the meaning of our symbols for each vertices whose coupling constants are collected in tables \ref{tab:SC3mesons}, \ref{tab:SCBBP}, \ref{tab:SCBBV}.
 %++++++++++++++++++++++++++++++++++++++++++++++++++++++++++++++++++++++++++++++++++++++++++++++++++++++++++++++++++++++
\begin{table}[!htbp]
  \centering
 \caption{Strong coupling constants of $VPP$ and $VVV$ vertices.}
 \label{tab:SC3mesons}
  \begin{tabular}{|c|c|c|c|c|c|c|c|c|c|c|c|}  %c
   \hline %
   vertex&g&vertex&g&vertex&g  &vertex&g&vertex&g&vertex&g\\
   \hline %
   $\rho^{+}\rightarrow\pi^{0}\pi^{+}$&6.05&$\rho^{0}\rightarrow\pi^{-}\pi^{+}$&6.05&$\rho^{+}\rightarrow K^{+}\overline{K}^{0}$&4.60
   &$\rho^{0}\rightarrow K^{0}\overline{K}^{0}$&-3.25&$\rho^{0}\rightarrow K^{+}K^{-}$&3.25&$\omega\rightarrow K^{+}K^{-}$&3.25\\
   \hline %
    $\phi\rightarrow K^{-}K^{+}$&4.60&$\overline{K}^{*0}\rightarrow\eta_8\overline{K}^{0}$&5.63&$\overline{K}^{*0}\rightarrow K^{-}\pi^{+}$&4.60
   &  $\overline{K}^{*0}\rightarrow\overline{K}^{0}\pi^{0}$&-3.25&$K^{*+}\rightarrow K^{+}\pi^{0}$&3.25 &$\phi\rightarrow \overline{K}^{0}K^{0}$&4.60\\
   \hline %
    $K^{*+}\rightarrow\eta_8 K^{+}$&5.63&$K^{*+}\rightarrow\pi^{+}K^{0}$&4.60&$K^{*0}\rightarrow\pi^{-}K^{+}$&4.60
   &   $K^{*0}\rightarrow K^{0}\eta_8$&5.63&$K^{*0}\rightarrow\pi^{0}K^{0}$&-3.25&$\omega\rightarrow K^{0}\overline{K}^{0}$&3.25\\
   \hline %
      \hline %
   $\rho^{+}\rightarrow\rho^{0}\rho^{+}$&7.38&$\rho^{0}\rightarrow\rho^{-}\rho^{+}$&7.38&$\rho^{+}\rightarrow K^{*+}\overline{K}^{*0}$&5.22
   &   $\rho^{0}\rightarrow K^{*+}K^{*-}$&3.69&$\omega\rightarrow K^{*+}K^{*-}$&3.69&$\rho^{0}\rightarrow K^{*0}\overline{K}^{*0}$&-3.69\\
   \hline %
    $\overline{K}^{*0}\rightarrow\phi\overline{K}^{*0}$&5.22&$\overline{K}^{*0}\rightarrow\overline{K}^{*0}\rho^{0}$ &-3.69&$\overline{K}^{*0}\rightarrow\overline{K}^{*0}\omega$&3.69
   &    $K^{*+}\rightarrow\rho^{+}K^{*0}$&5.22&$K^{*+}\rightarrow\phi K^{*+}$&5.22&$K^{*+}\rightarrow K^{*+}\rho^{0}$&3.69\\
   \hline %
    $K^{*+}\rightarrow\omega K^{*+}$&3.69&$K^{*0}\rightarrow\rho^{0}K^{*0}$&-3.69&$K^{*0}\rightarrow\omega K^{*0}$&3.69
   &   $K^{*0}\rightarrow K^{*0}\phi$&5.22&$\phi\rightarrow K^{*-}K^{*+}$&5.22&$\overline{K}^{*0}\rightarrow K^{*-}\rho^{+}$&5.22\\
   \hline %
   $\omega\rightarrow K^{*0}\overline{K}^{*0}$&3.69&$\phi\rightarrow \overline{K}^{*0}K^{*0}$&5.22& & &&&&&&\\
   \hline %
   \end{tabular}
\end{table}
 %++++++++++++++++++++++++++++++++++++++++++++++++++++++++++++++++++++++++++++++++++++++++++++++++++++++++++++++++++++++

 %++++++++++++++++++++++++++++++++++++++++++++++++++++++++++++++++++++++++++++++++++++++++++++++++++++++++++++++++++++++
\begin{table}[!htbp]
  \centering
 \caption{Strong coupling constants of ${\cal B}_{c3}{\cal B}_{c3}P$, ${\cal B}_{c3}{\cal B}_{c6}P$ and ${\cal B}_{c6}{\cal B}_{c6}P$ vertices.}
 \label{tab:SCBBP}
  \begin{tabular}{|c|c|c|c|c|c|c|c|c|c|c|c|}  %c
   \hline %
   vertex&g&vertex&g&vertex&g  &vertex&g&vertex&g&vertex&g\\
   \hline %
   $\Xi_{c}^{0}\rightarrow\Xi_{c}^{+}\pi^{-}$&0.99&$\Xi_{c}^{+}\rightarrow\Xi_{c}^{0}\pi^{+}$&0.99&$\Xi_{c}^{0}\rightarrow\Xi_{c}^{0}\pi^{0}$&-0.70  &$\Xi_{c}^{0}\rightarrow\Lambda_{c}^{+}K^{-}$&-0.90&$\Xi_{c}^{+}\rightarrow\Xi_{c}^{+}\pi^{0}$&0.70& $\Xi_{c}^{0}\rightarrow\Xi_{c}^{0}\eta_8$&-0.70\\
   \hline %
    $\Lambda_{c}^{+}\rightarrow\Lambda_{c}^{+}\eta_1$&0.75& $\Xi_{c}^{+}\rightarrow\Xi_{c}^{+}\eta_1$&0.07&$\Xi_{c}^{0}\rightarrow\Xi_{c}^{0}\eta_1$&0.07  &$\Lambda_{c}^{+}\rightarrow\Lambda_{c}^{+}\eta_8$&0.81&$\Xi_{c}^{+}\rightarrow\Xi_{c}^{+}\eta_8$& -0.70 &$\Xi_{c}^{+}\rightarrow\Lambda_{c}^{+}\overline{K}^{0}$&0.90 \\
   \hline %
  $\Sigma_{c}^{0}\rightarrow\Sigma_{c}^{+}\pi^{-}$&8.0&$\Xi_{c}^{'0}\rightarrow\Xi_{c}^{'0}\pi^{0}$ &-4.0&$\Xi_{c}^{'0}\rightarrow\Sigma_{c}^{0}\overline{K}^{0}$&9.0  &$\Sigma_{c}^{+}\rightarrow\Sigma_{c}^{0}\pi^{+}$ &8.0&$\Xi_{c}^{'0}\rightarrow\Sigma_{c}^{+}K^{-}$&6.4&$\Xi_{c}^{'0}\rightarrow\Xi_{c}^{'+}\pi^{-}$&5.7 \\
   \hline %
   $\Sigma_{c}^{0}\rightarrow\Sigma_{c}^{0}\pi^{0}$&-8.0 &$\Sigma_{c}^{0}\rightarrow\Xi_{c}^{'0}K^{0}$&9.0&$\Xi_{c}^{'+}\rightarrow\Xi_{c}^{'+}\pi^{0}$&4.0  &$\Xi_{c}^{'+}\rightarrow\Sigma_{c}^{++}K^{-}$&9.0&$\Xi_{c}^{'+}\rightarrow\Xi_{c}^{'0}\pi^{+}$ &5.7 &$\Sigma_{c}^{++}\rightarrow\Sigma_{c}^{+}\pi^{+}$&8.0 \\
   \hline %
   $\Sigma_{c}^{++}\rightarrow\Xi_{c}^{'+}K^{+}$&9.0&$\Sigma_{c}^{+}\rightarrow\Xi_{c}^{'+}K^{0}$&6.4&$\Sigma_{c}^{+}\rightarrow\Sigma_{c}^{++}\pi^{-}$&8.0  &$\Xi_{c}^{'+}\rightarrow\Omega_{c}^{0}K^{+}$&9.0&$\Xi_{c}^{'+}\rightarrow\Sigma_{c}^{+}\overline{K}^{0}$&6.4&$\Sigma_{c}^{+}\rightarrow\Xi_{c}^{'0}K^{+}$&6.4\\
   \hline %
   $\Omega_{c}^{0}\rightarrow\Xi_{c}^{'+}K^{-}$&9.0&$\Xi_{c}^{'0}\rightarrow\Omega_{c}^{0}K^{0}$&9.0&$\Sigma_{c}^{0}\rightarrow\Sigma_{c}^{0}\eta_1$ &-2.6    &$\Omega_{c}^{0}\rightarrow\Omega_{c}^{0}\eta_1$&-11.0&$\Omega_{c}^{0}\rightarrow\Xi_{c}^{'0}\overline{K}^{0}$&9.0&$\Sigma_{c}^{0}\rightarrow\Sigma_{c}^{0}\eta_8$ &4.6  \\
   \hline %
    $\Omega_{c}^{0}\rightarrow\Omega_{c}^{0}\eta_8$&-10.4 &$\Sigma_{c}^{+}\rightarrow\Sigma_{c}^{+}\eta_1$&-2.6&$\Xi_{c}^{'0}\rightarrow\Xi_{c}^{'0}\eta_1$&-2.6  &$\Xi_{c}^{'+}\rightarrow\Xi_{c}^{'+}\eta_1$&-2.6&$\Sigma_{c}^{+}\rightarrow\Sigma_{c}^{+}\eta_8$&4.6&$\Xi_{c}^{'0}\rightarrow\Xi_{c}^{'0}\eta_8$&-2.3\\
   \hline %
   $\Xi_{c}^{0}\rightarrow\Xi_{c}^{'+}\pi^{-}$&4.4&$\Xi_{c}^{'+}\rightarrow\Xi_{c}^{+}\pi^{0}$&3.1&$\Xi_{c}^{'+}\rightarrow\Xi_{c}^{0}\pi^{+}$&4.4  &$\Omega_{c}^{0}\rightarrow\Xi_{c}^{+}K^{-}$&6.5&$\Xi_{c}^{'+}\rightarrow\Lambda_{c}^{+}\overline{K}^{0}$&-4.6&$\Sigma_{c}^{+}\rightarrow\Lambda_{c}^{+}\pi^{0}$&6.5 \\
   \hline %
   $\Xi_{c}^{0}\rightarrow\Omega_{c}^{0}K^{0}$&6.5&$\Omega_{c}^{0}\rightarrow\Xi_{c}^{0}\overline{K}^{0}$&6.5&$\Sigma_{c}^{0}\rightarrow\Xi_{c}^{0}K^{0}$&-7.1   &$\Xi_{c}^{'+}\rightarrow\Lambda_{c}^{+}K^{0}$&-4.6&$\Xi_{c}^{+}\rightarrow\Xi_{c}^{'0}\pi^{+}$&4.4&$\Lambda_{c}^{+}\rightarrow\Xi_{c}^{'0}K^{+}$&4.6\\
     \hline %
   $\Xi_{c}^{0}\rightarrow\Sigma_{c}^{+}K^{-}$&-5.0&$\Xi_{c}^{'0}\rightarrow\Xi_{c}^{0}\pi^{0}$&-3.1&$\Xi_{c}^{+}\rightarrow\Xi_{c}^{'0}\pi^{+}$&4.4  &$\Xi_{c}^{'0}\rightarrow\Lambda_{c}^{+}K^{-}$&4.6&$\Xi_{c}^{'0}\rightarrow\Xi_{c}^{+}\pi^{-}$&4.4&$\Xi_{c}^{+}\rightarrow\Omega_{c}^{0}K^{+}$ &6.5 \\
     \hline %
    $\Sigma_{c}^{++}\rightarrow\Xi_{c}^{+}K^{+}$&-7.1&$\Xi_{c}^{+}\rightarrow\Sigma_{c}^{++}K^{-}$&-7.1&$\Sigma_{c}^{++}\rightarrow\Lambda_{c}^{+}\pi^{+}$&-6.5  &$\Xi_{c}^{0}\rightarrow\Sigma_{c}^{0}\overline{K}^{0}$&-7.1&$\Lambda_{c}^{+}\rightarrow\Sigma_{c}^{++}\pi^{-}$&-6.5&$\Lambda_{c}^{+}\rightarrow\Sigma_{c}^{0}\pi^{+}$&6.5\\
     \hline %
    $\Sigma_{c}^{+}\rightarrow\Xi_{c}^{0}K^{+}$&-5.0&$\Sigma_{c}^{0}\rightarrow\Lambda_{c}^{+}\pi^{-}$&6.5&$\Xi_{c}^{'+}\rightarrow\Xi_{c}^{+}\eta_8$&5.4  &$\Sigma_{c}^{+}\rightarrow\Xi_{c}^{+}K^{0}$&-5.0&$\Xi_{c}^{+}\rightarrow\Sigma_{c}^{+}\overline{K}^{0}$&-5.0&$\Xi_{c}^{'0}\rightarrow\Xi_{c}^{0}\eta_8$&5.4\\
     \hline %
    $\Lambda_{c}^{+}\rightarrow\Xi_{c}^{0}K^{+}$ &-0.90 &$\Xi_{c}^{+}\rightarrow\Lambda_{c}^{+}K^{0}$ &0.90 & $\Xi_{c}^{'+}\rightarrow\Xi_{c}^{'+}\eta_8$ &-2.3 &&&&&&\\
   \hline %
    \end{tabular}
\end{table}
 %++++++++++++++++++++++++++++++++++++++++++++++++++++++++++++++++++++++++++++++++++++++++++++++++++++++++++++++++++++++

 %++++++++++++++++++++++++++++++++++++++++++++++++++++++++++++++++++++++++++++++++++++++++++++++++++++++++++++++++++++++
\begin{table}[!htbp]
  \centering
 \caption{Strong coupling constants of ${\cal B}_{c3}{\cal B}_{c3}V$, ${\cal B}_{c3}{\cal B}_{c6}V$ and ${\cal B}_{c6}{\cal B}_{c6}V$ vertices.}
 \label{tab:SCBBV}
  \begin{tabular}{|c|c|c|c|c|c|c|c|c|c|c|c|}  %c
   \hline %
   vertex&$f_{1}$&$f_{2}$&vertex&$f_{1}$&$f_{2}$  &vertex&$f_{1}$&$f_{2}$&vertex&$f_{1}$&$f_{2}$\\
   \hline %
   $\Xi_{c}^{0}\rightarrow\Lambda_{c}^{+}K^{*-}$&-4.6&-6.0&$\Xi_{c}^{0}\rightarrow\Xi_{c}^{0}\rho^{0}$&-6.0&-7.5
   &$\Xi_{c}^{+}\rightarrow\Xi_{c}^{0}\rho^{+}$&8.5&10.6&$\Lambda_{c}^{+}\rightarrow\Xi_{c}^{0}K^{*+}$&-4.6&-6.0\\
   \hline %
   $\Xi_{c}^{0}\rightarrow\Xi_{c}^{0}\phi$&4.6&6.0&$\Xi_{c}^{0}\rightarrow\Xi_{c}^{0}\omega$&5.5&7.5
   &$\Lambda_{c}^{+}\rightarrow\Lambda_{c}^{+}\omega$&4.9&6.0&$\Xi_{c}^{0}\rightarrow\Xi_{c}^{+}\rho^{-}$&8.5&10.6\\
   \hline %
    $\Xi_{c}^{+}\rightarrow\Lambda_{c}^{+}\overline{K}^{*0}$&4.6&6.0&$\Xi_{c}^{+}\rightarrow\Xi_{c}^{+}\rho^{0}$&6.0&7.5
   &$\Xi_{c}^{+}\rightarrow\Xi_{c}^{+}\omega$&5.5&7.5&$\Xi_{c}^{+}\rightarrow\Xi_{c}^{+}\phi$&4.6&6.0\\
   \hline %
   $\Sigma_{c}^{0}\rightarrow\Xi_{c}^{'0}K^{*0}$&5.0&30.0&$\Xi_{c}^{'0}\rightarrow\Xi_{c}^{'0}\rho^{0}$&-2.5&-16.0
   &$\Xi_{c}^{'0}\rightarrow\Xi_{c}^{'+}\rho^{-}$&3.5&22.6&$\Xi_{c}^{'0}\rightarrow\Xi_{c}^{'0}\phi$&4.0&21.0\\
   \hline %
   $\Xi_{c}^{'+}\rightarrow\Xi_{c}^{'0}\rho^{+}$&3.5&22.6&$\Sigma_{c}^{+}\rightarrow\Xi_{c}^{'0}K^{*+}$&3.5&21.2
   &$\Xi_{c}^{'0}\rightarrow\Xi_{c}^{'0}\omega$&2.4&15.0&$\Xi_{c}^{'0}\rightarrow\Sigma_{c}^{+}K^{*-}$&3.5&21.2\\
   \hline %
   $\Xi_{c}^{'0}\rightarrow\Sigma_{c}^{0}\overline{K}^{*0}$&5.0&30.0&$\Sigma_{c}^{+}\rightarrow\Sigma_{c}^{++}\rho^{-}$&4.0&27.0
   &$\Sigma_{c}^{++}\rightarrow\Xi_{c}^{'+}K^{*+}$&5.0&30.0&$\Xi_{c}^{'+}\rightarrow\Sigma_{c}^{++}K^{*-}$&5.0&30.0\\
   \hline %
   $\Xi_{c}^{'+}\rightarrow\Sigma_{c}^{+}\overline{K}^{*0}$&3.5&21.2&$\Xi_{c}^{'+}\rightarrow\Xi_{c}^{'+}\rho^{0}$&2.5&16.0
   &$\Xi_{c}^{'+}\rightarrow\Xi_{c}^{'+}\phi$&4.0&21.0&$\Xi_{c}^{'+}\rightarrow\Xi_{c}^{'+}\omega$&2.4&15.0\\
   \hline %
    $\Sigma_{c}^{+}\rightarrow\Xi_{c}^{'+}K^{*0}$&3.5&21.2&$\Sigma_{c}^{++}\rightarrow\Sigma_{c}^{+}\rho^{+}$&4.0&27.0
   &$\Sigma_{c}^{0}\rightarrow\Sigma_{c}^{0}\rho^{0}$&-4.0&-27.0&$\Sigma_{c}^{0}\rightarrow\Sigma_{c}^{0}\omega$&3.5&24.0\\
   \hline %
   $\Sigma_{c}^{+}\rightarrow\Sigma_{c}^{0}\rho^{+}$&4.0&27.0&$\Sigma_{c}^{0}\rightarrow\Sigma_{c}^{+}\rho^{-}$&4.0&27.0
   &$\Xi_{c}^{'+}\rightarrow\Omega_{c}^{0}K^{*+}$&7.0&35.0&$\Omega_{c}^{0}\rightarrow\Xi_{c}^{'0}K^{*0}$&7.0&35.0\\
   \hline %
   $\Omega_{c}^{0}\rightarrow\Omega_{c}^{0}\phi$&11.0&52.0&$\Omega_{c}^{0}\rightarrow\Xi_{c}^{'+}K^{*-}$&7.0&35.0
   &$\Omega_{c}^{0}\rightarrow\Xi_{c}^{'0}\overline{K}^{*0}$&7.0&35.0&$\Sigma_{c}^{+}\rightarrow\Sigma_{c}^{+}\omega$&3.5&24.0\\
   \hline %
   $\Xi_{c}^{'0}\rightarrow\Xi_{c}^{0}\rho^{0}$&-1.5&-11.0&$\Xi_{c}^{'0}\rightarrow\Xi_{c}^{0}\phi$&-2.1&-13.0
   &$\Xi_{c}^{'0}\rightarrow\Xi_{c}^{0}\omega$&1.2&8.0&$\Xi_{c}^{'+}\rightarrow\Xi_{c}^{+}\omega$&1.5&11.0\\
   \hline %
   $\Lambda_{c}^{+}\rightarrow\Xi_{c}^{'0}K^{*+}$&2.3&14.1&$\Xi_{c}^{'0}\rightarrow\Lambda_{c}^{+}K^{*-}$&2.3&14.1
   &$\Xi_{c}^{+}\rightarrow\Xi_{c}^{'0}\rho^{+}$&2.1&15.6&$\Xi_{c}^{'0}\rightarrow\Xi_{c}^{+}\rho^{-}$&2.1&15.6\\
   \hline %
    $\Xi_{c}^{'+}\rightarrow\Lambda_{c}^{+}\overline{K}^{*0}$&-2.3&-14.1&$\Sigma_{c}^{+}\rightarrow\Xi_{c}^{0}K^{*+}$&-2.2&-13.0
   &$\Xi_{c}^{+}\rightarrow\Sigma_{c}^{+}K^{*0}$&-2.2&-13.0&$\Xi_{c}^{'+}\rightarrow\Xi_{c}^{0}\rho^{+}$&2.1&15.6\\
   \hline %
   $\Xi_{c}^{'+}\rightarrow\Xi_{c}^{+}\rho^{0}$&1.5&11.0&$\Xi_{c}^{'+}\rightarrow\Xi_{c}^{+}\phi$&-2.1&-13.0
   &$\Xi_{c}^{0}\rightarrow\Xi_{c}^{'+}\rho^{-}$&2.1&15.6&$\Sigma_{c}^{+}\rightarrow\Xi_{c}^{+}\overline{K}^{*0}$&-2.2&-13.0\\
   \hline %
   $\Sigma_{c}^{+}\rightarrow\Lambda_{c}^{+}\rho^{0}$&2.6&16.0&$\Xi_{c}^{+}\rightarrow\Omega_{c}^{0}K^{*+}$&3.3&20.0
   &$\Lambda_{c}^{+}\rightarrow\Xi_{c}^{'+}K^{*0}$&-2.3 &-14.1&$\Xi_{c}^{0}\rightarrow\Sigma_{c}^{+}K^{*-}$&-2.2&-13.0\\
   \hline %
   $\Sigma_{c}^{++}\rightarrow\Xi_{c}^{+}K^{*+}$&-3.1&-18.4&$\Xi_{c}^{+}\rightarrow\Sigma_{c}^{++}K^{*-}$&-3.1&-18.4
   &$\Sigma_{c}^{++}\rightarrow\Lambda_{c}^{+}\rho^{+}$&-2.6&-16.0&$\Lambda_{c}^{+}\rightarrow\Sigma_{c}^{++}\rho^{-}$&-2.6&-16.0\\
   \hline %
   $\Lambda_{c}^{+}\rightarrow\Sigma_{c}^{0}\rho^{+}$&2.6&16.0&$\Sigma_{c}^{0}\rightarrow\Lambda_{c}^{+}\rho^{-}$&2.6&16.0
   &$\Omega_{c}^{0}\rightarrow\Xi_{c}^{+}K^{*-}$&3.3&20.0&$\Xi_{c}^{0}\rightarrow\Sigma_{c}^{0}\overline{K}^{*0}$&-2.2&-13.0\\
   \hline %
    $\Xi_{c}^{0}\rightarrow\Omega_{c}^{0}K^{*0}$&3.3&20.0&$\Sigma_{c}^{0}\rightarrow\Xi_{c}^{0}K^{*0}$&-2.2&-13.0
   &$\Omega_{c}^{0}\rightarrow\Xi_{c}^{0}\overline{K}^{*0}$&3.3&20.0&$\Lambda_{c}^{+}\rightarrow\Xi_{c}^{+}K^{*0}$&4.6&6.0 \\
   \hline %
   \end{tabular}
\end{table}
 %++++++++++++++++++++++++++++++++++++++++++++++++++++++++++++++++++++++++++++++++++++++++++++++++++++++++++++++++++++++

%===========================================================================================

\end{document}